\documentclass[12pt]{withesis}
\usepackage{epsfig}
\usepackage{times}

\usepackage{atlasphysics}
\usepackage{dcolumn}
\usepackage{bm}
\usepackage{xspace}
\usepackage{multirow}
\usepackage{bigstrut}
\usepackage{amsmath}
\usepackage{subfigure}
\usepackage{color}
\usepackage{lineno}
\usepackage{graphicx}
\usepackage{caption}
\usepackage{amsmath}
\usepackage[para]{footmisc}
\usepackage{hyperref}
\usepackage{floatrow}
\usepackage{rotating}
\usepackage{cite,mcite}
\usepackage{afterpage}
\usepackage{amssymb}
\usepackage{geometry}
\usepackage{CJKutf8}
\usepackage{soul} 
\usepackage{mathrsfs}
\usepackage{booktabs}

\newcolumntype{L}[1]{>{\raggedright\let\newline\\\arraybackslash\hspace{0pt}}m{#1}}
\newcolumntype{C}[1]{>{\centering\let\newline\\\arraybackslash\hspace{0pt}}m{#1}}
\newcolumntype{R}[1]{>{\raggedleft\let\newline\\\arraybackslash\hspace{0pt}}m{#1}}

\newcommand{\thirteentev}{\mbox{\ensuremath{\sqrt{s}=13\TeV}}}

\newcommand{\ggF}{\ensuremath{gg\mathrm{F}}}
\newcommand{\VBF}{\ensuremath{\mathrm{VBF}}}
\newcommand{\VH}{\ensuremath{VH}}
\newcommand{\WH}{\ensuremath{WH}}
\newcommand{\ZH}{\ensuremath{ZH}}

\newcommand{\ttH}{\ensuremath{\ttbar H}}
\newcommand{\bbH}{\ensuremath{\bbbar H}}
\newcommand{\tH}{\ensuremath{tH}}

\newcommand{\ppggF}{\ensuremath{pp\to H}}
\newcommand{\ppVBF}{\ensuremath{pp\to qqH}}

\newcommand{\ppWH}{\ensuremath{pp\to WH}}
\newcommand{\ppZH}{\ensuremath{pp\to ZH}}
\newcommand{\ppttH}{\ensuremath{pp\to\ttbar H}}
\newcommand{\ppbbH}{\ensuremath{pp\to\bbbar H}}
\newcommand{\pptH}{\ensuremath{pp\to tH}}

\newcommand{\Hgg}{\ensuremath{H\to gg}}
\newcommand{\Hyy}{\ensuremath{H\to\gamma\gamma}}

\newcommand{\HZZ}{\ensuremath{H\to ZZ^{(*)}}}
\newcommand{\HWW}{\ensuremath{H\to WW^{(*)}}}
\newcommand{\Htt}{\ensuremath{H\to\tau\tau}}
\newcommand{\Hcc}{\ensuremath{H\to c\bar{c}}}

\newcommand{\Hmm}{\ensuremath{H\to \mu\mu}}
\newcommand{\HZg}{\ensuremath{H\to Z\gamma}}

\newcommand{\mmumu}{\ensuremath{m_{\mu\mu}}}

\newcommand{\monoHbb}{mono-$h$($b\bar{b}$)}
\newcommand{\zpthdm}{\ensuremath{Z^{\prime}\mathrm{-2HDM}}}
\newcommand{\thdma}{\ensuremath{\mathrm{2HDM+}a}}
\newcommand{\bbA}{\ensuremath{b\bar{b}A}}
\newcommand{\tanb}{\ensuremath{\tan\beta}}
\newcommand{\mh}{\ensuremath{m_h}}
\newcommand{\mHpm}{\ensuremath{m_{H^{\pm}}}}
\newcommand{\ma}{\ensuremath{m_a}}
\newcommand{\mchi}{\ensuremath{m_\chi}}
\newcommand{\mZp}{\ensuremath{m_{Z^\prime}}}
\newcommand{\tbtag}{2 $b$-tag}
\newcommand{\thbtag}{$\geq$3 $b$-tag}

\noappendixtables                
\noappendixfigures               

\begin{document}



\clearpage\pagenumbering{roman}  

\title{Investigation of Higgs Boson decaying to di-muon, Dark Matter produced in association with a Higgs boson decaying to \lowercase{$b$}-quarks and unbinned profiled unfolding}

\author{Chen-Hsun (Jay) Chan}
\advisorname{Sau Lan Wu}
\advisortitle{Professor}
\department{Physics}
\oralexamdate{05/08/2023}
\committeeone{Sau Lan Wu, Professor, Physics}
\committeetwo{Kevin Black, Professor, Physics}
\committeethree{Tulika Bose, Professor, Physics}
\committeefour{Snezana Stanimirovic, Professor, Astronomy}

\date{2023}
\newgeometry{left=1in,right=1in,bottom=1in,top=1in}
\maketitle


\copyrightpage


\begin{acknowledgments}
First and foremost, I would like to express my deepest gratitude to my advisor Professor Sau Lan Wu.
Sau Lan's guidance and support have been instrumental in my success throughout my PhD.
I am grateful to have joined her research group and had the opportunity to work on experiments such as the \Hmm, \monoHbb, Higgs combination, \thdma\ combination and ITk upgrade.
Sau Lan's mentorship and leadership have enabled me to make important contributions to these projects and stand out in the analysis teams.
In addition, she connected me to the Lawrence Berkeley National Lab, where I had an amazing experience towards the end of my PhD.
Sau Lan is a great mentor who truly cares about students' lives and careers. She not only taught me how to become a great physicist but also inspired me to become a hard-working and self-motivated person.

I am grateful to the Department of Physics at the University of Wisconsin-Madison for providing me with the opportunity to pursue my graduate studies.
I would also like to acknowledge my thesis defense committee members, including my advisor Sau Lan Wu, and Kevin Black, Tulika Bose, and Snezana Stanimirovic, for reading my thesis and providing valuable feedback.
Their comments and suggestions helped me to improve the quality of my work and make it a more comprehensive and coherent piece of scientific writing.

The Wisconsin ATLAS group is like a big family.
I am grateful to Chen Zhou for instructing me and giving me advice for all aspects of my research projects, including \Hmm, \monoHbb, Higgs combination, \thdma\ combination, and my ATLAS qualification task on ITk simulation.
Chen also helped me with all my presentations, funding proposals and even job applications.
He has been super helpful throughout my PhD. I cannot be more grateful for the everyday help and support from Sylvie Padlewski as well.
I also enjoyed the active discussions and company from Alex Wang, Rui Zhang, Alkaid Cheng, Tuan Pham, Wasikul Islam and Amy Tee.
Furthermore, the technical support from Shaojun Sun and Wen Guan has been incredible.

I would like to thank my ATLAS analysis collaborators, including Giacomo Artoni, Jan Kretzschmar, Fabio Cerutti, Hongtao Tang, Andrea Gabrielli, Gaetano Barone, Miha Zgubic, Yanlin Liu, Miha Muskinja, Alice Alfonsi, Brendon Bullard, and Hanna Borecka-Bielska in \Hmm, Spyridon Argyropoulos, Dan Guest, James Frost, Philipp Gadow, Andrea Matic, Anindya Ghosh, Eleni Skorda, Jon Burr, and Samuel Meehan in \monoHbb, Zirui Wang and Lailin Xu in the \thdma\ combination, Nicolas Morange, Kunlin Ran and Rahul Balasubramanian in the Higgs combination, and Ben Smart and Chen Zhou for supervising my ATLAS qualification task. It was a pleasure working with all these incredible people, and I learned a lot from each individual.

I also want to acknowledge my fellow PhD students Manuel Silva and Maria Veronica Prado for all the joyful activities in Wisconsin and Switzerland. While I was at CERN, I had lots of fun hanging out with Yaun-Tang Chou, Otto Lau, Yvonne Ng, and Bobby Tse.

The days I was based at Berkeley are among the happiest in my life. I would like to express my sincere appreciation to Maurice Garcia-Sciveres and Kevin Einsweiler for providing me the opportunity to work on the ITk pixel upgrade and to Elisabetta Pianori and Timon Heim for giving me day-to-day training and guidance.
I am also extremely grateful to have had the opportunity to work with Ben Nachman on the Machine Learning projects.
Ben taught me a lot about Machine Learning, and his innovative ideas continue to inspire and motivate me.
I also appreciate the support and company from other lab members, including Aleksandra Dimitrievska, Marija Marjanovic, Daniel Joseph Antrim, Hongtao Yang, Xiangyang Ju, Taisai Kumakura, Haoran Zhao, Miha Muskinja, Karol Krizka, Elodie Resseguie, Juerg Beringer, Marko Stamenkovic, Rebecca Carney, Zhicai Zhang, Mariel Pettee, Shuo Han, Emily Anne Thompson, Elham E Khoda, Ian Dyckes, Maria Giovanna Foti, Carl Haber, Elliot Reynolds, Vinicius Mikuni, Alessandra Ciocio, and Angira Rastogi.

I would also like to thank my mom Li-Chin Huang and my partner Gabriel Alcaraz for their love and support.
Their unwavering encouragement and belief in me have been crucial to my success. I am grateful for their presence in my life and their unwavering support throughout my PhD journey.

Finally, I want to express my heartfelt appreciation to all my friends who have been a part of my life in Wisconsin, Switzerland, and Berkeley. I am grateful for the joy and laughter that they brought into my life, and their unwavering support and encouragement have been invaluable to me.

\end{acknowledgments}

\tableofcontents
\listoftables
\listoffigures


\begin{abstract}
%
%
%

\noindent       
The discovery of the Standard Model (SM) Higgs boson by ATLAS and CMS at the LHC in 2012 marked a major milestone in particle physics.
However, many questions remain unanswered, which has led to an active research program to search for either rare SM phenomena or Beyond Standard Model (BSM) physics that involve the Higgs boson.
In this dissertation, I present two example searches involving the Higgs boson, using proton-proton ($pp$) collision data collected by the ATLAS detector.

\vspace*{0.5em}
\noindent       
The first search tackles the problem of how the SM Higgs couples to the second-generation fermions. It searches for the dimuon decay of the SM Higgs boson (\Hmm) using data corresponding to an integrated luminosity of 139 fb$^{-1}$ collected by the ATLAS detector in $pp$ collisions at $\sqrt{s} = 13\ \mathrm{TeV}$ at the LHC. To identify this rare decay, we train boosted decision trees to separate signal and background. We obtain an observed (expected) significance over the background-only hypothesis for a Higgs boson with a mass of 125.09 GeV of 2.0$\sigma$ (1.7$\sigma$). The observed upper limit on the cross section times branching ratio for $pp\to H\to\mu\mu$ is 2.2 times the SM prediction at 95\% confidence level, while the expected limit on a \Hmm\ signal assuming the absence (presence) of a SM signal is 1.1 (2.0). The best-fit value of the signal strength parameter, defined as the ratio of the observed signal yield to the one expected in the SM, is $\mu = 1.2 \pm 0.6$.

\vspace*{0.5em}
\noindent       
In the second search, we look for Dark Matter produced in association with a Higgs boson decaying to $b$-quarks.
This search uses the same dataset as the \Hmm\ search and targets events that contain large missing transverse momentum and either two $b$-tagged small-radius jets or a single large-radius jet associated with two $b$-tagged subjets.
We split events into multiple categories that target different phase spaces of the Dark Matter signals.
We do not observe a significant excess from the SM prediction.
We interpret the results using two benchmark models with two Higgs doublets extended by either a heavy vector boson $Z^\prime$ (referred to as \zpthdm) or a pseudoscalar singlet $a$ (referred to as \thdma) that provide a Dark Matter candidate $\chi$. For \zpthdm, the observed limits extend up to a $Z^\prime$ mass of 3.1 TeV at 95\% confidence level for a mass of 100 GeV for the Dark Matter candidate. For \thdma, we exclude masses of a up to 520 GeV and 240 GeV for $\tan\beta=1$ and $\tan\beta=10$, respectively, and for a Dark Matter mass of 10 GeV. Additionally, we set limits on the visible cross sections, which range from 0.05 fb to 3.26 fb, depending on the regions of missing transverse momentum and $b$-quark jet multiplicity.

\vspace*{0.5em}
\noindent       
In addition to the two physics analyses, unfolding is a key procedure in the high energy experiments, which corrects data for the detector effect.
I present a new unfolding method that allows to unfold data without having any artificial binning and is also able to profile nuisance parameters simultaneously.
This new method provides much higher flexibility and increases the reusability for different downstream tasks compared to the traditional approach.
It will benifit any future analyses including Higgs physics and Dark Matter searches.
\end{abstract}

\clearpage\pagenumbering{arabic} 
\chapter{Introduction}
\label{chap:introduction}
The Higgs boson \cite{PhysRevLett.13.321, HIGGS1964132, PhysRevLett.13.508, PhysRevLett.13.585, PhysRev.145.1156, PhysRev.155.1554} has played a crucial role in the Standard Model (SM) and the Higgs mechanism.
It is responsible for particles acquiring mass through their interactions with the Higgs boson.
According to the Higgs mechanism, the W and Z bosons acquire mass through the electroweak symmetry breaking, while fermions' mass is generated through the Yukawa coupling between fermions and the Higgs boson.
The SM Higgs boson was discovered by the ATLAS and CMS experiments at the LHC in 2012 \cite{ATLAS:2012yve, CMS:2012qbp, ATLAS:2012oga}.
This achievement, along with subsequent observations \cite{ATLAS:2015yey, ATLAS:2014aga, ATLAS:2018kot, ATLAS:2018ynr, ATLAS:2018mme}, has resolved a numbrt of puzzles in high energy physics.
However, many questions related to the Higgs boson remain unanswered, such as whether it interacts with fermions in the first and second generations and whether it can be a portal to Beyond Standard Model (BSM) particles such as extra Higgs bosons arising from the Two-Higgs-Doublet Model (2HDM) \cite{Lee:1973iz, Branco:2011iw}, and Dark Matter \cite{2013ApJS..208...19H, Planck:2018nkj, doi:10.1146/annurev.aa.25.090187.002233, Bertone:2004pz, Feng:2010gw}.
To answer these remaining questions, a broad program of searches for either rare SM phenomena or BSM that involve the Higgs boson have been proposed and initiated.
In this dissertation, I will give a detailed review of two example searches that involve the Higgs boson and use the proton-proton (p-p) collision data collected by the ATLAS detector \cite{Collaboration_2008, CERN-LHCC-2012-009}.
The focus is on strategies that enhance the sensitivity to the searched signals.

The first search is a search for the dimuon decay of the SM Higgs boson (\Hmm) \cite{Hmumu}.
It is the most promising channel for probing the couplings between the SM Higgs Boson and the second generation fermions at the LHC, thanks to excellent muon identification and reconstruction \cite{ATLAS:2020auj}.
However, due to the small decay branching ratio and a large amount of background dominated by the Drell-Yan process, it is a very challenging analysis with a signal-to-background ratio typically at the level of 0.1\%.
The small signal to background ratio leads to two major challenges.
Firstly, it is difficult to establish the signal without good separation between signal and background.
Secondly, measurements can be easily biased by statistical fluctuation and background mismodeling.
We thus designed special techniques and analysis strategies to address these two challenges, including a categorization using dedicated boosted decision trees to separate signal and background and enhance the signal sensitivity, and a background modeling procedure based on spurious signal tests which ensure that background distributions can be modeled by the chosen functions without significant bias.

The second search is to look for Dark Matter produced in association with a Higgs boson decaying to $b$-quarks \cite{monoHbb}.
In contrast with the \Hmm\ search, \Hbb\ has a much larger decay branching ratio and has been established prior to these analyses.
Such a dominant decay channel is particularly useful to search for BSM, such as Dark Matter, which can be produced in association with the Higgs boson.
This channel, which connects the Dark sector with the Higgs sector, has great sensitivity to DM models with the extended Higgs sector \cite{Branco:2011iw, Lee:1973iz}, such as \zpthdm\ \cite{zp2hdm} and \thdma\ \cite{LHCDarkMatterWorkingGroup:2018ufk, 2hdma}.
It targets events that contain large missing transverse momentum and a Higgs boson which decays to $b$-quarks (referred as \monoHbb).
The Higgs boson can be in low energy or highly boosted.
We thus designed different strategies to better reconstruct the Higgs boson.
Events are split by different kinematic variables into multiple categories which target different phase spaces of the DM signals.
Upper limits on signal cross-section in a model-independent context, which will be useful for future reinterpretation with new DM models, are also calculated.

It is interesting to point out the main differences between these two analyses.
Firstly, \Hmm\ search is a search for a single rare SM process, while the search of \monoHbb\ looks for multiple possible BSM models with indefinite signal parameters.
Secondly, \Hmm\ search uses a multivariate-variable-analysis (MVA) approach with machine learning to categorize the events and maximize the search sensitivity, while the search of \monoHbb\ adopts the traditional cut-based approach, which reduces model dependency.
Finally,\Hmm\ uses a data-driven approach with analytical functions to model the signal and background, while \monoHbb\ estimates the signal and background with Monte Carlo (MC) simulation templates.
These two analyses together serve as great examples that cover different perspectives of searches in high-energy particle experiments.

In additional to physics analyses, unfolding is an important procedure in particle physics experiments which corrects for detector effects and provides differential cross section measurements that can be used for a number of downstream tasks, such as extracting fundamental physics parameters.
Traditionally, unfolding is done by discretizing the target phase space into a finite number of bins and is limited in the number of unfolded variables.
Recently, there have been a number of proposals to perform unbinned unfolding with machine learning. However, none of these methods (like most unfolding methods) allow for simultaneously constraining (profiling) nuisance parameters.
We thus propose a new machine learning-based unfolding method, referred to as unbinned profiled unfolding (UPU) that results in an unbinned differential cross section and can profile nuisance parameters.
The machine learning loss function is the full likelihood function, based on binned inputs at detector-level.
We first demonstrate the method with simple Gaussian examples and then show the impact on a simulated Higgs boson cross section measurement.

The rest of the thesis is organized as follows: Chapter~\ref{chap:higgs} briefly introduces the properties of the SM Higgs boson, its interactions with other particles and the extended Higgs sector associated with DM;
Chapter~\ref{chap:detector} introduces the LHC and describes the ATLAS detector;
Chapter~\ref{chap:sample} describes the data and simulation samples used for \Hmm\ and \monoHbb\ analyses;
Chapter~\ref{chap:object} summarizes the physics object definiations used for \Hmm\ and \monoHbb\ analyses;
Chapter~\ref{chap:hmumu} details the analysis strategies for \Hmm\, while Chapter~\ref{chap:monoHbb} describes the analysis strategies for \monoHbb;
Chapter~\ref{chap:upu} presents the study of the newly proposed method of unbinned profiled unfolding;
Chapter \ref{chap:conclusion} provides a summary and outlook.                
\chapter{SM Higgs boson and the extended Higgs sector}
\label{chap:higgs}

The SM Higgs boson \cite{Higgs:1964pj, Higgs:1966ev, Englert:1964et, Guralnik:1964eu, Kibble:1967sv} is a massive scalar boson with spin-0, CP-even, charge-neutral, and colorless properties.
Its mass has been measured to be 125.09 GeV \cite{ATLAS:2015yey}.
The Higgs boson ($H$) can interact with quarks ($q$), charged-leptons ($\ell^{\pm}$), $Z$, $W^{\pm}$ as well as itself (self-coupling).
The strength of these interactions is determined by the mass of the interacting particle.
The Lagrangian of the Higgs couplings can be expressed as:
\begin{equation}
\label{eq:HCouplingLagrangian}
\mathcal{L}\supset -g_{Hff}f\bar{f}H + \delta_V V_\mu V^\mu \left( g_{HVV}H + \frac{g_{HHVV}}{2} H^2 \right) + \frac{g_{HHH}}{6} H^3 + \frac{g_{HHHH}}{24} H^4,
\end{equation}
where $f \in \left\{q,\, \ell^{\pm}\right\}$, $V \in \left\{Z,\, W^{\pm}\right\}$, and $\delta_W = 1$, $\delta_Z = \frac{1}{2}$.
The $g$'s in Eq.~\ref{eq:HCouplingLagrangian} represent the coupling strength to each particle:
\begin{equation}
  \label{eq:HCouplingStrength}
  g_{Hff} = \frac{m_f}{\nu},\ g_{HVV} = \frac{2 m_V^2}{\nu},\ g_{HHVV} = \frac{2 m_V^2}{\nu^2},\ g_{HHH} = \frac{3 m_H^2}{\nu},\ g_{HHHH} = \frac{3 m_H^2}{\nu^2}.\ 
\end{equation}
Specifically, $g_{Hff}$ is proportional to the mass of the fermion, while $g_{HVV}$ and $g_{HHVV}$ are proportional to the square of the mass of the $Z$ and $W^\pm$ bosons, respectively.
The coupling strength to the Higgs boson itself is represented by $g_{HHH}$ and $g_{HHHH}$, both of which are proportional to the square of the Higgs boson mass, \mH.
The vacuum expectation value of the Higgs field, $\nu = \left(\sqrt{2}G_F\right)^{-\frac{1}{2}} \approx 246$ GeV, is used in Eq.~\ref{eq:HCouplingStrength} to calculate the coupling strengths.

In the following sections, I will briefly discuss the production and decay of the Higgs boson at the LHC, as well as the extended Higgs sector that includes additional particles that bridge the Dark sector.

\afterpage{\clearpage}


\section{SM Higgs boson production at the LHC}
\label{chap:higgs:sec:production}
\begin{figure}[h!]
  \centering
  \subfigure[]{\includegraphics[width=0.49\textwidth]{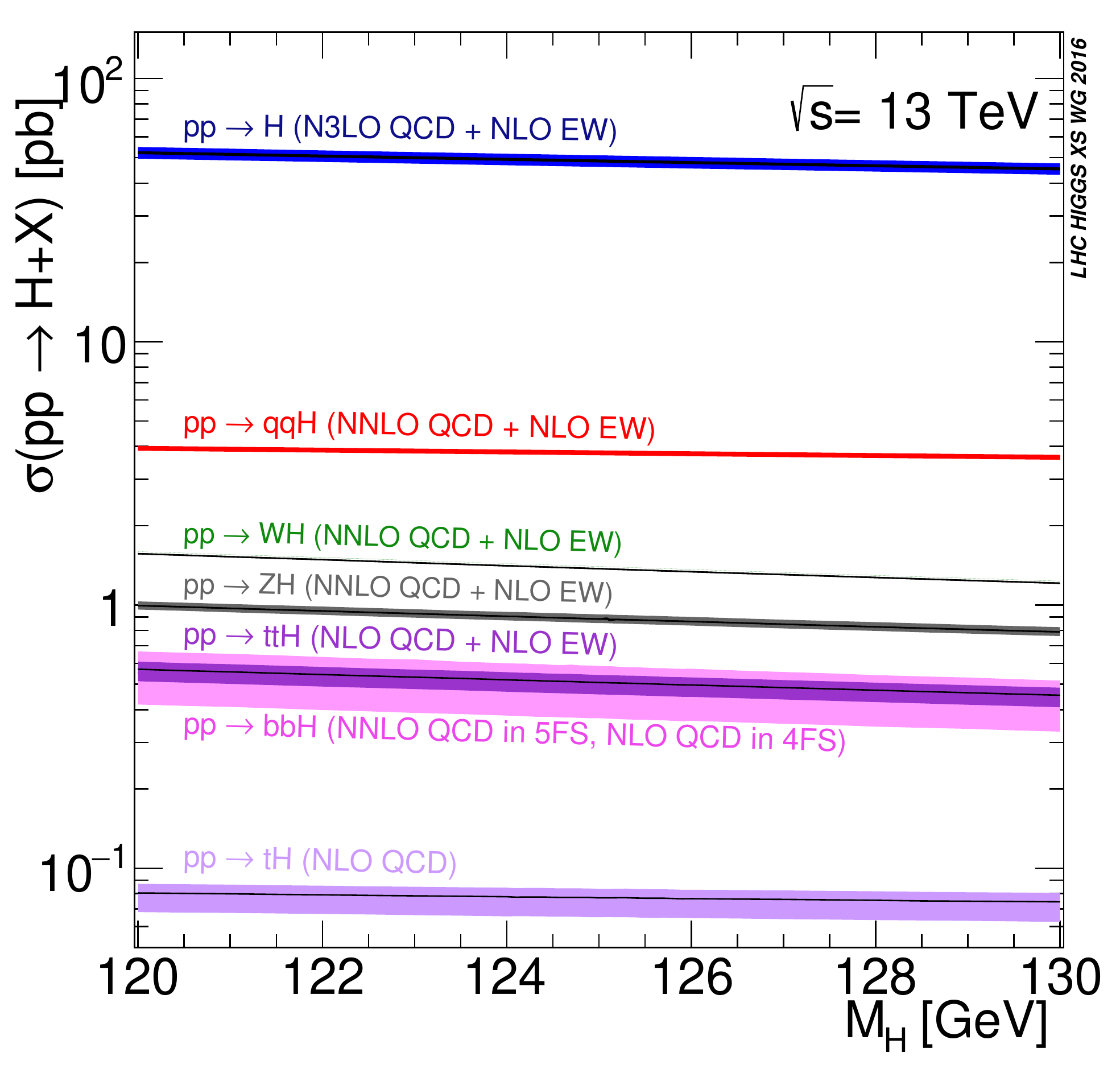}}
  \subfigure[]{\includegraphics[width=0.49\textwidth]{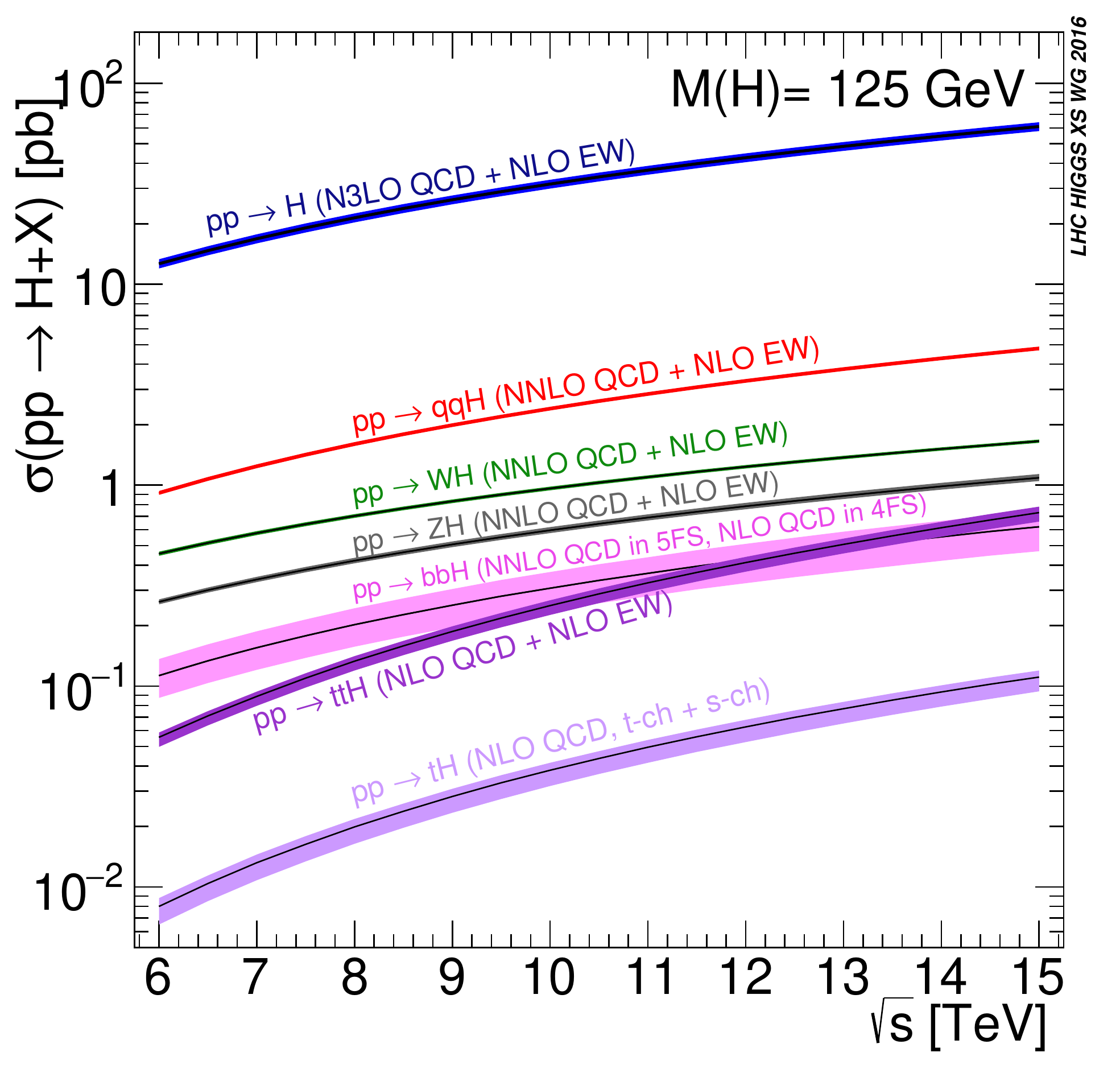}}
  \caption{
    The production cross sections of the SM Higgs boson
    (a) as a function of \mH\ at \thirteentev\ and
    (b) as a function of $\sqrt{s}$ with the experimental measured $m_{H} = 125.09$ GeV.
    The central values are depicted by solid lines, while the colored bands illustrate the theoretical uncertainties.
  }
  \label{fig:higgs_xs}
\end{figure}

The SM Higgs boson can be produced through several processes via $pp$ collision at the LHC. The main production modes are as follows:
\begin{itemize}
\item Gluon fusion production (\ggF) \ppggF
\item Vector boson fusion production (\VBF) \ppVBF
\item Associated production with a $W$ boson (\ppWH) or a $Z$ boson (\ZH) \ppZH
\item Associated production with a pair of top quarks (\ttH) \ppttH,
  or a pair of bottom quarks (\bbH) \ppbbH
\item Associated production with a single top quark (\tH) \pptH
\end{itemize}
The production cross-sections for each mode depend on the Higgs boson mass \mH\ and the center-of-mass energy ($\sqrt{s}$) of the $pp$ collision.
Fig.~\ref{fig:higgs_xs}~(a) summarizes the production cross-sections as a function of \mH\ at $\sqrt{s}=13$ TeV, while Fig.~\ref{fig:higgs_xs}~(b) shows the cross-sections as a function of $\sqrt{s}$ with the experimental measured $\mH = 125.09$ GeV.

\begin{figure}[h!]
  \centering
  \subfigure[]{\includegraphics[width=0.4\textwidth]{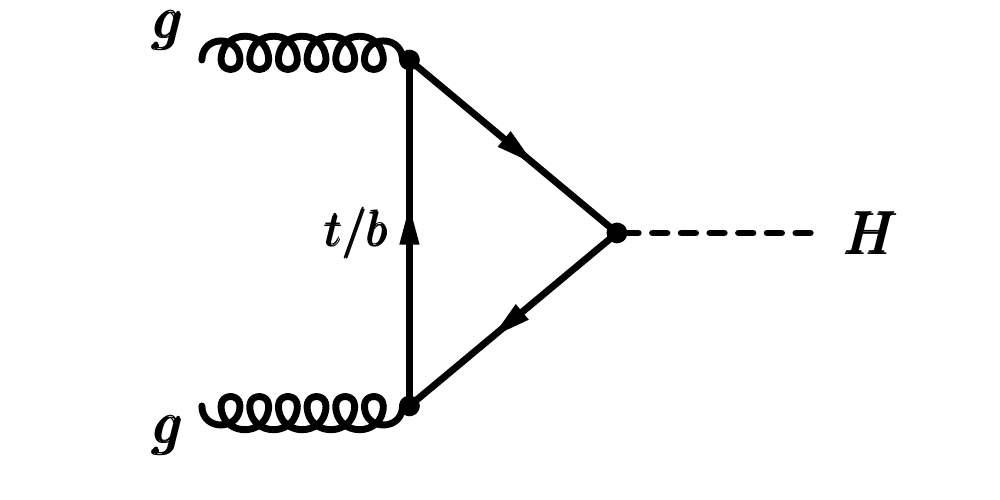}}
  \subfigure[]{\includegraphics[width=0.4\textwidth]{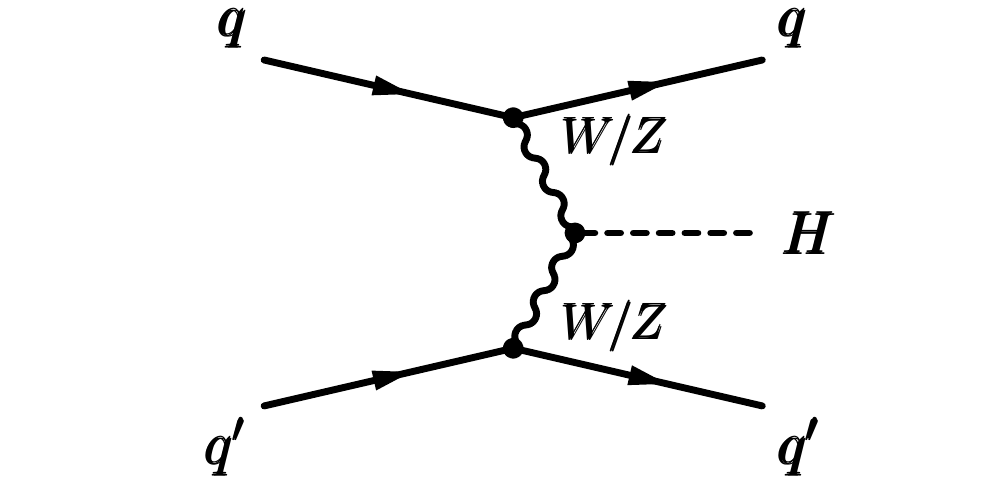}}
  \caption{Feynman diagrams showing the LO production of Higgs via (a) \ggF\ and (b) \VBF\ processes.}
  \label{fig:feyn_ggF}
\end{figure}

The \ggF\ production is expected to dominate as it involves the strong interaction between gluons, which is very strong at the LHC energy scales.
As gluons are massless, their interaction with the Higgs boson must be mediated through a heavy quark loop.
Fig.~\ref{fig:feyn_ggF}~(a) shows the leading-order (LO) Feynman diagram of the \ggF\ production.

The \VBF\ production has the second largest production cross-section.
This process is characterized by two forward jets, which are crucial for its identification.
Since the Higgs boson is produced through interactions with vector bosons, this process can be used to measure the Higgs couplings to vector bosons.
Fig.~\ref{fig:feyn_ggF}~(b) shows the LO Feynman diagram of the \VBF\ production.

\begin{figure}[h!]
  \centering
  \subfigure[]{\includegraphics[width=0.31\textwidth]{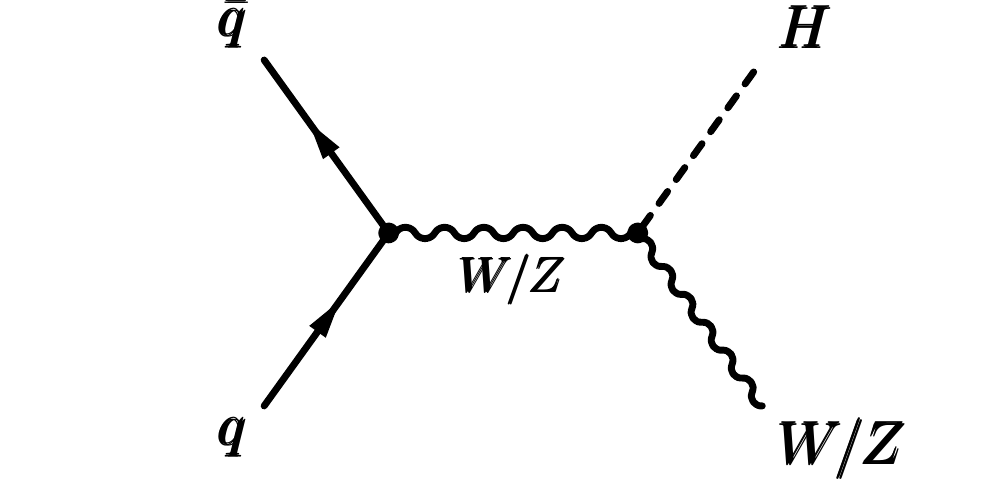}}
  \subfigure[]{\includegraphics[width=0.31\textwidth]{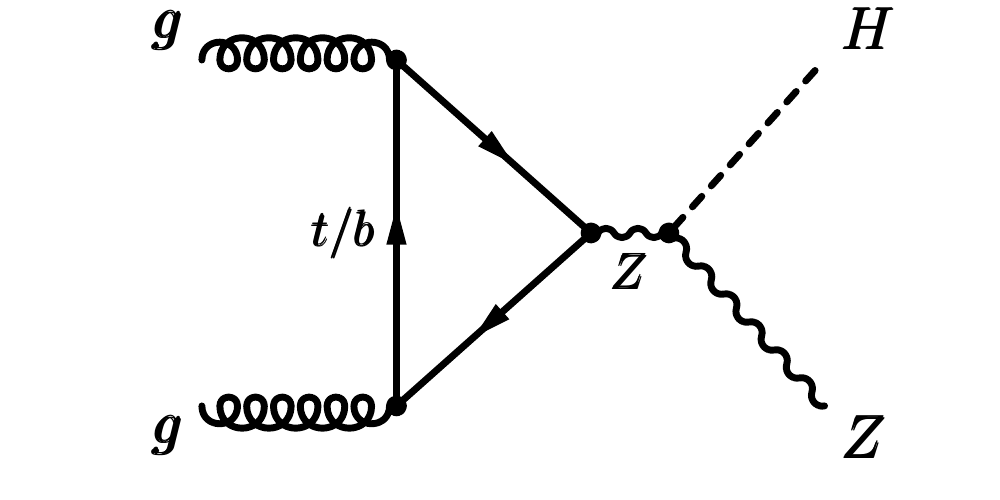}}
  \subfigure[]{\includegraphics[width=0.31\textwidth]{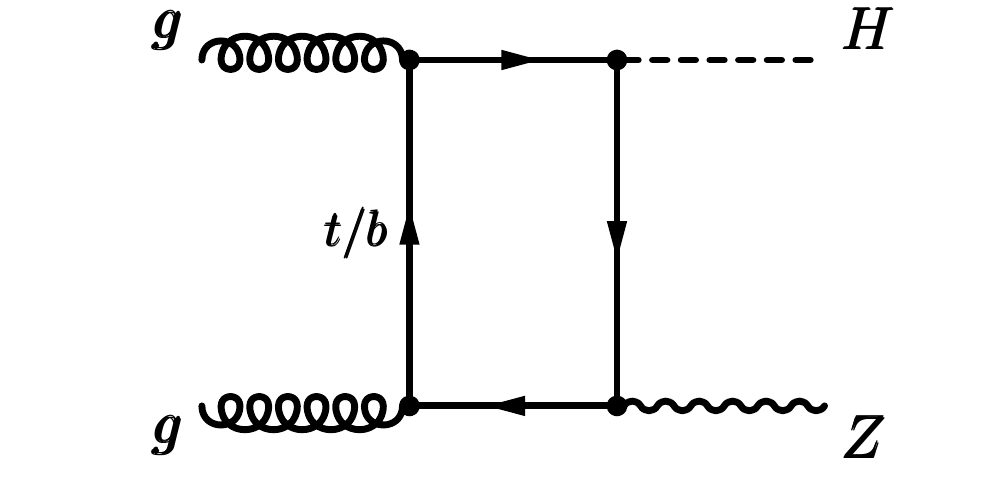}}
  \caption{Examples of LO Feynman diagrams for Higgs production via \VH, where the Higgs boson is produced in association with either a $W$ boson (\WH) or a $Z$ boson (\ZH).}
  \label{fig:feyn_VH}
\end{figure}

The next main production mode is the \VH\ production.
The Higgs boson is produced in association with either a $W$ boson (\WH) or a $Z$ boson (\ZH).
The vector boson can decay leptonically or hadronically, but the leptonic decay channels are usually easier to identify due to good reconstruction and resolution of the leptons.
The LO Feynman diagrams of the \VH\ production are shown in Fig.~\ref{fig:feyn_VH}.

\begin{figure}[h!]
  \centering
  \subfigure[]{\includegraphics[width=0.31\textwidth]{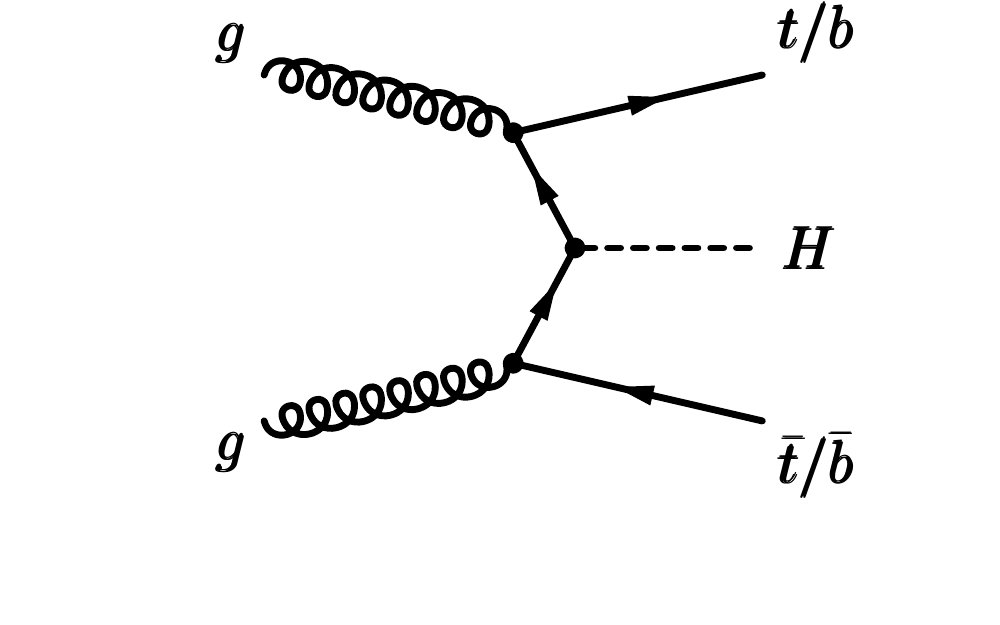}}
  \subfigure[]{\includegraphics[width=0.31\textwidth]{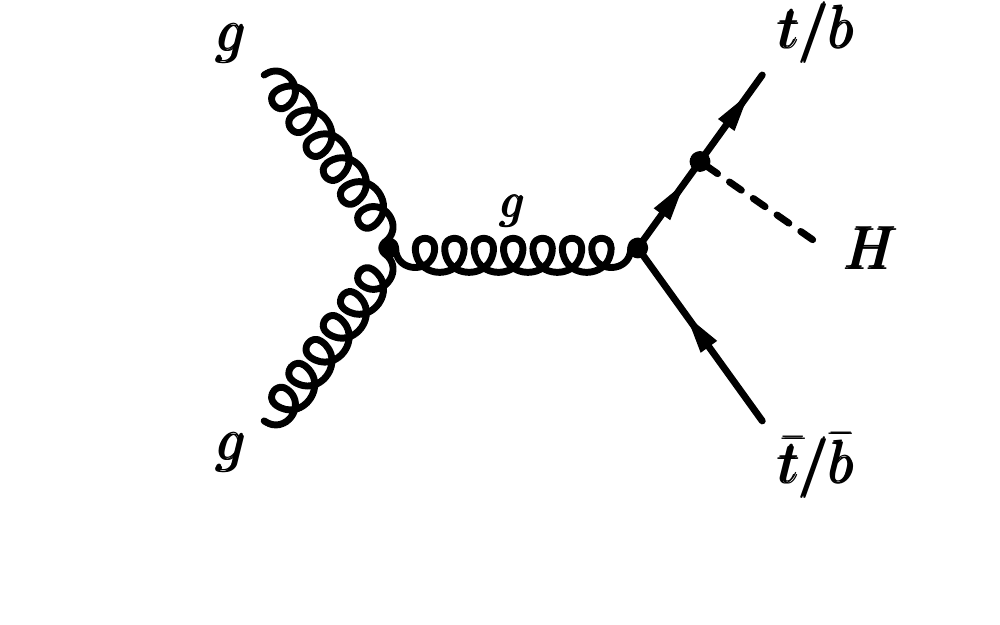}}
  \subfigure[]{\includegraphics[width=0.31\textwidth]{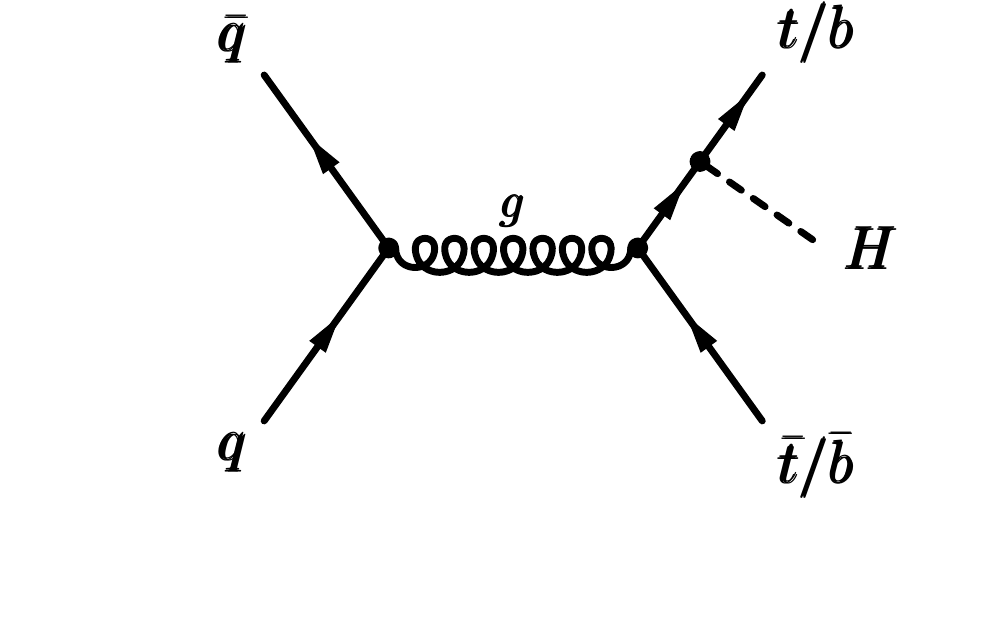}}
  \caption{Examples of LO Feynman diagrams for Higgs boson production via the \ttH\ and \bbH\ processes.}
  \label{fig:feyn_ttH}
\end{figure}

\begin{figure}[h!]
  \centering
  \subfigure[]{\includegraphics[width=0.24\textwidth]{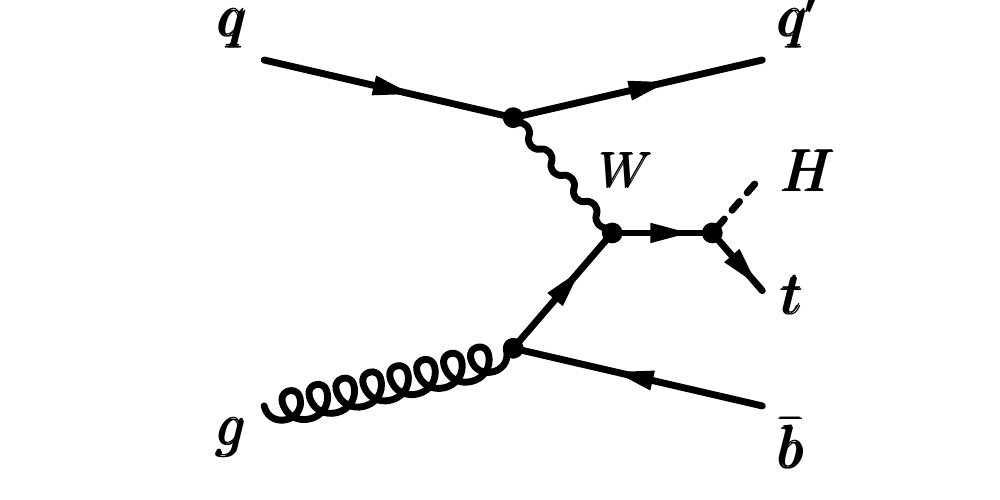}}
  \subfigure[]{\includegraphics[width=0.24\textwidth]{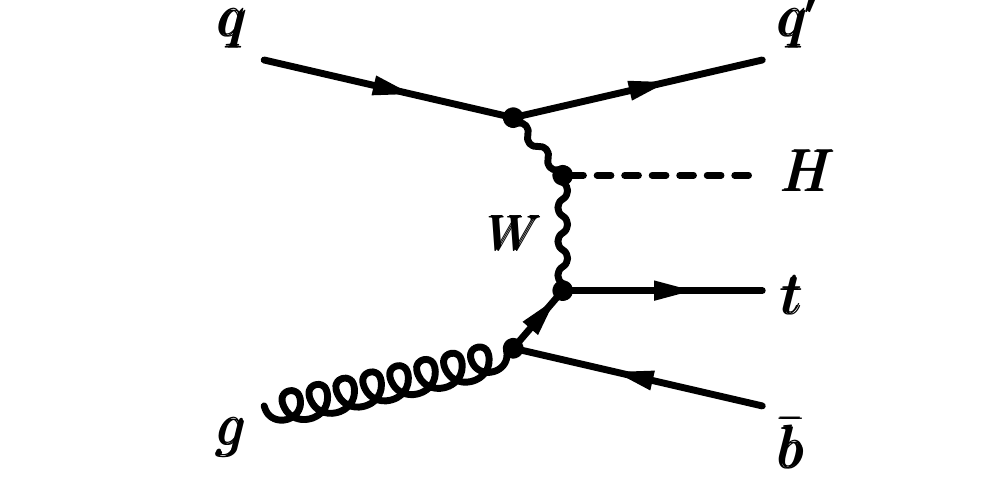}}
  \subfigure[]{\includegraphics[width=0.24\textwidth]{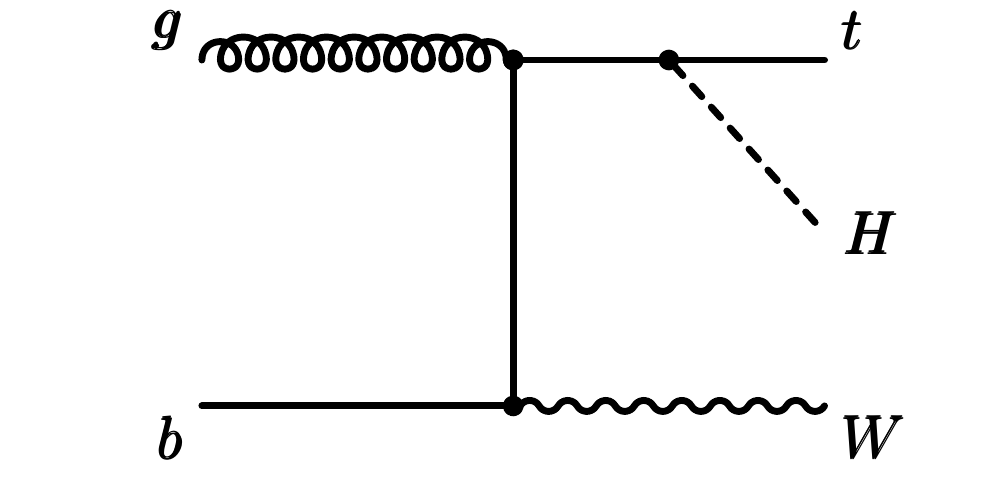}}
  \subfigure[]{\includegraphics[width=0.24\textwidth]{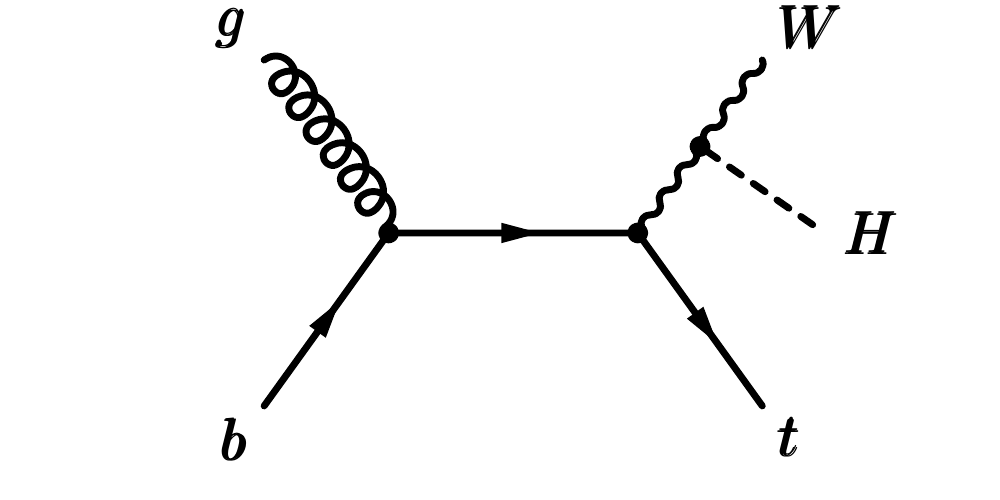}}
  \caption{Examples of LO Feynman diagrams for Higgs boson production via the \tH\ process.}
  \label{fig:feyn_tH}
\end{figure}

Although the \ttH, \bbH\ and \tH\ productions have relatively small cross-sections, they are important for direct measurements of the Yukawa couplings to the third-generation fermions as the Higgs boson is produced via the interactions with these fermions.
Among them, \ttH\ has been extensively explored to probe the large top-Higgs Yukawa coupling.
The LO Feynman diagrams for \ttH\ and \bbH\ productions are shown in Fig.~\ref{fig:feyn_ttH}, while the LO Feynman diagrams of \tH\ are shown in Fig.~\ref{fig:feyn_tH}.

\section{SM Higgs boson decay at the LHC}
\label{chap:higgs:sec:decay}

\begin{figure}[htbp]
\begin{center}
\includegraphics[width=0.49\columnwidth]{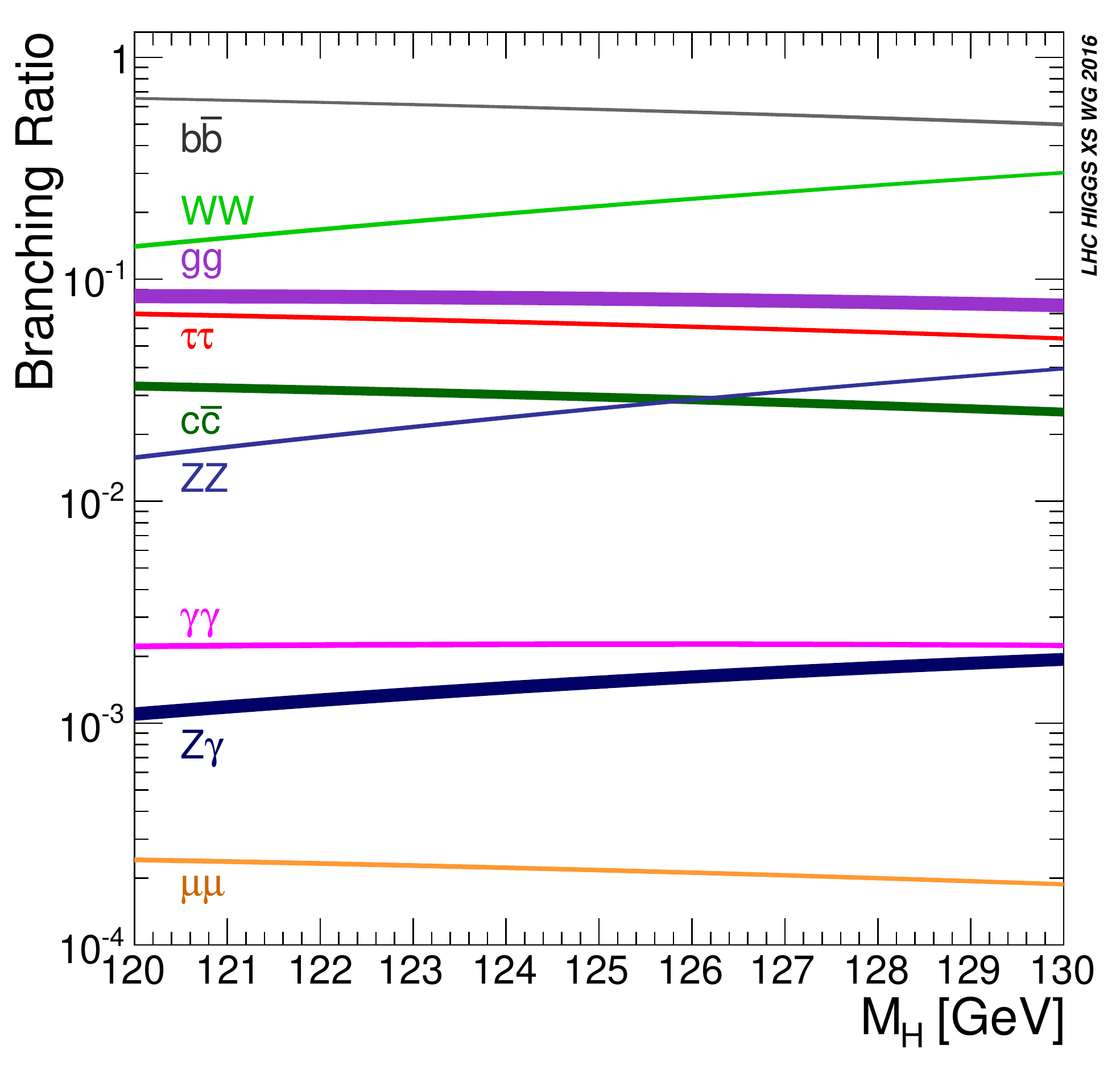}
\caption{The branching ratios of the primary decay channels of the Higgs boson as a function of \mH.}
\label{fig:Higgs_BR}
\end{center}
\end{figure}

The Higgs boson is a short-lived particle with a lifetime predicted to be approximately $10^{-22}$ s, corresponding to a width of approximately 4 \MeV.
Therefore, the Higgs boson can only be observed indirectly through its decay products at the LHC.
Fig.~\ref{fig:Higgs_BR} displays the branching ratios of the primary decay channels of the Higgs boson as a function of \mH.
The branching ratios of the Higgs decays at $m_{H} = 125.09\ \mathrm{GeV}$ are listed in Tab.~\ref{tab:Higgs_BR}.
At around $m_{H} = 125\ \mathrm{GeV}$, \Hbb\ is the dominant decay channel, with a branching ratio of 57.5\%, whereas $H\to \mu\mu$ has a relatively small branching ratios due to the very small value of the muon mass.

\begin{table*}[h]
\begin{center}
\begin{tabular}{cc}
\toprule
Decay channel  &  Branching ratio [\%]\\
\midrule
\Hbb & $57.5 \pm 1.9$    \\
\HWW & $21.6 \pm 0.9$    \\
\Hgg & $8.56 \pm 0.86$   \\
\Htt & $6.30 \pm 0.36$   \\
\Hcc & $2.90 \pm 0.35$   \\
\HZZ & $2.67 \pm 0.11$   \\
\Hyy & $0.228 \pm 0.011$   \\
\HZg & $0.155 \pm 0.014$   \\
\Hmm & $0.022 \pm 0.001$   \\
\bottomrule
\end{tabular}
\caption{The branching ratios of the SM Higgs decays at $\mH = 125.09\ \mathrm{GeV}$.}
\label{tab:Higgs_BR}
\end{center}
\end{table*}



\section{Extended Higgs sector --- type II Two-Higgs-Doublet Model}
\label{chap:higgs:sec:extended}

The SM assumes that the Higgs sector is an SU(2) doublet, which can be extended with a second Higgs doublet to create the Two-Higgs-Doublet Model (2HDM) \cite{Lee:1973iz, Branco:2011iw}.
The 2HDM is a key component of many well-motivated BSM theories, such as supersymmetry \cite{Haber:1984rc} and axion models \cite{Kim:1986ax, Peccei:1977hh}, and is motivated by the flexibility it provides in the scalar mass spectrum and the existence of additional sources of CP violation \cite{Trodden:1998qg}.
It also enables the connection to the Dark sector through extra particles.
Here, the focus will be on the type II 2HDM (2HDM II), where one of the Higgs doublets couples to up-type quarks, while the other couples to down-type quarks and leptons.
The Yukawa couplings are expressed by:
\begin{equation}
\mathcal{L} \supset - \left(\bar{Q}Y_{u} \tilde{\Phi}_{u} u_R + \bar{Q}Y_{d} \tilde{\Phi}_{d} d_R + \bar{L}Y_{\ell}\Phi_{d}\ell_{R} + \mathrm{h.c.}     \right),
\end{equation}
where $Y_{f}$ are the Yukawa couplings, $Q$ and $L$ are the left-handed quark and lepton doublets, and $f_{R}$ are the right-handed fermions.
$\Phi_u$ and $\Phi_d$ are the two Higgs doublets, which are parametrized as:
\begin{equation}
\begin{split}
&\Phi_u = \frac{1}{\sqrt{2}} \left( \begin{matrix}
\cos \beta H^{+} \\
\nu_u +\cos\alpha h + \sin \alpha H + i \cos \beta A
\end{matrix} \right), \\
&\Phi_d = \frac{1}{\sqrt{2}} \left( \begin{matrix}
-\sin \beta H^{+} \\
\nu_d - \sin\alpha h + \cos \alpha H - i \sin \beta A,
\end{matrix} \right) \\
\end{split}
\end{equation}
where $h$, $H$ are neutral CP-even Higgs bosons, $H^{\pm}$ are charged Higgs bosons, and $A$ is a neutral CP-odd Higgs boson (pseudo-scalar).
$\nu_f$ are the vacuum expectation values of the two Higgs doublets, which satisfy $\nu =  \sqrt{\nu_u^2 + \nu_d^2} \approx 246\ \mathrm{GeV}$.

The electroweak (EW) symmetry breaking leads to interactions between the CP-even mass eigenstates and the EW gauge bosons, which can be written as:
\begin{equation}
\mathcal{L} \supset \delta_V V_\mu V^\mu g_{HVV} \left( \sin \left( \beta - \alpha \right) h +  \cos \left( \beta - \alpha \right) H \right),
\end{equation}
where $\alpha$ is the mixing angle between $h$ and $H$, and $\beta$ is defined with $\tan\beta \equiv \frac{\nu_u}{\nu_d}$.
To simplify the model, it is well-motivated to enforce the alignment limit $\alpha = \beta - \pi / 2$, such that $h$ is consistent with the SM Higgs boson (and therefore $m_h \approx 125\ \mathrm{GeV}$).
Note that $H$ here denotes the other Higgs boson (as opposed to the SM Higgs boson), which is likely to be heavier under experimental constraints.
This notation will be used throughout the context of this model.

\subsection{\zpthdm}

\begin{figure}[h!]
  \centering
  \includegraphics[width=0.4\textwidth]{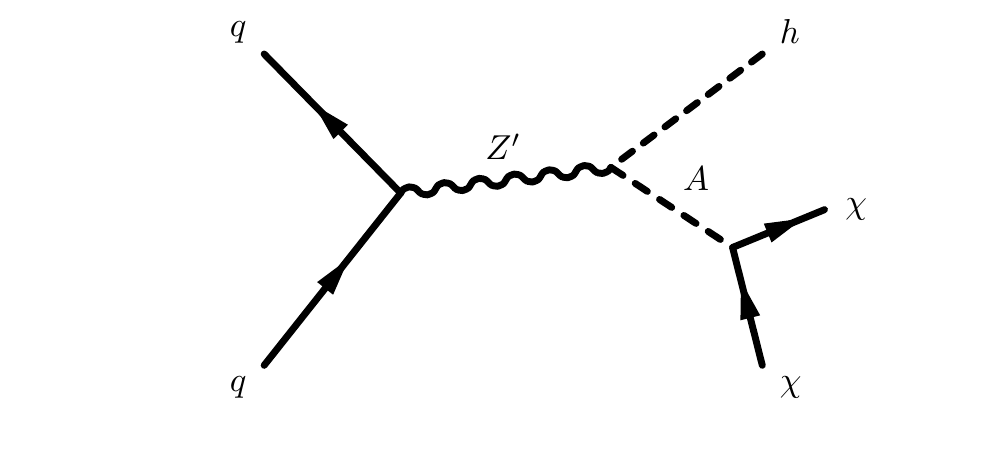}
  \vspace{-5mm}
  \caption{An example Feynman diagram of \zpthdm\ in the \monoHbb search.}
  \label{fig:feyn_zp2hdm}
\end{figure}

The \zpthdm\ \cite{zp2hdm} assumes that the pseudo-scalar $A$ in the 2HDM II is the sole mediator of the interaction between SM and DM.
It extends the 2HDM II with an additional vector boson, $Z^\prime$, which mixes with the $Z$ boson.
By coupling the DM to the pseudo-scalar, $A$, constraints from DM coupling to SM Higgs boson or vector bosons are avoided.
The additional vector boson $Z^\prime$, on the other hand, allows for resonant production of $Z^\prime \to h + A$ (and $A \to \chi\chi$, where $\chi$ represents the DM particle).
This makes it an interesting signal process to look for in the \monoHbb\ search, as the Feynman diagram shown in Fig.~\ref{fig:feyn_zp2hdm}.
Furthermore, the model assumes that $Z^\prime$ couples only to right-handed quarks and the down-type Higgs doublet $\Phi_d$, which avoids potentially stringent limits from dilepton resonance searches.
It is assumed that the coupling between $A$ and $\chi$ is strong, such that the $A\to \chi\chi$ decay has a very large branching ratio, as long as the $A\to t\bar{t}$ decay is kinematically forbidden.

\vspace{-0.5mm}

The \zpthdm\ involves eight parameters, namely \tanb, \mH, \mHpm\ and \mA\ from the 2HDM II; $g_Z$ (the gauge coupling of $Z^\prime$), \mZp (the mass of $Z^\prime$), \mchi (the mass of the DM particle $chi$), and $g_\chi$ (the coupling between $A$ and $\chi$).
To simplify the analysis, we assume $\mH = \mHpm = \mA$ and set \tanb\ and $g_\chi$ to unity, and $g_Z$ to 0.8.
Since these parameters have negligible impact on the kinematic distributions, but only on the production cross-sections and decay branching ratios, the results can be easily scaled for different values of these parameters.
The choices of these values are made to have the most interesting phenomenology as recommended by the LHC Dark Matter Working Group \cite{LHCDarkMatterWorkingGroup:2018ufk}.
Consequently, in the interpretation of the \monoHbb\ analysis, only \mZp\ and \mA\ are the free parameters .

%

\subsection{\thdma}

The \thdma\ \cite{LHCDarkMatterWorkingGroup:2018ufk,2hdma} introduces an extra pseudo-scalar, denoted by $a$, to the 2HDM II, which mixes with the pseudo-scalar $A$ and serves as a mediator between the DM and SM sectors.
The interaction between $a$ and the DM particle $\chi$ is expressed by:
\begin{equation}
\mathcal{L} \supset - i g_\chi a \bar{\chi} \gamma_5 \chi,
\end{equation}
where $g_\chi$ is the coupling strength between $a$ and $\chi$.
Meanwhile, the relevant couplings between Higgs bosons and SM fermions are given by:
\begin{equation}
\begin{split}
\mathcal{L} \supset & - g_{htt} \bar{t} \left[ h + \xi_t H - i \xi_t \left( \cos\theta A - \sin\theta a \right) \gamma_5 \right] t \\
& - \sum_{f=b,\tau} g_{hff} \bar{f} \left[ h + \xi_f H + i \xi_f \left( \cos\theta A - \sin\theta a \right) \gamma_5 \right] f \\
& - g_{htt} V_{tb} \xi_t H^+ \bar{t_R} b_L + g_{hbb} V_{tb} \xi_b H^+ \bar{t_L} b_R + \mathrm{h.c.},
\end{split}
\end{equation}
where $g_{hff}$ are the SM Yukawa couplings defined in Eq.~\ref{eq:HCouplingStrength}, $V_{ij}$ are the elements of the Cabibbo-Kobayashi-Maskawa matrix elements, and $\xi_f$ are given by:
\begin{equation}
\xi_t = -\cot\beta, \ \xi_b = \xi_\tau = \tanb.
\label{eq:2hdma_tb_coupling}
\end{equation}

\begin{figure}[h!]
  \centering
  \subfigure[]{\includegraphics[width=0.4\textwidth]{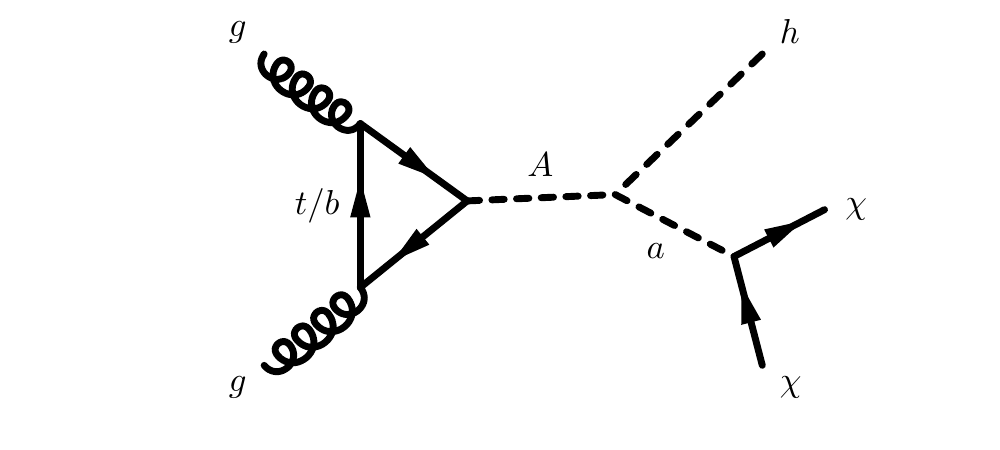}}
  \subfigure[]{\includegraphics[width=0.4\textwidth]{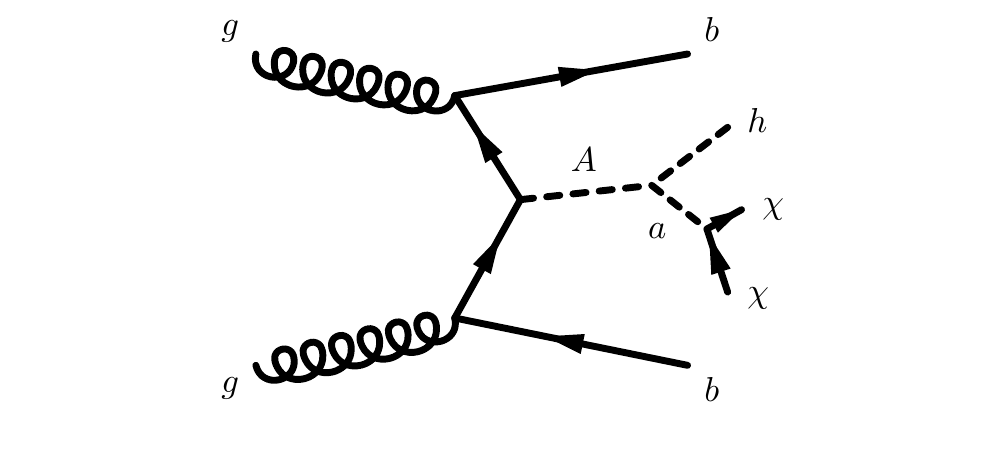}}
  \caption{Example diagrams for \thdma\ with (a) \ggF\ and (b) \bbA\ productions.}
  \label{fig:feyn_2hdma}
\end{figure}

Notably, large values of $\cot\beta$ result in stronger couplings between $t$ and the additional Higgs bosons, while larger values of $\tan\beta$ lead to stronger couplings between $b (\tau)$ and the additional Higgs bosons.
The model yields rich phenomenology with various signatures and complementarity among different experimental searches.
As depicted in Fig.~\ref{fig:feyn_2hdma}, there are two primary ways of producing the $A \to ha \to h\chi\chi$ process, namely, gluon fusion production (\ggF) and associated production with a pair of bottom quarks (\bbA), which give rise to the \monoHbb\ signature of interest.

The \thdma\ can be fully characterized by eleven parameters: \tanb, \mH, \mHpm\ and \mA\ from the 2HDM II; the mixing angle between the two pseudo-scalars $\theta$; the quartic couplings between the Higgs bosons $\lambda_3$, $\lambda_{a1}$, $\lambda_{a2}$; the mass of the additional pseudo-scalar \ma; the coupling to the Dark sector $g_\chi$; and the mass of the DM particle \mchi.
To simplify the analysis, we set $\mA = \mHpm = \mH$ to avoid constraints from EW precision measurements.
Furthermore, we set $\lambda_3 = \lambda_{a1} = \lambda_{a2} = 3$ to ensure the stability of the Higgs potential and maximize the trilinear couplings between CP-even and CP-odd Higgs bosons.
Lastly, we set $g_\chi = 1$ to obtain a large branching ratio of $a\to\chi\chi$, provided that the the decay is kinematically allowed.
Like \zpthdm, these values are chosen to have the most interesting phenomenology as recommended by the LHC Dark Matter Working Group \cite{LHCDarkMatterWorkingGroup:2018ufk}.
These considerations reduce the number of free parameters to five, namely \mA, \ma, \tanb, $\theta$ and \mchi.

\clearpage

\chapter{The Large Hadron Collider and the ATLAS detector}
\label{chap:detector}

\section{The Large Hadron Collider}

The Large Hadron Collider (LHC) \cite{Evans:2008zzb} is a particle accelerator designed to study proton-proton collisions at unprecedented energy levels.
It is located at the European Organization for Nuclear Research (CERN) in Geneva, Switzerland and is the largest and most powerful particle accelerator in the world.

\begin{figure}[htbp]
  \begin{center}
  \includegraphics[width=0.95\columnwidth]{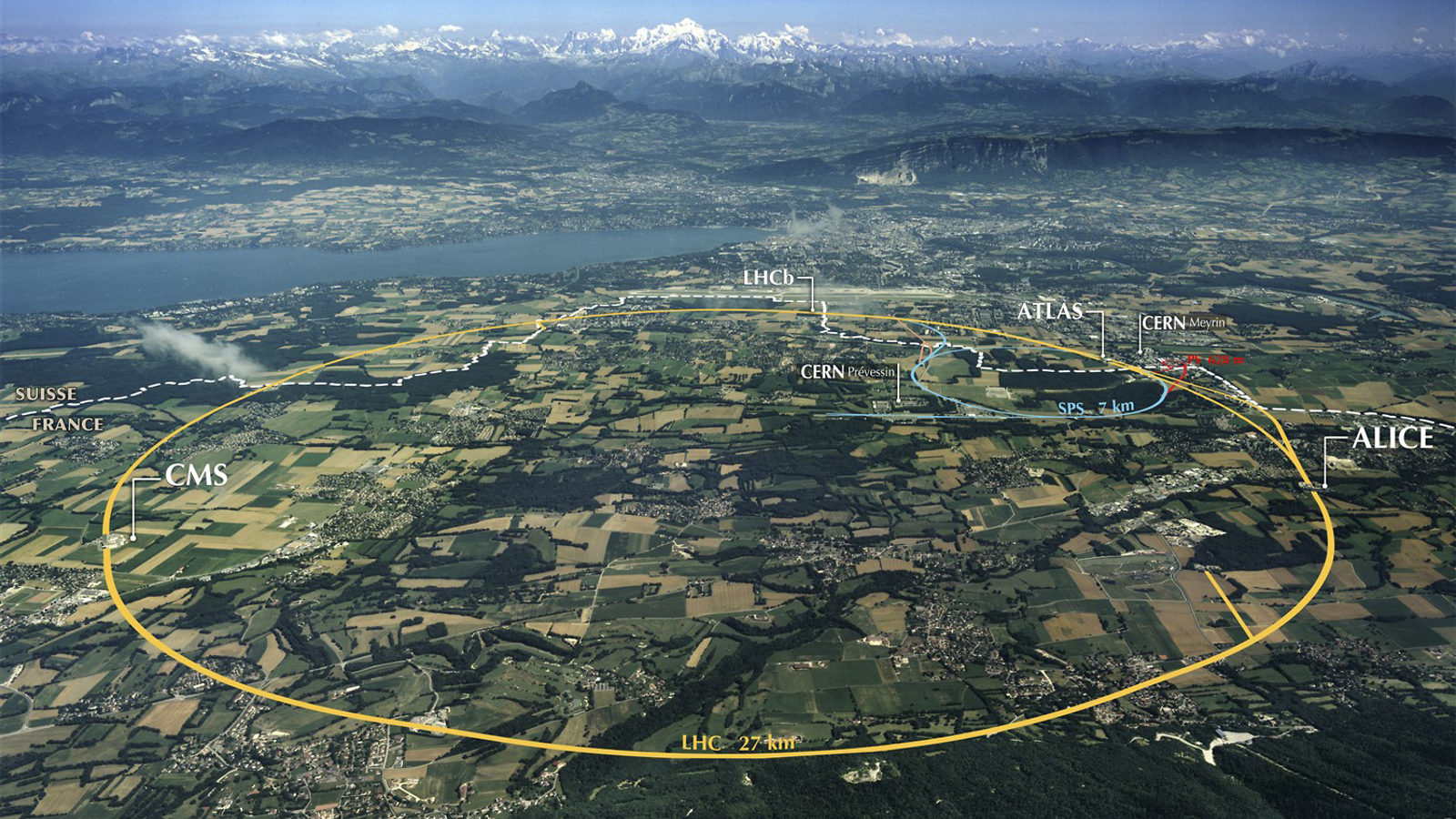}
  \caption{The large hadron collider illustrated in the map of the Geneva area in Switzerland.}
  \label{fig:lhc}
  \end{center}
  \end{figure}

As illustrated in Fig.~\ref{fig:lhc}, the LHC is a circular accelerator with a circumference of 27 kilometers (17 miles) and is situated underground, at a depth of 100 meters (328 feet).
The protons are accelerated to nearly the speed of light and then made to collide at four different points around the LHC, where large detectors such as ATLAS \cite{Collaboration_2008, CERN-LHCC-2012-009} and CMS \cite{CMS:2008xjf} are located.

The energy scale of the LHC is immense, with proton-proton collisions occurring at center-of-mass energies up to 13 TeV.
This is over 100 times the energy achieved by the previous largest accelerator, the Tevatron \cite{Holmes:2011ey} at Fermilab in the United States.

The LHC has undergone several upgrades since it began operations in 2008.
The second run of the LHC, known as Run 2, began in 2015 and lasted until the end of 2018.
During this run, the LHC collided protons at higher energies and intensities than ever before, allowing scientists to study the properties of the Higgs boson in greater detail and search for new particles beyond the standard model of particle physics.

The current phase of the LHC is Run 3, which began in 2022.
The main goal of Run 3 is to collect data on proton-proton collisions at a higher luminosity, which is the number of collisions per second.
This will enable scientists to study rare processes and phenomena that occur at higher energy scales than previously explored.
The Run 3 is expected to continue until 2024, followed by a period of maintenance and upgrade in preparation for the High-Luminosity LHC project.

The High-Luminosity LHC (HL-LHC) project involves upgrading the accelerator and detectors to increase the collision rate and improve the accuracy of the measurements.
The HL-LHC is currently planned to start operation in 2028 and will further increase the collision rate and luminosity of the LHC, opening new avenues of research in particle physics.

\section{The ATLAS detector}
\begin{figure}[h!]
  \centering
  \includegraphics[width=0.9\textwidth]{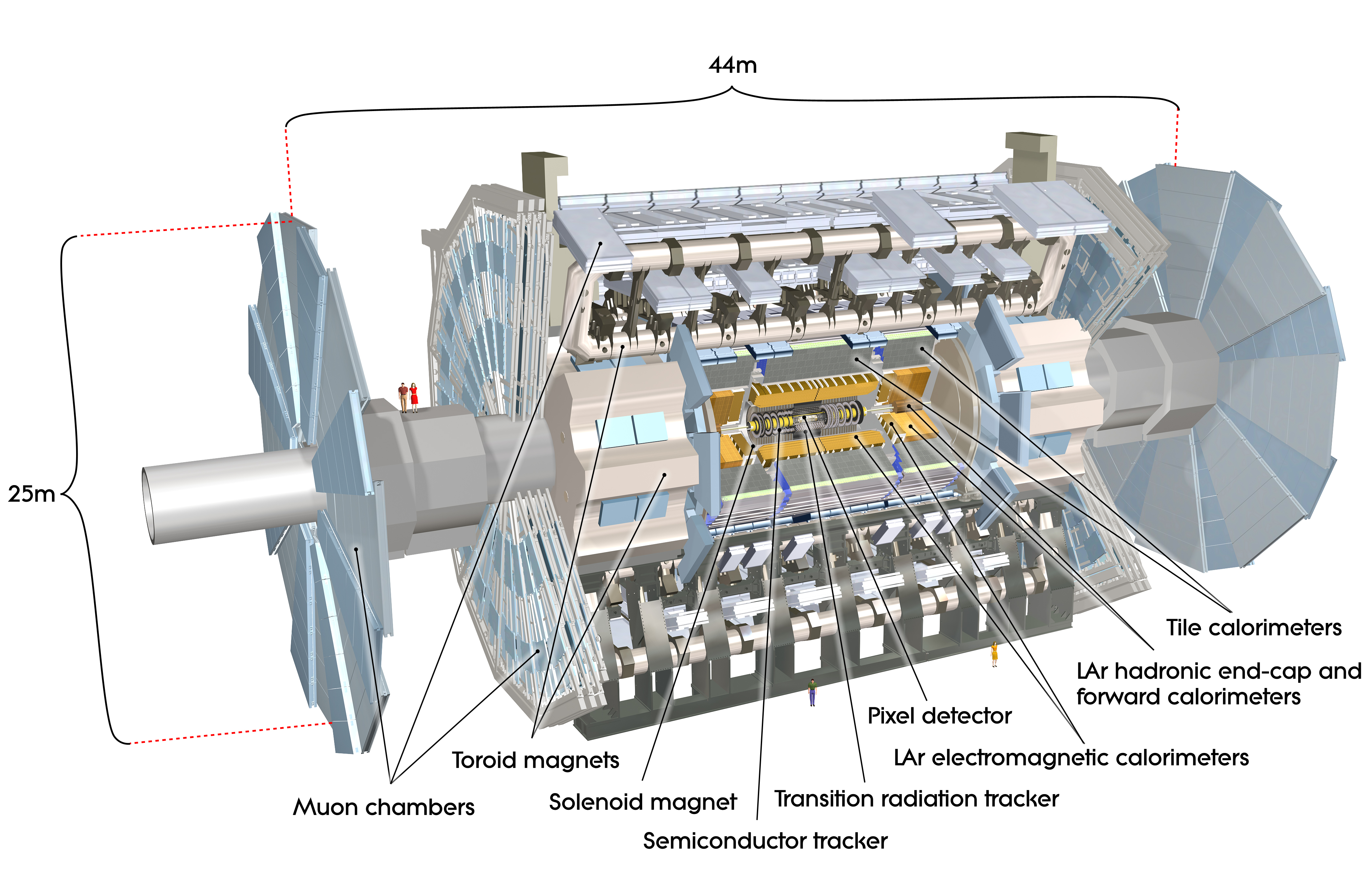}
  \caption{Computer generated image of the entire ATLAS detector.}
  \label{fig:atlas}
\end{figure}

The ATLAS experiment \cite{Collaboration_2008, CERN-LHCC-2012-009} is a versatile detector that features a forward-backward symmetric cylindrical geometry and nearly complete solid angle coverage of 4$\pi$.
As shown in Fig.~\ref{fig:atlas}, the ATLAS detector consists of an inner tracking detector, a thin superconducting solenoid, electromagnetic and hadronic calorimeters, and a muon spectrometer that includes three large superconducting toroidal magnet systems.

The inner-detector system (ID) is subject to a 2 T axial magnetic field that bends charged particles in the $r-\phi$ plane and provides tracking capabilities within the pseudorapidity range $|\eta| < 2.5$.
The high-granularity silicon pixel detector covers the vertex region and typically provides four position measurements (hits) per track, with the first hit usually in the insertable B-layer installed before Run 2.
It is followed by the silicon microstrip tracker (SCT), which usually provides eight measurements per track.
These silicon detectors are complemented by the transition radiation tracker (TRT), which enables radially extended track reconstruction up to $|\eta| = 2.0$.

The calorimeter system has approximately 188,000 cells and covers the range $|\eta| < 4.9$.
Within the region $|\eta| < 3.2$, electromagnetic calorimetry is provided by barrel and endcap high-granularity lead/liquid-argon (LAr) sampling calorimeters (ECAL), with an additional thin LAr presampler covering $|\eta| < 1.8$ to correct for energy loss in material upstream of the calorimeters.
The ECAL has a depth between 24 and 27 radiation lengths ($X_0$), and its granularity in the barrel in terms of $\Delta\eta \times \Delta\phi$ is typically $0.025 \times \pi/128$, with variations in segmentation with layer and $|\eta|$ as described in Ref.~\cite{ATLAS:2016krp}.

Hadronic calorimetry is provided by the steel/scintillator-tile calorimeter (HCAL), which is segmented into three barrel structures within $|\eta| < 1.7$, and two copper/LAr hadronic endcap calorimeters.
The solid angle coverage is completed with forward copper/LAr and tungsten/LAr calorimeter modules (FCAL) optimized for electromagnetic (FCAL1) and hadronic (FCAL2 and FCAL3) measurements, respectively.
The combined depth of the calorimeters for hadronic energy measurements is more than 10 nuclear interaction lengths nearly everywhere across the full detector acceptance ($|\eta| < 4.9$).
The granularity is as fine as $0.1 \times \pi/32$, again with variations in segmentation with layer and $|\eta|$ as described in Ref.~\cite{ATLAS:2016krp}.

The muon spectrometer (MS) includes two types of chambers: trigger and high-precision tracking chambers.
These chambers measure the deflection of muons in the $r$-$z$ plane due to a magnetic field produced by superconducting air-core toroids.
The field integral of the toroids ranges from 2.0 to 6.0 $\mathrm{T} \cdot \mathrm{m}$ throughout most of the detector.
Precision chambers cover the region $|\eta| < 2.7$ with three stations of monitored drift tube (MDT) chambers.
In the $|\eta| > 2.0$ region, where the background is higher, the innermost MDT station is replaced with cathode-strip chambers (CSCs).
Each MDT chamber provides six to eight $\eta$ measurements along the muon track, while the CSCs provide four simultaneous measurements of $\eta$ and $\phi$.
The nominal single-hit resolution of the MDTs and CSCs is about 80 $\mathrm{\mu m}$ and 60 $\mathrm{\mu m}$, respectively, in the bending plane.
The chambers are accurately aligned using optical sensors to achieve a 10\% transverse momentum resolution for 1 TeV muons.
The muon trigger system covers the range $|\eta| < 2.4$ with resistive-plate chambers (RPCs) in the barrel, consisting of three doublet stations for $|\eta| < 1.05$, and thin-gap chambers (TGCs) in the endcap regions, consisting of one triplet station followed by two doublets for $1.0 < |\eta| < 2.4$.
The RPCs and TGCs provide tracking information complementary to the precision chambers, especially improving the determination of the track coordinate in the non-bending direction, referred to as the second coordinate.
The typical spatial resolution for the position measurements in the RPCs and TGCs is 5-10 mm in both the bending plane and non-bending direction.

To select interesting events, a two-level trigger system \cite{ATLAS:2016wtr} is used, consisting of a hardware-based first-level (Level-1, L1) and a software-based high-level trigger (HLT).
The L1 trigger accepts events from the 40 MHz bunch crossings at a rate below 100 kHz, which the HLT reduces in order to record events to disk at about 1 kHz.

\chapter{Data and simulated event samples}
\label{chap:sample}

\section{Data samples}

Both the \Hmm\ and \monoHbb\ analyses utilize $pp$ collision data collected by the ATLAS detector at $\sqrt{s} = 13$ TeV from 2015 to 2018.
To ensure the proper functioning of all components of the ATLAS detector, basic data quality requirements were imposed on all events.
The full dataset corresponds to an integrated luminosity of 139 fb$^{-1}$, with an uncertainty of 1.7\%.
For the \Hmm analysis, events must satisfy a combination of single-muon triggers with transverse momentum (\pT) thresholds up to 26 GeV for isolated muons and 50 GeV for muons without any isolation requirement imposed, allowing for the recovery of some inefficiency introduced by the isolation requirement at the trigger level for high momentum muons.
For the \monoHbb analysis, events must pass the most efficient available \MET\ trigger, which reaches full efficiency by an offline \MET\ value of approximately 200 GeV.
In addition, \monoHbb\ also employs events passing single-electron or single-muon triggers \cite{ATLAS:2020gty,ATLAS:2019dpa} to construct 2-lepton control regions (described in section~\ref{chap:monoHbb:sec:categorization:ssec:cr}).

\afterpage{\clearpage}

\section{Simulated event samples}

Simulated Monte Carlo (MC) samples are generated to optimize event selections and model signal and background processes.
The simulation procedure involves generating parton-level events, processing them for parton showering, hadronization, underlying events, etc., and processing them through the ATLAS detector simulation.
For all samples, the detector simulation is based on \textsc{Geant}4 \cite{ATLAS:2010arf, AGOSTINELLI2003250} (referred to as full-simulation samples), except for an additional sample for the Drell-Yan process ($Z/\gamma^{*}\to\mu\mu$), where experimental effects are approximated using parametrizations (referred to as fast-simulation samples) to speed up the process.

The simulation also includes the effects of multiple $pp$ collisions in the same or neighboring bunch crossings (pile-up) by overlaying inelastic $pp$ interactions produced using \textsc{Pythia} 8 \cite{Sjostrand:2014zea} with the NNLPDF2.3LO set of parton distribution functions (PDFs) \cite{Ball:2012cx} and the A3 set of tuned parameters \cite{ATL-PHYS-PUB-2016-017}.
Events are reweighted to match the distribution of the average number of interactions per bunch crossing observed in data.
Simulated events are corrected to reflect the momentum scales and resolutions, as well as the triggers, reconstruction, identification, and isolation efficiencies measured in data for all physics objects used in the analyses.

\subsection{Simulated event samples for $H\to\mu\mu$}

The signal for the $H\to\mu\mu$ analysis is the \Hmm\ process. The \Hmm\ samples are generated for the main Higgs boson production modes, including $gg$F, VBF, $VH$ and $t\bar{t}H$.
The Higgs boson mass is set to 125 GeV and the corresponding decay width is $\Gamma_H = 4.07$ MeV \cite{LHCHiggsCrossSectionWorkingGroup:2013rie}.
The samples are normalized using the latest available theoretical calculations of the corresponding SM production cross sections \cite{LHCHiggsCrossSectionWorkingGroup:2016ypw}, as well as the $H\to\mu\mu$ branching ratio of $2.17 \times 10^{-4}$ calculated with HDECAY \cite{Djouadi:1997yw,Spira:1997dg,Djouadi:2006bz,DJOUADI2019214} and PROPHECY4F \cite{Bredenstein:2006ha,Bredenstein:2006rh,Bredenstein:2006nk}.
The $gg$F sample is simulated using the \textsc{Powheg} NNLOPS program \cite{Nason:2004rx,Frixione:2007vw,Alioli:2010xd,Campbell:2012am,Hamilton:2012np,Hamilton:2012rf,Hamilton:2013fea,Hamilton:2015nsa} with the PDF4LHC15 set of PDFs \cite{Butterworth:2015oua}.
The simulation achieves next-to-next-to-leading-order (NNLO) accuracy in QCD for inclusive observables after reweighting the Higgs boson rapidity spectrum \cite{Catani:2007vq}.
The parton-level events are processed by \textsc{Pythia} 8 for the Higgs decay, parton showering, final-state radiation (QED FSR), hadronization, and underlying event using the AZNLO set of tuned parameters \cite{ATLAS:2014alx}.
The sample is normalized to a next-to-next-to-next-to-leading-order QCD calculation with next-to-leading-order (NLO) electroweak corrections \cite{Aglietti:2004nj,Actis:2008ug,Actis:2008ts,Anastasiou:2008tj,Pak:2009dg,Harlander:2009bw,Harlander:2009mq,Harlander:2009my,Anastasiou:2015vya,Anastasiou:2016cez,Dulat:2018rbf,Bonetti:2018ukf}.
The VBF and $q\bar{q}/qg \to VH$ samples are generated at NLO accuracy in QCD using the \textsc{Powheg}-\textsc{Box} program \cite{Nason:2009ai,Cullen:2011ac,Luisoni:2013cuh}. The loop-induced $gg\to ZH$ sample is generated at leading order (LO) using \textsc{Powheg}-\textsc{Box}.
The same settings for the PDF set and \textsc{Python} 8 as for the \ggF\ sample are adopted for the VBF and V$H$ samples.
The VBF sample is then normalized to an approximate-NNLO QCD cross-section with NLO EW corrections \cite{Ciccolini:2007jr,Ciccolini:2007ec,Bolzoni:2010xr}.
The \VH\ samples are normalized to NNLO QCD cross-section with NLO electroweak corrections for $q\bar{q}/qg \to VH$ and NLO + next-to-leading-logarithm (NNLL) accuracy QCD for $gg\to ZH$ \cite{Ciccolini:2003jy,Brein:2003wg,Brein:2011vx,Altenkamp:2012sx,Denner:2014cla,Brein:2012ne,Harlander:2014wda,Harlander:2018yio}.
The $t\bar{t}H$ sample is generated by \textsc{Mad}\textsc{Graph}5\_aMC@NLO \cite{Alwall:2014hca,Artoisenet:2012st} at NLO accuracy in QCD with the NNPDF3.0NLO PDF set \cite{NNPDF:2014otw} and interfaced to \textsc{Pythia} 8 using the A14 set of tuned parameters \cite{ATL-PHYS-PUB-2014-021}.
The sample is normalized to NLO QCD cross-section with NLO EW corrections \cite{Beenakker:2002nc,Dawson:2003zu,Zhang:2014gcy,Frixione:2015zaa}.

The background processes comprise Drell-Yan (DY) $Z/\gamma^{*}\to\mu\mu$ process, diboson ($WW$, $WZ$ and $ZZ$) process, $t\bar{t}$, single-top-quark productions, as well as $t\bar{t}V$ process.
The full-simulated DY and diboson samples are simulated by \textsc{Sherpa} 2.2.1 \cite{Sherpa:2019gpd}.
Specifically, the DY events are produced using NLO-accurate matrix elements for up to two partons and LO-accurate matrix elements for up to four partons calculated with the Comix \cite{Gleisberg:2008fv} and OpenLoops \cite{Cascioli:2011va,Denner:2016kdg} libraries and the NNPDF3.0 NNLO set.
They are matched to the \textsc{Sherpa} parton shower \cite{Schumann:2007mg} using the MEPS@NLO prescription \cite{Hoeche:2011fd,Hoeche:2012yf,Catani:2001cc,Hoeche:2009rj}.
The $t\bar{t}$ and single-top-quark samples are generated at NLO accuracy with \textsc{Powheg}-\textsc{Box} \cite{Frixione:2007nw,Re:2010bp} using the NNPDF3.0NLO PDF set interfaced to \textsc{Pythia} 8 for parton showering and hadronization using the A14 parameter set.
Additionally, the $Wt$ process undergoes the diagram removal scheme \cite{Frixione:2008yi} to remove the overlap with the $t\bar{t}$ production.
The $t\bar{t}V$ sample is generated using \textsc{Mad}\textsc{Graph}5\_aMC@NLO \cite{Alwall:2014hca} at NLO in a set-up similar to the $t\bar{t}H$ sample.

To provide enough statistics for background modeling with low statistical uncertainties, an additional fast-simulated DY samples is produced.
This sample is generated using \textsc{Sherpa} 2.2.4 \cite{Hoche:2019flt} with LO matrix elements and up to three additional partons.
It uses the CT14 NNLO PDF set \cite{Dulat:2015mca}.
The parton-level events are then processed with \textsc{Pythia} 8 for QED and QCD parton showering and hadronization.
The CKKW-L algorithm \cite{Lonnblad:2001iq} is used to remove the double-counted QCD emissions with a merging scale of 20 GeV.
The experimental effects are parametrized based on the full-simulation samples or directly from ATLAS data.
This reproduces the reconstruction and selection efficiencies of detector-level objects using event weighting and models the resolution of the ATLAS detector with predetermined probability distributions.
Detailed descriptions are used for the muon momentum resolution and muon trigger and selection efficiencies, photons from QED FSR, hadronic jets from primary interaction, pile-up events in terms of kinematics and the number of associated ID tracks, and for the effect of pile-up and the underlying event on the measurement of the missing transverse momentum $E^\mathrm{miss}_\mathrm{T}$.

\subsection{Simulated event samples for mono-$h$($b\bar{b}$)}

The signal for the \monoHbb\ analysis comprises \zpthdm\ and \thdma. Both models are simulated using \textsc{Mad}\textsc{Graph}5\_aMC@NLO v2.6.5 \cite{Alwall:2014hca} at LO in QCD, interfaced with \textsc{Pythia} 8 using the A14 set of tuned parameters.
\zpthdm\ and the \ggF\ production for \thdma\ are simulated in the 5-flavor scheme, while the \bbA\ production of \thdma\ is simulated in the 4-flavor scheme.
Multiple parameters values are used for the parameter scans of the interpretations.

The background consists of single vector-boson productions ($Z$ and $W$), diboson productions, \ttbar, single top-quark, $t\bar{t}V$, \ttH\ and \VH production.
The $Z$ and $W$ samples are generated using \textsc{Sherpa} 2.2.1 with the same settings as for the DY sample used for the $H\to\mu\mu$ analysis.
The diboson samples are simulated using \textsc{Sherpa} 2.2.2 with NLO-accurate matrix elements for up to one parton and LO-accurate matrix elements for up to three partons for the $q\bar{q}$-initiated process and LO-accurate matrix elements for up to one parton for the $gg$-initiated process.
$Z$, $W$ and diboson samples are matched with the \textsc{Sherpa} parton shower using the MEPS@NLO prescription.
The \ttbar, single top-quark and \ttH\ samples are generated using \textsc{Powheg}\textsc{Box} with the same settings as in \Hmm\ analysis.
The \VH\ samples are generated by \textsc{Powheg}\textsc{Box} with the same settings as $H\to\mu\mu$, and the $t\bar{t}V$ sample is generated using \textsc{Mad}\textsc{Graph}5\_aMC@NLO at NLO with the same settings as in $H\to\mu\mu$.
All samples are interfaced with \textsc{Pythia} 8 for the parton shower and hadronization with the same corresponding settings as used for $H\to\mu\mu$.

\clearpage

\chapter{Physics object definitions}
\label{chap:object}

In this chapter, the physics objects utilized to characterize and select events for the two analyses will be described.
These physics objects are reconstructed from the ID tracks, calorimeter energy deposits, and MS tracks for each event.

\section{Primary vertex}

To enhance the quality of the event reconstruction, the vertices must meet the requirement of having at least two ID tracks with \pT\ greater than 500 MeV.
The primary vertex of the hard interaction is identified as the vertex with the highest sum of $\pT^2$ of tracks.

\section{Jets}

Different event topologies require various approaches to reconstruct jets using the anti-kt algorithm \cite{Cacciari:2008gp,Cacciari:2011ma}.
The reconstruction methods include small-radius (small-$R$) jets for both the \Hmm\ and \monoHbb\ analyses, large-radius (large-$R$), and variable-radius (variable-$R$) track-jets for \monoHbb.

Small-$R$ jets are reconstructed with $R = 0.4$ from `particle flow' objects formed by combining ID tracks and calorimeter energy clusters \cite{ATLAS:2017ghe}.
These jets must have $|\eta| < 4.5$ and $\pT > 25\ \GeV$ (for \Hmm) or $\pT > 20\ \GeV$ (for \monoHbb) for $|\eta| < 2.5$ and $\pT > 30\ \GeV$ for $2.5 < |\eta| < 4.5$.
For small-$R$ jets with $|\eta| < 2.5$ and $\pT < 60\ \GeV$, the Jet Vertex Tagger (JVT) algorithm \cite{ATLAS:2015ull} is used to confirm their origin from the primary vertex, by computing a multivariate likelihood from tracking information.

Large-$R$ jets are reconstructed with $R = 1.0$ from calorimeter energy clusters calibrated using the local hadronic cell weighting (LCW) scheme \cite{ATLAS:2016krp}.
This choice of $R$ captures all jets produced in the decay of a boosted heavy object, such as a Higgs boson.
To mitigate the impact of pile-up, large-$R$ jets are trimmed by removing any $R = 0.2$ subjets that have less than 5\% of the original jet energy \cite{Krohn:2009th}.

Variable-$R$ track-jets are used to identify subjets that originate from $b$-hadrons within large-$R$ jets.
These are reconstructed from ID tracks using an $R$ that shrinks as the proto-jet's \pT\ increases \cite{Krohn:2009zg} and are matched to the large-$R$ jets by ghost association \cite{Cacciari:2008gn}.
The $R$ of variable-$R$ track-jets is set to $R = 30\ \GeV/\pT$, with minimum and maximum values of 0.02 and 0.4, respectively.
The algorithm can reconstruct separate jets from closely spaced $b$-hadrons, such as in highly boosted \Hbb\ decays.

Central small-$R$ jets ($|\eta| < 2.5$) and variable-$R$ track-jets containing $b$-hadrons ($b$-tagged jets) are identified using the `MV2c10` \cite{ATLAS:2018sgt,ATLAS:2019bwq} (used in \Hmm\ for central small-$R$ jets) or `DL1' tagger (used in \monoHbb\ for both central small-$R$ jets and variable-$R$ track-jets) \cite{ATL-PHYS-PUB-2017-013}.
In \monoHbb, the $b$-tagging working point (WP), which corresponds to a 77\% efficiency (77\% WP) in \ttbar\ events, is used for both central small-$R$ jets and variable-$R$ track-jets.
In \Hmm, two $b$-tagging WPs, which correspond to a 60\% (60\% WP) and a 85\% (85\% WP) efficiency in $t\bar{t}$ events, are used for central small-$R$ jets.

Since the decays of $b$-hadrons can produce muons, which are vetoed when building particle-flow objects and excluded from the energies of either the small-$R$ or large-$R$ jets, in \monoHbb\ the four-momenta of non-isolated muons falling inside the b-tagged jet cones can be added to correct for the total jet momentum.
The correction is done for the muon (two muons) closeset to the jet axis of small-$R$ (large-$R$) jets.
This correction is used in \monoHbb\ to calculate the mass of the Higgs boson candidate (\mh)

\section{Leptons}

Muon reconstruction is performed by combining tracks in the ID and MS.
For \Hmm, additional muon candidates are considered to improve the efficiency.
In the region of $|\eta| < 0.1$, muon candidates are identified by matching a reconstructed ID track to either an MS track segment or a calorimetric energy deposit consistent with a minimum-ionizing particle.
In the region $2.5 < |\eta| < 2.7$, where the ID does not cover, additional muons are reconstructed from an MS track with hits in the three MS layers and combined with forward ID hits.

Muon candidates must satisfy the `Loose' or `Medium' criteria defined in Ref.~\cite{ATLAS:2016lqx}, depending on the analysis.
The criteria require a transverse momentum $\pT > 6\ \GeV$ and $|\eta| < 2.7$.
Muons with an associated ID track must also be matched to the primary vertex with $z_0 \sin\theta < 0.5$ mm and $|d_0|/\sigma\left(d_0\right) < 3$, where $z_0$ is the longitudinal impact parameter, $\theta$ is the polar angle of the track, $d_0$ is the transverse impact parameter calculated relative to the measured beam-line position, and $\sigma\left(d_0\right)$ is the uncertainty on $d_0$.
Isolation criteria \cite{ATLAS:2020rej} are imposed to suppress non-prompt muons produced from hadron decays.
The isolation criteria utilize ID track and calorimeter energy deposit information in a range of $\Delta R < 0.2$ around the muon.

Muons can lose a significant amount of energy through QED final-state-radiation (FSR), which can reduce the signal dimuon mass and deteriorate the $m_{\mu\mu}$ resolution.
To improve the resolution of the signal kinematic spectrum for \Hmm, up to one final-state photon candidate is included as part of the dimuon system for each event.
Photon candidates are reconstructed \cite{ATLAS:2020rej} from energy cluster found in the EM calorimeter with the requirement that $f_1 > 0.2$, where $f_1$ is the energy fraction of the cluster in the first layer of the EM calorimeter.
Additionally, only photons with $\Delta R \left(\gamma, \mu\right) < 0.2$ are considered. To reduce background from pile-up interactions, photons are also required to pass a $p_\mathrm{T}^\gamma$ threshold, which increases linearly from 3 GeV at $\Delta R = 0$ to 8 GeV at $\Delta R = 0.2$.
If there are more than one photon candidates, the photon with the highest $p_\mathrm{T}^\gamma$ is selected.
About 5\% of the events include a QED FSR photon, and the total width of the signal $m_{\mu\mu}$ spectrum is reduced by approximately 3\%.

Electrons are reconstructed by matching clusters of energy in the EM calorimeter to ID tracks.
The identification criteria for electrons include the `Medium' criterion used in \Hmm\ and the `Loose` criterion used in \monoHbb\ \cite{ATLAS:2019qmc}.
Additionally, electrons must have $p_\mathrm{T} > 7$ GeV (used in both \Hmm\ and \monoHbb), or 27 GeV (used only in \monoHbb), $|\eta| < 2.47$ and not $1.37 < |\eta| < 1.52$.
Like muons, electrons must be isolated and be matched to the primary vertex with $z_0 \sin\theta < 0.5$ mm and $|d_0|/\sigma\left(d_0\right) < 5$.

The reconstruction of hadronically decaying $\tau$-leptons is initiated from small-$R$ jets created using LCW-calibrated clusters \cite{ATL-PHYS-PUB-2015-045}.
As hadronic $\tau$-lepton decays can produce either one or three charged pions, the jets are required to have one or three tracks within $\Delta R = 0.2$ of the jet axis.
A recurrent neural network (RNN) classifier is employed to identify the $\tau$-lepton \cite{ATL-PHYS-PUB-2019-033}, with inputs constructed from the clusters and tracks associated with the $\tau$-lepton.
All $\tau$-leptons are required to pass the `VeryLoose` WP \cite{ATL-PHYS-PUB-2019-033} and have $|\eta| < 2.4$ and $p_\mathrm{T} > 20$ GeV.

\section{Missing transverse momentum}

Particles that escape from the detector, such as neutrinos and Dark Matter, result in missing transverse momentum \MET.
\MET\ is the magnitude of the negative vector sum of the transverse momenta of all physics objects, including muons, electrons, and jets, as well as the ID tracks not linked to any physics objects (soft term).

\section{Overlap removal}

To prevent the same detector signals from being interpreted as different objects, an overlap removal procedure is implemented.
If any object is rejected at one stage, it is not considered in subsequent stages.
The procedure proceeds as follows: first, if two electrons share a track, the electron with lower \pT\ is discarded.
Next, any $\tau$-lepton within $\Delta R = 0.2$ of an electron or muon is removed.
Then, any electron that shares a track with a muon is removed.
Any small-$R$ jet is within $\Delta R = 0.2$ of an electron is then discarded, followed by any electron within a cone of \pT-dependent size around a small-$R$ jet.
If a small-$R$ jet with less than three tracks is linked to a muon or is within $\Delta R = 0.2$ of one, it is removed, and any muon within a cone of \pT-dependent size around a small-$R$ jet is discarded.
Next, any small-$R$ jet within $\Delta R = 0.2$ of a $\tau$-lepton is removed.
Finally, any large-$R$ jet within $\Delta R = 1.0$ of an electron is removed. Track-jets are not included in the overlap removal process as they are soely employed for $b$-tagging.

\clearpage

\chapter{Search for \Hmm}
\label{chap:hmumu}

\section{Event selections and categorization}
\label{chap:hmumu:sec:categorization}

\begin{figure}[h!]
  \centering
  \includegraphics[width=0.6\textwidth]{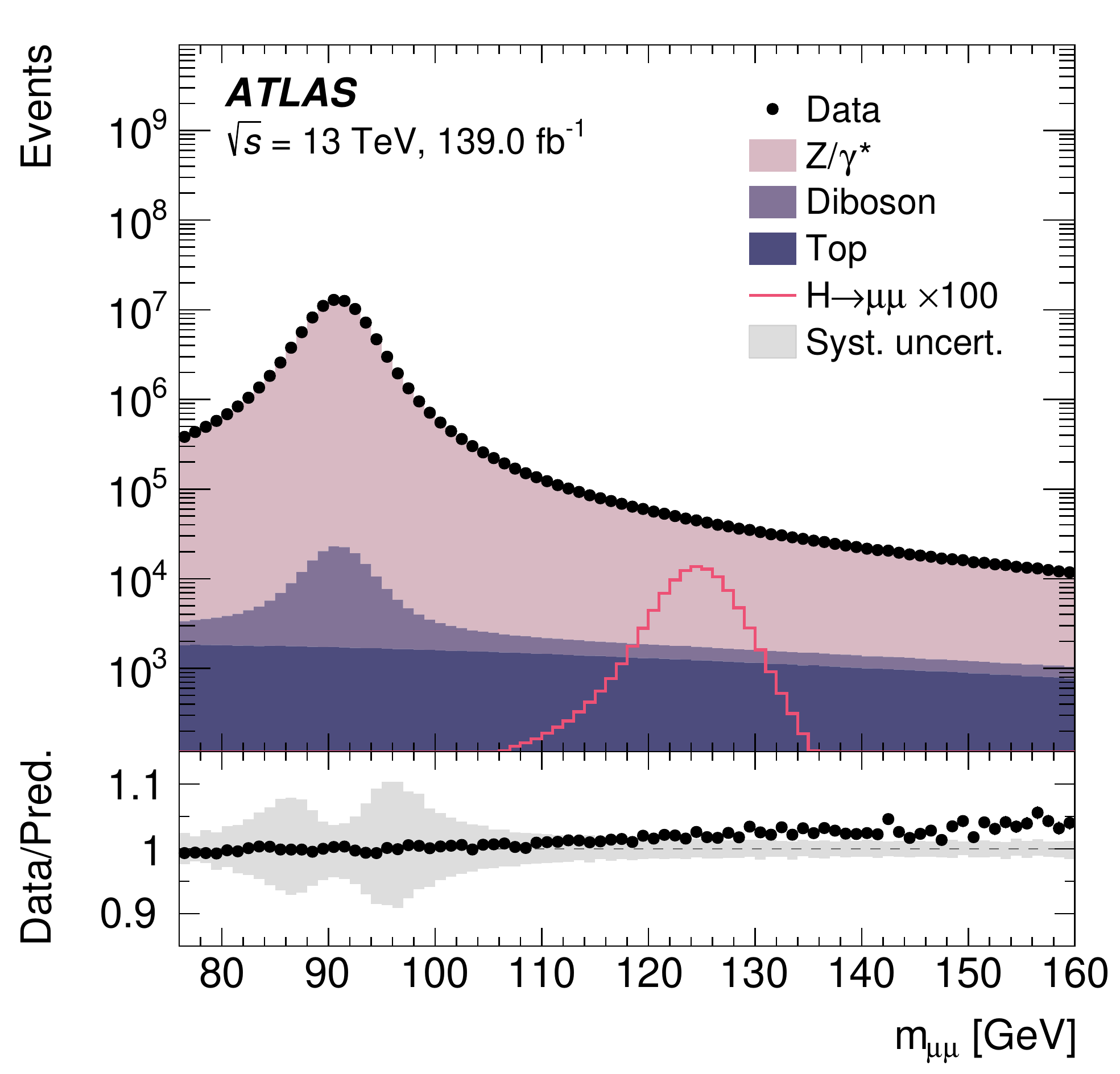}
  \caption{Dimuon invariant mass \mmumu\ in the range 76–160 GeV for all events passing the event selections. Data in points are compared to a full set of fully simulated background processes and the total background prediction is scaled to the integrated data yield. The shaded region in the bottom panel shows the impact of the systematic uncertainty on the muon momentum scale, resolution and muon trigger and reconstruction efficiencies. The \Hmm\ signal shown is the sum of the ggF, VBF, VH and ttH as open line, normalised to one hundred times the SM prediction for visibility.}
  \label{fig:hmm_inc}
\end{figure}

To select events in the interesting kinematic phase space for \Hmm\ signals, a number of event selections are applied.
Firstly, events must contain a primary vertex and at least a pair of opposite-charge muons.
The leading muon must have $\pT > 27\ \GeV$, while the subleading muon must have $\pT > 15\ \GeV$ in all categories except for the \VH\ 3-lepton category (defined in the following subsections).
Further selections are required in regions targeting different production modes, as detailed in the following subsections.
Tab.~\ref{tab:hmumu_evtsel} summarizes the main event selections, and Fig.~\ref{fig:hmm_inc} shows the distribution of selected events in the \mmumu\ spectrum.
The final statistical fitting is performed in the range of $110 < \mmumu < 160$ GeV, where \mmumu\ is the invariant mass of the dimuon system.
Approximately 450,000 data events are selected in this \mmumu\ range, and the mass window $120 < \mmumu < 130$ GeV is expected to contain about 868 \Hmm\ signal events, which corresponds to about 85\% of the total selected signals.
The total efficiency times acceptance of about 52\% is with respect to all \Hmm\ with \ggF, \VBF, \VH, and \ttH\ productions.

\begin{table}[h!]
  \begin{center}
    \caption{Summary of the main event selection criteria applied to
      all events as well as the the selection of jets.
      The bottom sections give the basic requirements on leptons and $b$-tagged
      jets for the analysis categories targeting different Higgs boson production modes.
      The subleading muon momentum threshold is $15\,\GeV$ in all categories except the \VH\
      3-lepton categories, where it is lowered to $10\,\GeV$.}
    \label{tab:hmumu_evtsel}
    \small
    \begin{tabular}{lc}
      \toprule
      & Selection \\
      \midrule
      & Primary vertex\\
      Common preselection & Two opposite-charge muons\\
      & Muons: $|\eta| < 2.7$, $\pT^\mathrm{lead} > 27\GeV$, $\pT^\mathrm{sublead} > 15\GeV$ (except $VH$ 3-lepton)\\
      \midrule
      Fit Region & $110< \mmumu< 160\GeV$\\
      \midrule
      \multirow{2}{*}{Jets} & $\pT > 25\GeV$ and $|\eta|<2.4$\\
      & or with $\pT > 30\GeV$ and $2.4 < |\eta|< 4.5$ \\
      \midrule
      \ttH Category & at least one additional $e$ or $\mu$ with $\pT > 15\GeV$, at least one $b$-jet ($85\%$ WP)\\
      $VH$ 3-lepton Categories & $\pT^\mathrm{sublead} > 10\GeV$, one additional $e$ ($\mu$) with $\pT > 15(10)\GeV$, no $b$-jets ($85\%$ WP)\\
      $VH$ 4-lepton Category & at least two additional $e$ or $\mu$ with $\pT > 8, 6\GeV$, no $b$-jets ($85\%$ WP)\\
      \ggF+VBF Categories & no additional $\mu$, no $b$-jets ($60\%$ WP)\\
      \bottomrule
    \end{tabular}
  \end{center}
\end{table}

The selected events are classified into 20 categories.
The \ttH\ category targets the \ttH\ production mode of the \Hmm\ signals, three \VH\ categories target \VH, four \VBF\ categories target \VBF, and twelve \ggF\ categories target \ggF.
The categories are defined by various kinematic variables and boosted decision trees (BDT) using the XGBoost package.
The categorization procedure separates events into regions with different signal-to-background ratios, enhancing the total signal sensitivity.

The remainder of the section describes the general strategy of BDT training and optimization of the BDT score boundaries between each category, as well as details the categorization for \ttH, \VH, \VBF, and \ggF\ categories.

\subsection{General BDT training strategy}

The overall approach is to employ a BDT as a classifier to differentiate between the targeted signal and the background.
These BDTs take a reduced set of kinematic variables as input to minimize unnecessary systematic uncertainties while maintaining a comparable performance.

All samples are initially weighted by their cross-sections and then normalized such that the mean weights of the training signal and background samples are equal to 1.
To prevent bias from tuning on the test samples, we adopt a four-fold training method (outlined in Tab.~\ref{tab:4foldTraining}).
The signal and background samples are first divided into four sections based on the remainder of event numbers divided by 4.
In each fold of the training, two of the four sections (50\% of the full samples) are used for the training set, one of the four sections (25\% of the full samples) is utilized as the validation set, and the remaining section (25\% of the full samples) is employed for the test set.
The training, validation, and test sets are rotated among the four folds of the training.
The hyperparameters of the BDT are optimized by maximizing the ROC AUC on the validation set using Bayesian Optimization \cite{NIPS2012_05311655}.
After the training, the BDT for each fold is used to classify the corresponding test set.

\begin{table*}[h!]
\begin{center}
\begin{tabular}{ccccc}
\toprule
Event number \% 4	 &  0 & 1 & 2 & 3 \\
\midrule
Fold 0  	& Test set		&  Validation set	& Trainin set		& Training set	\\
Fold 1	& Training set	& Test set			& Validation set 	& Training set	\\
Fold 2	& Trainin set	& Training set		& Test set			& Validation set	\\
Fold 3	& Validation set	& Trainin set		& Training set		& Test set		\\
\bottomrule
\end{tabular}
\caption{Summary of four-fold training method.}
\label{tab:4foldTraining}
\end{center}
\end{table*}

To ensure a similar shape of the BDT distribution in each fold of the training, the BDT outputs in each fold are transformed separately such that the unweighted signal sample in each fold has a uniform BDT distribution.
The \textsc{QuantileTransformer} function of the scikit-learn package \cite{scikit-learn} is used to perform this transformation of the BDT outputs.

\subsection{General strategy of BDT boundary optimization}

The BDT boundaries are optimized simultaneously to maximize the total number counting significance, denoted as $Z_{\mathrm{tot}}$:
\begin{equation}
Z_{\mathrm{tot}} = \sqrt{\sum_i Z_i^2},
\end{equation}
where $Z_i$ is the number counting significance in each BDT category and is calculated as:
\begin{equation}
Z_i = \sqrt{2 \left( \left(s_i + b_i\right) \log\left(\frac{s_i+b_i}{b_i}\right) - s_i\right)},
\end{equation}
where $s_i$ is the signal yield inside the dimuon mass window, $120 \leq \mmumu \leq 130\ \mathrm{GeV}$, and $b_i$ the background yield inside the dimuon mass window extrapolated from the yield of the data sideband, $110 \leq \mmumu \leq 120\ \mathrm{GeV}$ or $130 \leq \mmumu \leq 180\ \mathrm{GeV}$, by a scale factor of $0.2723$.
The scale factor is estimated from the inclusive observed dataset by taking the ratio between the center (dimuon mass 120-130 GeV) and the sideband (dimuon mass 110-180 excluding 120-130 GeV) data in the inclusive region of the dimuon selections.
To ensure sufficient statistics for the fitting in the later stage, each BDT category requires at least five background events within the dimuon mass window $120 \leq \mmumu \leq 130\ \mathrm{GeV}$.

During the stage of optimizing the categorization, only the ``test set score'' is used.
Specifically, the BDT score of each event used at this stage is obtained from a BDT model that is completely independent of the BDT models that have taken that event as an input (training event).
Therefore, any bias from the training or hyperparameter tuning will not show up in the samples used for the optimization.
The same samples used for optimizing the BDT boundaries are also used for the final sensitivity evaluation.
A validation using a 4-fold categorization method has been performed and showed a negligible difference in the sensitivity.

\subsection{\ttH\ category}

The \ttH\ category aims to identify the \ttH\ production mode of the Higgs boson, where the \ttbar\ pair undergoes dileptonical or semileptonic decay.
To be inluded in this category, events must have at least 1 lepton (either $e$ or $\mu$) with $\pT > 15$ GeV, in addition to the opposite-charge muon pair, and at least 1 $b$-jet tagged at the 85\% WP.
The two highest-\pT\ opposite-charge muons are used to reconstruct the Higgs boson and to calculate the dimuon mass $m_{\mu\mu}$.

A BDT is trained using simulated $t\bar{t}H\to\mu\mu$ events as signal and simulated SM background processes as background, with a preselection of $100 < \mmumu < 200$ GeV.
The BDT uses 12 kinematic variables as input features, which are listed below:

\begin{itemize}
\item $\pT^{\mu\mu}$: transverse momentum of the reconstructed Higgs boson (dimuon system).
\item $\cos\theta^*$: cosine of the lepton decay angle in the Collins-Soper frame.
\item $\pT^{\ell_{1\left(2\right)}}$: transverse momenta of the additional leptons.
\item $n_{j_c}$: number of central jets ($|\eta| < 2.5$).
\item $n_b$: number of $b$-jets.
\item $H_\mathrm{T}$: scalar sum of the transverse momenta of all jets.
\item $m_{\mathrm{Lep-}t}$: transverse mass of the leptonic top-quark system, consisting of the third lepton, $E^\mathrm{miss}_\mathrm{T}$ and a $b$-jet. If an event contains more than two $b$-jets, the $b$-jet is selected to minimize the difference between the reconstructed value and the SM value for $m_t$ and $m_W$.
\item $m_{\mathrm{Lep-}W}$: transverse mass of the leptonic $W$-boson system, consisting of the third lepton and $E^\mathrm{miss}_\mathrm{T}$.
\item $m_{\mathrm{Had-}t}$: invariant mass of the hadronic top-quark system, consisting of a $b$-jet and two non-$b$-jets. If an event contains more than one possible combination of the hadronic top-quark system, the combination is chosen by minimizing the $\chi$-square differences between the reconstructed value and the SM value for $m_t$ and $m_W$, where the reconstructed $m_W$ value is calculated from the system composed of two non-$b$-jets.
\item $m_{\ell\ell}$: invariant mass of the two additional leptons, where the two leptons must have opposite signs and the same flavor.
\item $m_{\mu\mu_3}$: invariant mass of the system composed of the third-highest \pT\ muon and one of the two leading muons with opposite charge to the third-highest \pT\ muon.
\end{itemize}

\begin{figure}[h!]
  \centering
  \subfigure[]{\includegraphics[width=0.40\textwidth]{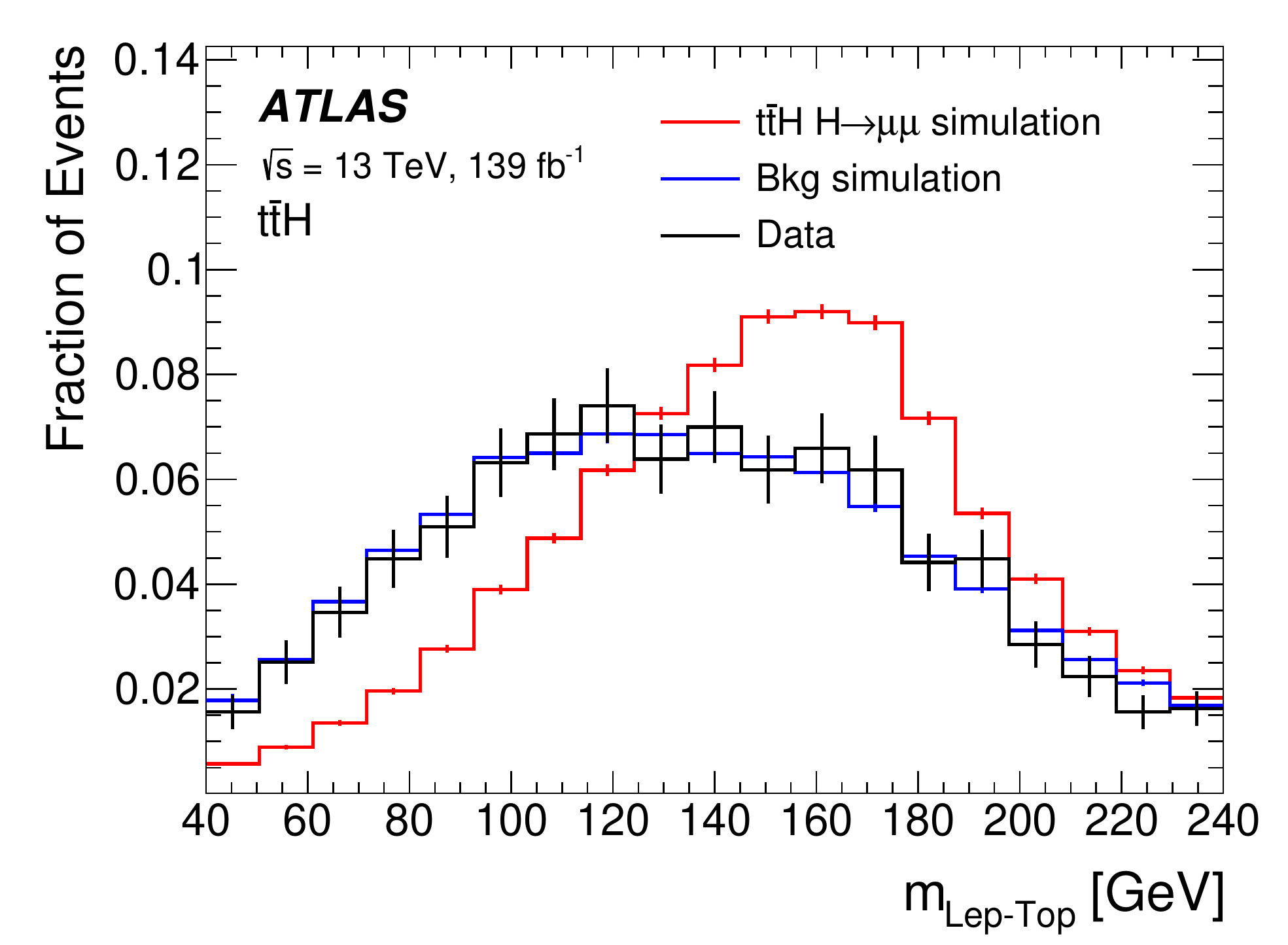}}
  \subfigure[]{\includegraphics[width=0.40\textwidth]{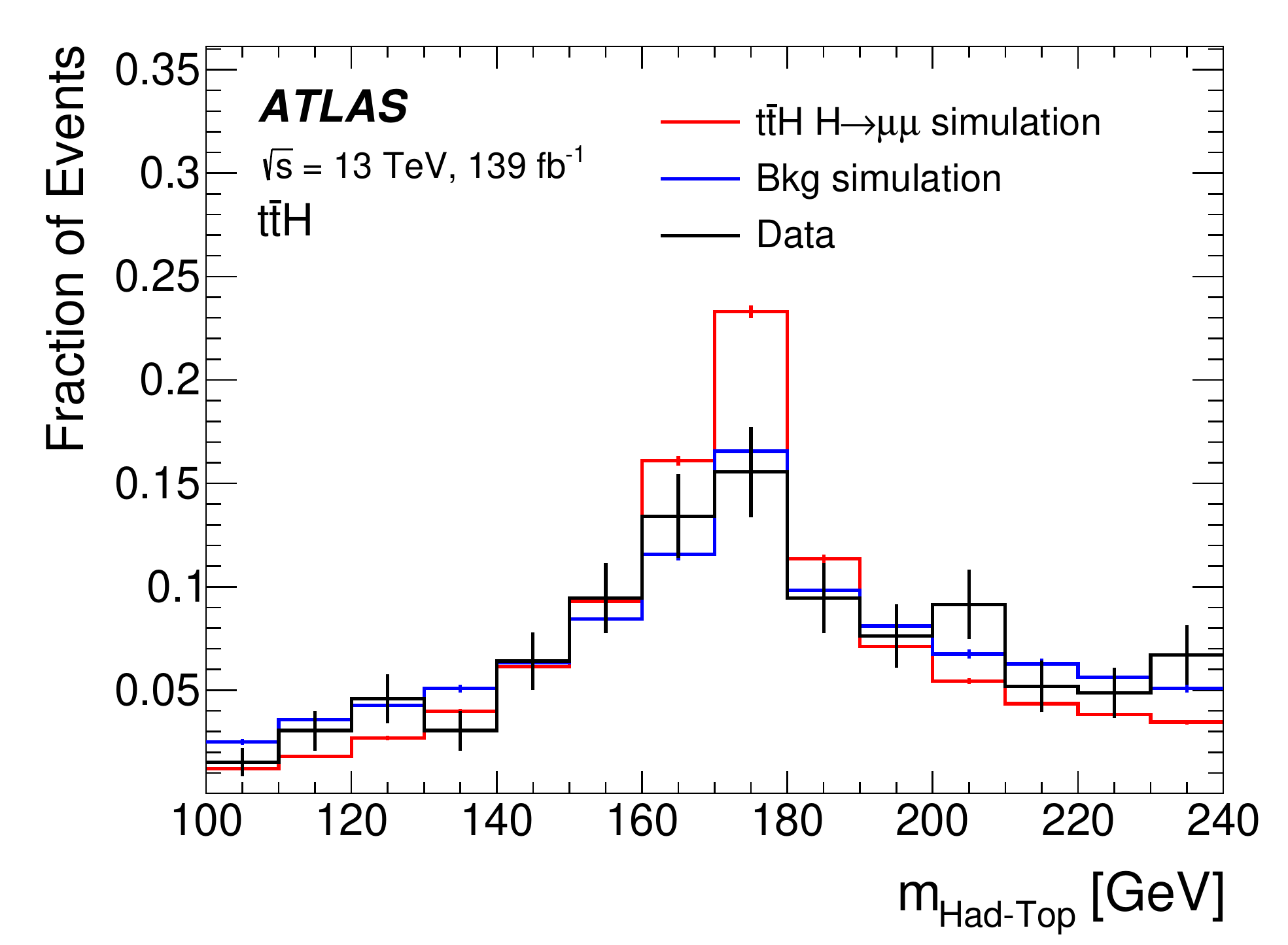}}
  \subfigure[]{\includegraphics[width=0.40\textwidth]{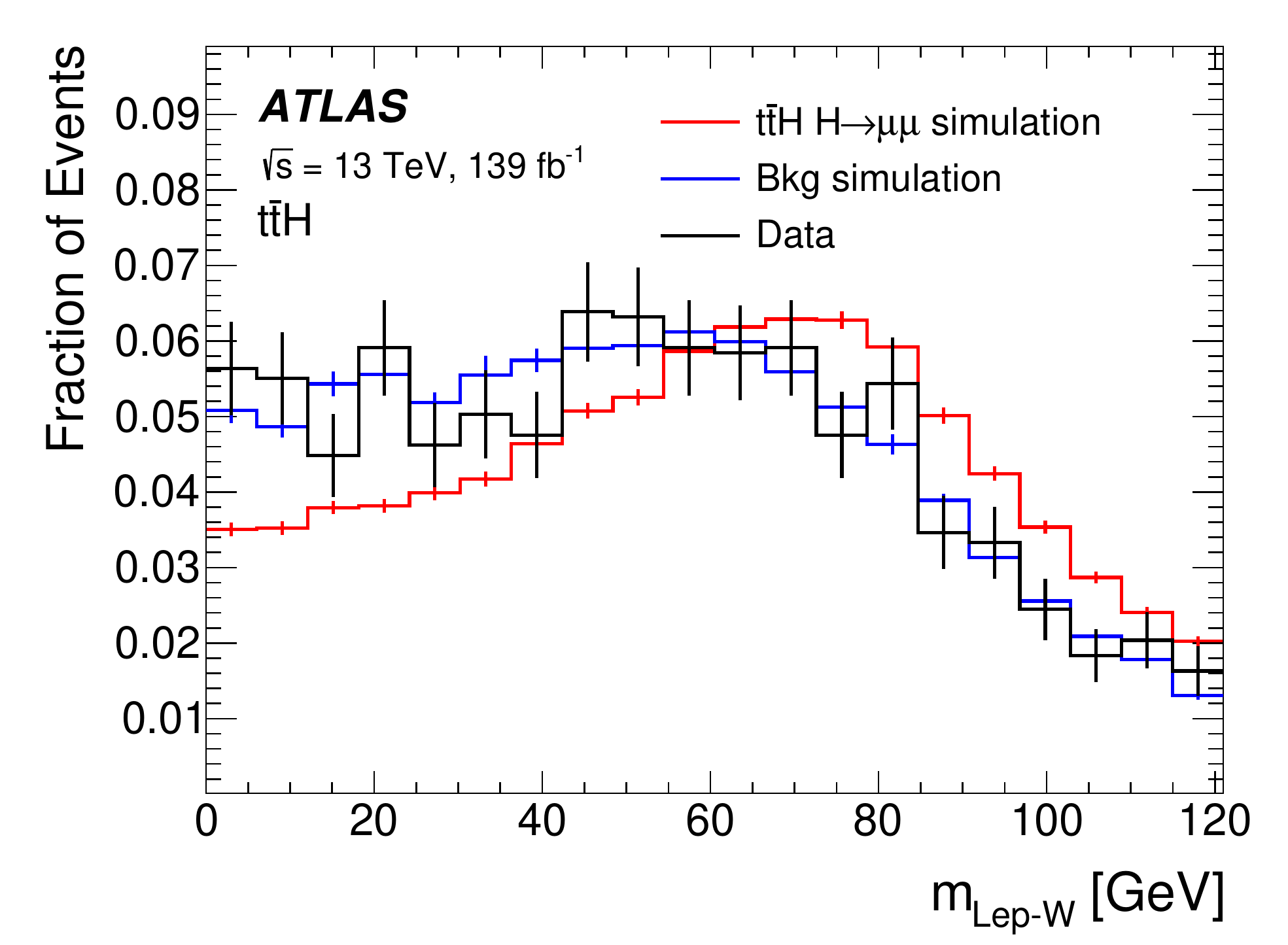}}
  \subfigure[]{\includegraphics[width=0.40\textwidth]{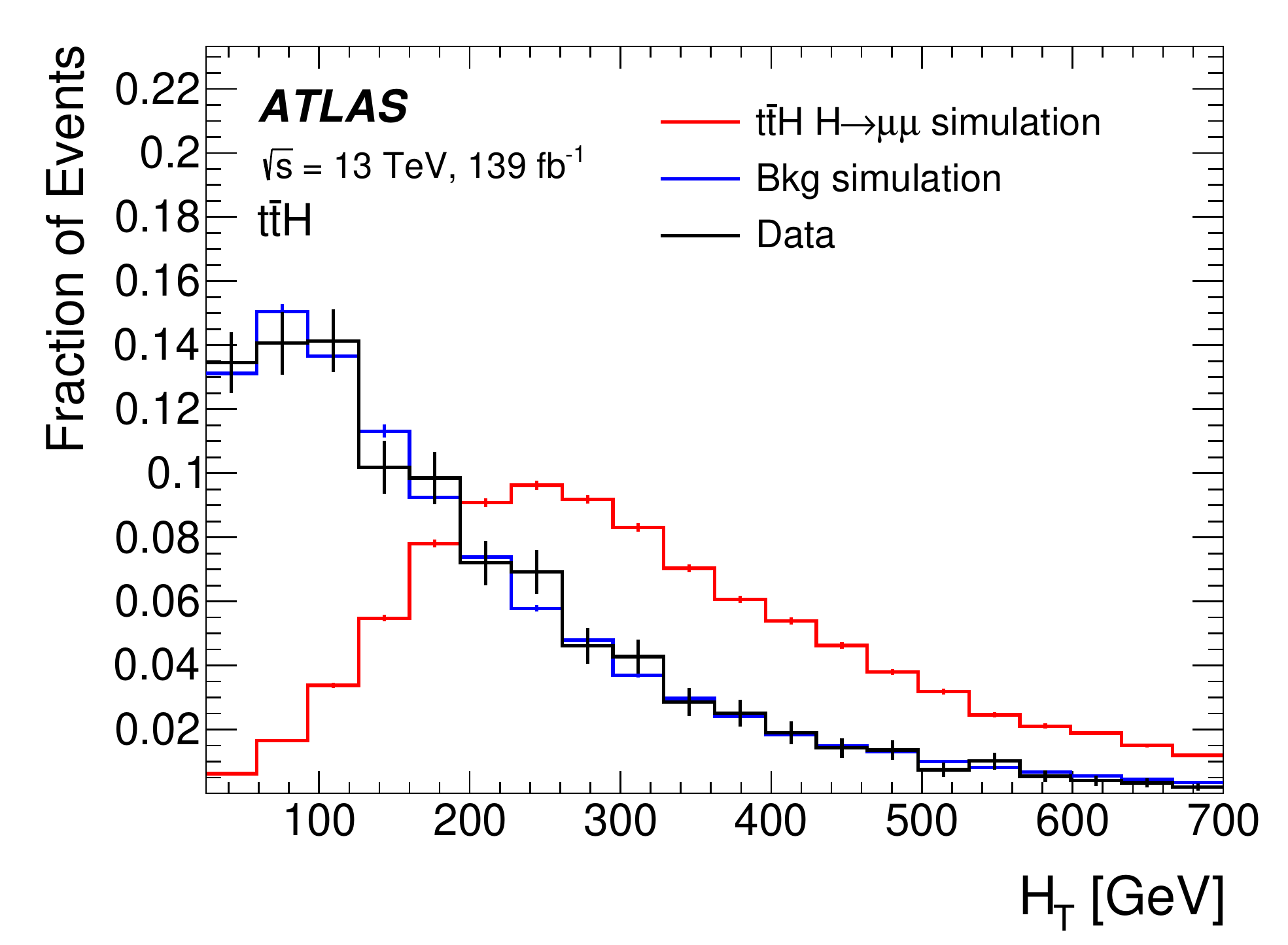}}
  \subfigure[]{\includegraphics[width=0.40\textwidth]{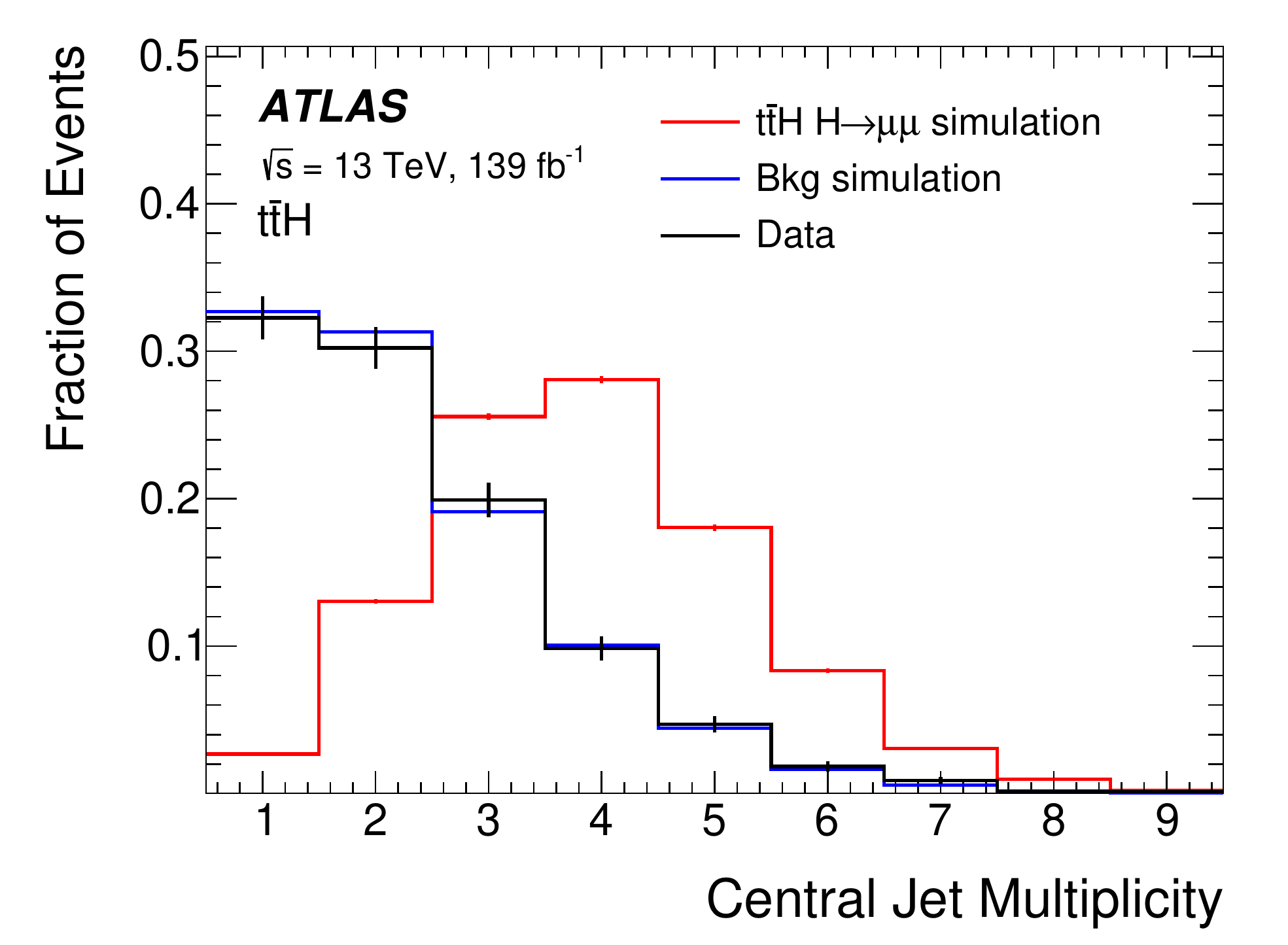}}
  \subfigure[]{\includegraphics[width=0.40\textwidth]{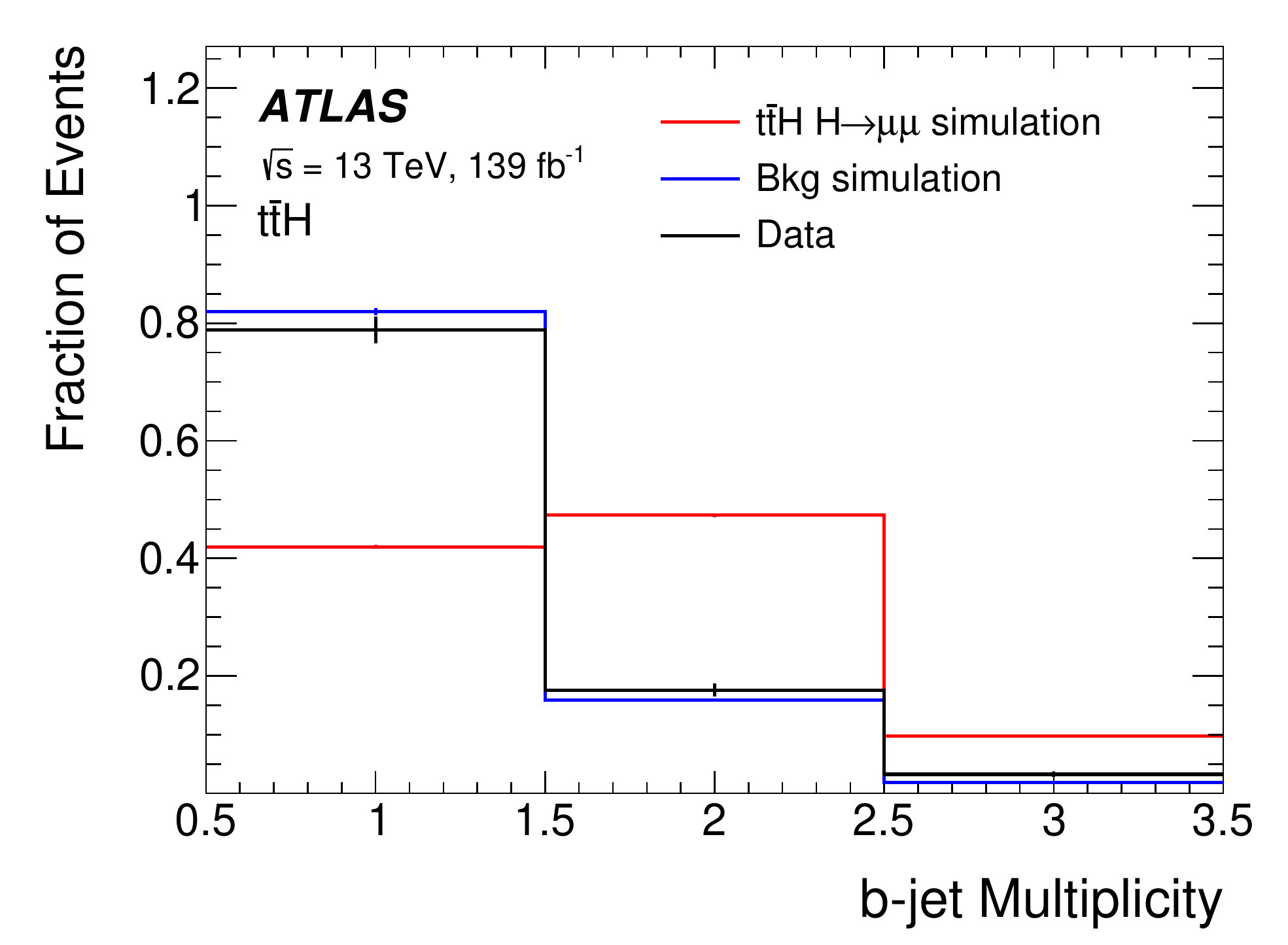}}
  \caption{Distributions of each training variable used for the \ttH\ category (part 1).}
  \label{fig:ttH_variables1}
\end{figure}

\begin{figure}[h!]
  \centering
  \subfigure[]{\includegraphics[width=0.40\textwidth]{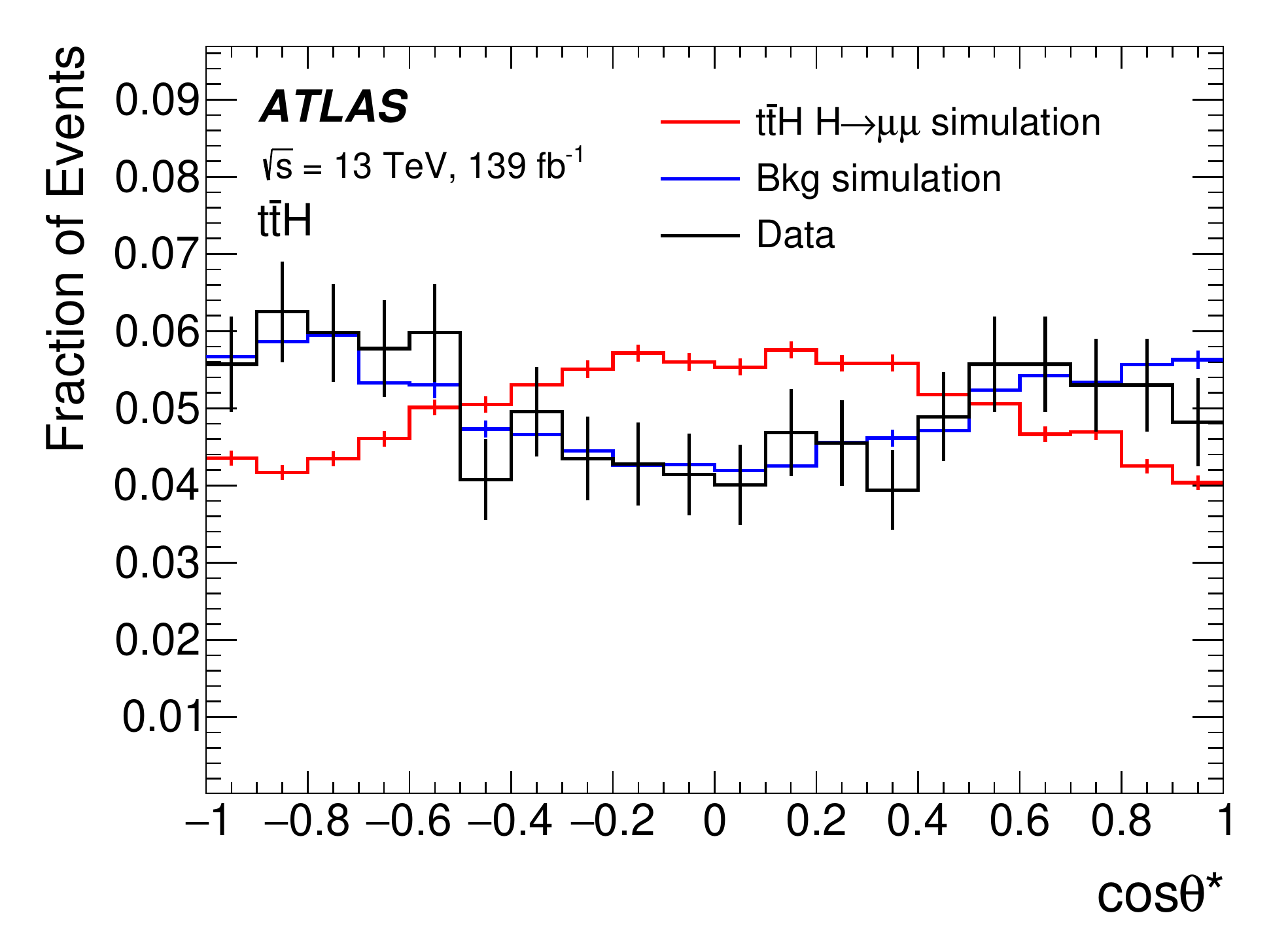}}
  \subfigure[]{\includegraphics[width=0.40\textwidth]{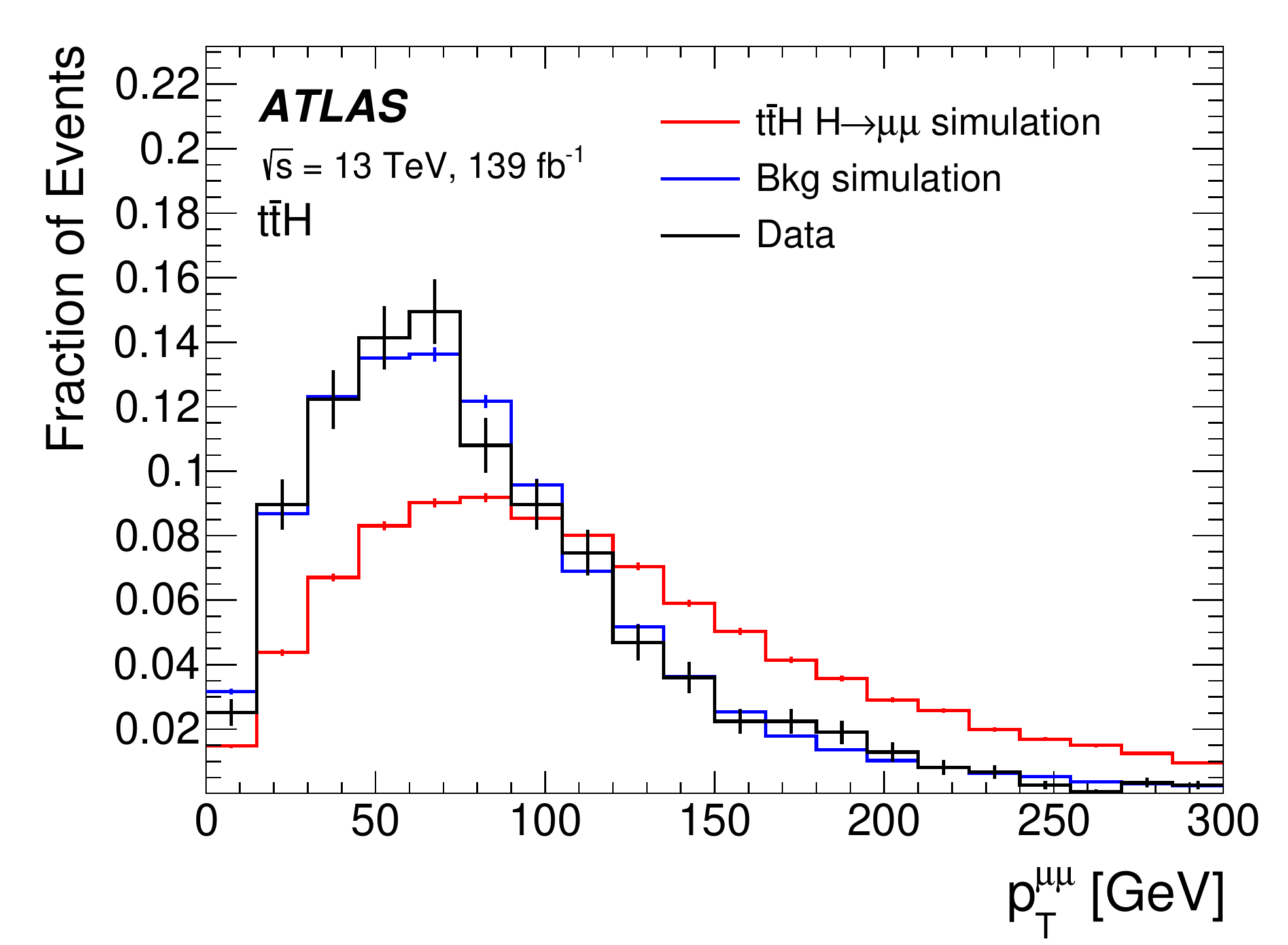}}
  \subfigure[]{\includegraphics[width=0.40\textwidth]{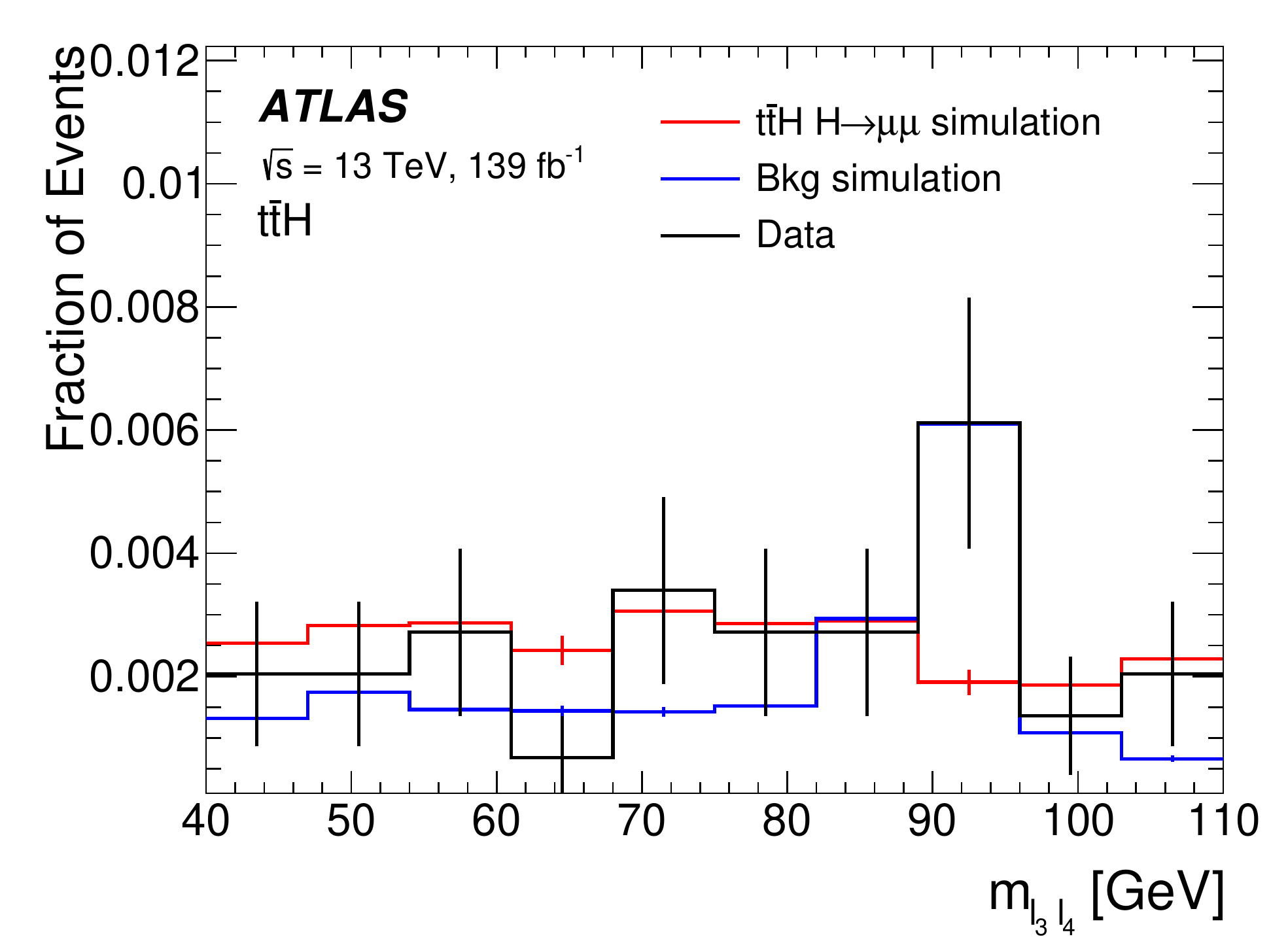}}
  \subfigure[]{\includegraphics[width=0.40\textwidth]{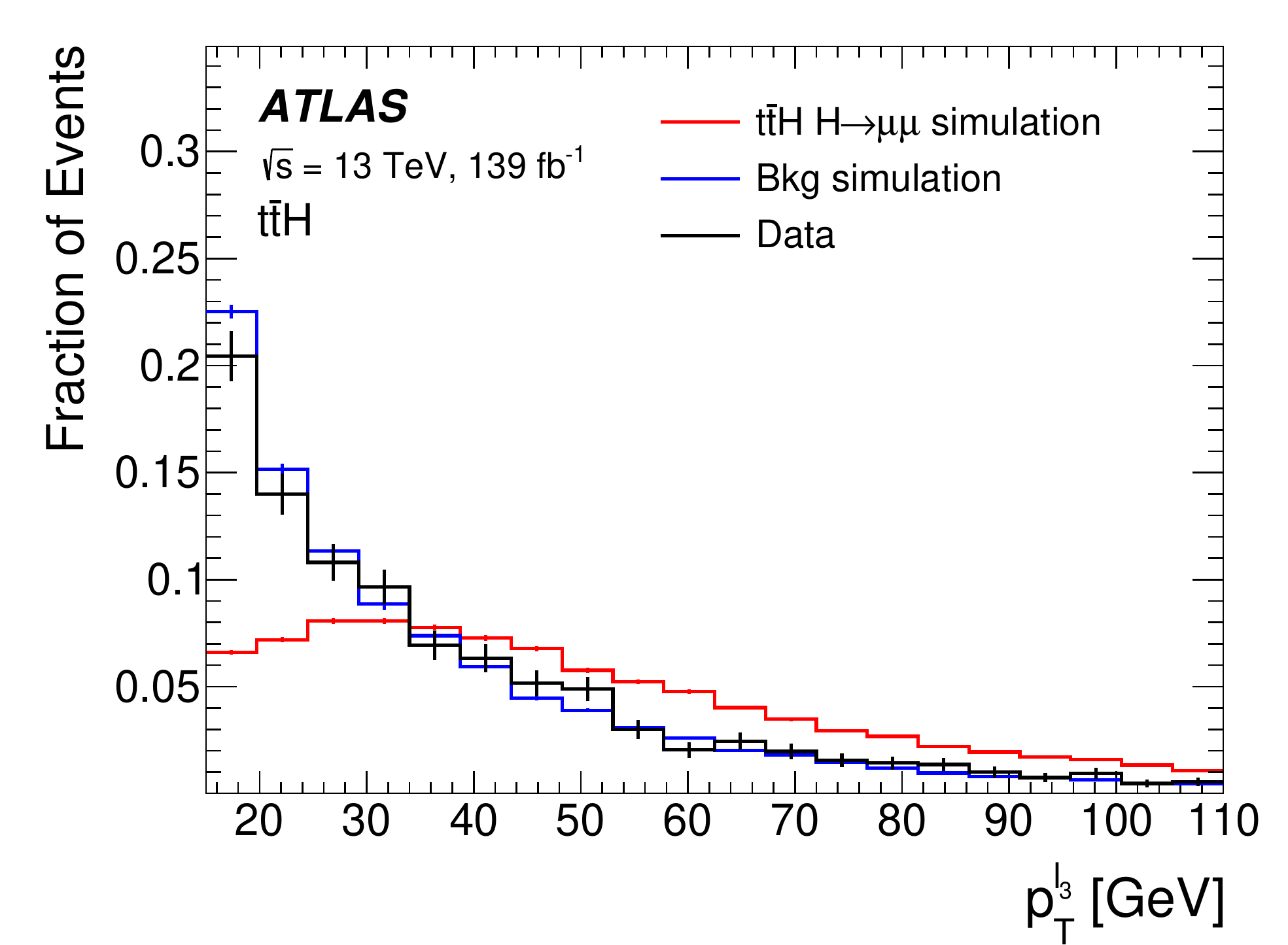}}
  \subfigure[]{\includegraphics[width=0.40\textwidth]{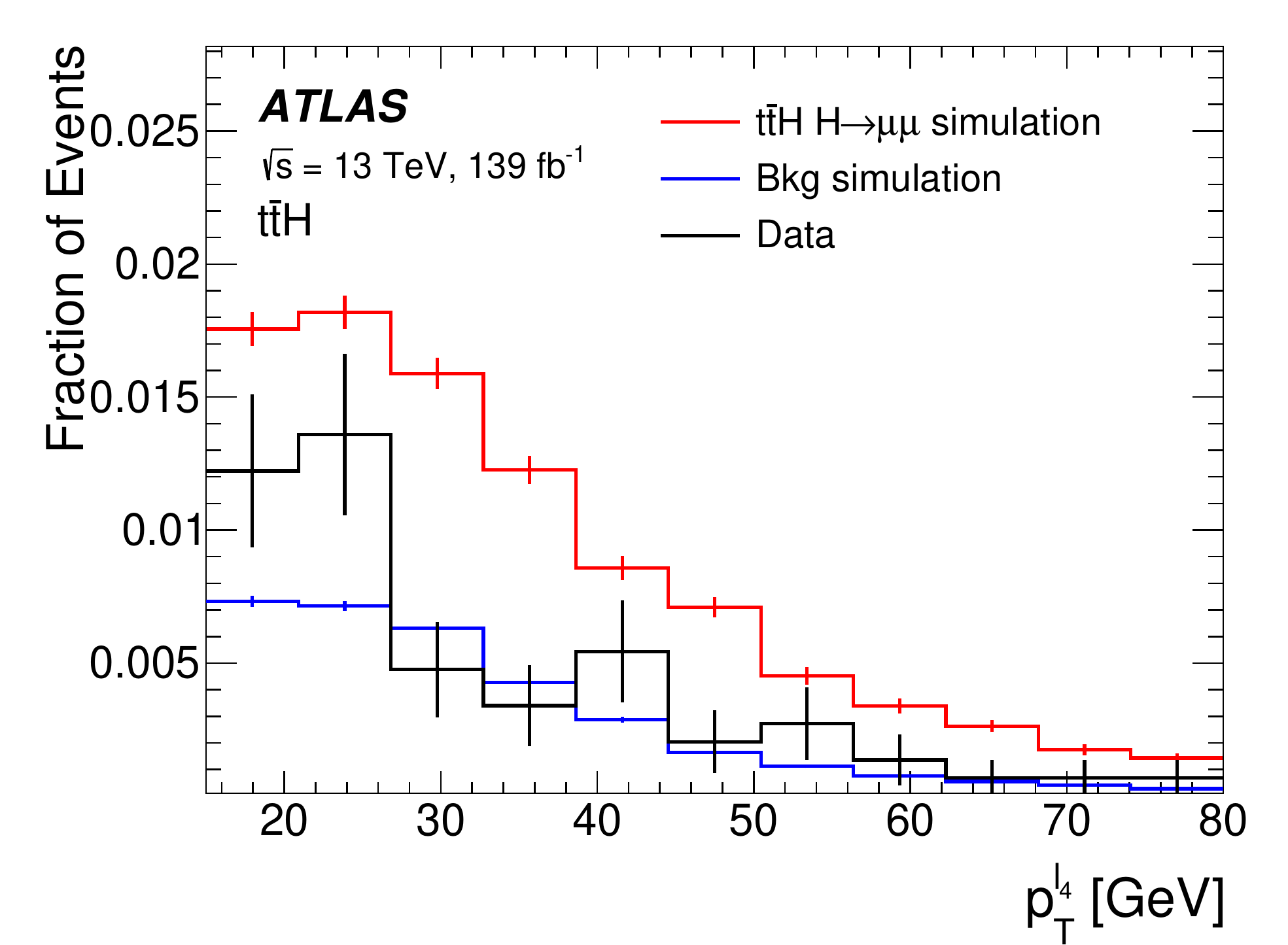}}
  \subfigure[]{\includegraphics[width=0.40\textwidth]{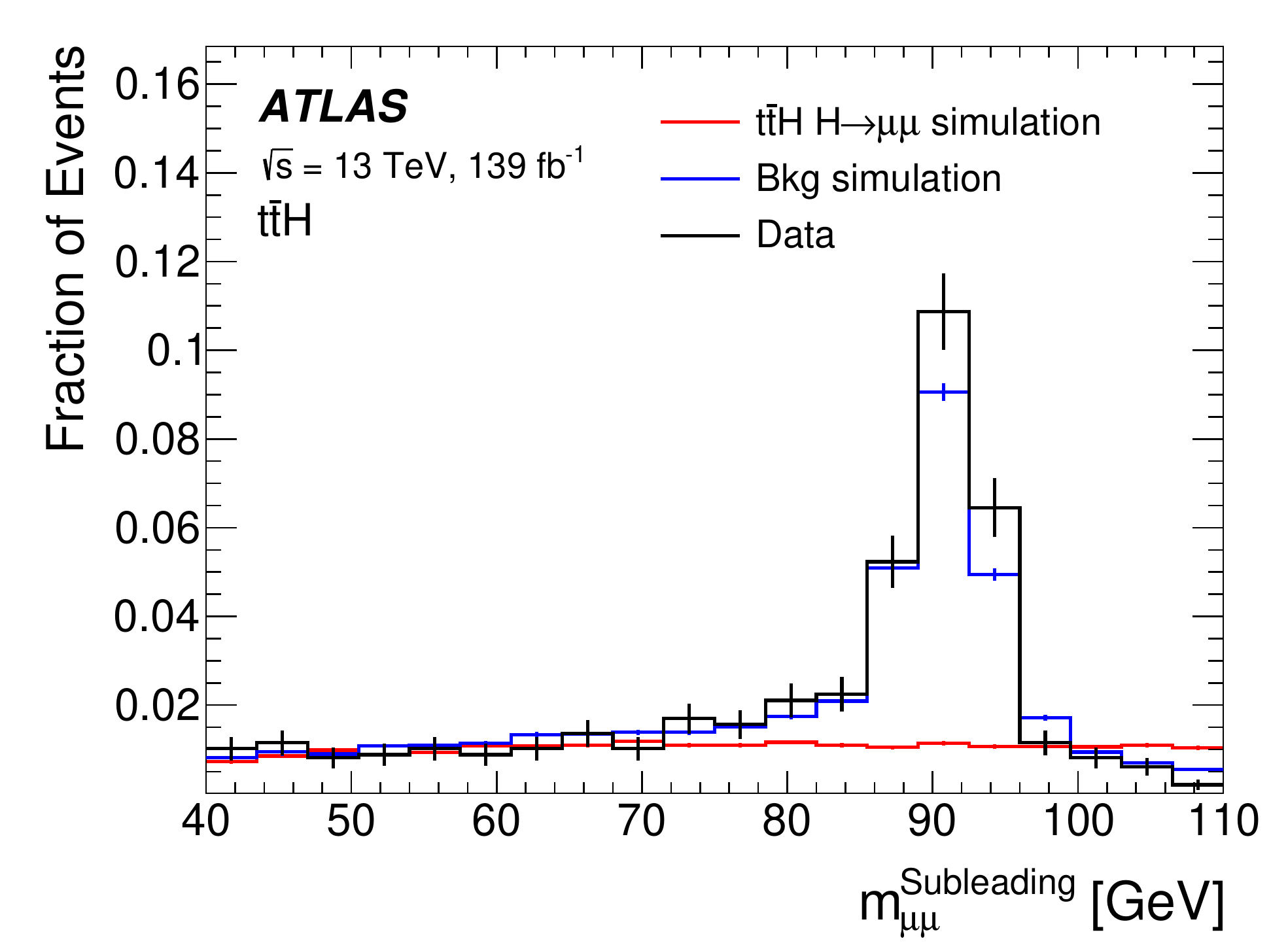}}
  \caption{Distributions of each training variable used for the \ttH\ category (part 2).}
  \label{fig:ttH_variables2}
\end{figure}

\begin{figure}[h!]
  \centering
  \includegraphics[width=0.6\textwidth]{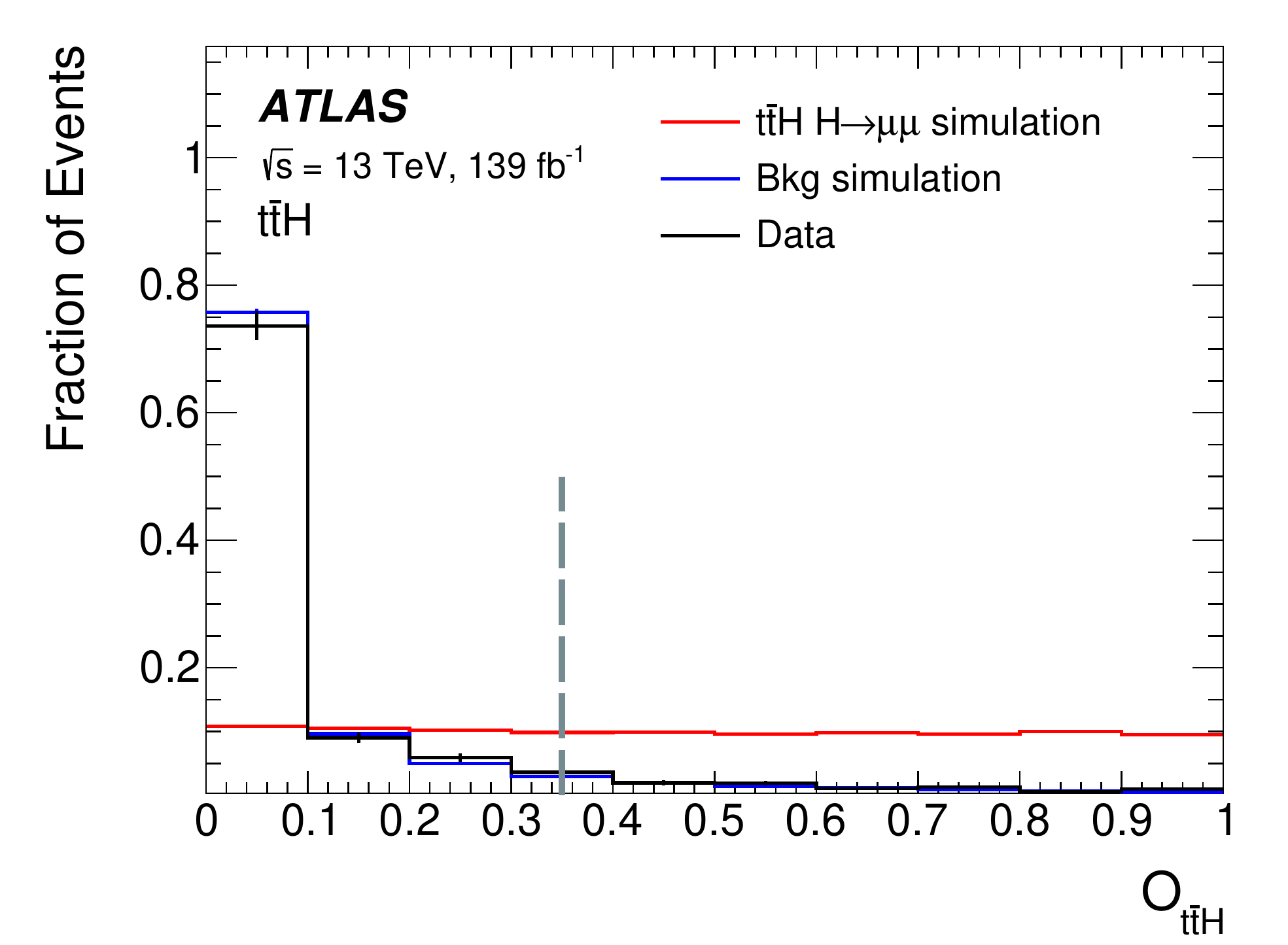}
  \caption{The \ttH\ BDT score distributions for signal and background. The vertical grey dashed line represents the BDT score threshold of 0.35 used to select events in the \ttH\ category.}
  \label{fig:bdt_ttH}
\end{figure}

Note that if an event does not contain enough objects to define the variables described above, the variables will be assigned with an unphysical arbitrary value.
Fig.~\ref{fig:ttH_variables1} and \ref{fig:ttH_variables2} show the distributions of the training variables used in the \ttH\ category.
The distributions of the \ttH\ BDT score for signal and background are shown in Fig.~\ref{fig:bdt_ttH}.

The \ttH\ category is defined based on the BDT score, with the dominant backgrounds including $t\bar{t}Z$ process, $t\bar{t}$ and dibosons.
Assuming the SM, the expected signal yield in this category is 1.2, with a high purity of 98\% for the \ttH\ production relative to the other Higgs boson production modes.
In the mass window $120 < \mmumu < 130$ GeV, the signal-to-background ratio is 8\% .

\subsection{\VH\ categories}

The \VH\ categories aim to select events in the \VH\ production mode, where the vector boson decays leptonically ($W\to\ell\nu$ or $Z\to\ell\ell$).
These events must not have any $b$-jets tagged at the 85\% WP and should have at least one lepton (muon or electron) in addition to the opposite-charge muon pair.
The events are further categorized into the 3-lepton channel and 4-lepton channel.

In the 3-lepton channel, events are required to have exactly three leptons, including the sub-leading muon with $p_\mathrm{T} > 10$ GeV and the additional muon (electron) with $p_\mathrm{T} > 10\ (15)$ GeV.
To reduce the main background $Z\to\mu\mu$ events, events containing an opposite-charge muon pair with an invariant mass between 80 and 105 GeV are vetoed.

In the 4-lepton channel, events must have at least four leptons, and two additional muons or electrons are required to have $\pT > 8$ and 6 GeV, respectively.
There should be at most one opposite-charge muon pair with an invariant mass between 80 and 105 GeV.

If there are more than two possible combinations of opposite-charge muon pairs, the assignments of the muons to either \Hmm\ or $Z\to\mu\mu$ are made based on the minimization of the $\chi^2$ criterion.
The criterion considers the difference between the reconstructed and the expected masses of $H$ and $Z$ with respect to the expected experimental resolutions.
The formula for $\chi^2$ is shown below:

\begin{equation}
\chi^2 =
\begin{cases}
\frac{\left(m_{\mu_1\mu_2} - m_H \right)^2}{\sigma\left(m_H\right)^2} + \frac{\left(m_\mathrm{T}^{\left(\mu_3,MET\right)} - m_W \right)^2}{\sigma\left(m_W\right)^2}, & \text{if } n_\mu = 3\\
\frac{\left(m_{\mu_1\mu_2} - m_H \right)^2}{\sigma\left(m_H\right)^2} + \frac{\left(m_{\mu_3\mu_4} - m_Z \right)^2}{\sigma\left(m_Z\right)^2}, & \text{if } n_\mu \geq 4
\end{cases}
,
\end{equation}
where $m_{\mu_1\mu_2}$ is the invariant mass of the opposite-charge dimuon system forming \Hmm\ (also referred to as \mmumu), $m_\mathrm{T}^{\left(\mu_3,MET\right)}$ is the transverse mass of the system consisting of the additional muon and \MET , and $m_{\mu_3\mu_4}$ is the invariant mass of the additional opposite-charge dimuon system forming $Z\to\mu\mu$.
Using this pairing algorithm achieves an accuracy of about 93\% and 97\% in the 3-lepton and 4-lepton channels, respectively.
If the additional leptons are electrons, they are matched to either $W\to e\mu$ or $Z\to ee$ without ambiguity.

A BDT is trained in the 3-lepton (4-lepton) channel in the range of $110 < \mmumu < 160$ GeV, using simulated $WH \to \mu\mu$ ($ZH \to \mu\mu$) as the training signal and simulated Standard Model (SM) background as the training background.
In the 3-lepton channel, the nine training variables used for the BDT are:

\begin{itemize}
\item $\Delta\phi_{\mu\mu,MET}$: azimuthal separation between $H\to\mu\mu$ and $E^\mathrm{miss}_\mathrm{T}$.
\item $p_\mathrm{T}^W$: transverse momentum of the $W$ boson.
\item $p_\mathrm{T}^{\ell_W}$: transverse momentum of the $W$ lepton.
\item $m_\mathrm{T}^W$: transverse mass of the $W$ boson system.
\item $\Delta\phi_{\mu\mu,W}$: azimuthal separation between $H\to\mu\mu$ and $W$.
\item $\Delta\eta_{\mu\mu,W}$: separation in pseudorapidity between $H\to\mu\mu$ and $W$.
\item $E^\mathrm{miss}_\mathrm{T}$: missing transverse momentum.
\item $p_\mathrm{T}^j$: transverse momentum of the leading jet if present.
\item $n_{j}$: number of jets. 
\end{itemize}

In the 4-lepton channel, the eight training variables used for the BDT are:

\begin{itemize}
\item $\Delta\phi_{\ell_Z,\ell_Z}$: azimuthal separation between the two leptons from $Z\to\ell\ell$.
\item $m_\mathrm{T}^Z$: invariant mass of the $Z$ boson system.
\item $\Delta\phi_{\mu\mu,Z}$: azimuthal separation between $H\to\mu\mu$ and $Z$.
\item $\Delta\eta_{\mu\mu,Z}$: separation in pseudorapidity between $H\to\mu\mu$ and $Z$.
\item $E^\mathrm{miss}_\mathrm{T}$: missing transverse momentum.
\item $\pT^{j_1}$: transverse momentum of the leading jet if present.
\item $\pT^{j_2}$: transverse momentum of the subleading jet if present.
\item $n_{j}$: number of jets. 
\end{itemize}

\begin{figure}[h!]
  \centering
  \subfigure[]{\includegraphics[width=0.40\textwidth]{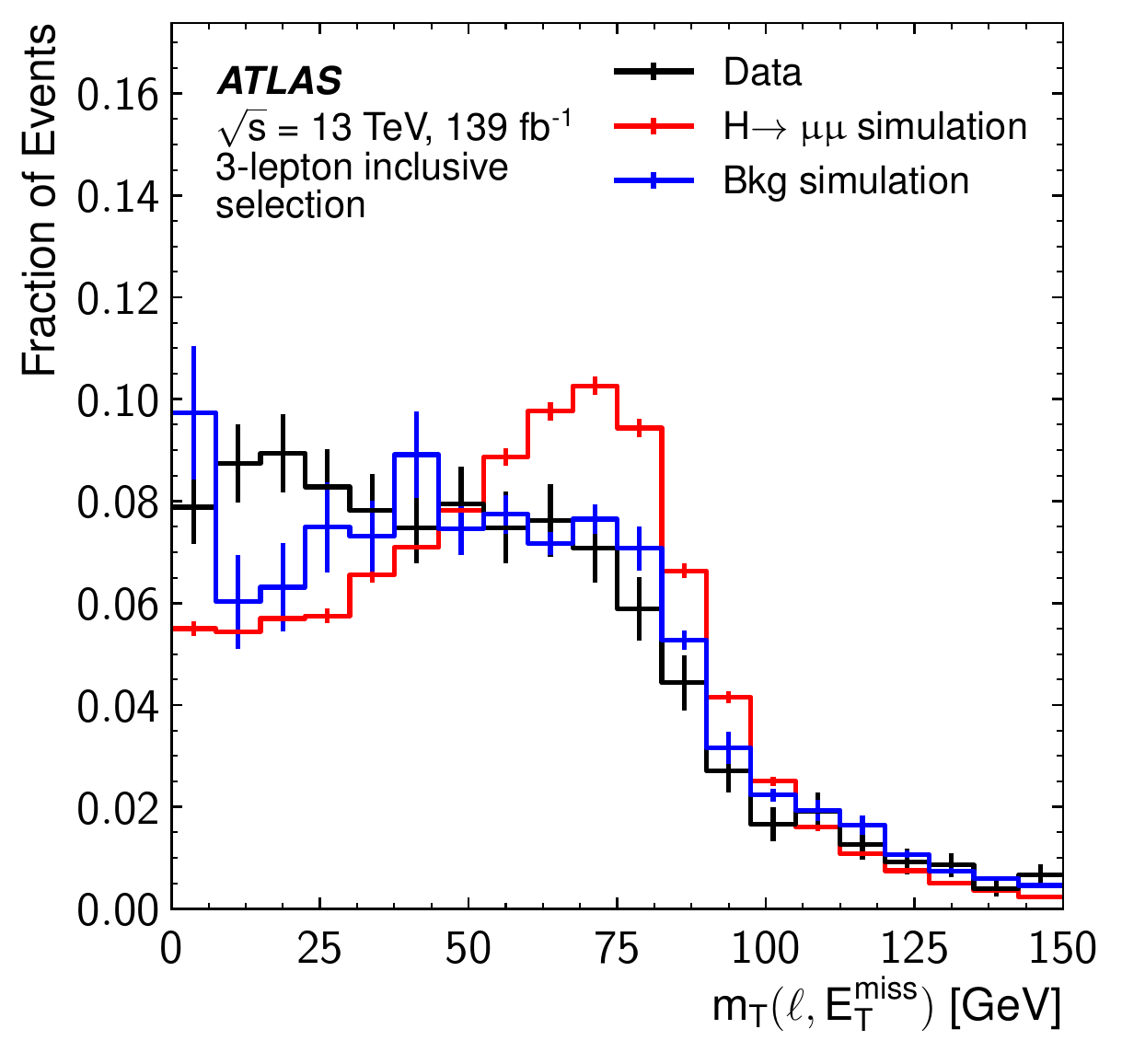}}
  \subfigure[]{\includegraphics[width=0.40\textwidth]{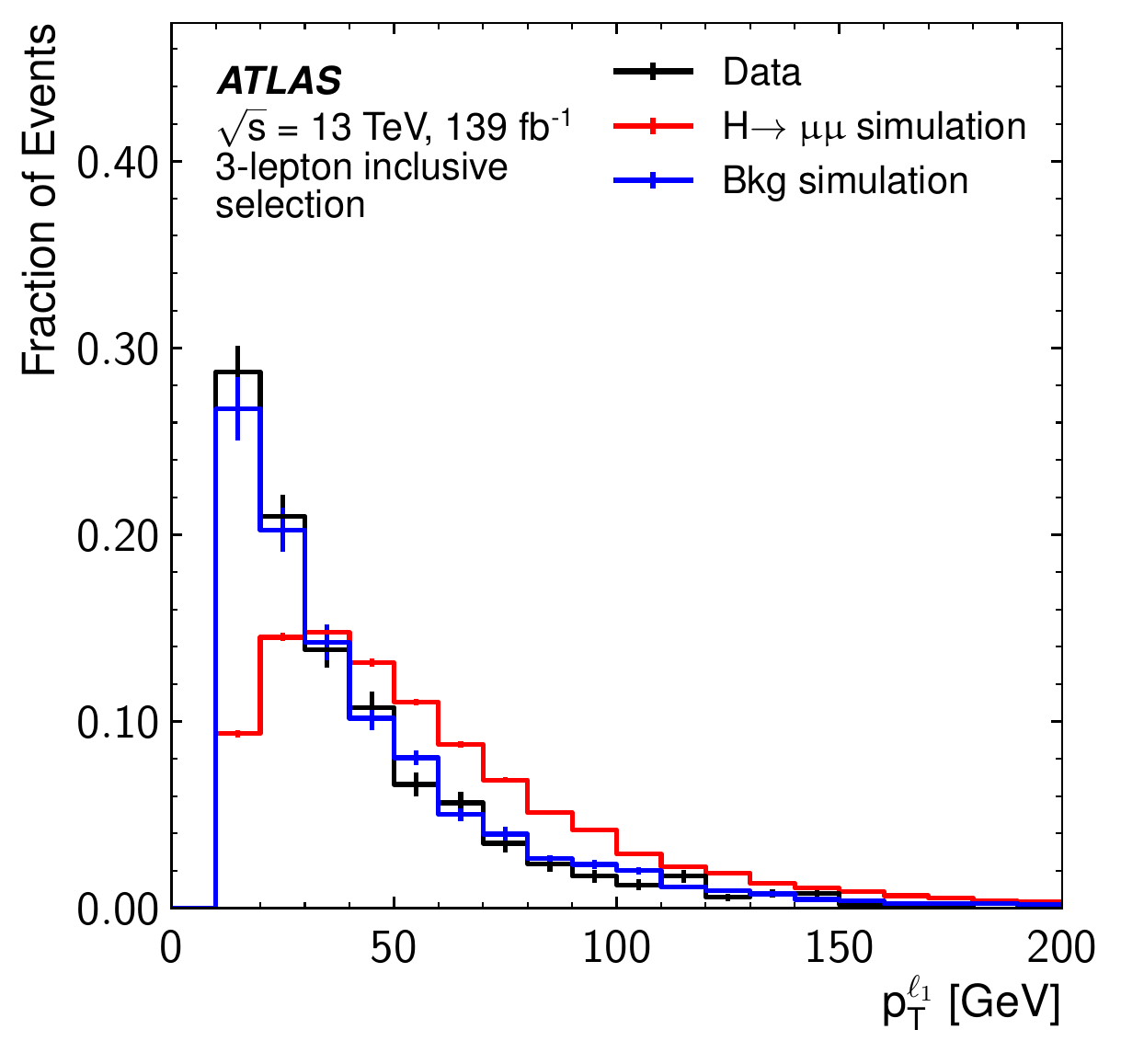}}
  \subfigure[]{\includegraphics[width=0.40\textwidth]{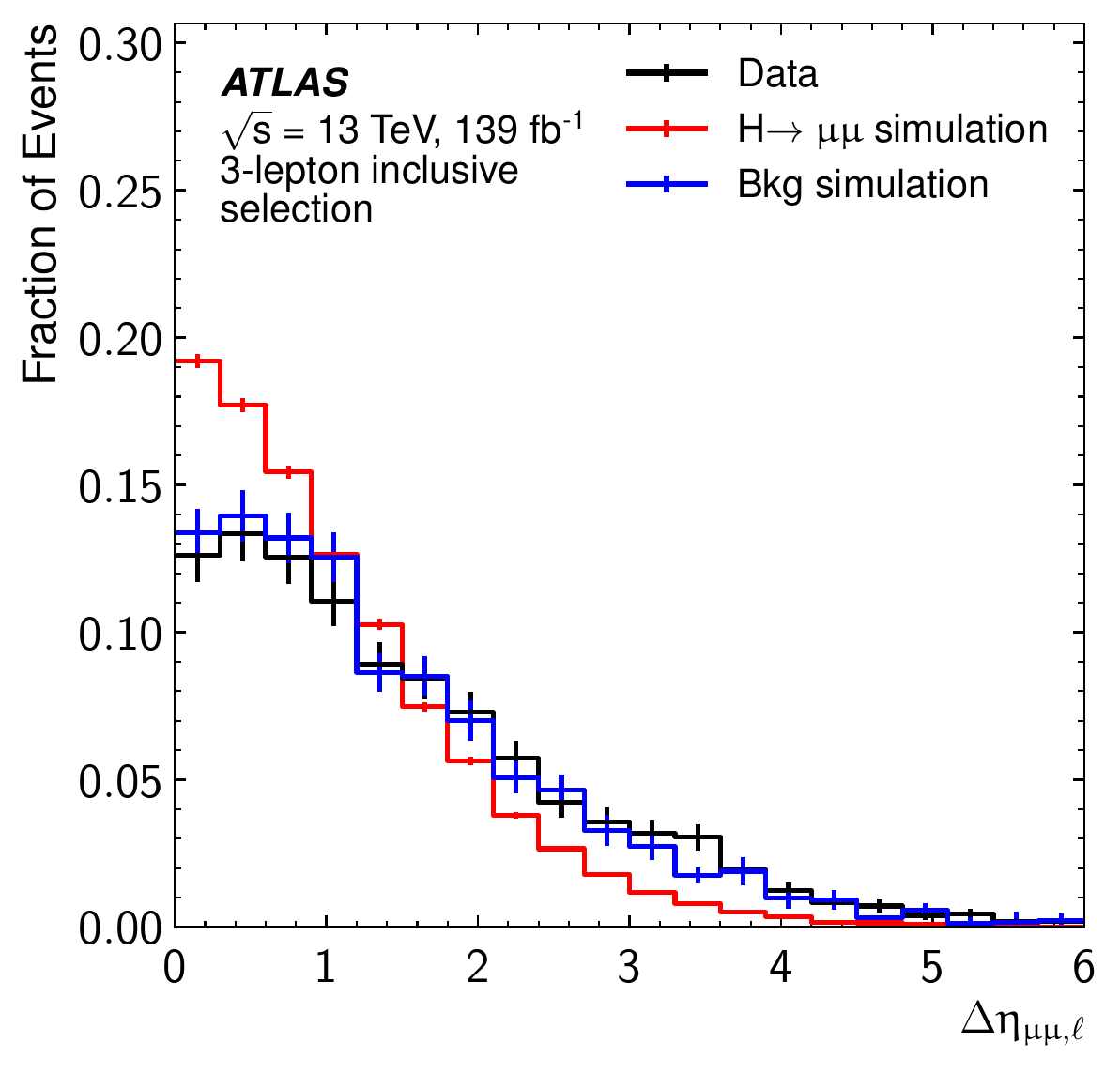}}
  \subfigure[]{\includegraphics[width=0.40\textwidth]{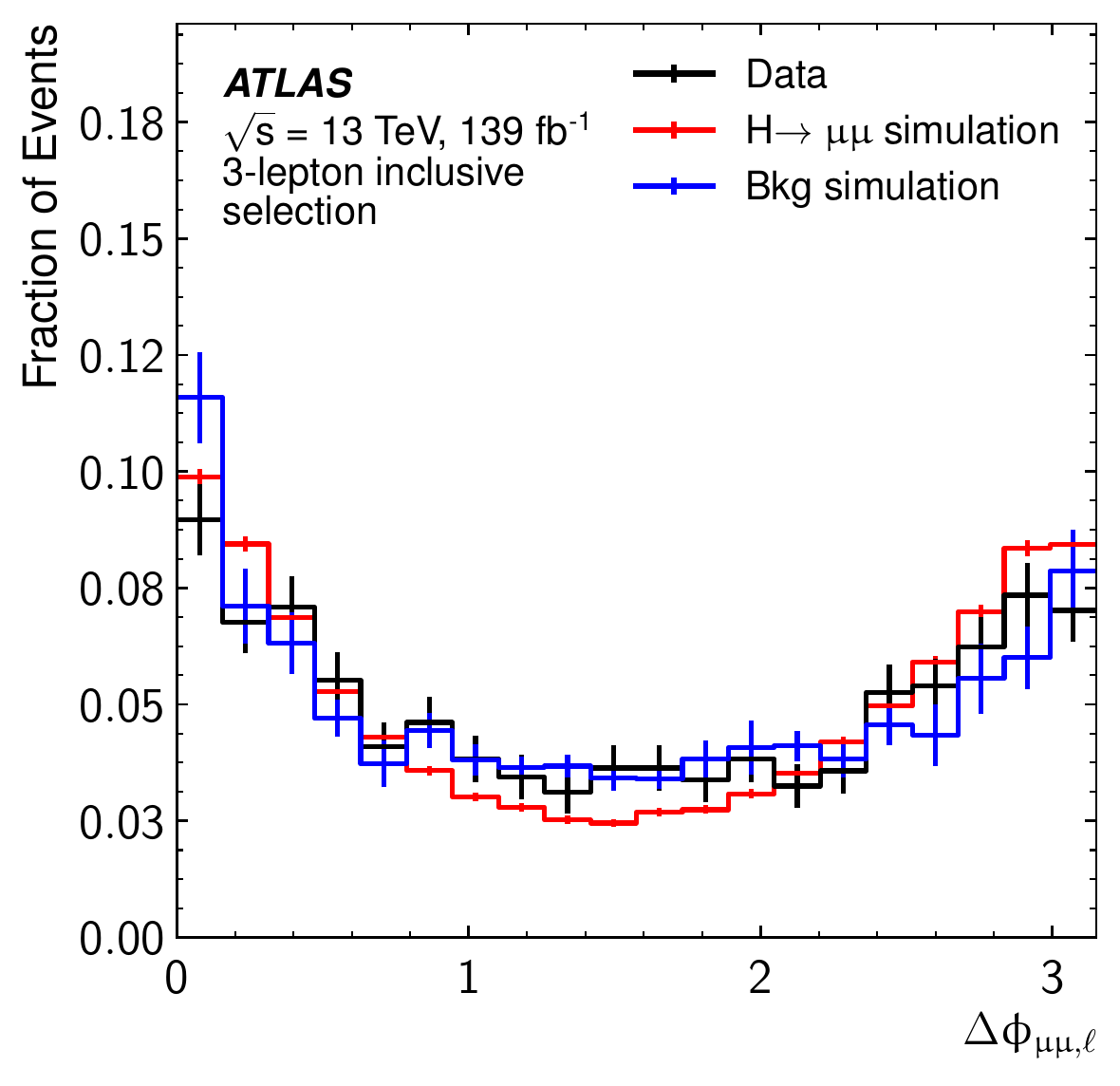}}
  \subfigure[]{\includegraphics[width=0.40\textwidth]{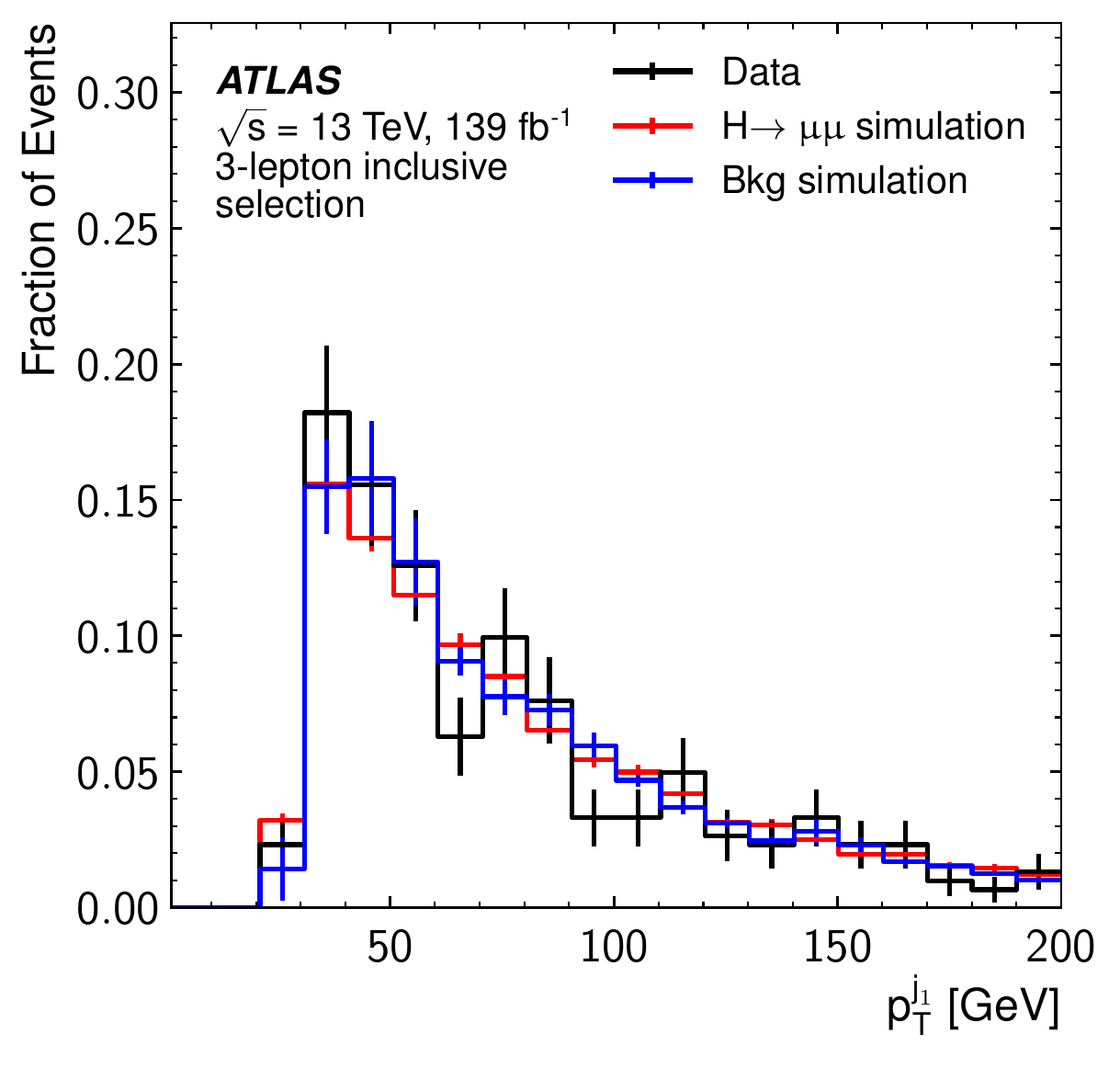}}
  \subfigure[]{\includegraphics[width=0.40\textwidth]{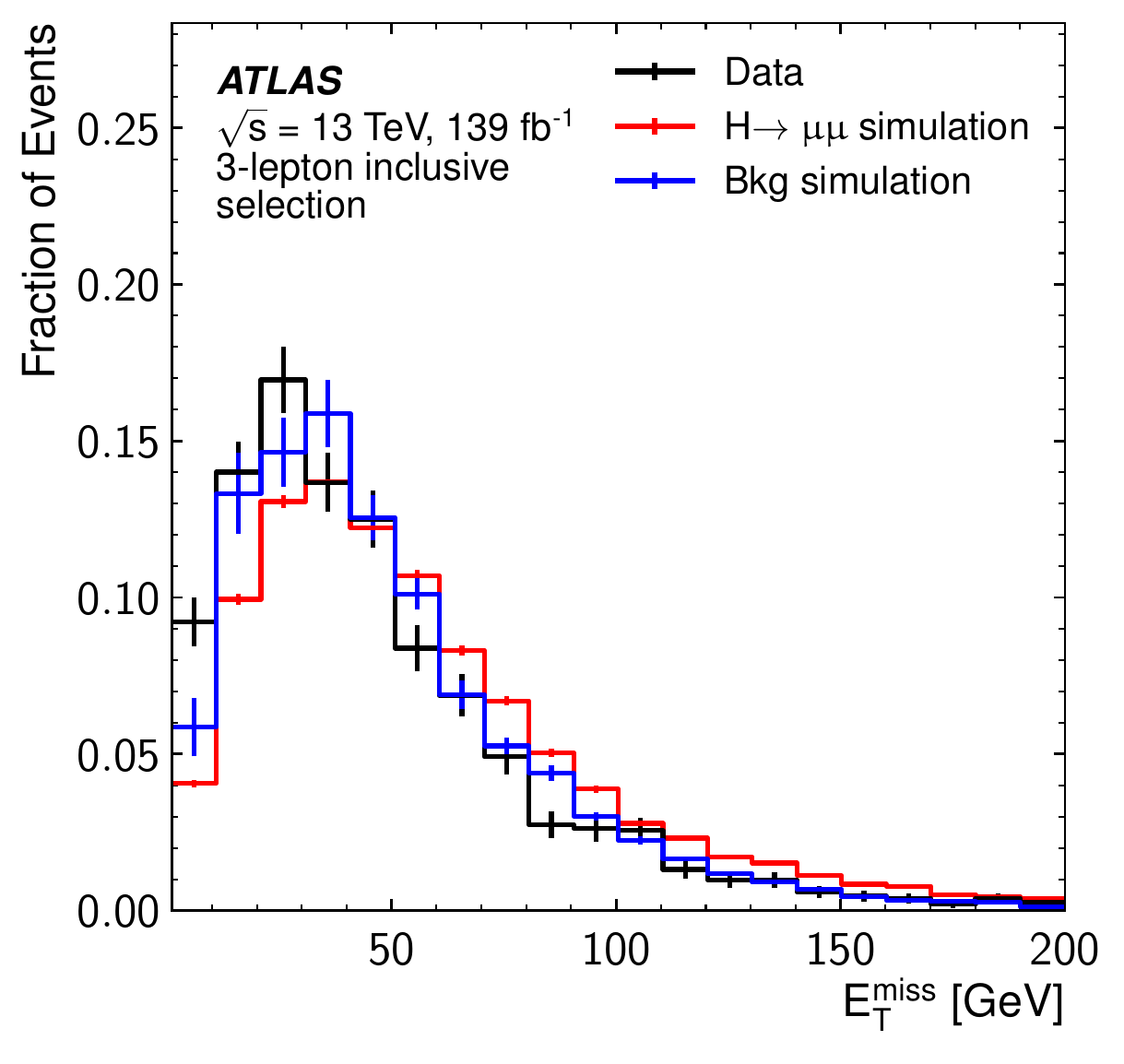}}
  \caption{Distributions of each training variable used for the \VH\ 3-lepton channel (part 1).}
  \label{fig:VH3lep_variables1}
\end{figure}

\begin{figure}[h!]
  \centering
  \subfigure[]{\includegraphics[width=0.40\textwidth]{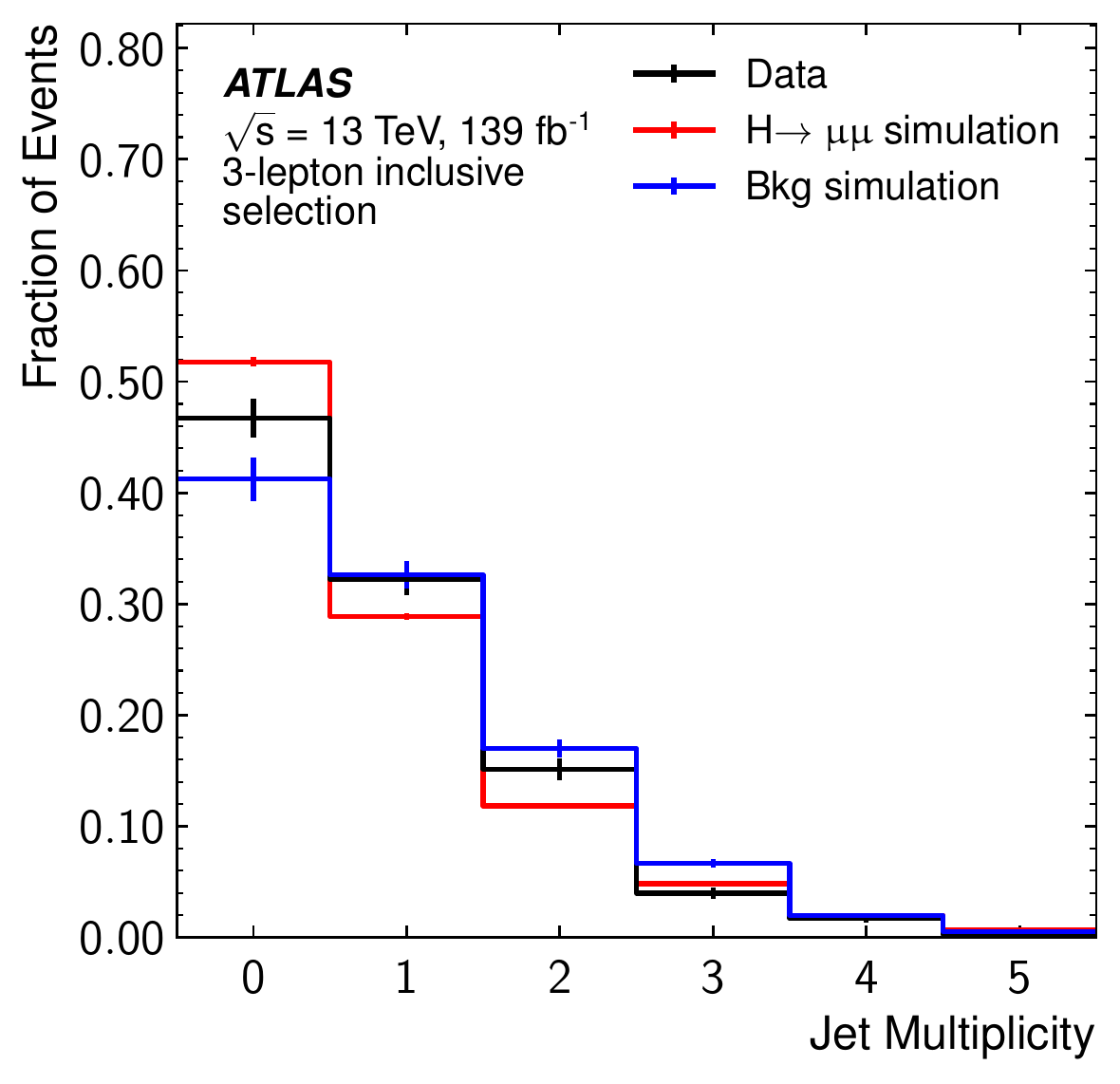}}
  \subfigure[]{\includegraphics[width=0.40\textwidth]{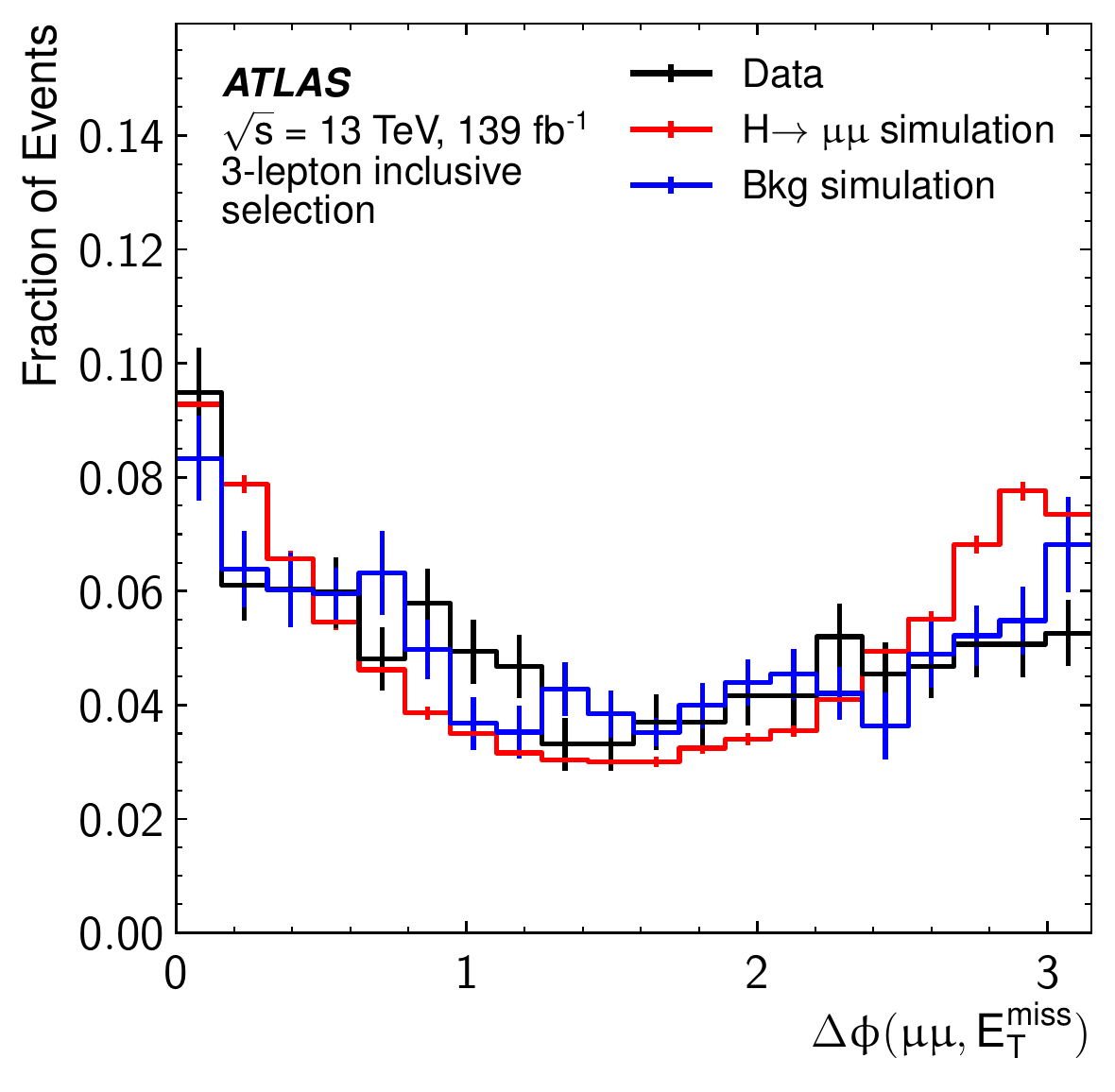}}
  \subfigure[]{\includegraphics[width=0.40\textwidth]{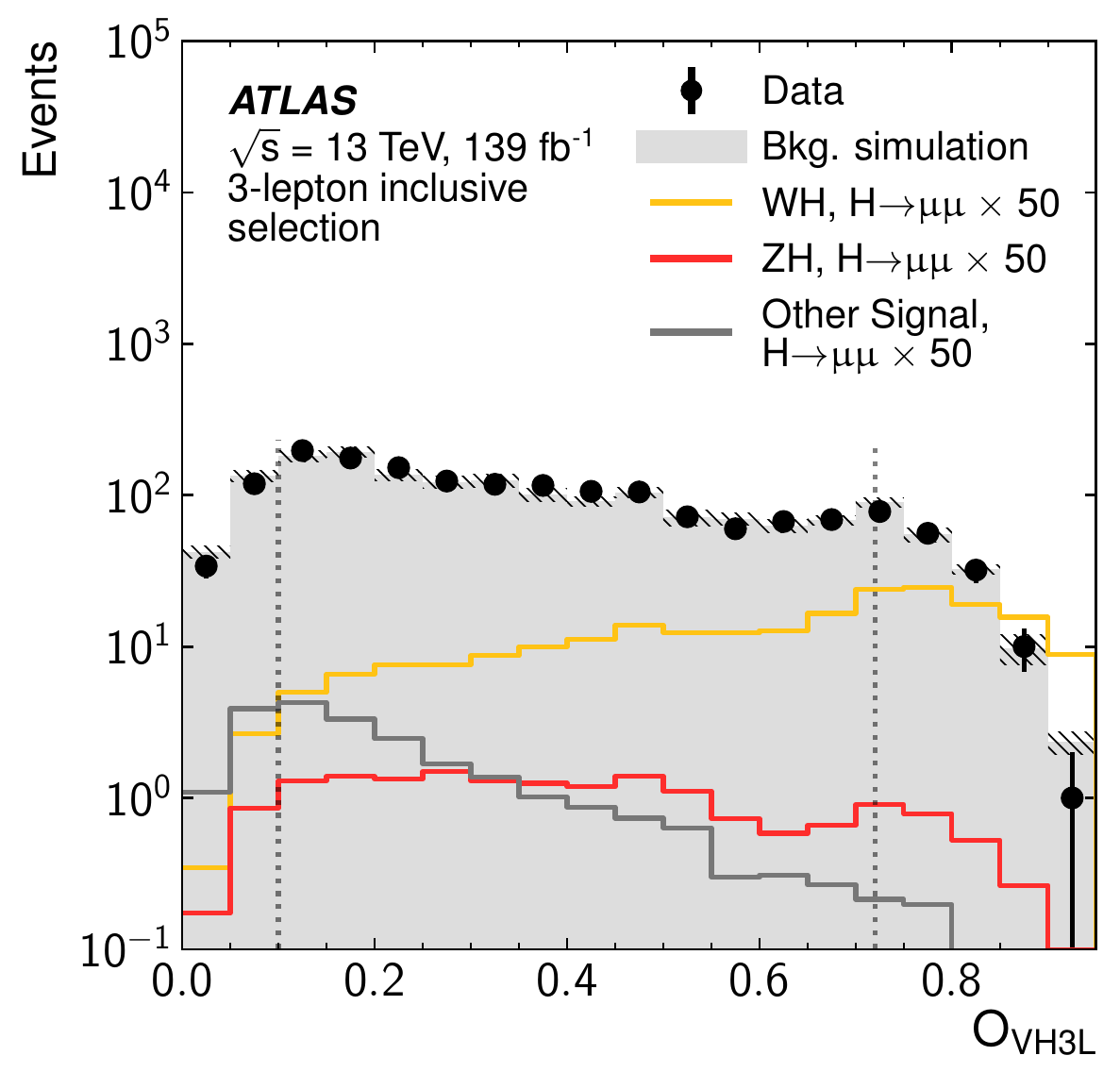}}
  \caption{Distributions of each training variable used for the \VH\ 3-lepton channel (part 2).}
  \label{fig:VH3lep_variables2}
\end{figure}

\begin{figure}[h!]
  \centering
  \subfigure[]{\includegraphics[width=0.40\textwidth]{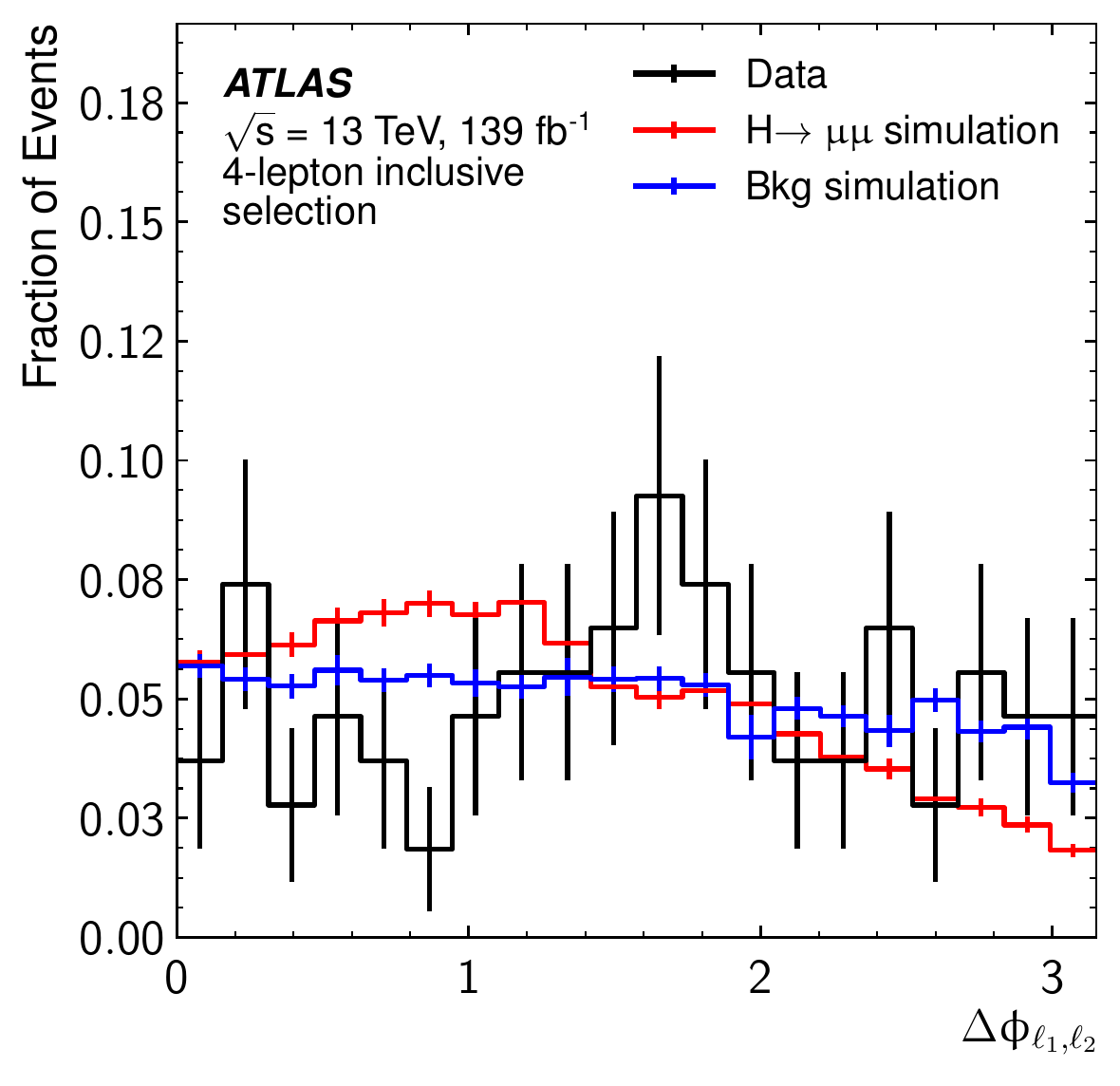}}
  \subfigure[]{\includegraphics[width=0.40\textwidth]{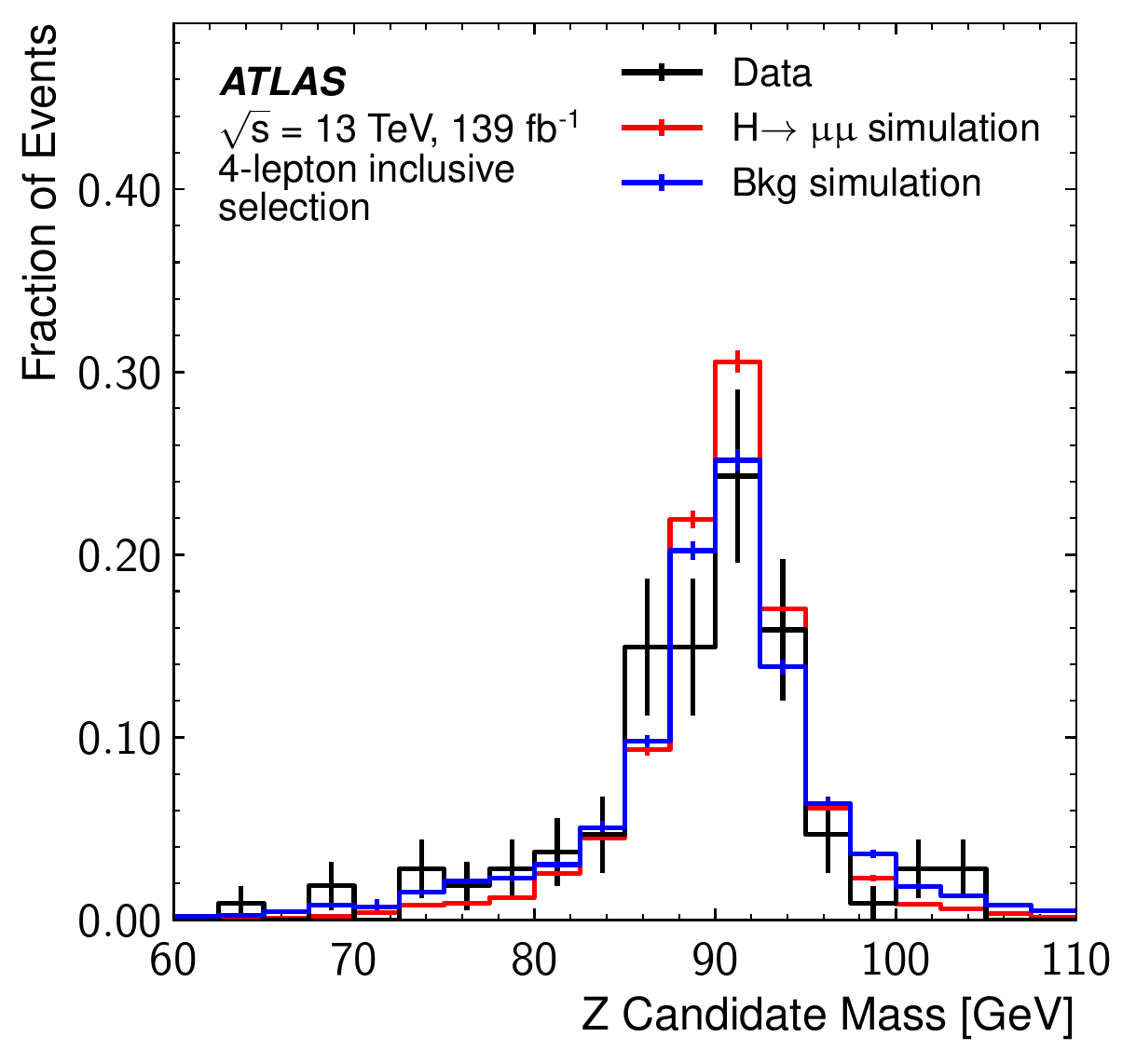}}
  \subfigure[]{\includegraphics[width=0.40\textwidth]{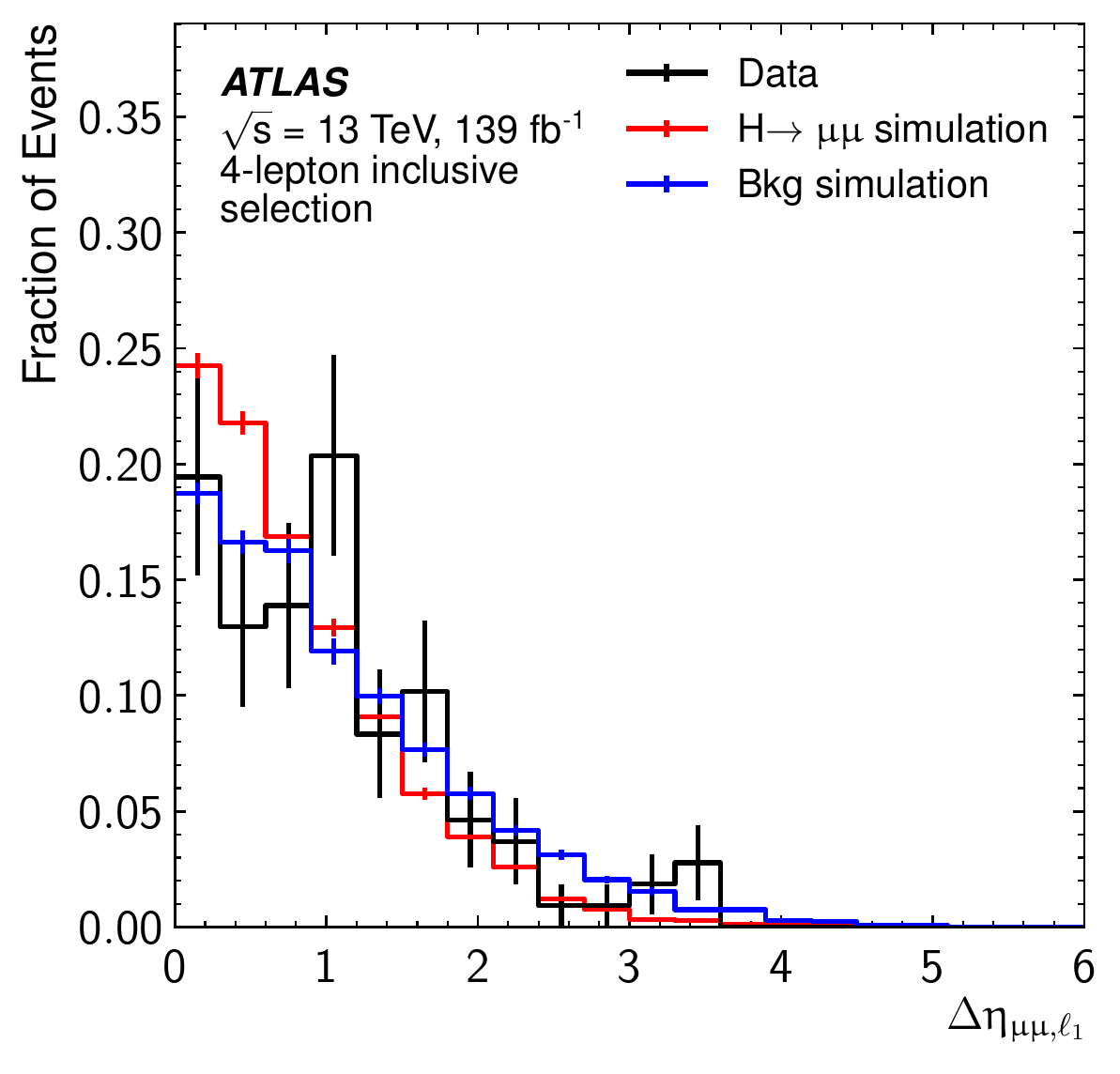}}
  \subfigure[]{\includegraphics[width=0.40\textwidth]{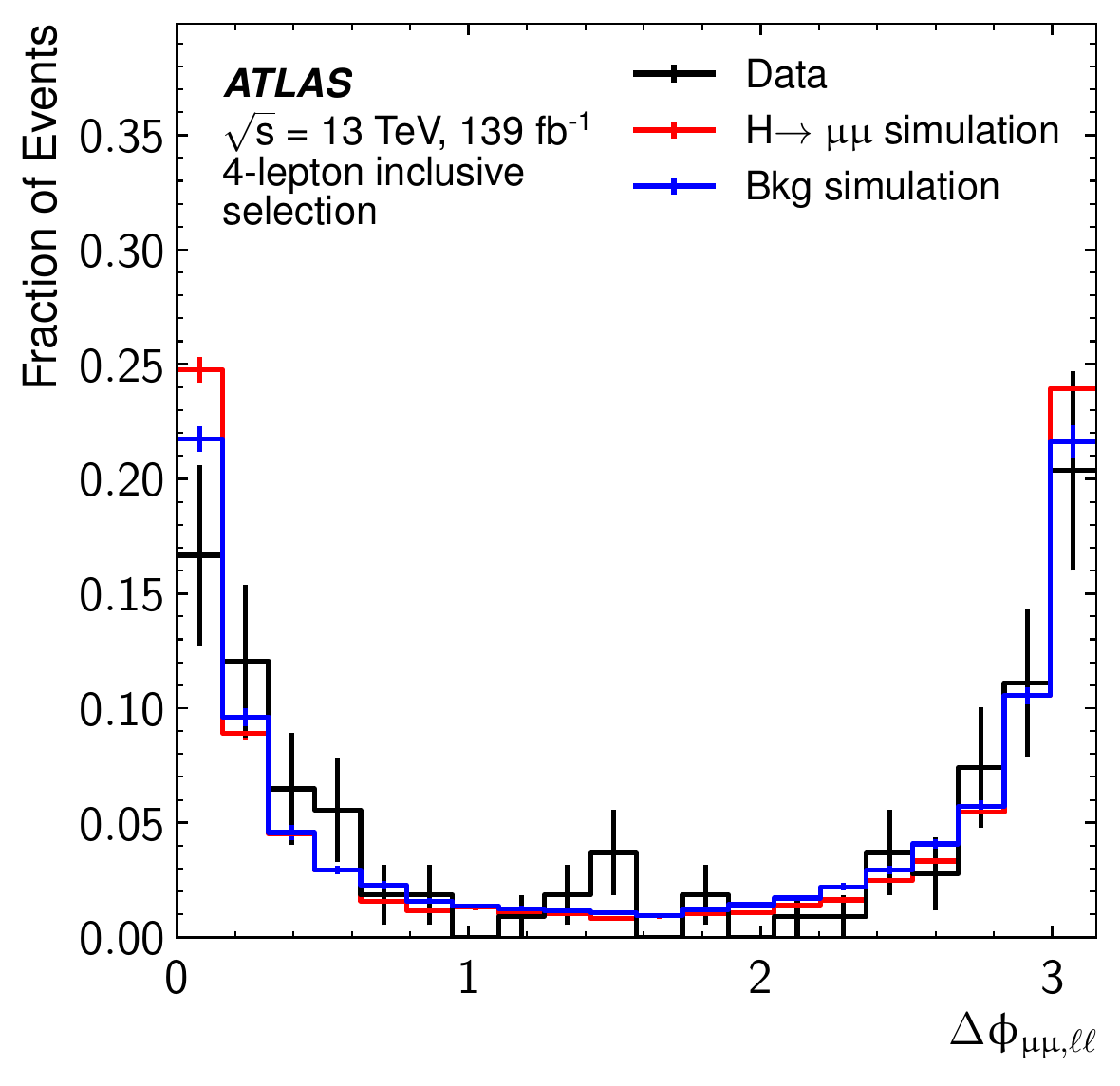}}
  \subfigure[]{\includegraphics[width=0.40\textwidth]{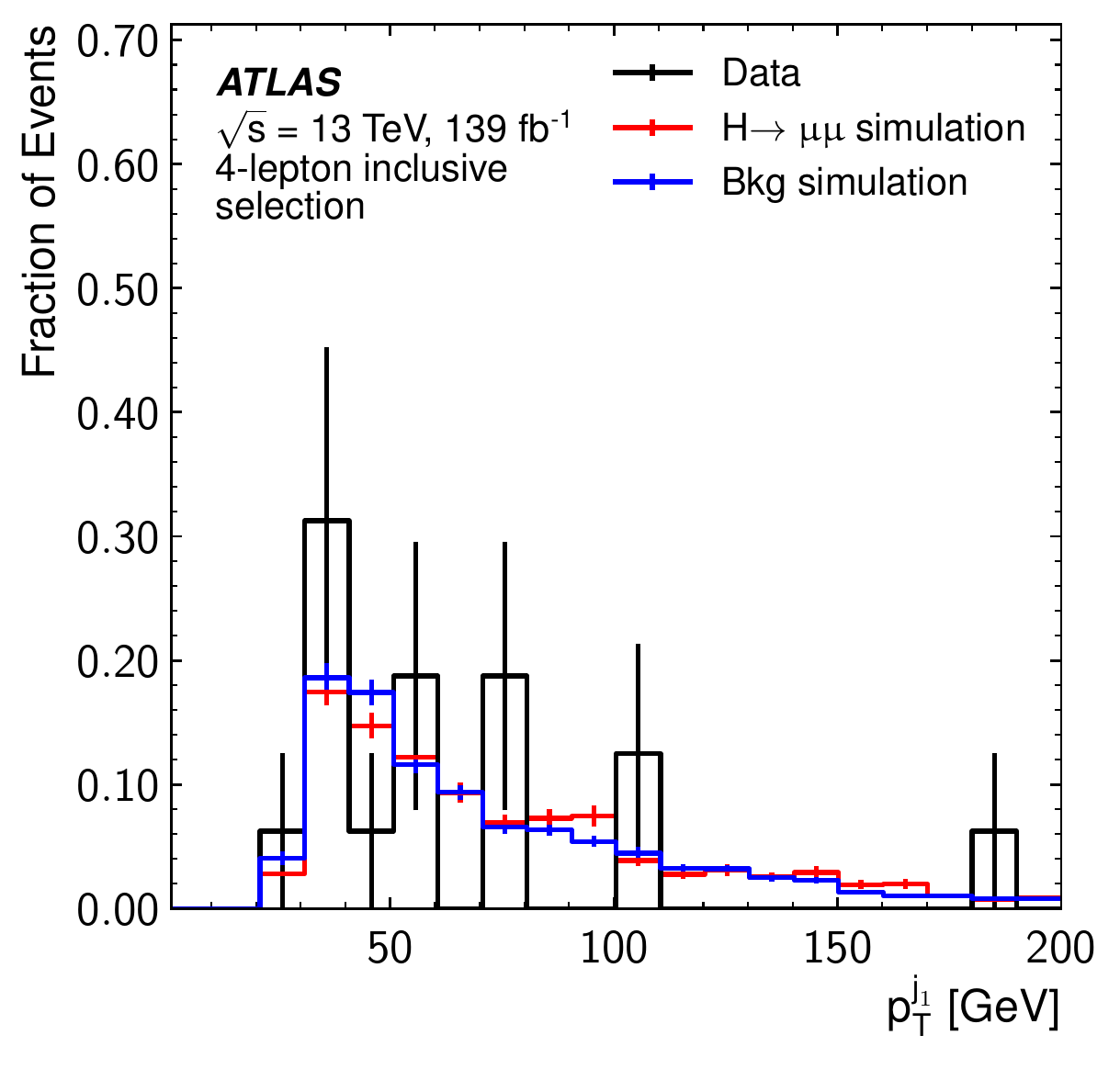}}
  \subfigure[]{\includegraphics[width=0.40\textwidth]{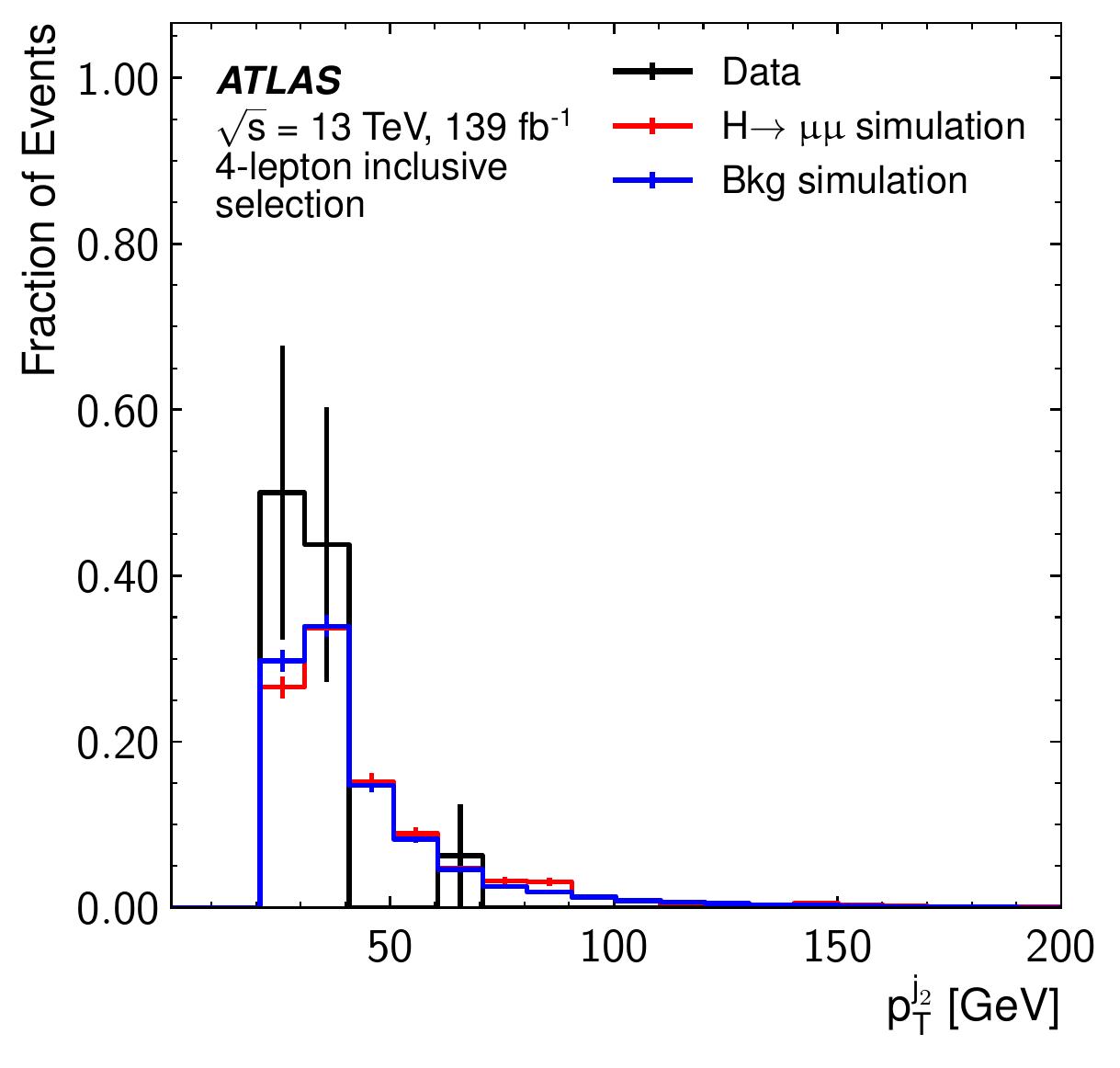}}
  \caption{Distributions of each training variable used for the \VH\ 4-lepton channel (part 1).}
  \label{fig:VH4lep_variables1}
\end{figure}

\begin{figure}[h!]
  \centering
  \subfigure[]{\includegraphics[width=0.40\textwidth]{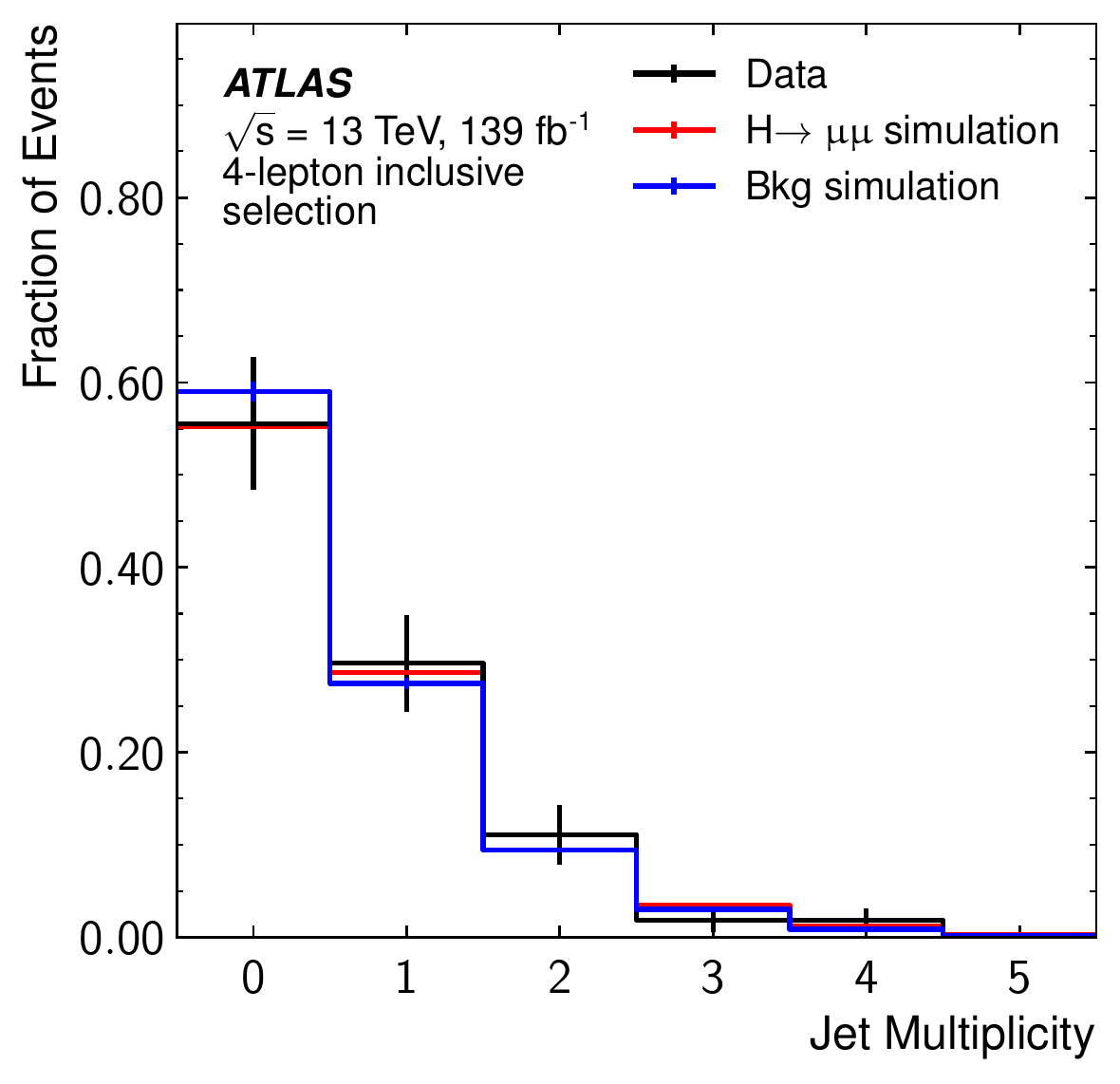}}
  \subfigure[]{\includegraphics[width=0.40\textwidth]{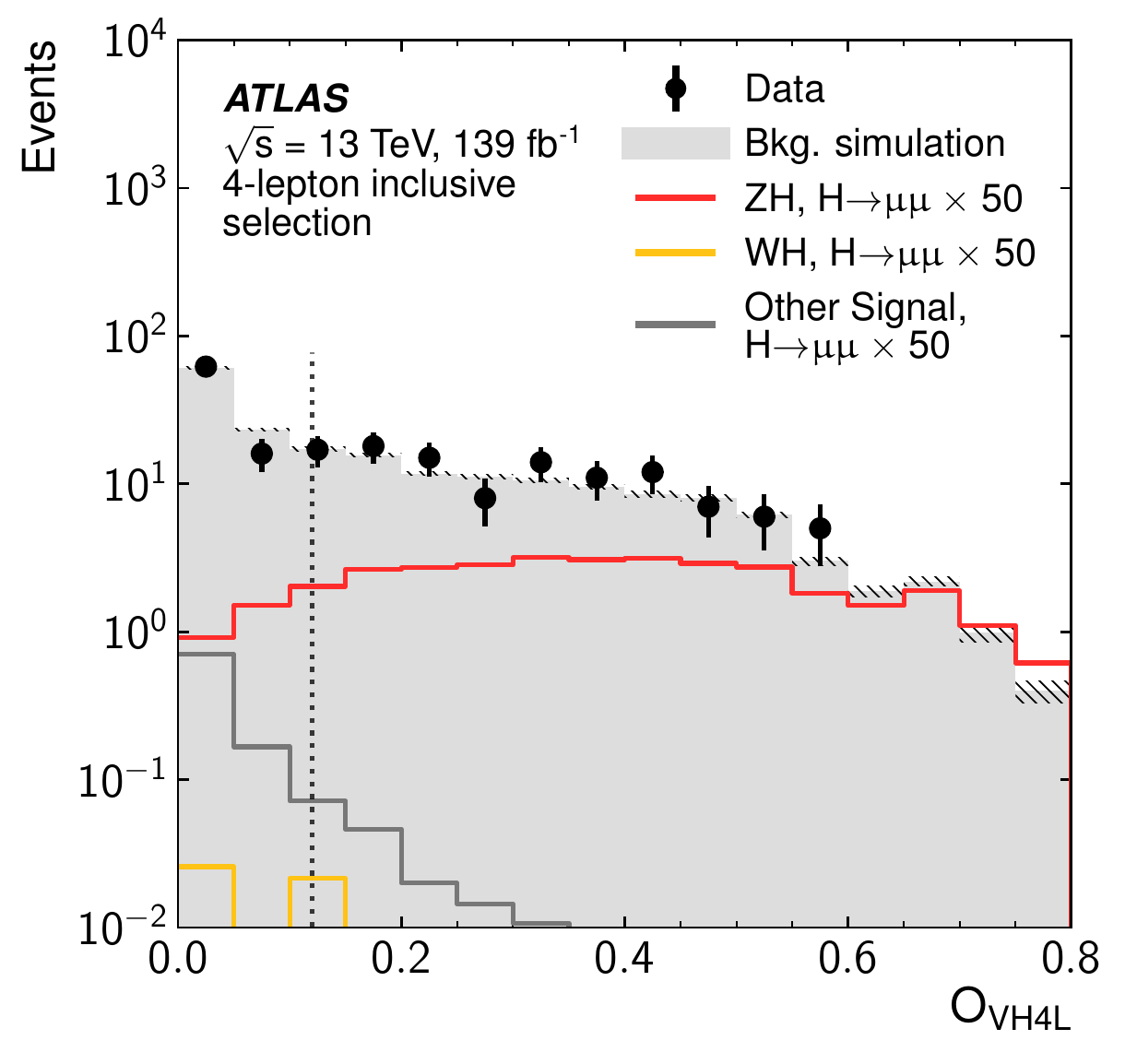}}
  \caption{Distributions of each training variable used for the \VH\ 4-lepton channel (part 2).}
  \label{fig:VH4lep_variables2}
\end{figure}

\begin{figure}[h!]
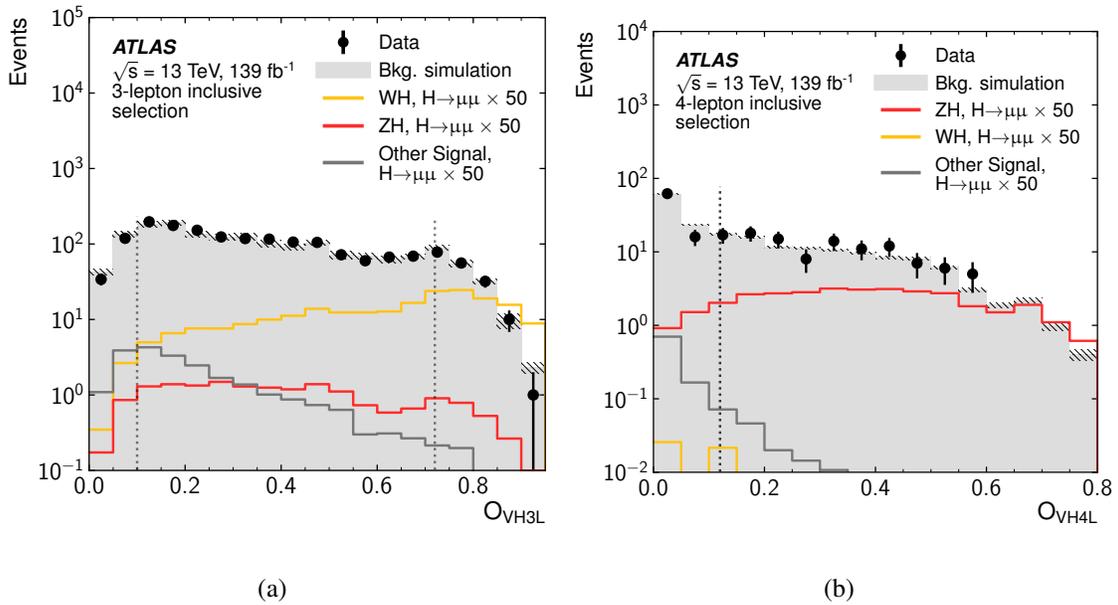

  \centering
  \subfigure[]{\includegraphics[width=0.45\textwidth]{figure/hmumu/figaux_09i.pdf}}
  \subfigure[]{\includegraphics[width=0.45\textwidth]{figure/hmumu/figaux_10h.pdf}}
  \caption{The \VH\ BDT score distributions for signal and background in (a) 3-lepton channel and (b) 4-lepton channel. The grey vertical dashed lines indicate the BDT score boundary define each \VH\ categories.}
  \label{fig:VH_BDT}
\end{figure}

Like the \ttH\ category, to handle events that do not have enough objects to define certain variables, an unphysical arbitrary value will be assigned to those variables in the BDT for the \VH\ category.
Fig.~\ref{fig:VH3lep_variables1} and \ref{fig:VH3lep_variables2} show the distributions of the training variables used for the \VH\ 3 lepton channels, and Fig~\ref{fig:VH4lep_variables1} and \ref{fig:VH4lep_variables2} show the distributions of the training variables used for the \VH\ 4 lepton channels.
The score distributions of the BDT for both signal and background in the \VH\ 3-lepton and 4-lepton channels are illustrated in Fig.~\ref{fig:VH_BDT}.

The BDT scores define three \VH\ categories, with two categories, ``VH3LH'' and ``VH3LM'', defined in the 3-lepton channel using the 3-lepton BDT and one category, ``VH4L'', defined in the 4-lepton channel using the 4-lepton BDT.
The dominant background in the \VH\ categories includes diboson, \ttbar, and DY processes, with the diboson process contributing about 70\% (55\%) of the total background in the VH3LH (VH3LM) category, while in the VH4L category, about 98\% of the background is from the $ZZ$ process.
The expected signal yield based on the SM assumption is 1.4, 2.8, and 0.5 in the VH3LH, VH3LM, and VH4L categories, respectively.
The \VH\ production is expected to have a signal purity of 89\% in the VH3LM category and more than 99\% in the VH3LH and VH4L categories relative to other Higgs boson production modes.

\subsection{\ggF\ and \VBF\ categories}

The remaining events that are not selected by the \ttH\ and \VH\ categories are classified into the \ggF\ and \VBF\ categories.
The events are divided into three channels based on jet multiplicities: 0-jet, 1-jet, and 2-jet, where 2-jet includes events with the number of jets $n_j \geq 2$.
Events containing $b$-jets tagged at the 60\% efficiency working point or a third muon with $\pT > 15$ GeV are rejected.

In total four BDTs are trained for the \ggF\ and \VBF\ categories. One of them (\VBF\ classifier) focuses on differentiating \VBF\ production in 2-jet channel, while the other three (\ggF\ classifiers) target all \Hmm\ signals in 0-jet, 1-jet, 2-jet each. For the \VBF\ classifier, \VBF\ $H\to\mu\mu$ events with $n_j \geq 2$ are used as the training signal, while the SM background with $n_j \geq 2$ is used as the training background. For the \ggF\ classifiers, all \Hmm\ signal events with $n_j = 0,\,1,\,\geq2$ are used as the training signal, and the SM background with $n_j = 0,\,1,\,\geq2$ is used as the training background, for 0-jet, 1-jet, 2-jet, respectively. All of the training events are required to be within the range of $120 < \mmumu < 130$ GeV. Seventeen training variables used for the BDTs are listed below:

In total, four BDTs are trained for the \ggF\ and \VBF\ categories.
One of them (the \VBF\ classifier) focuses on differentiating \VBF\ production in the 2-jet channel, while the other three (\ggF\ classifiers) target all \Hmm\ signals in the 0-jet, 1-jet, and 2-jet channels, respectively.
For the \VBF\ classifier, \VBF\ \Hmm\ events with $n_j \geq 2$ are used as the training signal, while the SM background with $n_j \geq 2$ is used as the training background.
For the \ggF\ classifiers, all \Hmm\ signal events with $n_j = 0,\,1$ and $\geq2$ are used as the training signal, and the SM background with $n_j = 0,\,1$ and $\geq2$ is used as the training background for the 0-jet, 1-jet, and 2-jet channels, respectively.
All of the training events are required to be within the range of $120 < \mmumu < 130$ GeV.
Seventeen training variables are used for the BDTs and are listed below:

\begin{itemize}
\item $p_\mathrm{T}^{\mu\mu}$: transverse momentum of the dimuon system.
\item $y_{\mu\mu}$: rapidity of the dimuon system.
\item $\cos\theta^*$: cosine of the lepton decay angle in the Collins-Soper frame.
\item $p_\mathrm{T}^{j_{1\left(2\right)}}$: transverse momentum of the leading (subleading) jet if present.
\item $\eta_{j_{1\left(2\right)}}$: pseudorapidity of the leading (subleading) jet if present.
\item $\Delta\phi_{\mu\mu,j_{1\left(2\right)}}$: azimuthal separation between $H\to\mu\mu$ and the leading (subleading) jet if present.
\item $N_{\mathrm{track}}^{j_{1\left(2\right)}}$: number of ID tracks with $p_{\mathrm{T}} > 0.5$ GeV associated with the leading (subleading) jet if present with $p_{\mathrm{T}} > 50$ GeV and $|\eta| < 2.1$.
\item $p_\mathrm{T}^{jj}$: transverse momentum of the dijet system formed by the two leading jets if present.
\item $y_{jj}$: rapidity of the dijet system if present.
\item $\Delta\phi_{\mu\mu,jj}$: azimuthal separation between $H\to\mu\mu$ and the dijet system if present.
\item $m_{jj}$: invariant mass of the dijet system if present.
\item $E^\mathrm{miss}_\mathrm{T}$: missing transverse momentum.
\item $H_\mathrm{T}$: scalar sum of the transverse momenta of all jets.
\end{itemize}

\begin{figure}[h!]
  \centering
  \subfigure[]{\includegraphics[width=0.45\textwidth]{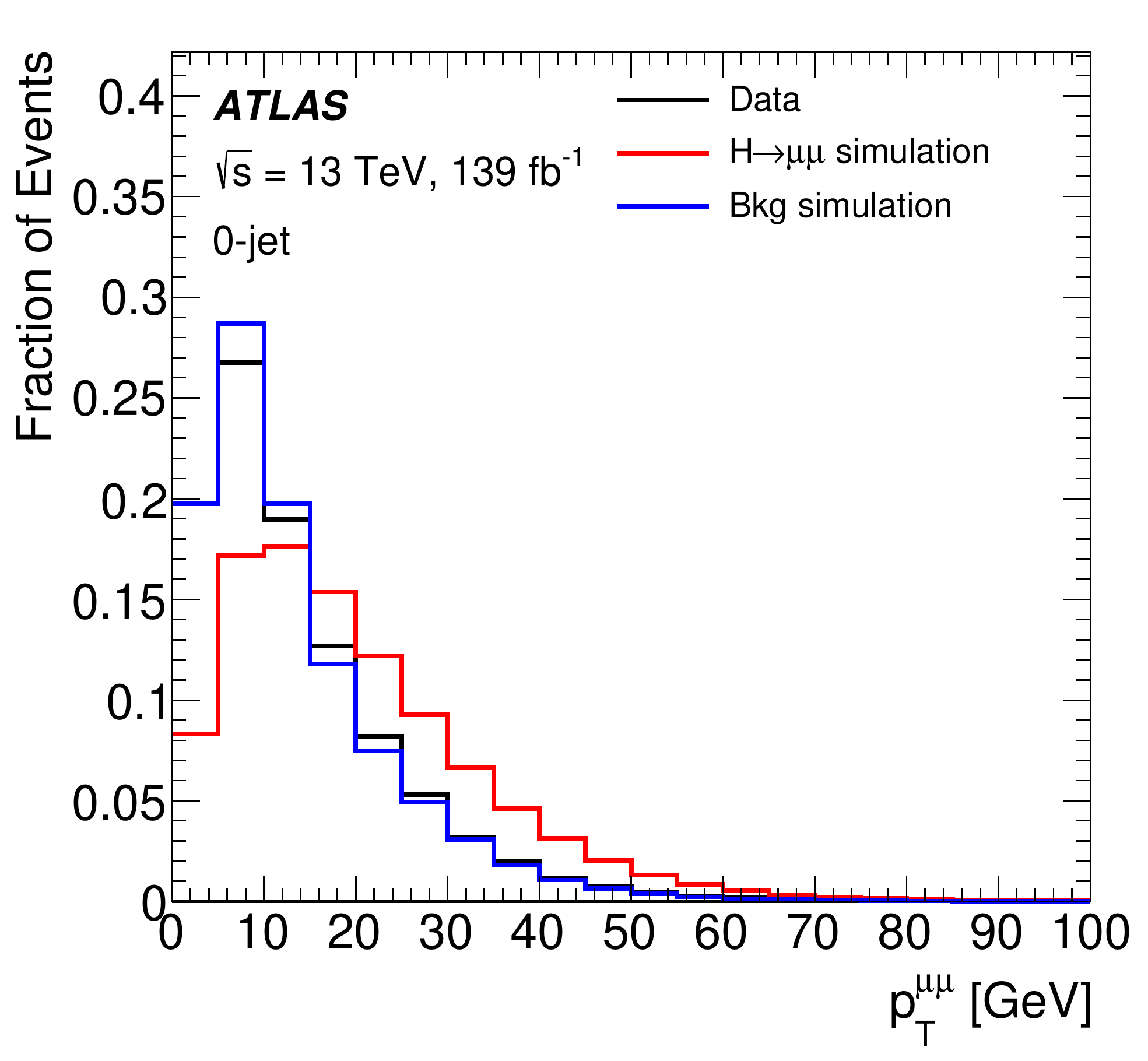}}
  \subfigure[]{\includegraphics[width=0.45\textwidth]{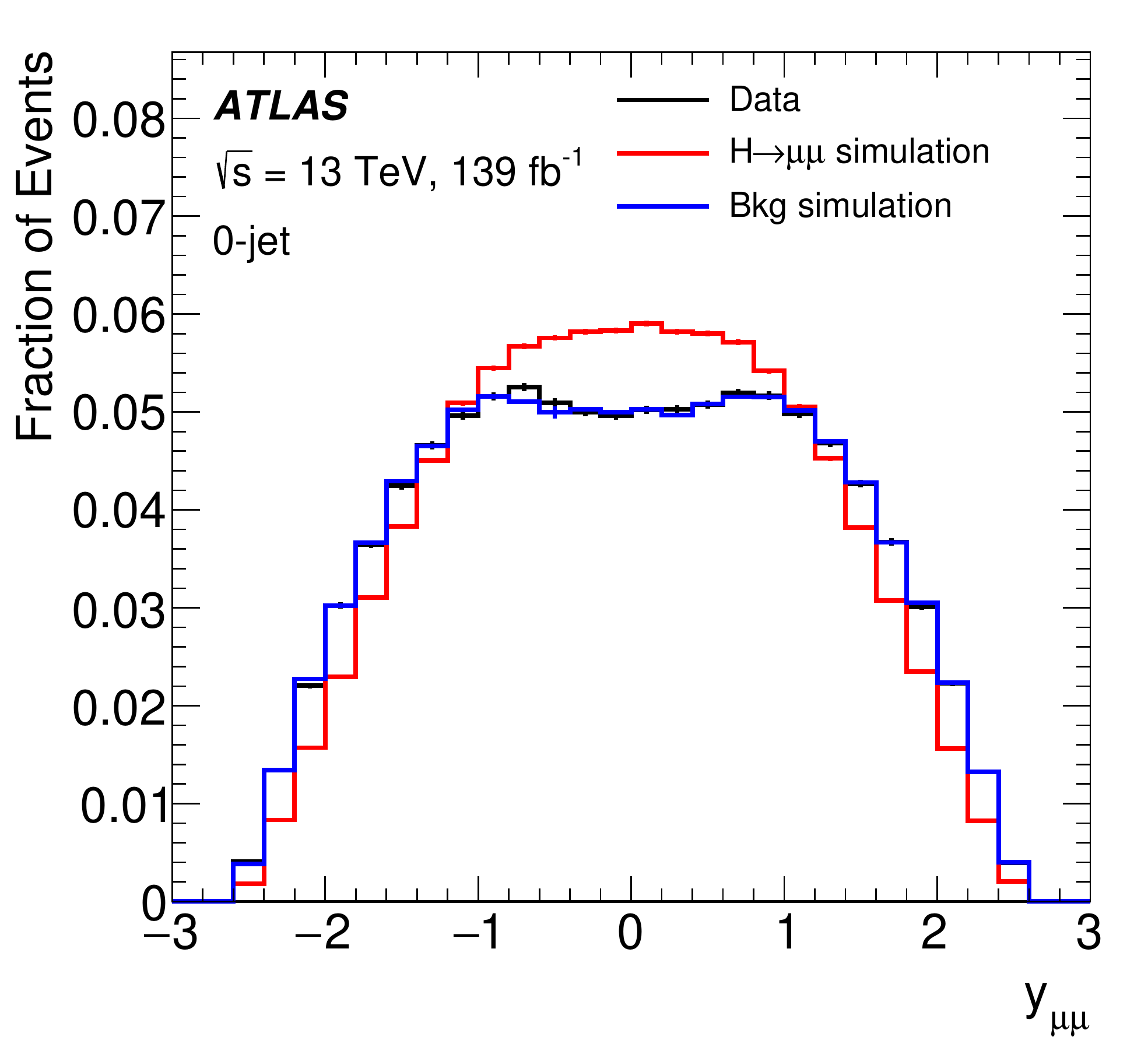}}
  \subfigure[]{\includegraphics[width=0.45\textwidth]{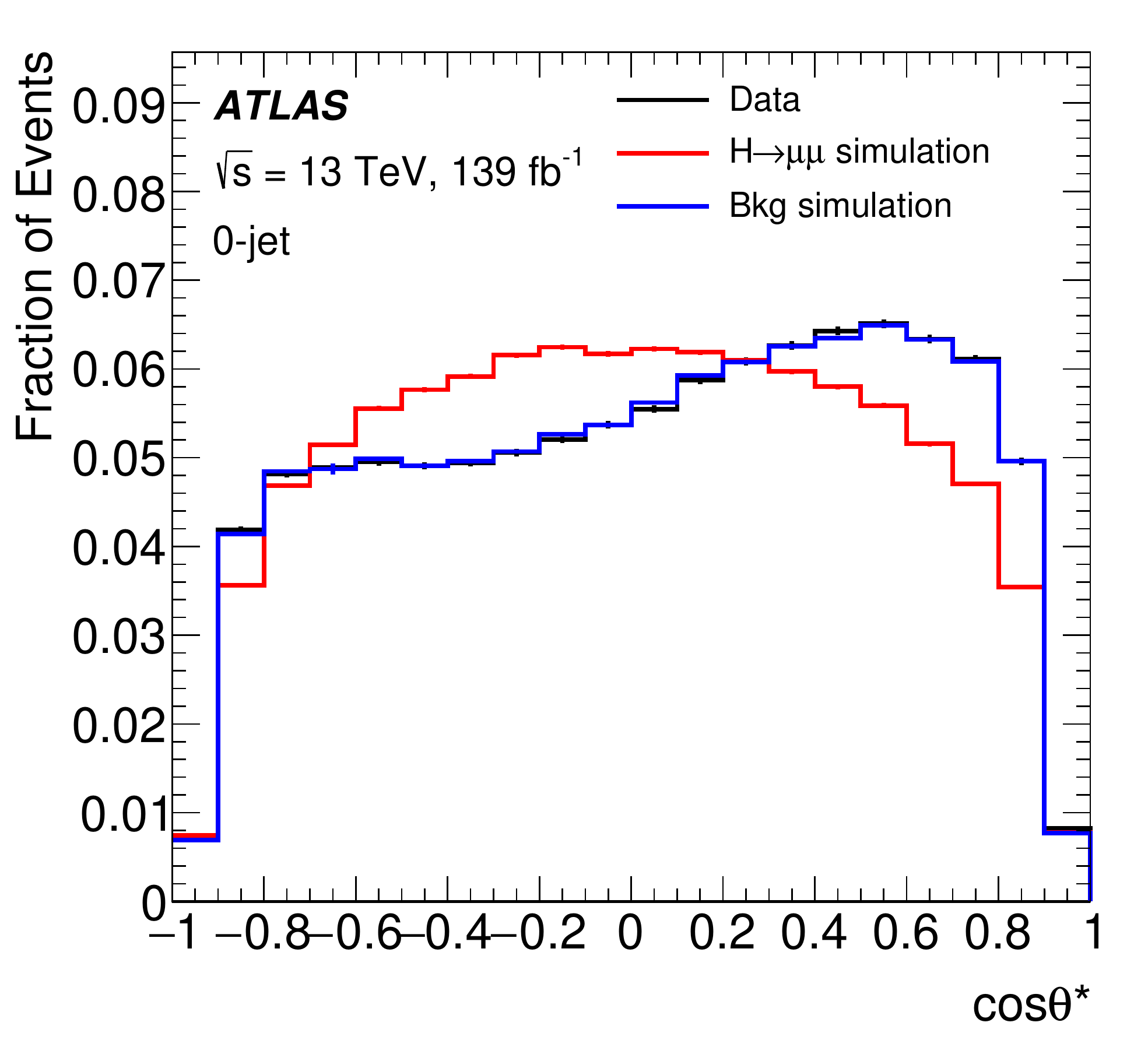}}
  \caption{Distributions of each training variable in the 0-jet channel.}
  \label{fig:0jet_variables}
\end{figure}

\begin{figure}[h!]
  \centering
  \subfigure[]{\includegraphics[width=0.45\textwidth]{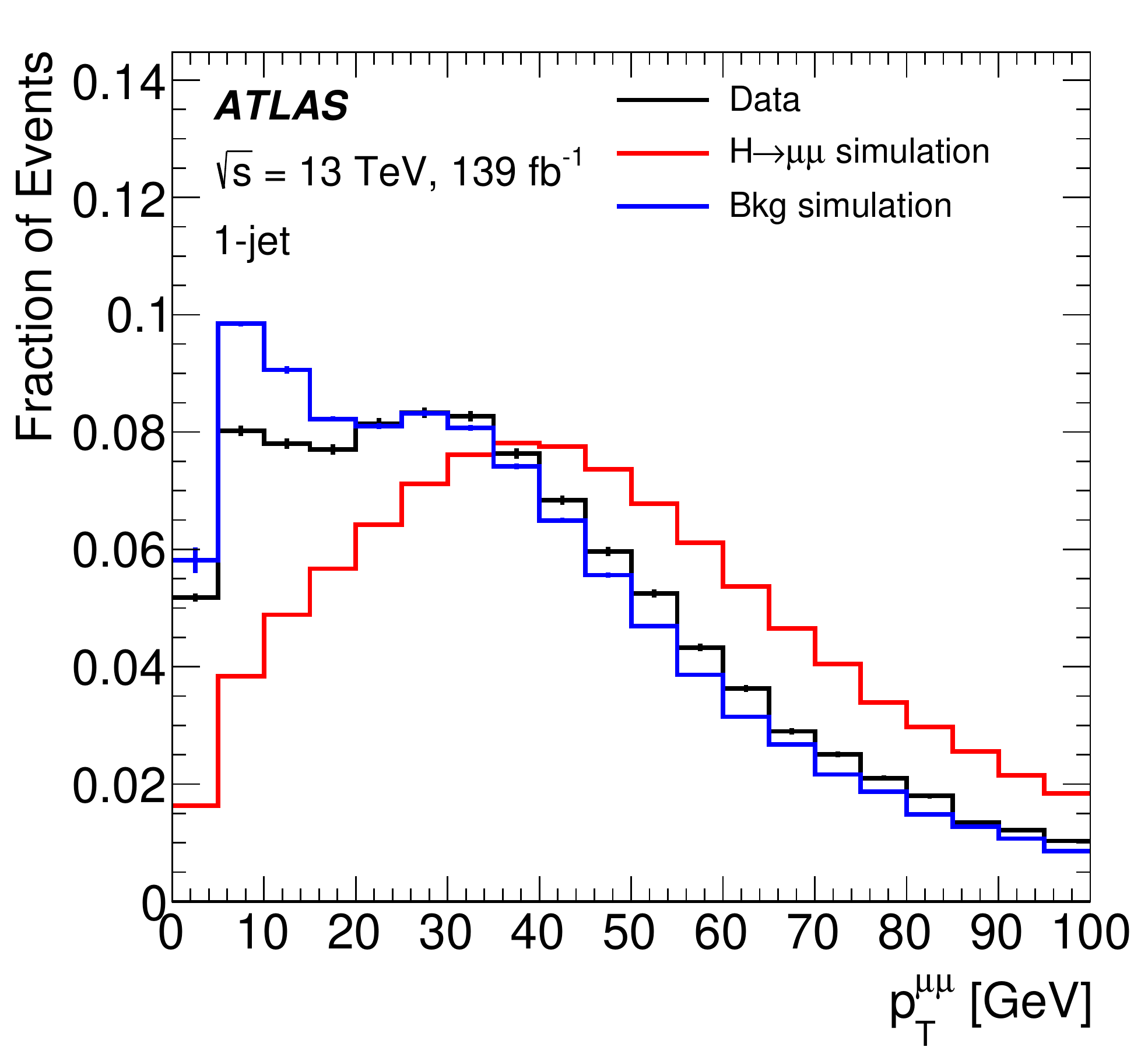}}
  \subfigure[]{\includegraphics[width=0.45\textwidth]{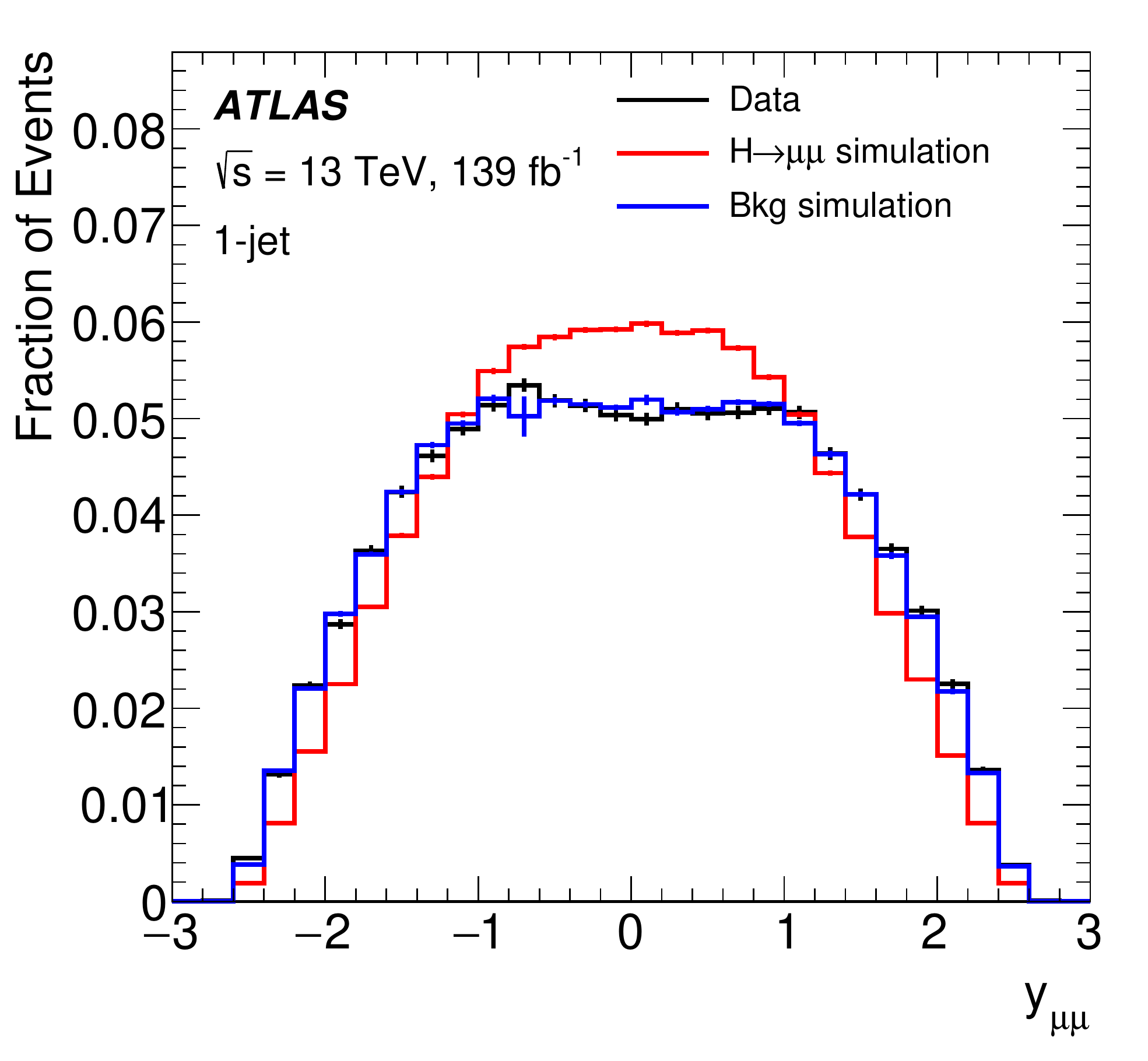}}
  \subfigure[]{\includegraphics[width=0.45\textwidth]{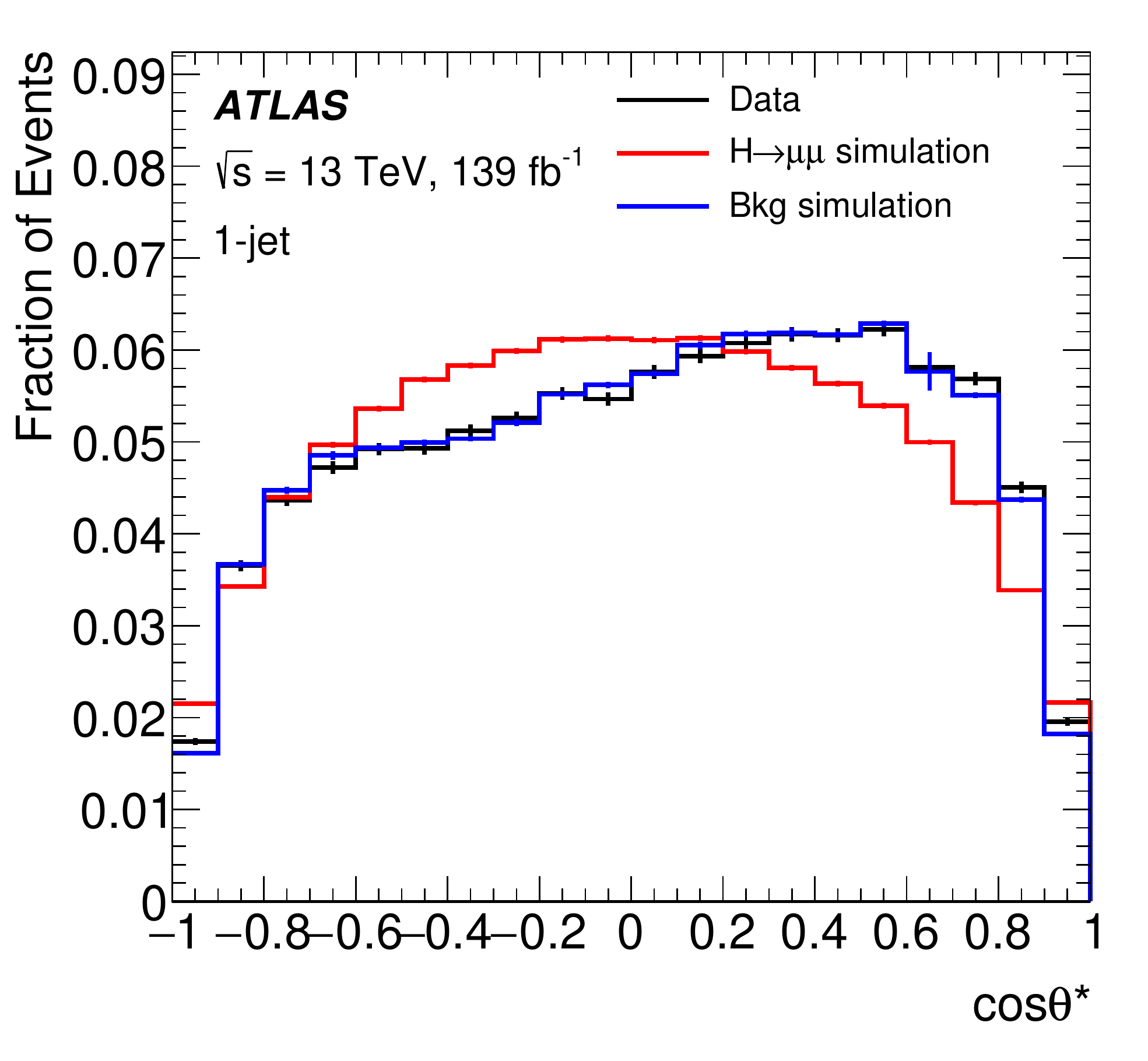}}
  \subfigure[]{\includegraphics[width=0.45\textwidth]{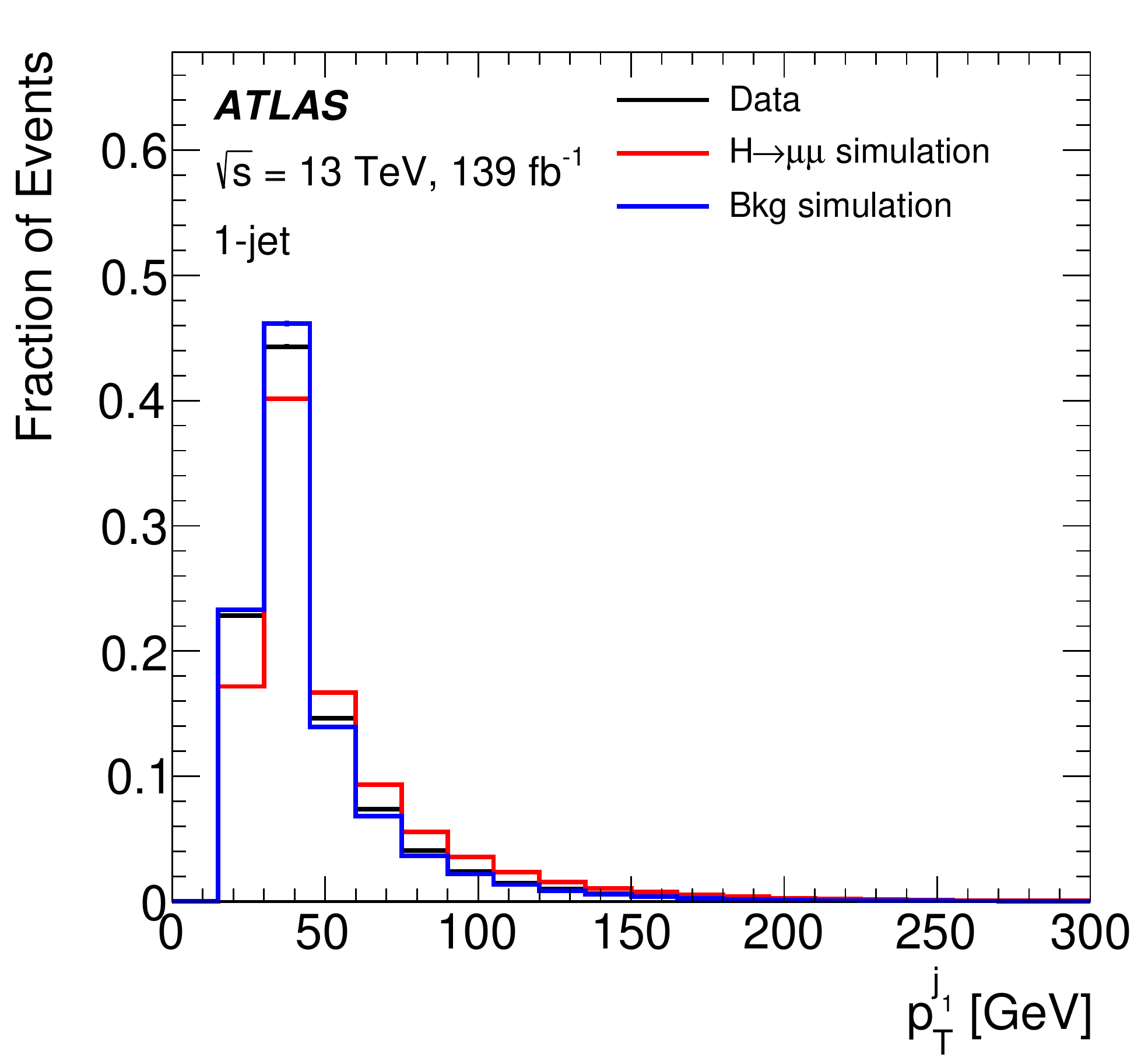}}
  \caption{Distributions of each training variable in the 1-jet channel (part 1).}
  \label{fig:1jet_variables1}
\end{figure}

\begin{figure}[h!]
  \centering
  \subfigure[]{\includegraphics[width=0.45\textwidth]{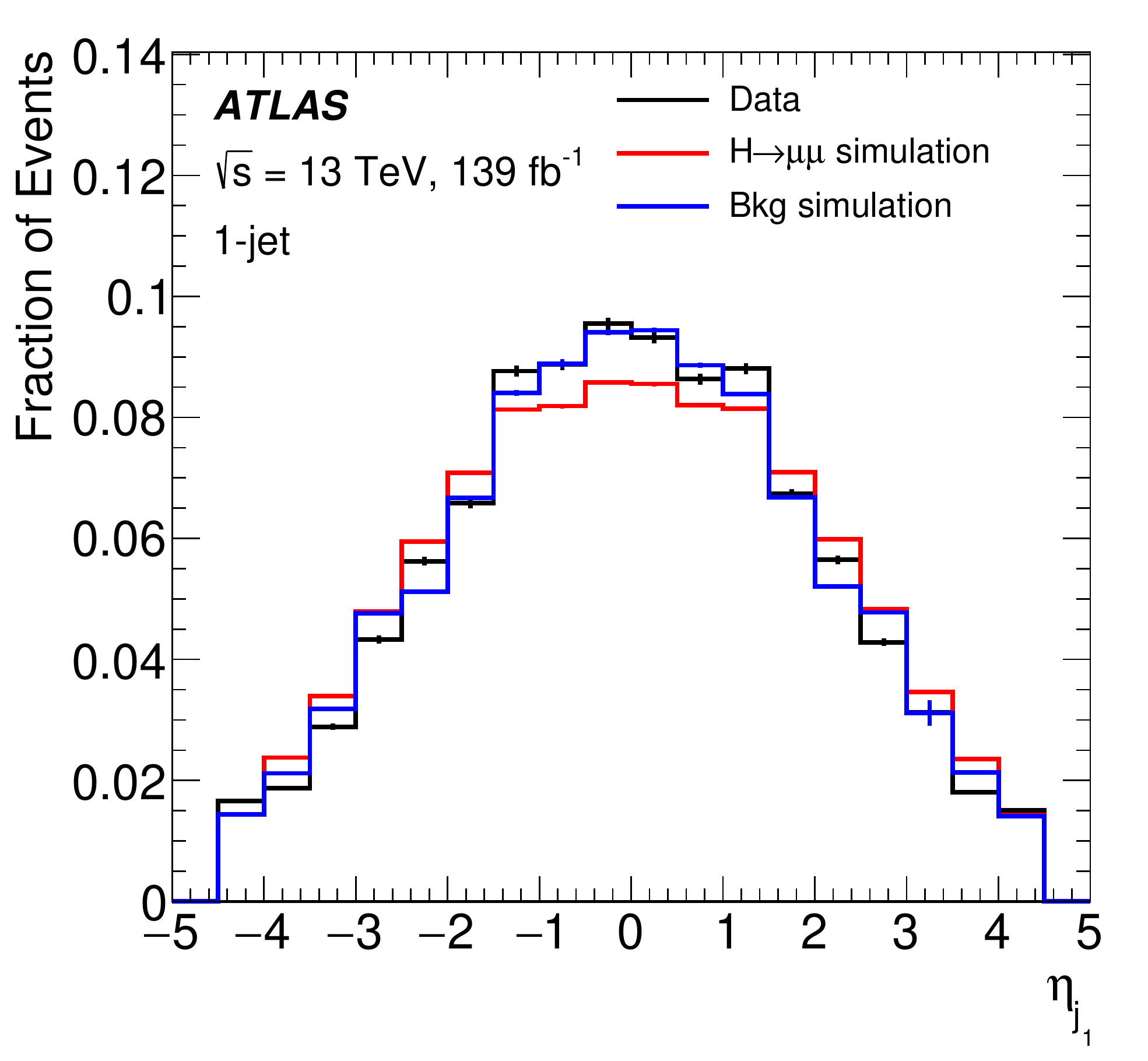}}
  \subfigure[]{\includegraphics[width=0.45\textwidth]{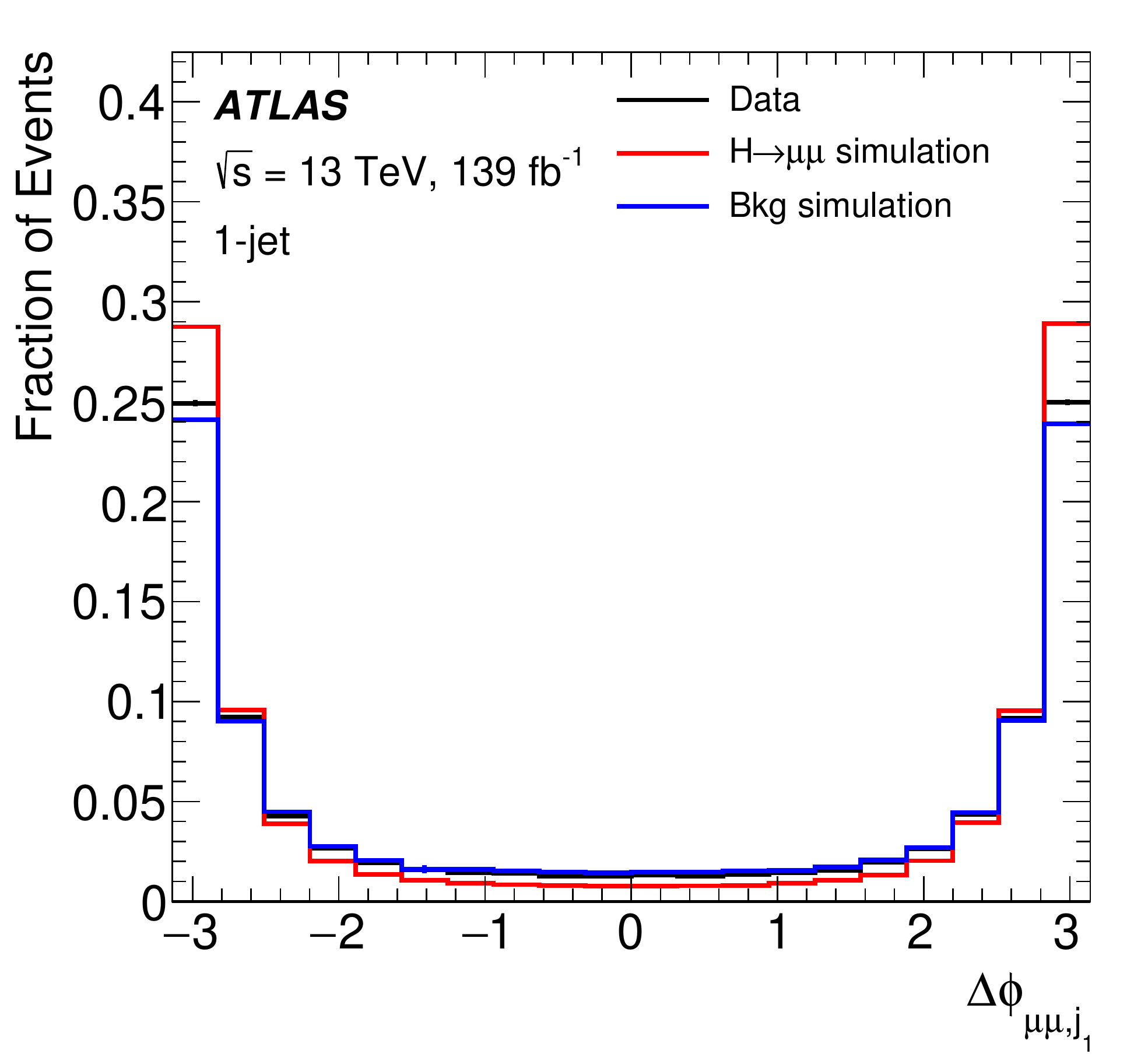}}
  \subfigure[]{\includegraphics[width=0.45\textwidth]{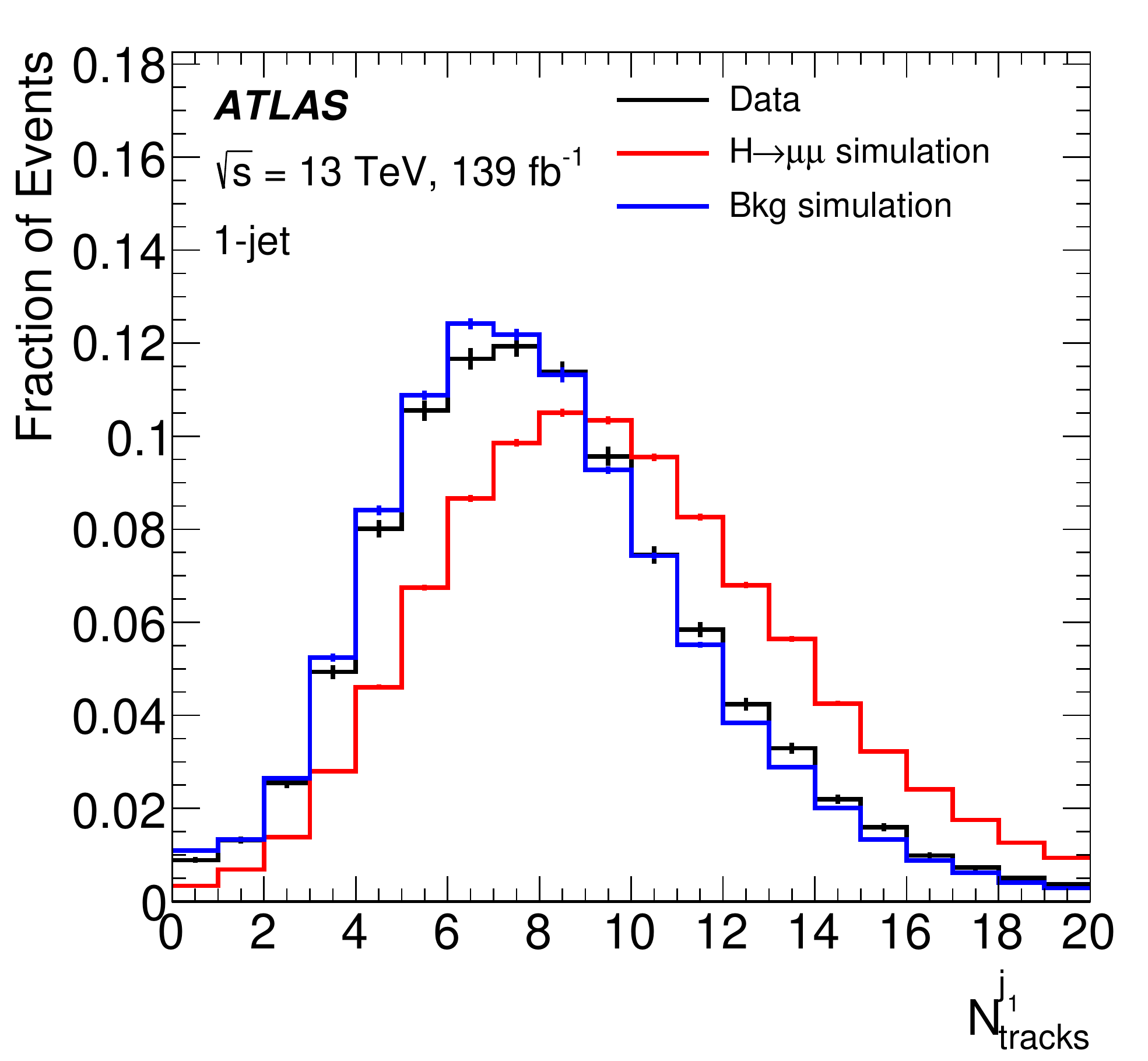}}
  \caption{Distributions of each training variable in the 1-jet channel (part 2).}
  \label{fig:1jet_variables2}
\end{figure}

\begin{figure}[h!]
  \centering
  \subfigure[]{\includegraphics[width=0.40\textwidth]{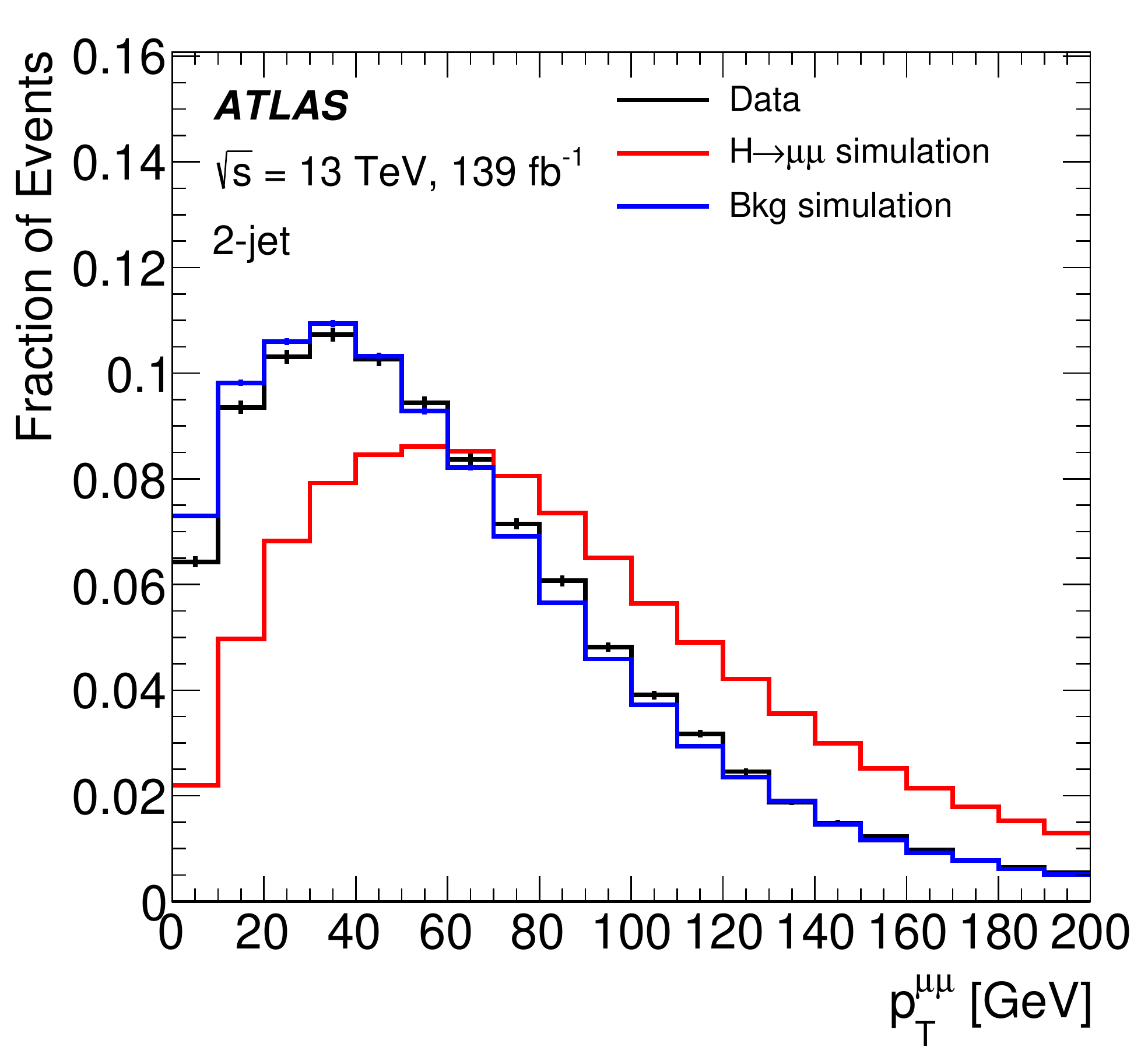}}
  \subfigure[]{\includegraphics[width=0.40\textwidth]{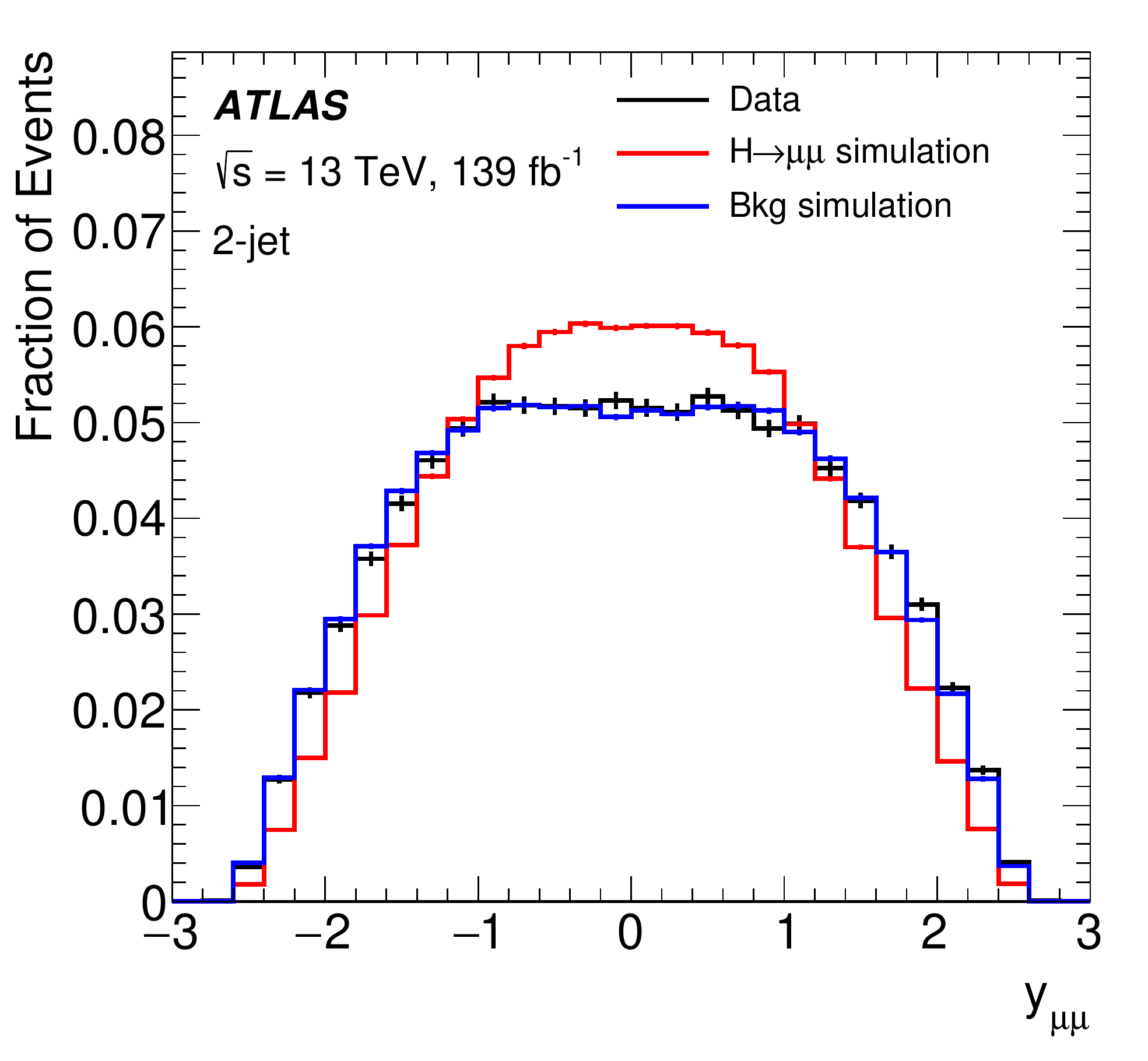}}
  \subfigure[]{\includegraphics[width=0.40\textwidth]{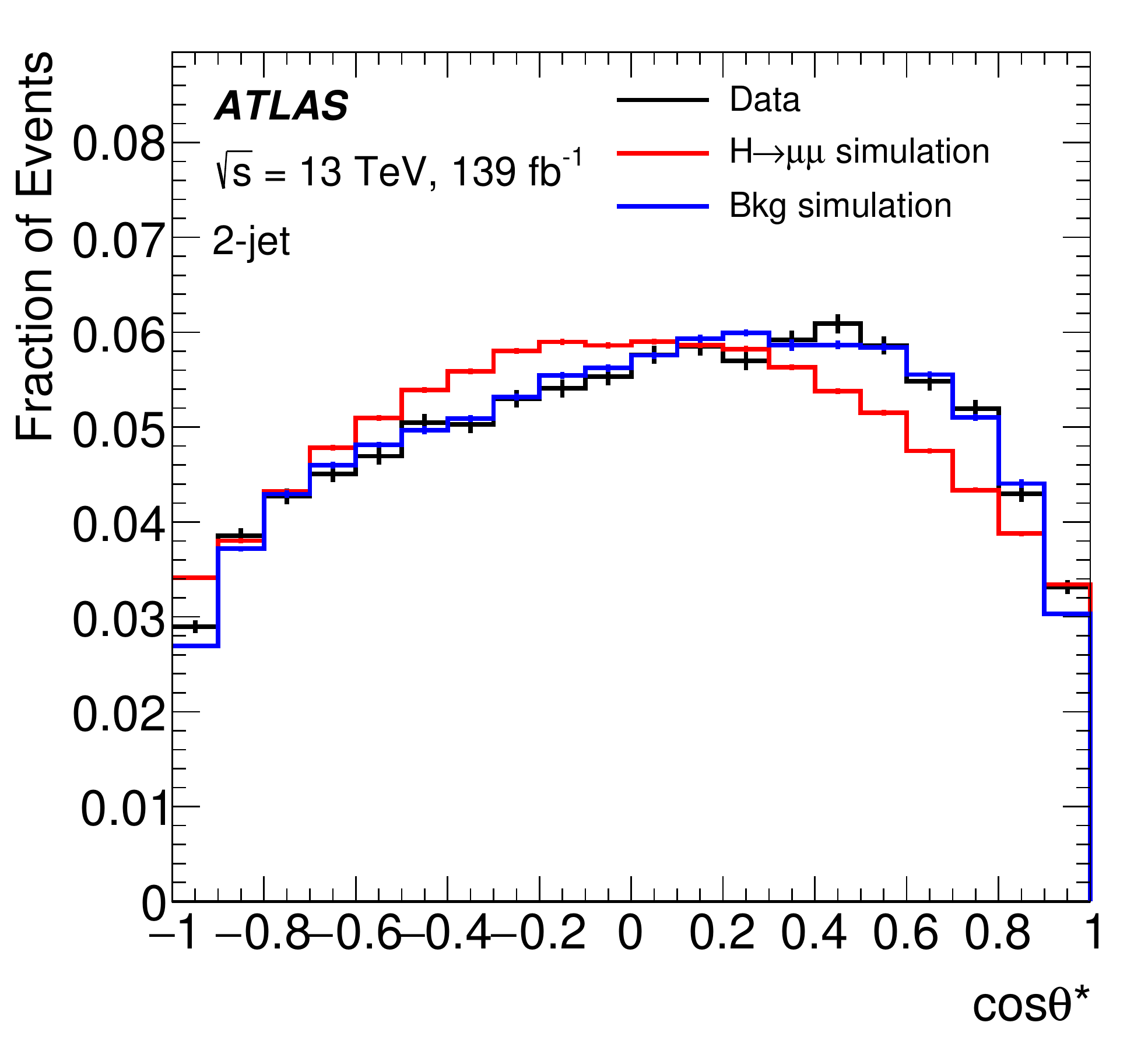}}
  \subfigure[]{\includegraphics[width=0.40\textwidth]{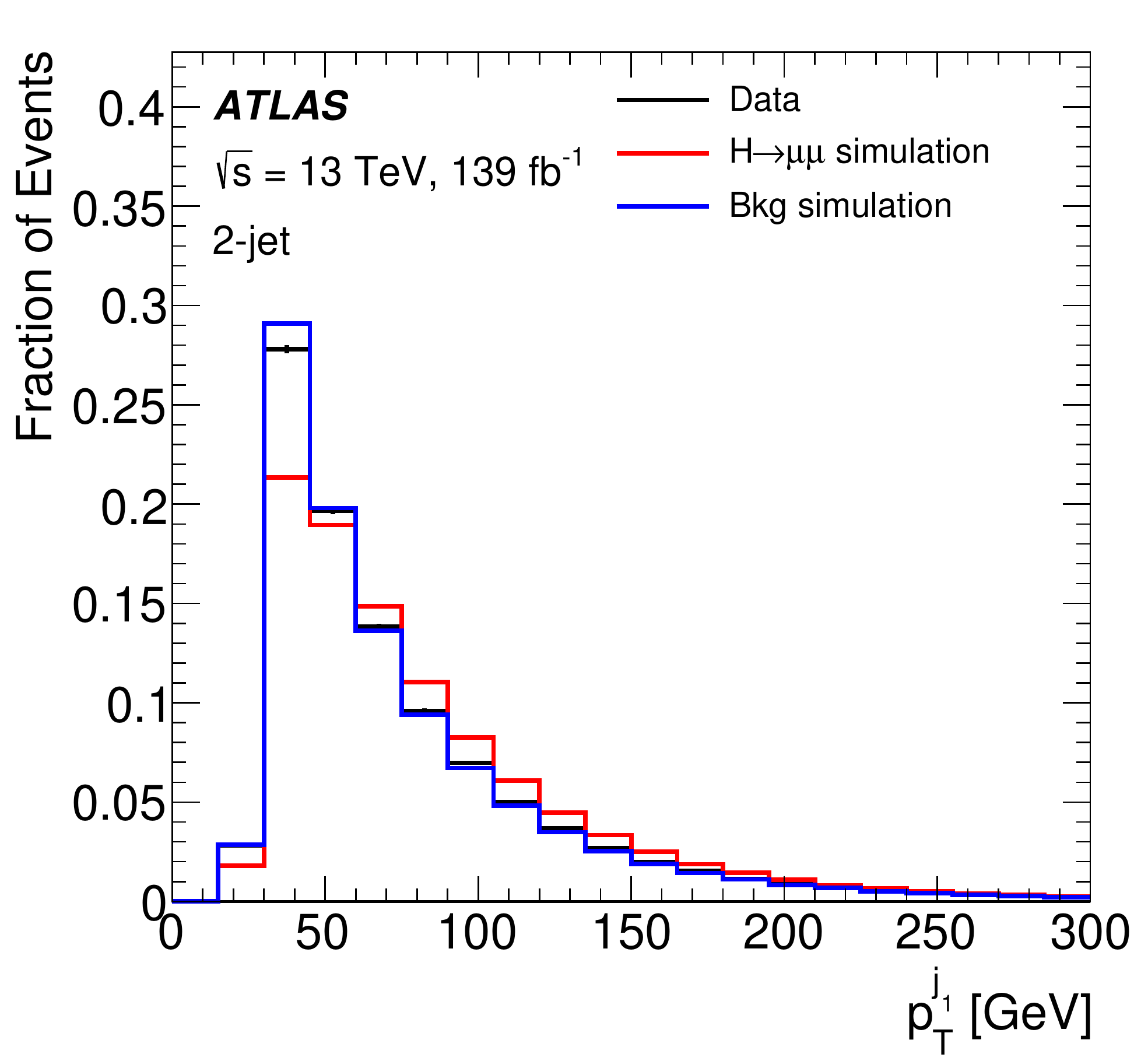}}
  \subfigure[]{\includegraphics[width=0.40\textwidth]{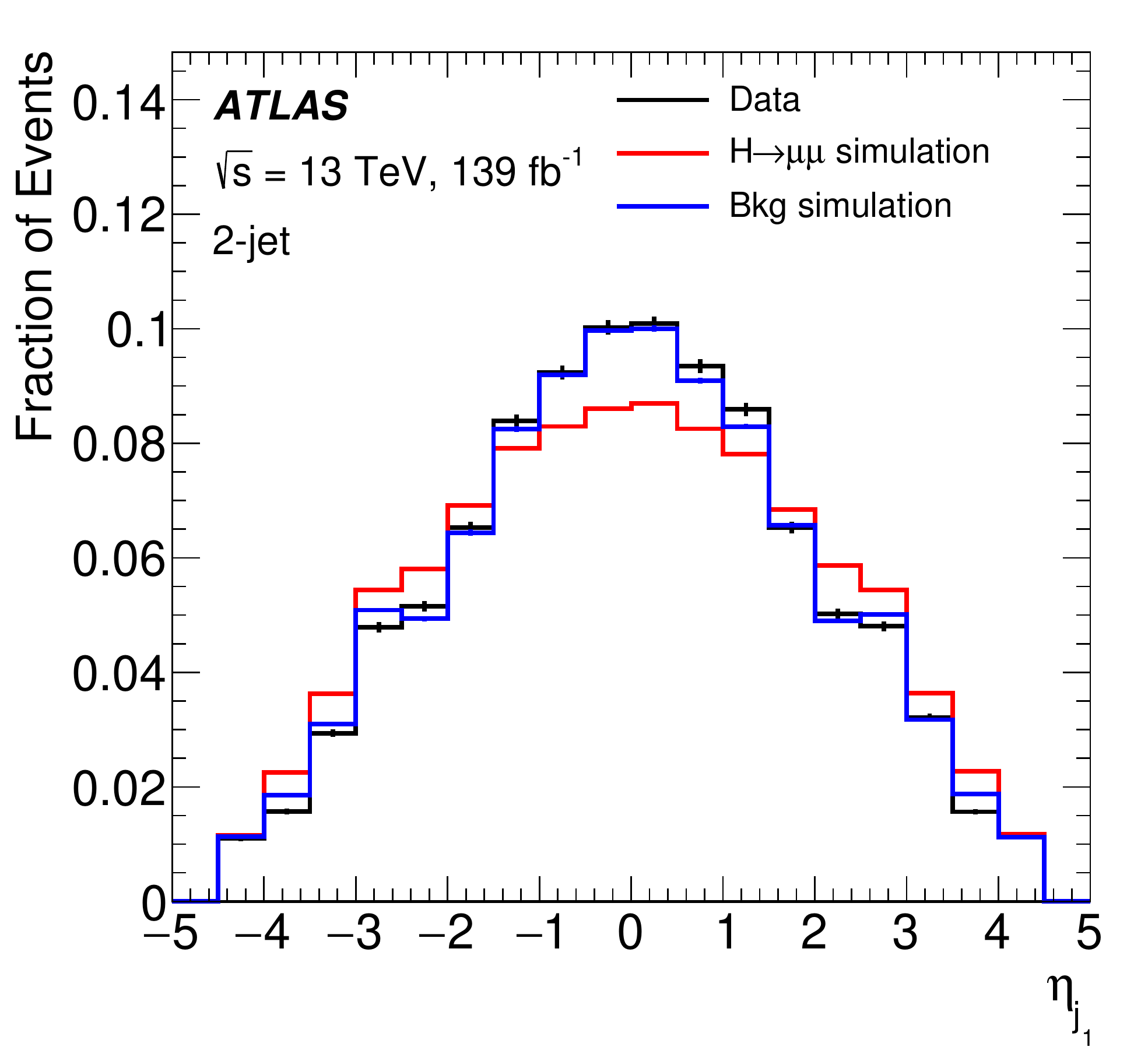}}
  \subfigure[]{\includegraphics[width=0.40\textwidth]{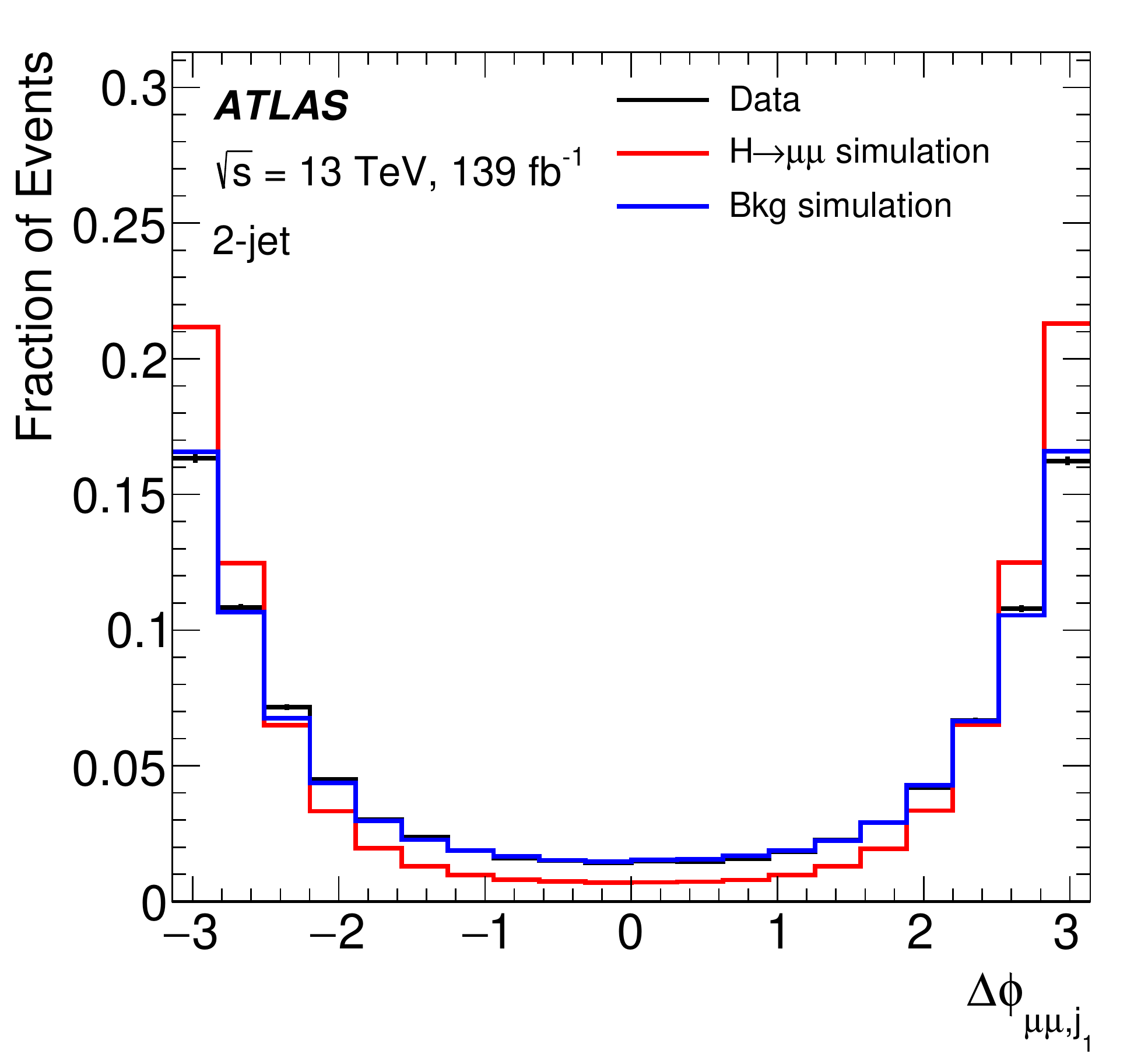}}
  \caption{Distributions of each training variable in the 2-jet channel (part 1).}
  \label{fig:2jet_variables1}
\end{figure}

\begin{figure}[h!]
  \centering
  \subfigure[]{\includegraphics[width=0.40\textwidth]{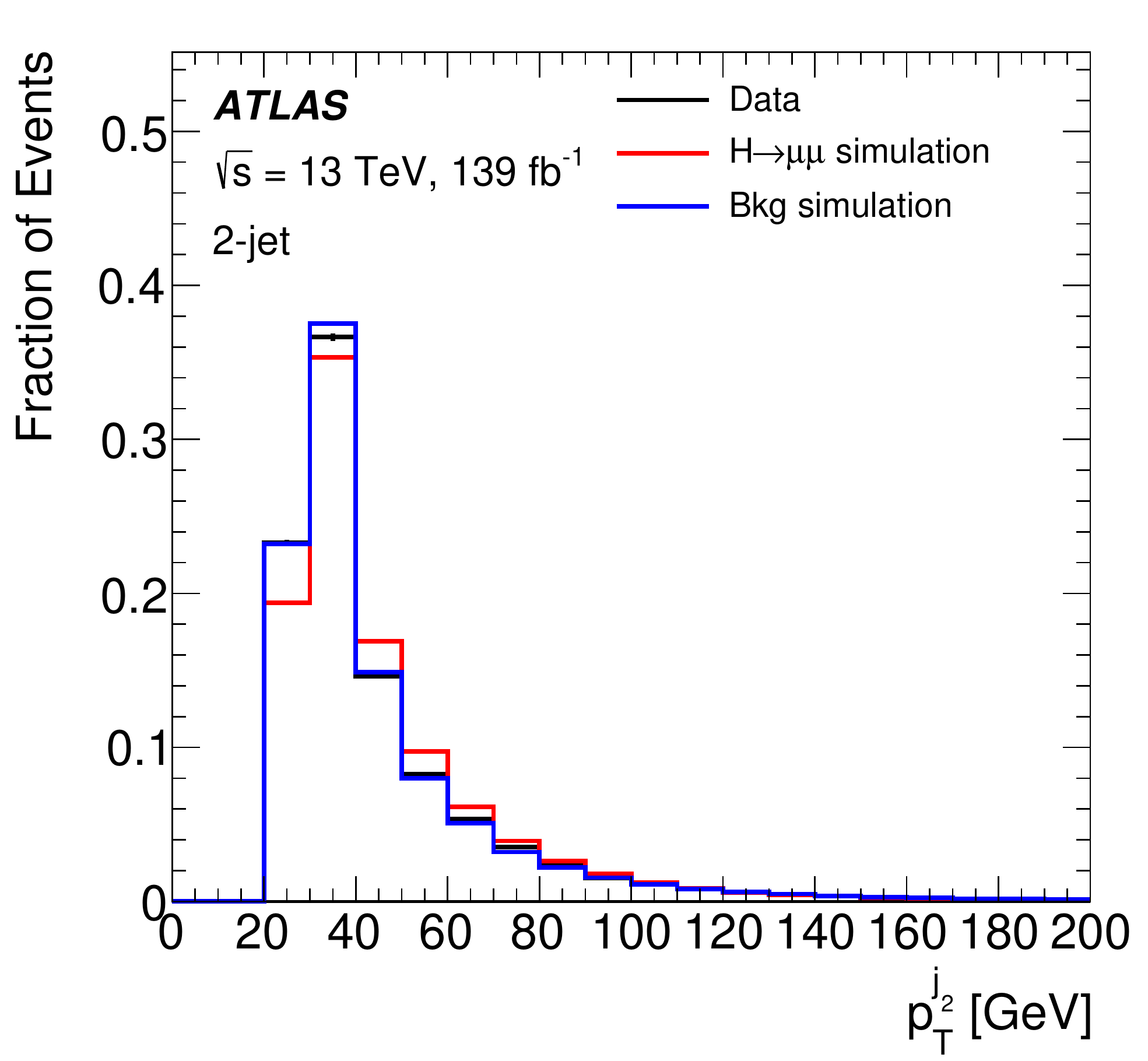}}
  \subfigure[]{\includegraphics[width=0.40\textwidth]{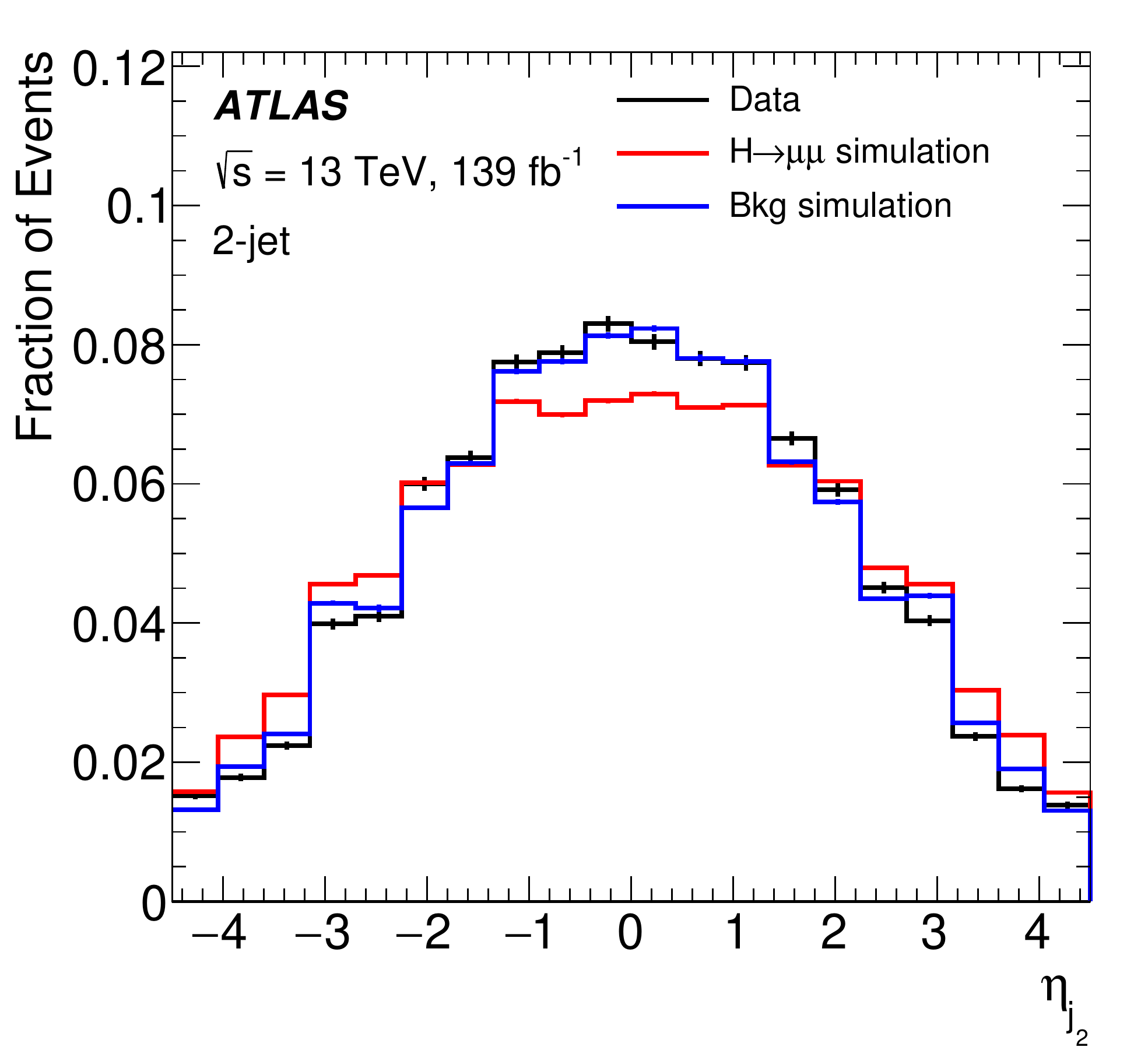}}
  \subfigure[]{\includegraphics[width=0.40\textwidth]{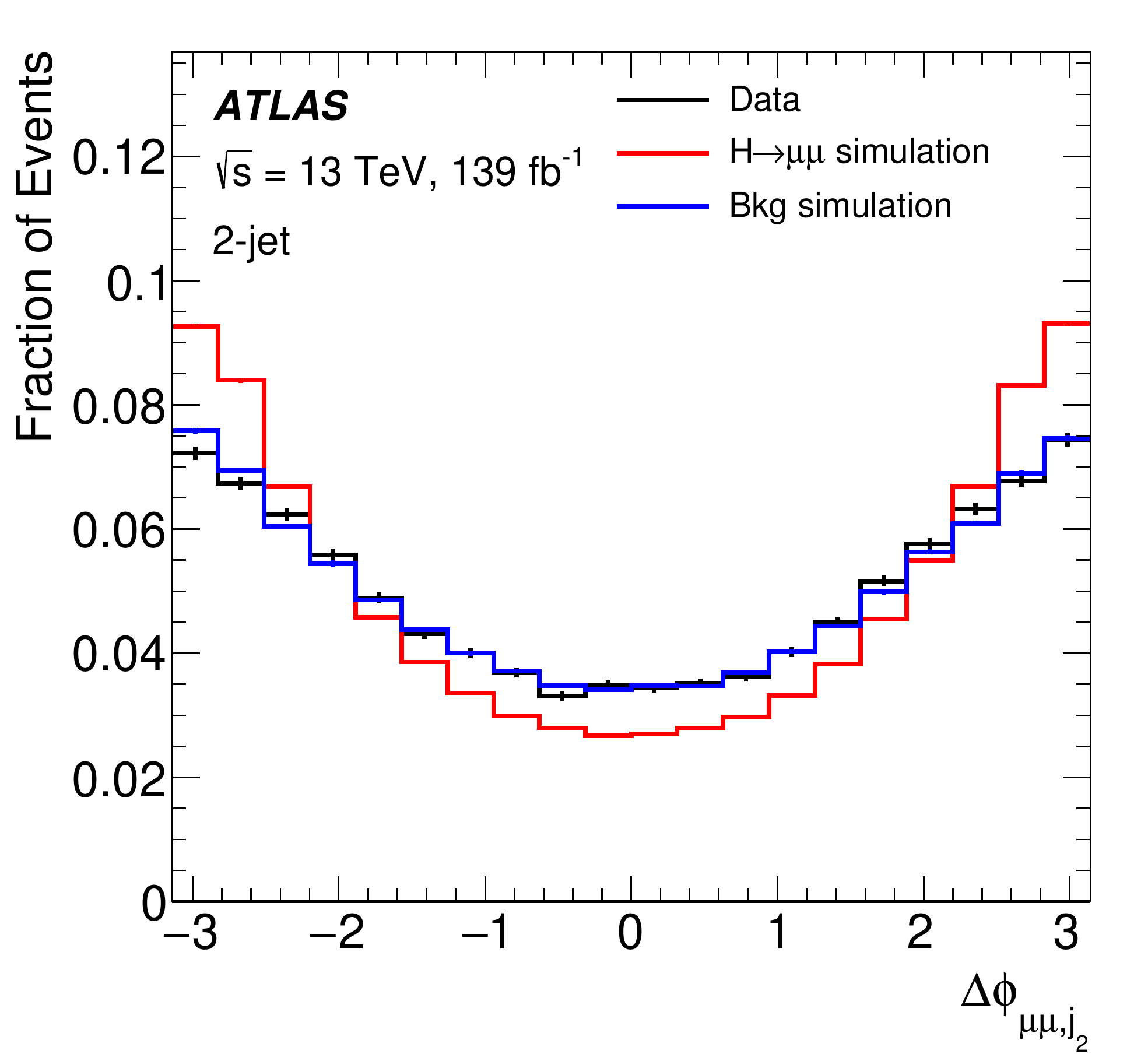}}
  \subfigure[]{\includegraphics[width=0.40\textwidth]{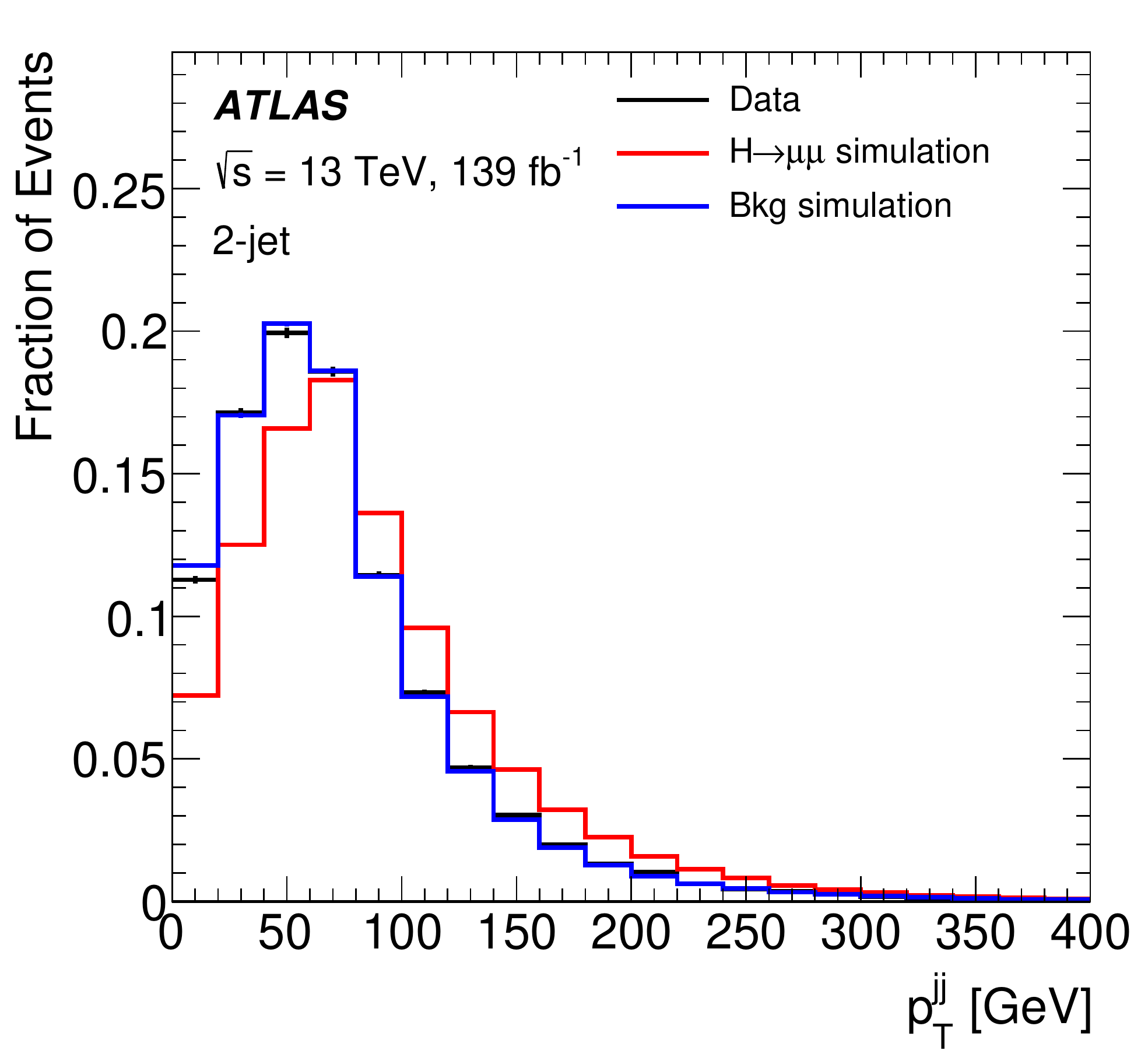}}
  \subfigure[]{\includegraphics[width=0.40\textwidth]{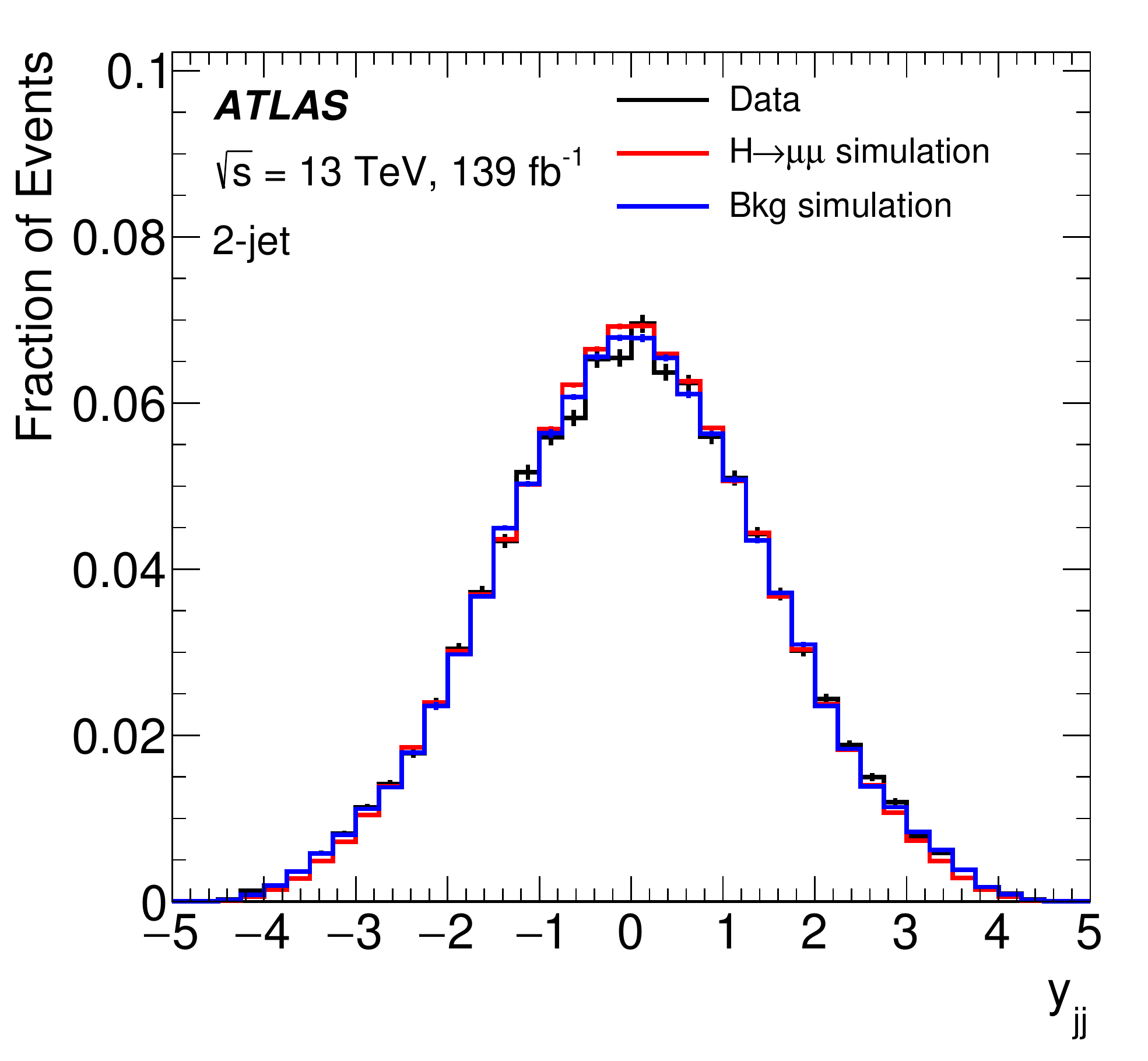}}
  \subfigure[]{\includegraphics[width=0.40\textwidth]{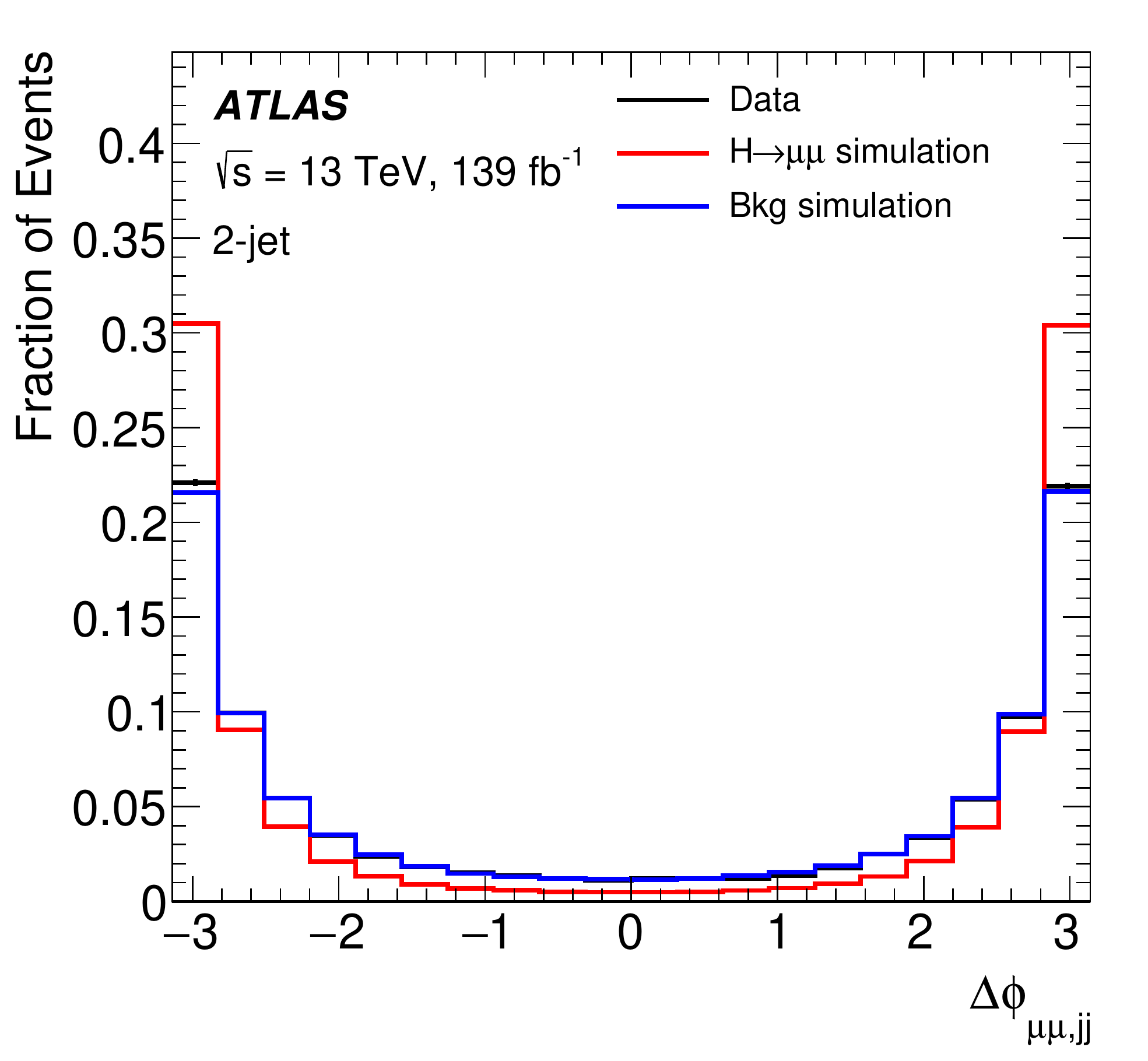}}
  \caption{Distributions of each training variable in the 2-jet channel (part 2).}
  \label{fig:2jet_variables2}
\end{figure}

\begin{figure}[h!]
  \centering
  \subfigure[]{\includegraphics[width=0.45\textwidth]{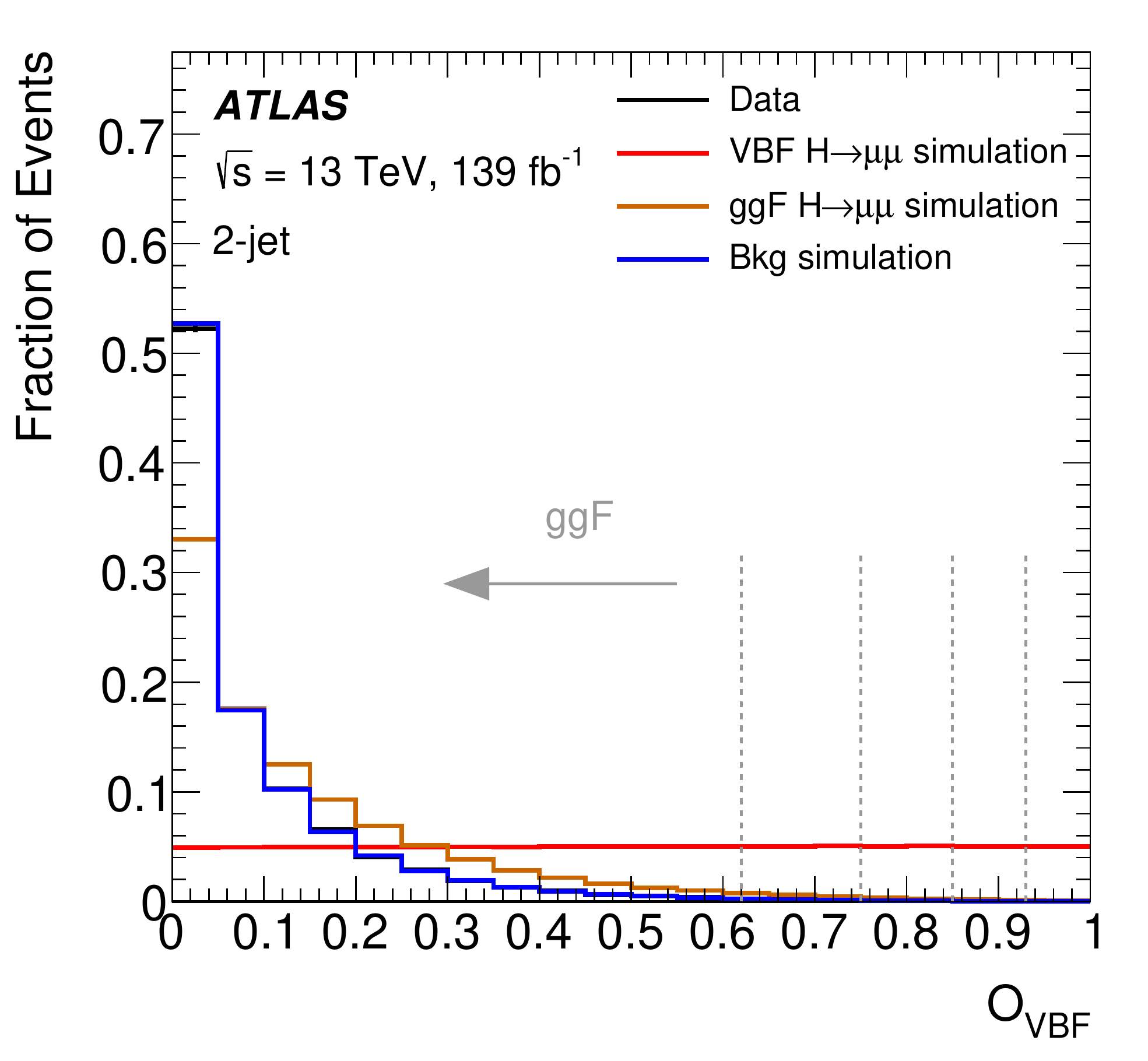}}
  \subfigure[]{\includegraphics[width=0.45\textwidth]{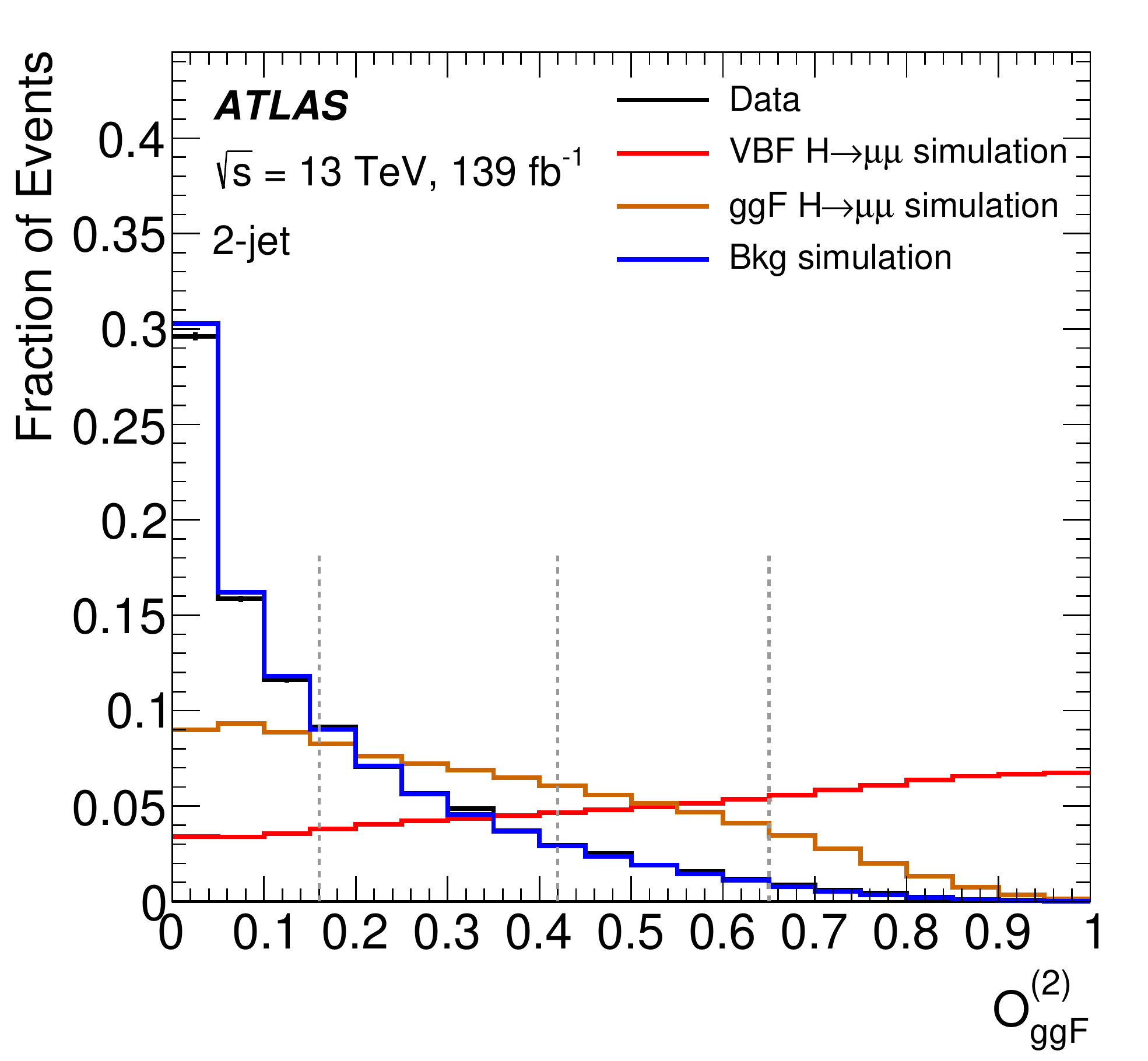}}
  \subfigure[]{\includegraphics[width=0.45\textwidth]{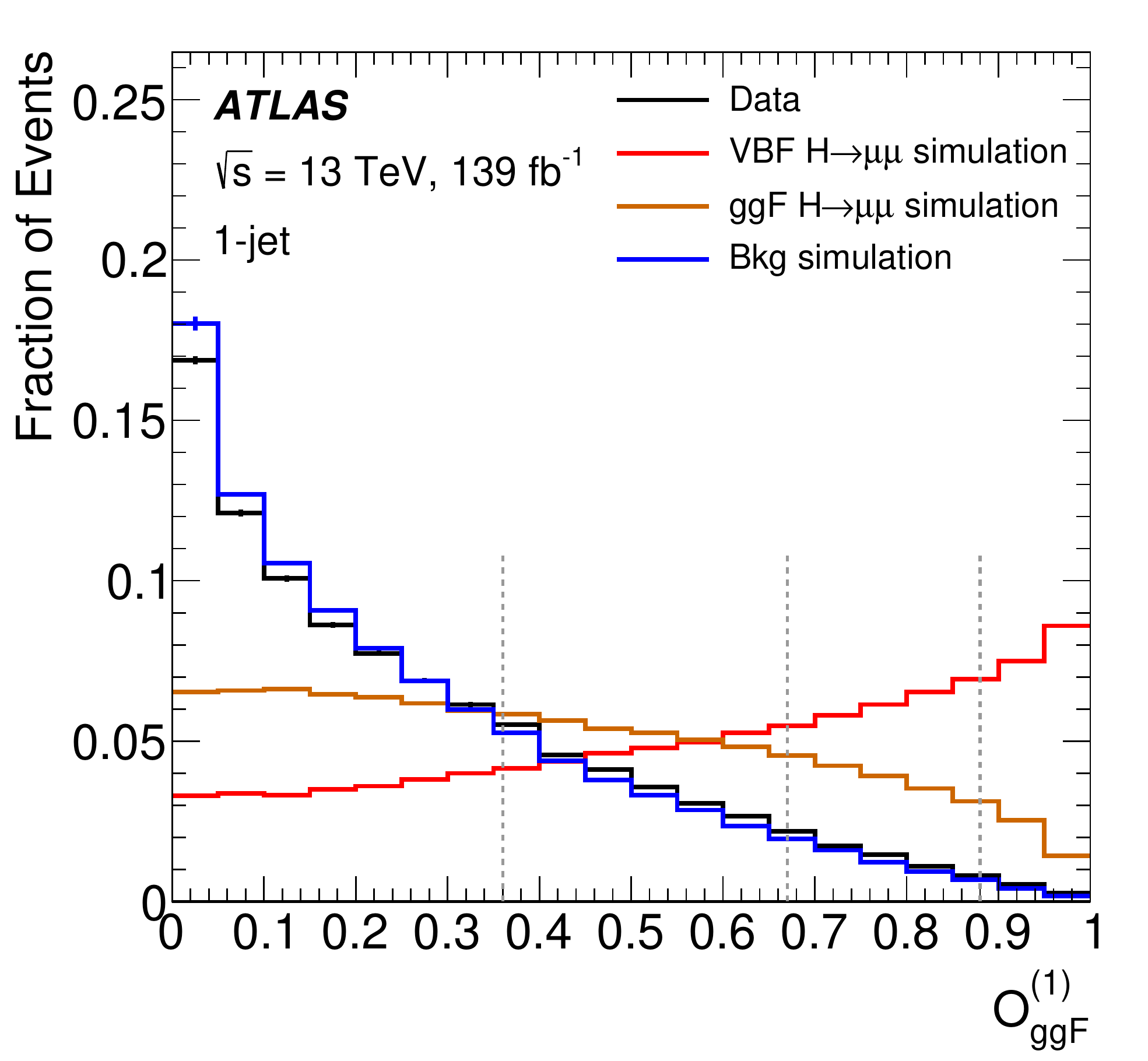}}
  \subfigure[]{\includegraphics[width=0.45\textwidth]{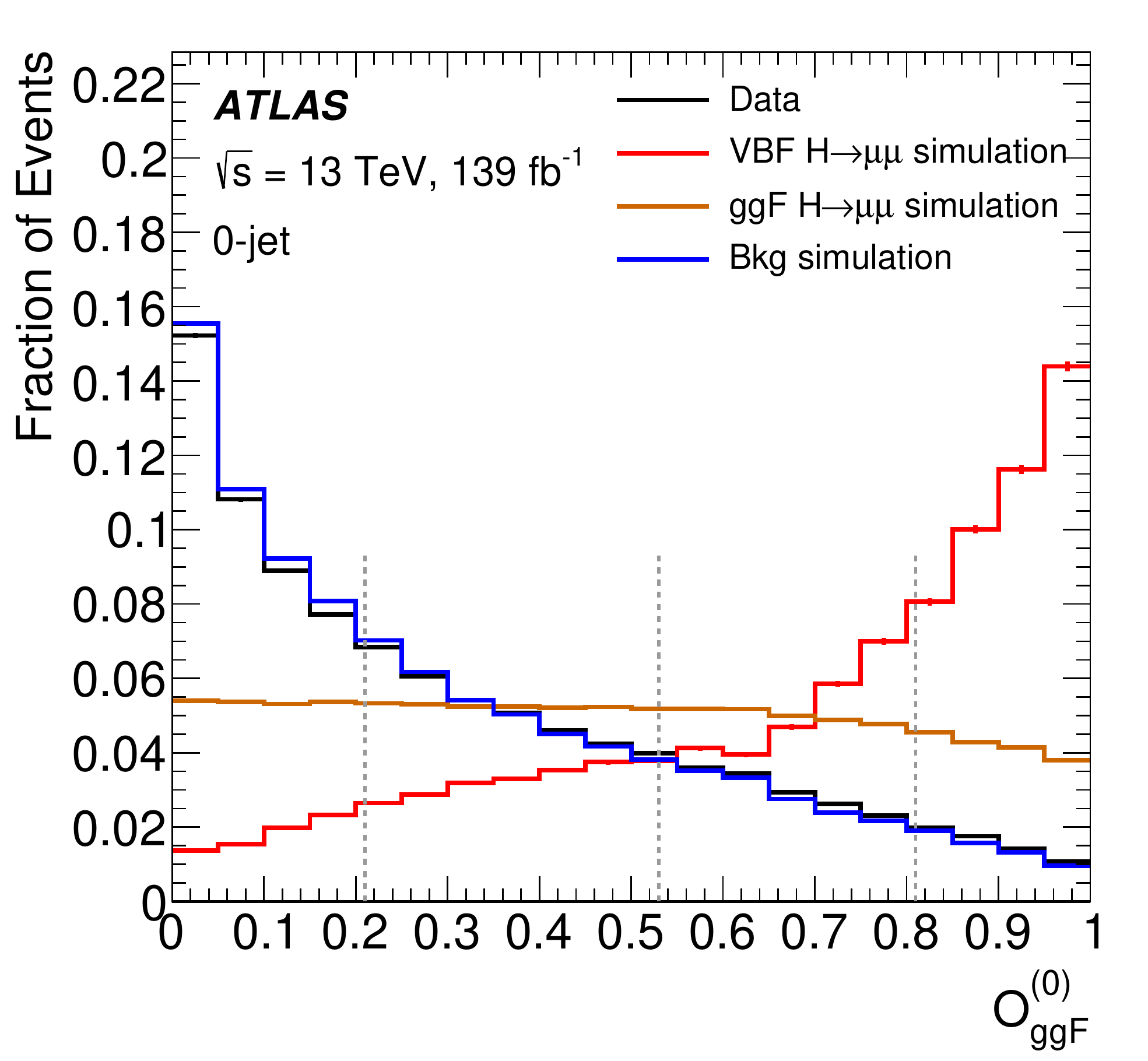}}
  \caption{The \VBF\ (a) and \ggF\ (b)-(d) BDT score distributions for signal and background. The grey vertical dashed lines indicate the BDT score boundary define each categories.}
  \label{fig:ggFVBF_BDT}
\end{figure}

It is worth noting again that variables will be assigned an unphysical arbitrary value if an event does not contain enough objects to define such variables.
Fig.~\ref{fig:0jet_variables}, \ref{fig:1jet_variables1}, \ref{fig:1jet_variables2}, \ref{fig:2jet_variables1} and \ref{fig:2jet_variables2} show the distributions of each training variable in the 0-jet, 1-jet and 2-jet channels, respectively.
Fig.~\ref{fig:ggFVBF_BDT} shows the BDT score distributions in the \ggF\ and \VBF\ categories for both signal and background.

Due to low statistics of sideband data with high VBF classifier output score $O_\mathrm{VBF}$, a function is fit to the VBF classifier output score distribution of sideband data for $O_\mathrm{VBF} \geq 0.5$ in order to reduce the statistical fluctuation when optimizing the BDT boundaries. The function used for the fit is Epoly2, i.e. $\exp(a_1 O_\mathrm{VBF} + a_2 O_\mathrm{VBF}^2)$. The BDT boundaries for the \VBF\ categories in $n_j \geq 2$ are then optimized by maximizing the number counting significance based on the fit function. Note that the raw event yields are being used outside the fit range.

In order to reduce statistical fluctuations when optimizing BDT boundaries for the \VBF\ categories in $n_j \geq 2$, a function is fit to the \VBF\ classifier output score distribution of sideband data with high \VBF\ classifier output score $O_\mathrm{VBF} \geq 0.5$.
The function used for the fit is Epoly2, i.e. $\exp(a_1 O_\mathrm{VBF} + a_2 O_\mathrm{VBF}^2)$.
Raw event yields are used outside the fit range.
The BDT boundaries for the \VBF\ categories are then optimized by maximizing the number counting significance based on the fit function.

The events in the 2-jet channel are sorted into four \VBF\ categories (``VBF Very High'', ``VBF High'', ``VBF Medium'' and ``VBF Low'') based on the \VBF\ classifier output score.
The remaining events in the 2-jet channel are sorted into four 2-jet \ggF\ categories (``2-jet Very High'', ``2-jet High'', ``2-jet Medium'' and ``2-jet Low'') based on the 2-jet \ggF\ classifier output score.
The events in the 1-jet channel are sorted into four 1-jet \ggF\ categories based on the 1-jet \ggF\ classifier output score.
Similarly, the events in the 0-jet channel are sorted into four 0-jet \ggF\ categories based on the 0-jet \ggF\ classifier output score.

\subsection{Summary of categorization}

\begin{figure}[h!]
  \centering
  \includegraphics[width=0.7\textwidth]{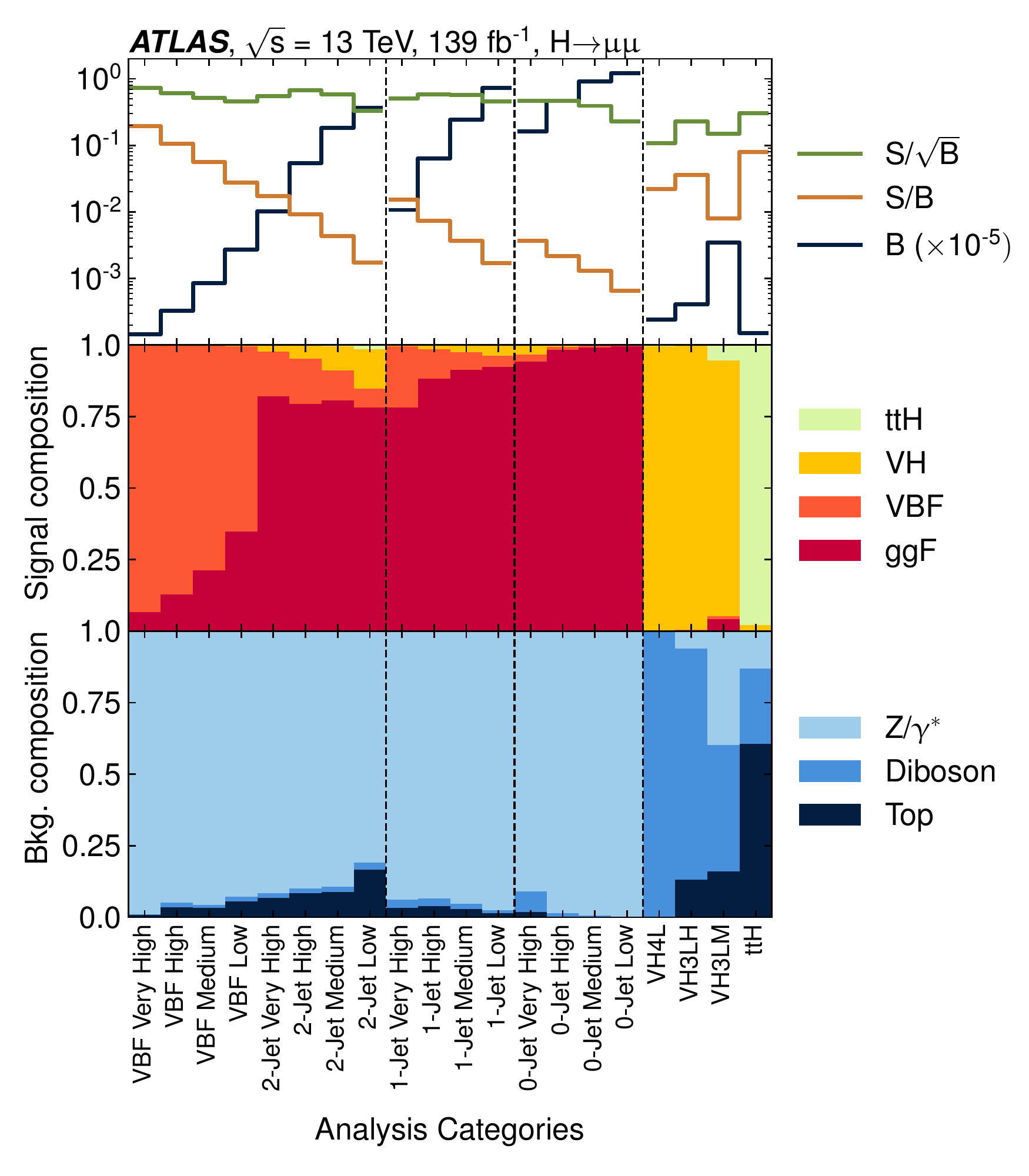}
  \caption{Summary of the signal and background composition in the twenty categories in the mass region $120 \GeV < \mumu\ < 130 \GeV$. The top panel shows the number of background events ($B$, multiplied by a factor of 10-5), the expected signal-to-background ratio for a SM \Hmm\ signal ($S/B$), as well as the number counting significance $S/\sqrt{B}$. The middle panel shows the expected contributions of different Higgs boson production modes to the total signal in each category. The bottom panel shows the background composition from DY, diboson and top-quark processes in simulation.}
  \label{fig:hmumu_categorization}
\end{figure}

Fig.~\ref{fig:hmumu_categorization} provides a summary of all twenty categories, each of which exhibits a signal-to-background ratio ranging from $10^{-3}$ to 0.2.
The DY process serves as the main background in all \ggF\ and \VBF\ categories, whereas the diboson (top) process dominates the background in the \VH\ (\ttH) categories.
Notably, each production mode is clearly separated within its corresponding category.

\section{Statistical analysis}
\label{chap:hmumu:sec:modeling}

To extract the \Hmm\ signal from the observed data, a binned maximum profile likelihood fit is performed simultaneously in all twenty categories using the \mmumu\ spectrum \cite{stat1,stat2}.
The likelihood function is constructed as a product of Poisson probability functions based on the expected signal and background yields in every category and bin considered in the fit, multiplied by Gaussian or log-normal priors which control the nuisance parameters (NP) $\theta$:

\begin{equation}
  \prod_{i \in \mathrm{bins}} \mathrm{Pois}\left( n_i | \mu S_i + B_i \right) \prod_{\theta_j} P_j\left( \theta_j \right),
  \label{eq:likelihood}
\end{equation}
where $n_i$ is the observed number of events, $S_i$ is the expected signal yield and $B_i$ is the expected background yield in bin $i$. $\mu$ is the signal strength, representing the ratio of the observed signal yield to the expected signal yield. In other words, the observed signal yield in bin $i$ is equal to $\mu S_i$. $S_i$ and $B_i$ are controlled by a set of NPs $\theta_j$, which can be correlated or uncorrelated with one another. The prior of the nuisance parameters $P_j$ is either a Gaussian or a log-normal distribution for the constrained case, and a uniform distribution for the unconstrained case. Both $S_i$ and $B_i$ are modeled by analytical functions (functions of \mmumu) in each category.

where $n_i$ is the observed number of events, $S_i$ is the expected signal yield, and $B_i$ is the expected background yield in bin $i$.
The signal strength $\mu$ represents the ratio of the observed signal yield to the expected signal yield.
In other words, the observed signal yield in bin $i$ is equal to $\mu S_i$.
$S_i$ and $B_i$ are controlled by a set of NPs $\theta_j$, which can be either correlated or uncorrelated with one another.
The prior of the nuisance parameters $P_j$ is either a Gaussian or a log-normal distribution for the constrained case, and a uniform distribution for the unconstrained case.
Both $S_i$ and $B_i$ are modeled by analytical functions, which are functions of \mmumu, in each category.

\subsection{Signal modeling}

In SM, the Higgs boson is predicted to be a narrow resonance with a width of 4.1MeV for $m_H = 125.09$ GeV. The observed signal shape is thus determined by detector resolution effects on the muon momentum measurement. Given the assyemtric feature of the detector resolution effects, we parametrize the signal \mmumu\ distribution as a double-sided crystal-ball function \cite{ATLAS:2014jdv} of \mmumu\ in each category:

In SM, the Higgs boson is predicted to have a narrow resonance with a width of 4.1 MeV for $m_H = 125.09$ GeV.
As a result of detector resolution effects on the muon momentum measurement, the observed signal shape is parametrized as a double-sided crystal-ball function of \mmumu\ in each category.
This parametrization is necessary due to the asymmetric nature of detector resolution effects.

\begin{equation}
			f_s(\mmumu;\mu, \sigma, \alpha_{L}, n_{L}, \alpha_{R}, n_{R}) =  \begin{cases}
            A_{L} \cdot (B_{L} - \frac{x - \mu}{\sigma})^{-n},
             & \mbox{for }\frac{x - \mu}{\sigma} < -\alpha_{L} \\
            \exp(- \frac{(x - \mu)^2}{2 \sigma^2}),
            & -\alpha_{L} \leq \mbox{for}\frac{x - \mu}{\sigma} \leq \alpha_{R} \\
            A_{R} \cdot (B_{R} - \frac{x - \mu}{\sigma})^{-n},
             & \mbox{for }\frac{x - \mu}{\sigma} > \alpha_{R}
            \end{cases},
           \label{eq:signal}
\end{equation}
where
\begin{equation}
\begin{split}
			A_{L/R} &= \left(\frac{n_{L/R}}{\left| \alpha_{L/R} \right|}\right)^n_{L/R} \cdot
            \exp\left(- \frac {\left|\alpha_{L/R} \right|^2}{2}\right),\\
            B_{L/R} &= \frac{n_{L/R}}{\left| \alpha_{L/R} \right|}  - \left| \alpha_{L/R} \right|,
\end{split}
\end{equation}
where the signal shape parameters $\mu$, $\sigma$, $\alpha_{L}$, $n_{L}$, $\alpha_{R}$ and $n_{R}$ are determined and fixed to a constant when statistical fitting is performed.

\begin{figure}[h!]
  \centering
  \includegraphics[width=0.6\textwidth]{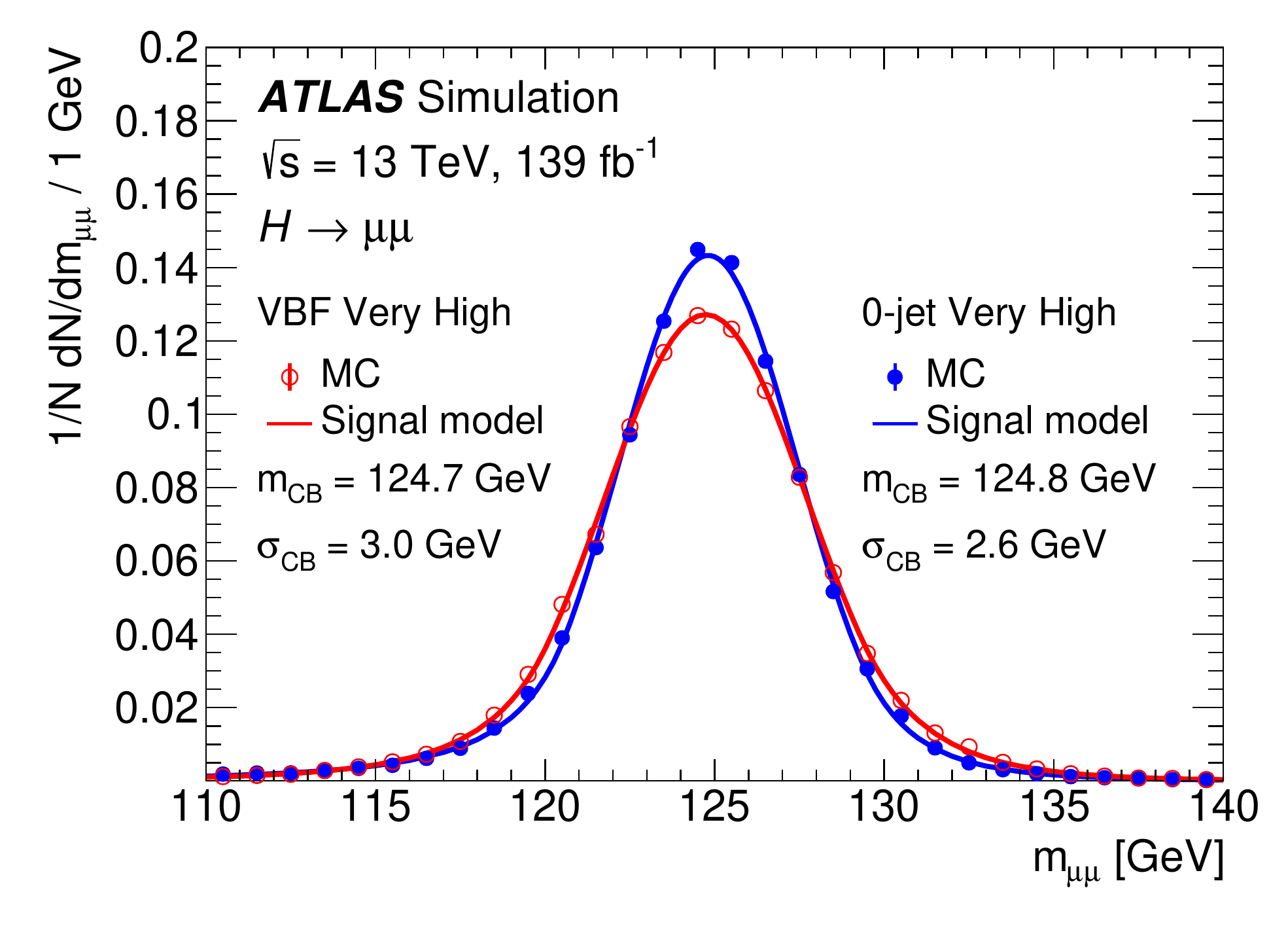}
  \caption{Eexamples of signal parametrization in the categories ``VBF Very High'' and ``2-jet Low''.}
  \label{fig:signal_modeling}
\end{figure}

The parametrization is done inclusively for all production models since no significant difference is found between the signal shapes of different production modes.
To determine the signal shape parameters, $f_s$ is fitted to the signal MC spectra summed over all production modes, assuming the relative normalizations as predicted by the SM.
Examples of signal parametrization in ``VBF Very High'' and ``2-jet Low'' are shown in Fig.~\ref{fig:signal_modeling}.
The Gaussian width varies between 2.6 and 3.2 GeV in all categories.
To test for potential biases in the extracted signal yields due to the analytic parameterizations, a signal injection procedure is used.
In a signal-plus-background fit to pseudo-data constructed from the expected signal and background distributions, the extracted signal yields agree with those injected within the statistical accuracy of about 0.3\%.

The uncertainties in signal modeling are represented by constrained nuisance parameters $\theta_j$.
Both theoretical and experimental sources of uncertainties are considered.
Theoretical uncertainties in signal production impact the number of expected signal events in each category.
For the main production modes, \ggF\ and \VBF, uncertainties include missing higher-order QCD corrections, PDFs, underlying event, and hadronization.
Uncertainties in \ggF\ signal are derived using an approach described in Ref.~\cite{LHCHiggsCrossSectionWorkingGroup} that considers QCD scales, renormalization, resummation, and migration between jet-multiplicity regions \cite{Liu:2013hba,Boughezal:2013oha,Campbell:2010ff,Stewart:2011cf,Gangal:2013nxa,Frederix:2016cnl,deFlorian:2012mx,Grazzini:2013mca}.
Uncertainties in \ggF\ Higgs boson transverse momentum, migration between different kinematic regions, and treatment of the top-quark mass in loop corrections are also accounted for.
Additionally, uncertainties are assigned for \ggF\ signal acceptance in \VBF\ topologies.
Uncertainties in predicted SM branching ratio and Higgs boson production cross sections are included according to Ref.~\cite{LHCHiggsCrossSectionWorkingGroup:2016ypw}.
Modeling uncertainties related to underlying event and parton showering are estimated by \textsc{Pythia} 8 systematic eigentune variations and comparison with \textsc{Herwig} 7 \cite{Bahr:2008pv,Bellm:2015jjp}.
The impact of theoretical uncertainties on predicted signal acceptance ranges between a few per mill and 15\% for \ggF\ production and a few per mill and 7\% for VBF production.
For \VH\ and \ttH\ categories, the impact ranges between a few per mill and about 18\%.

Experimental uncertainties arise from systematic uncertainties related to reconstructed physics objects used in the analysis, affecting expected signal yields in each category.
Systematic uncertainties in muon momentum scale and resolution also affect signal \mmumu\ distribution and shape parameters.
Experimental uncertainties include muon reconstruction and identification efficiencies, efficiencies due to trigger, isolation, and impact parameter requirements, muon momentum scale and resolution \cite{ATLAS:2016lqx,ATLAS:2020gty}, \MET\ soft term determination \cite{ATLAS:2018txj}, b-tagging efficiency \cite{ATLAS:2019bwq}, uncertainty in number of tracks associated with jets \cite{ATL-PHYS-PUB-2017-009}, pile-up modeling \cite{ATL-PHYS-PUB-2016-017}, uncertainties in electron reconstruction and identification efficiency \cite{ATLAS:2019qmc}, jet reconstruction efficiency, energy scale and resolution \cite{ATLAS:2017bje}.
The impact of experimental uncertainties on predicted signal yields and modeling in different categories is mainly due to uncertainties in jet energy scale and resolution and muon momentum resolution, affecting signal yields by up to about 10\% in some 2-jet categories.
Muon momentum resolution uncertainty has an impact on fitted yields ranging between 1\% and 6\% depending on the category.
The uncertainty of 240 MeV in the assumed Higgs mass from \cite{ATLAS:2015yey} is also considered, shifting the signal \mmumu\ distribution and changing $\mu$ in Eq.~\ref{eq:signal}.
Additionally, the uncertainty of 1.7\% in the combined 2015-2018 integrated luminosity is accounted for, derived from calibration of the luminosity scale using $x$-$y$ beam-separation scans \cite{ATLAS-CONF-2019-021}, obtained using the LUCID-2 detector \cite{Avoni_2018} for primary luminosity measurements.

\subsection{Background modeling}

The background shape is modeled using a core function multiplied by an empirical function.
The core function is a physics-motivated rigid shape that focuses on the non-trivial background shape, while the empirical function is a flexible function with free parameters that are determined by the observed data spectrum.
For all categories, the core function is the LO DY line-shape, as described in the appendix of Ref.~\cite{Hmumu}, convolved with the muon resolution, which is parametrized as a Gaussian based on the Drell-Yan full simulation sample.
The empirical functional form is selected from power-law function and exponential of polynomials families, as defined in Tab.~\ref{tab:empirical_functions}.
The selection is performed separately for each category based on the ``Spurious signal test'' and background-only fits.

\begin{table}[htb]
  \centering
  \caption{List of tested empirical functional forms for the background modelling.}
  \label{tab:empirical_functions}
  \begin{tabular}{l c}
    \toprule
    {{Function}} & {{Expression}} \\
    \midrule
    {PowerN}      &    $  \mmumu^{(a_{0}  + a_{1} \mmumu+  a_{2} \mmumu^2 + ... +a_{N} \mmumu^N) }  $ \\
    {EpolyN}      &    $ \exp(a_{1} \mmumu + a_{2} \mmumu^2 + ... + a_{N}  \mmumu^N )  $ \\
    \bottomrule
  \end{tabular}
\end{table}

The spurious signal test assesses the level of modeling bias by fitting a signal-plus-background model, as defined in Eq.~\ref{eq:likelihood}, to a background-only template.
For \ttH\ and \VH\ categories, the background-only template is the spectrum from the full simulation of all SM background, while for \VBF\ and \ggF\ categories, the DY fast simulation sample is used instead.
The resulting signal from the signal+background fit is the spurious signal, which should be less than 20\% of the expected data statistical uncertainties.
Background-only fits (Eq.~\ref{eq:likelihood} with $\mu$ fixed to 0) are performed on the background-only template and the sideband ($\mmumu < 120$ GeV or $\mmumu > 130$ GeV) of the observed data.
To pass the criteria, each $\chi^2$ p-value from the background-only fits is required to be greater than 1\%.
To determine the empirical function for each category, we select the function with the least degree of freedom that passes all the criteria.
The spurious signal yield is also taken as a modeling systematic uncertainty for each category.
We implement it as an additional signal (with the same shape as the actual signal shape) with yield associated with the spurious signal yield.
The spurious signal uncertainties are uncorrelated across categories, ranging from a few percent up to about 20\% of the expected data statistical uncertainties in the \VBF\ and \ggF\ categories and up to about 30\% in the \VH\ and \ttH\ categories, which have less statistical precision in their background simulated samples.

\subsection{Results}

\begin{figure}[h!]
  \centering
  \subfigure[]{\includegraphics[width=0.45\textwidth]{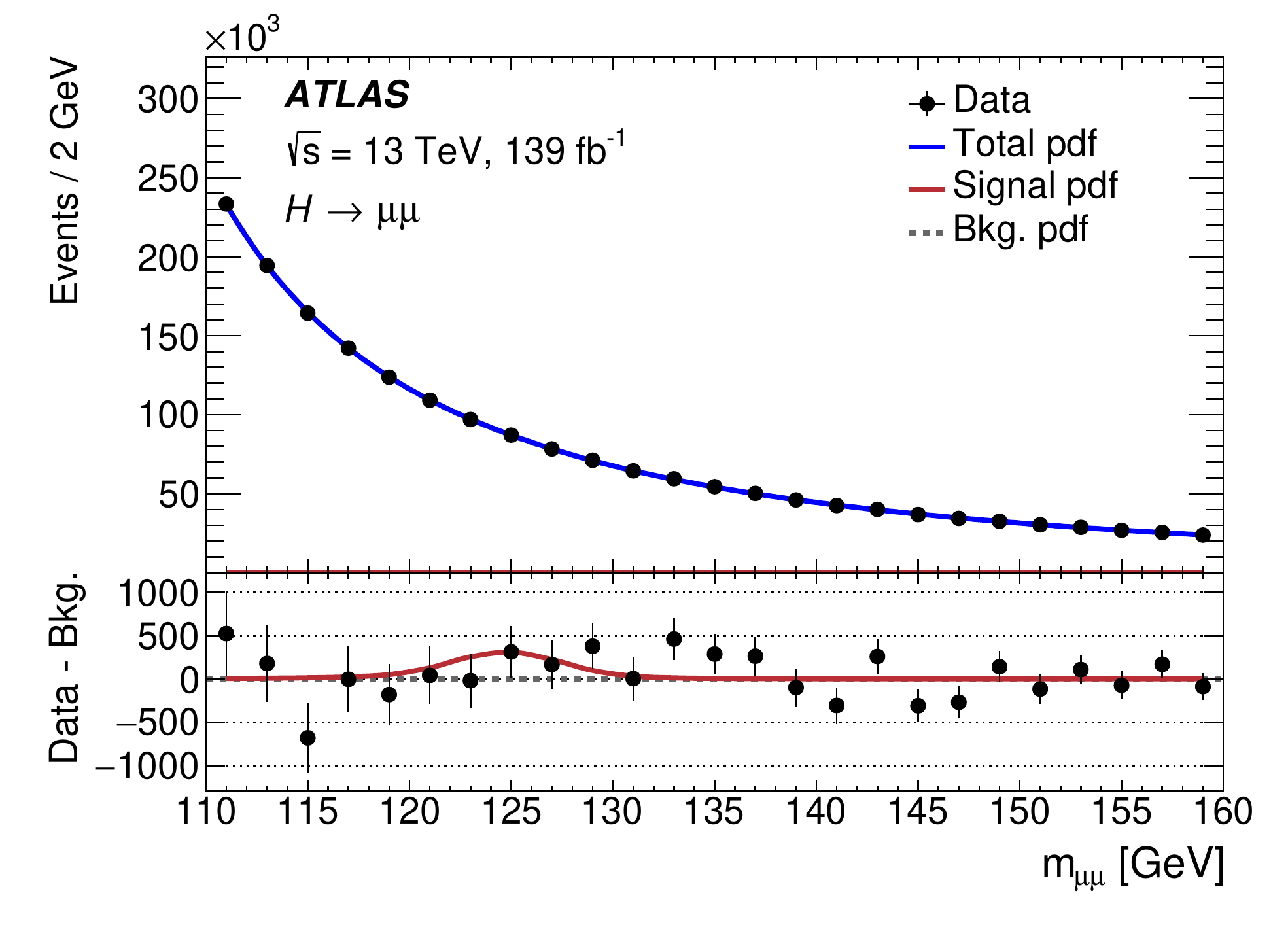}}
  \subfigure[]{\includegraphics[width=0.45\textwidth]{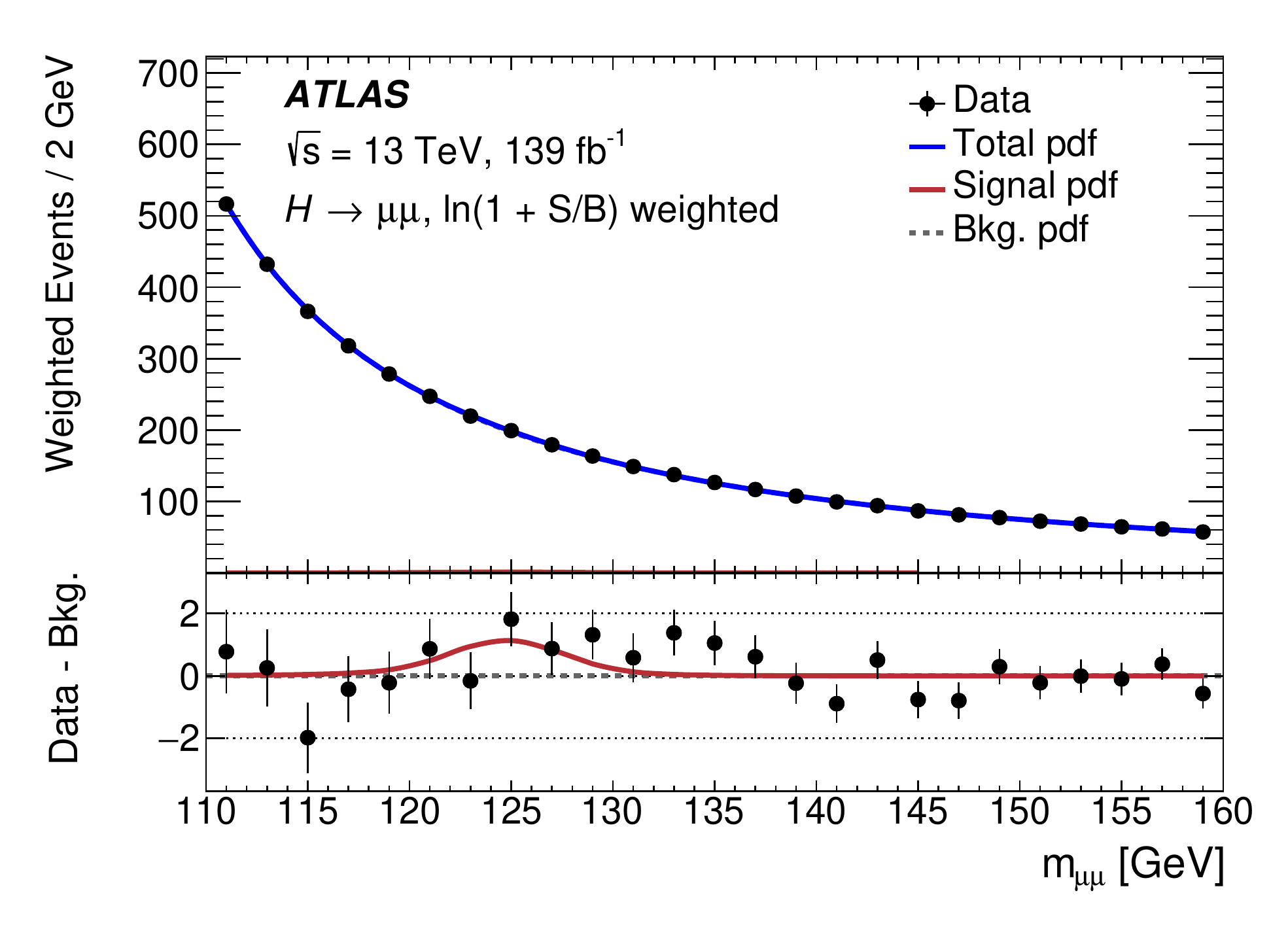}}
  \caption{Dimuon invariant mass spectrum in all the analysis categories observed in data. In (a) the unweighted sum of all events and signal plus background probability density functions (pdf) are shown, while in (b) events and pdfs are weighted by $\ln\left(1 + S/B\right)$, where S are the observed signal yields and B are the background yields derived from the fit to data in the \mmumu\ = 120–130 GeV window. The background and signal pdf are derived from the fit to the data, with S normalised to its best-fit value. The lower panels compare the fitted signal pdf, normalised to the signal best-fit value, to the difference between the data and the background model. The error bars represent the data statistical uncertainties. }
  \label{fig:fit_result}
\end{figure}

The analysis performs a simultaneous binned maximum profile likelihood fit over the range of 110-160 GeV for \mmumu, using a bin size of 0.1 GeV.
The confidence intervals are determined using the profile-likelihood-ratio test statistic \cite{Cowan:2010js}.
The best-fit signal strength parameter is $\mu = 1.2 \pm 0.6$, corresponding to an observed (expected) significance of 2.0$\sigma$ (1.7$\sigma$) with respect to the null hypothesis of no \Hmm\ signal.
The spectra of the dimuon invariant mass for all analysis categories after the signal-plus-background fit are presented in Fig.~\ref{fig:fit_result}.
In Fig.~\ref{fig:fit_result}(b), the events are weighted by $\ln\left(1 + S/B\right)$ to better reflect the total sensitivity to the \Hmm\ signal, where $S$ and $B$ are the observed signal yields and background yields derived from the fit to data in the $\mmumu = 120$-130 GeV window.
The values for $S$, $B$, and other key quantities are listed in Tab.~\ref{tab:Results}.
A goodness-of-fit test is performed using the saturated model technique \cite{Cousins2013GeneralizationOC}, which returns a probability of 10\%.
The uncertainty in the signal strength is dominated by the statistical error of $\pm0.58$.
The systematic uncertainties on the signal strength have an impact of $^{+0.18}_{-0.13}$, with contributions from signal theory uncertainties accounting for $^{+0.13}_{-0.08}$, signal experimental uncertainties accounting for $^{+0.07}_{-0.03}$, and spurious-signal uncertainties accounting for $\pm0.10$.

\begin{table}[htbp]
  \centering
  \caption{Number of events observed in the $\mmumu = 120$--$130\GeV$ window in data, the number of signal events expected in the SM ($S_\mathrm{SM}$), and
   events from signal ($S = \mu \times S_\mathrm{SM}$) and background ($B$) as derived from the combined fit to the data with a signal strength parameter of $\mu = 1.2$.
   The uncertainties in $S_\mathrm{SM}$ correspond to the systematic uncertainty of the SM prediction,
   the uncertainty in $S$ is given by that in $\mu$, and the uncertainty in $B$
   is given by the sum in quadrature of the statistical uncertainty from the fit and the spurious signal uncertainty.
   In addition the observed number of signal events divided by the square root of the number of background events ($S/\sqrt{B}$) 
   and the signal-to-background ratio ($S/B$) in \% for each of the 20 categories described in the text are displayed.}
\label{tab:Results}
  \small
  \begin{tabular}{ l  r  c c c c c c}
    \toprule
    Category & Data &  $S_\mathrm{SM}$  & $S$        & $B$       & $S/\sqrt{B}$ & $S/B$ [\%] &  $\sigma ~[\GeV] $\\
    \midrule
    VBF Very High     & 15    & 2.81 $\pm$ 0.27 & 3.3 $\pm$ 1.7  & 14.5    $\pm$ 2.1 & 0.86 & 22.6~~  & 3.0\\
    VBF High          & 39    & 3.46 $\pm$ 0.36 & 4.0 $\pm$ 2.1  & 32.5    $\pm$ 2.9 & 0.71 & 12.4~~  & 3.0\\
    VBF Medium        & 112   & 4.8  $\pm$ 0.5  & 5.6 $\pm$ 2.8  & 85      $\pm$ 4   & 0.61 & 6.6     & 2.9 \\
    VBF Low           & 284   & 7.5  $\pm$ 0.9  & 9   $\pm$ 4    & 273     $\pm$ 8~~   & 0.53 & 3.2   & 3.0 \\
    2-jet Very High   & 1030  & 17.6 $\pm$ 3.3~~  & 21  $\pm$ 10   & 1024    $\pm$ 22~~  & 0.63 & 2.0 & 3.1\\
    2-jet High        & 5433  & 50   $\pm$ 8~~    & 58  $\pm$ 30   & 5440    $\pm$ 50~~  & 0.77 & 1.0 & 2.9\\
    2-jet Medium      & 18\,311 & 79   $\pm$ 15   & 90  $\pm$ 50   & 18\,320   $\pm$ 90~~~~~  & 0.66 & 0.5 & 2.9 \\
    2-jet Low         & 36\,409 & 63   $\pm$ 17   & 70  $\pm$ 40   & 36\,340   $\pm$ 140~~~ & 0.37 & 0.2   & 2.9\\
    1-jet Very High   & 1097  & 16.5 $\pm$ 2.4~~  & 19  $\pm$ 10   & 1071    $\pm$ 22~~  & 0.59 & 1.8      & 2.9 \\
    1-jet High        & 6413  & 46   $\pm$ 7~~    & 54  $\pm$ 28   & 6320    $\pm$ 50~~  & 0.69 & 0.9 & 2.8 \\
    1-jet Medium      & 24\,576 & 90   $\pm$ 11   & 100 $\pm$ 50~~   & 24\,290   $\pm$ 100~~~ & 0.67 & 0.4 & 2.7 \\
    1-jet Low         & 73\,459 & 125  $\pm$ 17~~   & 150 $\pm$ 70~~   & 73\,480   $\pm$ 190~~~ & 0.53 & 0.2 & 2.8 \\
    0-jet Very High   & 15\,986 & 59   $\pm$ 11   & 70  $\pm$ 40   & 16\,090   $\pm$ 90~~~~~  & 0.55 & 0.4   & 2.6 \\
    0-jet High        & 46\,523 & 99   $\pm$ 13   & 120 $\pm$ 60~~   & 46\,190   $\pm$ 150~~~ & 0.54 & 0.3   & 2.6 \\
    0-jet Medium      & 91\,392 & 119  $\pm$ 14~~   & 140 $\pm$ 70~~   & 91\,310   $\pm$ 210~~~ & 0.46 & 0.2 & 2.7 \\
    0-jet Low         & 121\,354 & 79   $\pm$ 10   & 90  $\pm$ 50   & 121\,310  $\pm$ 280~~~~~ & 0.26 & 0.1  & 2.7 \\
    VH4L              & 34    & 0.53 $\pm$ 0.05 & 0.6 $\pm$ 0.3  & 24      $\pm$ 4   & 0.13 & 2.6 & 2.9 \\
    VH3LH             & 41    & 1.45 $\pm$ 0.14 & 1.7 $\pm$ 0.9  & 41      $\pm$ 5   & 0.27 & 4.2 & 3.1 \\
    VH3LM             & 358   & 2.76 $\pm$ 0.24 & 3.2 $\pm$ 1.6  & 347     $\pm$ 15  & 0.17 & 0.9 & 3.0 \\
    ttH              & 17    & 1.19 $\pm$ 0.13 & 1.4 $\pm$ 0.7  & 15.1    $\pm$ 2.2 & 0.36 & 9.2 & 3.2 \\ 
    \bottomrule
  \end{tabular}
\end{table}

A modified frequentist CLs method \cite{Cowan:2010js, ALRead_2002} is used to compute an upper limit on the signal strength parameter $\mu$ at 95\% confidence level (CL).
The observed upper limit on $\mu$ is found to be 2.2, with an expected limit of 1.1 for the absence of an \Hmm\ signal and an expected limit of 2.0 for an \Hmm\ signal at SM strength.
Assuming the SM cross section for Higgs boson production, the corresponding upper limit on the branching ratio is $B\left(\Hmm\right) < 4.7 \times 10^{-4}$ at 95\% CL.
This result is an improvement of approximately 2.5 times in expected sensitivity compared to the previous ATLAS publication \cite{ATLAS:2017eix}.
The increase in sensitivity is mainly due to a larger dataset and more advanced analysis techniques, resulting in an additional 25\% improvement.
\chapter{Search for \monoHbb}
\label{chap:monoHbb}

\section{Event selections and categorization}
\label{chap:monohbb:sec:categorization}

In the \monoHbb\ analysis, we aim to detect a Higgs boson decaying into two $b$-quarks along with a large missing transverse momentum \MET.
To achieve this, we categorize events into non-overlapping regions, namely signal regions and control regions, based on their expected signal or background content.
The control regions require the presence of one or two leptons, while the signal regions veto events containing loose leptons.

Reconstructing the two $b$-quarks as separate jets becomes challenging when the Higgs boson is significantly boosted, as the angle between the two b-jets is inversely proportional to the \pT\ of the Higgs boson.
To overcome this challenge, we split the analysis into ``resolved'' regions, where the decay products of the Higgs boson are reconstructed as two separate jets, and ``merged'' regions, where the entire Higgs boson decay is reconstructed as a single jet.

In the case of $b$-associated production within the \thdma\ benchmark model, the Higgs boson and DM particles are produced with an additional pair of $b$-quarks from gluon splitting.
Hence, to enhance sensitivity to these models, we further divide all regions into those requiring exactly two $b$-jets and those requiring three or more $b$-jets, which we refer to as ``\tbtag'' and ``\thbtag'' regions, respectively.
As \MET\ is often highly correlated with the mediator mass (e.g. \mZp\ in the \zpthdm\ and \mA\ in the \thdma), \MET\ is very sensitive to different signal models and parameters.
All regions are thus also divided by \MET\ in both resolved and merged regions. The binning of the \MET\ splitting varies between \tbtag, \thbtag, signal regions and control regions, depending on the sample size available.

\subsection{Common selections}

To ensure the reliability of events, we require the presence of a reconstructed primary vertex and exclude fake jets originating from beam-induced backgrounds, cosmic-ray showers, or noisy calorimeter cells \cite{ATLAS-CONF-2015-029}.
Additionally, events must satisfy $\MET > 150$ GeV and exclude $\tau$-lepton.
To further suppress background from $\tau$-lepton decays, we apply the ``extended $\tau$-lepton veto'', which involves vetoing events with small-$R$ jets that have a track multiplicity between 1 and 4 and $\Delta\phi\left(\mathrm{jet}, \MET\right) < 22.5^{\circ}$.
To reject events with \MET\ arising from leptonic heavy-flavour decays or severely mismeasured jets, we reject events where any of the leading up to three small-$R$ jets have $\Delta\phi\left(\mathrm{jet}, \MET\right) < 20^{\circ}$.
Control regions replace \MET\ with $E^\mathrm{miss}_{\mathrm{T, lep. invis.}}$ to mimic the signal regions' effect, where leptons are considered invisible in the calculation of missing transverse momentum.
For the final fit, we use the mass of the Higgs boson candidate (\mh) as the discriminating variable, where $50\ \mathrm{GeV} < \mh < 280$ GeV for the resolved regions and $50\ \mathrm{GeV} < \mh < 270$ for the merged regions.
The lower limit is the lowest calibrated large-$R$ jet mass, and the upper limit is set to be significantly larger than the Higgs boson mass, determined by the \mh\ binning used in the fit, which depends on the sample size available.

\subsection{Signal regions}

Tab.~\ref{tab:selections.summary} summarizes the signal region selections for both the merged and resolved regions.
To reduce the contribution of SM processes producing \MET\ through the decay $W\to\ell\nu$, events containing any loose electron or muon are rejected.

\begin{table}
    \centering
    \caption{
        Summary of selections used to define the signal regions used in the analysis.
        The kinematic variables are defined in the text.
    } 
    \vspace{0.2cm}   
    \begin{tabular}{c|c}
        \toprule
        \textbf{Resolved} & \textbf{Merged} \\
        \midrule
        \multicolumn{2}{c}{Primary \MET\ trigger} \\
        \midrule
        \multicolumn{2}{c}{Data quality selections} \\
        \midrule
        \multicolumn{2}{c}{$\MET\ > 150$ \GeV} \\
        \midrule
        \multicolumn{2}{c}{Lepton veto \& extended $\tau$-lepton veto} \\
        \midrule
        \multicolumn{2}{c}{$\Delta\phi(\mathrm{jet}_{1,2,3}, \MET) > 20\degree$} \\
        \midrule
        $\MET\ < 500$ \GeV & $\MET\ > 500$ \GeV \\
        \midrule
        At least 2 small-$R$ jets & At least 1 large-$R$ jet \\
        \midrule
        At least 2 $b$-tagged small-$R$ jets & At least 2 $b$-tagged associated variable-$R$ track-jets \\
        \midrule
        ${\pT}_h > 100\,\GeV$ if $\MET < 350$ \GeV & \multirow{2}{*}{---} \\
        ${\pT}_h > 300\,\GeV$ if $\MET > 350$ \GeV & \\
        \midrule
        $m_\mathrm{T}^{b,\mathrm{min}} > 170\,\GeV$ & --- \\
        \midrule
        $m_\mathrm{T}^{b,\mathrm{max}} > 200\,\GeV$ & --- \\
        \midrule
        $S > 12$ & --- \\
        \midrule
        $N_{\textrm{small-}R\,\textrm{jets}} \leq 4$ if 2 $b$-tag & \multirow{2}{*}{---} \\
        $N_{\textrm{small-}R\,\textrm{jets}} \leq 5$ if $\geq3$ $b$-tag &\\
        \midrule
        $50\,\GeV < m_h < 280\,\GeV$ & $50\,\GeV < m_h < 270~\GeV$ \\
        \bottomrule
    \end{tabular}
    \label{tab:selections.summary}
\end{table}

\subsubsection{Resolved regions}

The resolved regions select events with $\MET < 500\,\GeV$ and at least two $b$-tagged small-$R$ jets, where the two with the highest \pT\ form the Higgs boson candidate.
The combined \pT\ of this two-jet system (${\pT}_h$) must be greater than 100 GeV, and its mass is corrected for nearby muons, as described in Chap~\ref{chap:object}, to form \mh.

The dominant background in the resolved region is \ttbar\ production, where one top quark decays leptonically, but the lepton is either not reconstructed or not correctly identified.
In such cases, all the \MET\ in the event (beyond that from mismeasurement) comes from the decay of one of the two $W$ bosons, and therefore the transverse mass of the \MET and the corresponding $b$-jet should be approximately bounded from above by the top-quark mass.
We define the transverse mass as:

\begin{equation}
    m_\mathrm{T}^{b,\mathrm{min}/\mathrm{max}} = \sqrt{2\pT^{b,\mathrm{min}/\mathrm{max}}\MET(1-\mathrm{cos}\Delta\phi(\pT^{b,\mathrm{min}/\mathrm{max}}, \MET))}
\end{equation}
where $\pT^{b,\mathrm{min}}$ and $\pT^{b,\mathrm{max}}$ are defined as the \pT\ of the $b$-jet closest to (min) or furthest from (max) the \MET\ in $\phi$.
Events must satisfy $m_\mathrm{T}^{b,\mathrm{min}} > 170~\GeV$ and $m_\mathrm{T}^{b,\mathrm{max}} > 200~\GeV$.

To reduce contributions from multijet backgrounds, the signal region selections require an object-based \MET\ significance \cite{ATLAS-CONF-2018-038} $S > 12$, which assesses the likelihood that \MET\ is due to invisible particles rather than mismeasurements.
Data-driven estimates of the remaining multijet contributions were found to be smaller than the expected statistical uncertainty of the data, so the impact of multijet processes is not included in the background estimation.

Events are split to \tbtag\ and \thbtag\ regions. The \tbtag regions require events with exactly two $b$-tagged jets and at most four small-$R$ jets, counted only for central jets.
For the \thbtag\ regions, at most five small-$R$ jets are allowed to ensure a sufficient sample size in the corresponding control regions.
The \tbtag\ and \thbtag\ regions are split into three \MET\ bins: $150\,\GeV < \MET < 200\,\GeV$, $200\,\GeV < \MET < 350\,\GeV$ and $350\,\GeV < \MET < 500\,\GeV$.
This leads to six resolved signal regions: one for each combination of EmissT bin and number of b-tagged jets.
In the highest \MET\ bin, the requirement on ${\pT}_h$ is tightened to be greater than 300 GeV.

The \MET\ triggers become fully efficient at an offline \MET\ value close to 200 GeV but are also used for events in the range $150\,\GeV < \MET < 200~\GeV$.
To correct for MC mismodelling of the \MET\ trigger response, trigger efficiencies are measured in both data and simulation, and scale factors are calculated to correct the simulation.
The scale factors are calculated as a function of $E_{\mathrm{T,~lep,~invis}}^{\mathrm{miss}}$ in a region whose selection matches the $150\,\GeV < \MET < 200~\GeV$ and \tbtag\ region except that all \MET\ selections are dropped, exactly one $b$-tagged jet is required, and exactly one signal muon is required.
Events containing electrons are still vetoed.
The scale factors have values in the range 0.95–1.0.

\subsubsection{Merged regions}

The merged regions are defined by selecting events with $\MET > 500~\GeV$.
At least one large-$R$ jet is required, and the two leading variable-$R$ track-jets associated with the leading large-$R$ jet are required to be $b$-tagged.
This large-$R$ jet, defined to be the Higgs boson candidate, has its mass corrected for nearby muons, as described in Chapter~\ref{chap:object}, to form \mh.
Events are separated into two categories: those with no additional $b$-tagged variable-$R$ track-jets (referred to as \tbtag\ region) and those with at least one such $b$-tagged variable-$R$ track-jet not associated with the Higgs boson candidate (referred to as \thbtag\ region).
The merged 2 $b$-tag region is divided into two \MET\ bins,  $500\,\GeV < \MET < 750~\GeV$ and $\MET > 750~\GeV$. There is no further binning for the \thbtag\ region.

\subsection{Control regions}
\label{chap:monoHbb:sec:categorization:ssec:cr}

We define control regions to better constrain the normalization of main background processes where specific background processes dominate and negligible signal is expected.
1-muon control regions are defined in order to constrain the $W$ (produced in association with jets) and \ttbar\ processes.
$W$ and \ttbar\ processes contribute in the signal regions if leptons in the decays are either not identified or outside the kinematic acceptance.
The main contribution arises from hadronically decaying $\tau$-leptons.
As the shape of event variables is the same for all lepton flavors, the kinematic phase space of the signal region can be approximated closely by control regions requiring a medium muon (1-muon control regions).
In this case, $E_{\mathrm{T,~lep,~invis}}^{\mathrm{miss}}$ is used as a proxy for \MET\ to better approximate the signal regions, which veto the presence of any leptons.
Additionally, any other variable using \MET\ in its calculation, such as \MET\ significance and $m_\mathrm{T}^{\mathrm{b,min/max}}$, is constructed using $E_{\mathrm{T,~lep,~invis}}^{\mathrm{miss}}$.
This ensures that the \MET-related quantities in the control regions correspond to those in the signal regions.
Otherwise, the 1-muon control regions are defined by the same criteria as the signal regions.
In 1-muon \tbtag\ control regions, the sign of the muon charge is used as the fit discriminant, which separates $W$ and \ttbar.
In 1-muon \thbtag\ control regions, no further binning other than \MET is considered due to limited data statistics.

On the other hand, 2-lepton control regions are defined to constrain the $Z\to\nu\nu$ process produced in association with jets.
As the momentum of the $Z$ boson does not depend on its decay mode, $Z\to\nu\nu$ in the signal region can be closely modeled by $Z \to \ell^{+}\ell^{-}$ events by requiring exactly two loose electrons or muons with opposite charge.
These events are collected using single-electron or single-muon triggers.
Furthermore, one of the electrons or muons is required to be a medium electron or muon with $\pT > 27~\GeV$ or $\pT > 25~\GeV$, respectively.
The invariant mass of the di-lepton system is required to be consistent with the mass of the $Z$ boson within 10 GeV.
While keeping all other criteria of the signal regions, an additional criterion of $S < 5$ is imposed to suppress a remaining contribution of \ttbar\ processes.
To be similar to the signal regions, $E_{\mathrm{T,~lep,~invis}}^{\mathrm{miss}}$ is used as proxy for \MET\ and in the calculation of any other variable using \MET.
In 2-lepton control regions, no further binning other than \MET\ is considered.

All control regions are split to four \MET\ bins, including the resolved regions: $150\,\GeV < \MET < 200\,\GeV$, $200\,\GeV < \MET < 350\,\GeV$ $350\,\GeV < \MET < 500\,\GeV$, and merged region: $\MET > 500~\GeV$.

\section{Statistical analysis}
\label{chap:monoHbb:sec:stat}

The same binned profile likelihood function equation~\ref{eq:likelihood} is used in \monoHbb\ to perform statistical fitting in all signal regions and control regions simultaneously and extract signal yields. $S_i$ and $B_i$ are modeled by histograms derived from MC samples.

The analysis is split to a model-specific scenario and a model-independent scenario. In the model-specific scenario, one global signal strength $\mu$ is used for all signal and control regions. The invariant mass of the Higgs boson condidate \mh\ is used as the fit disriminant in the signal regions. The bin size of \mh\ is 5~\GeV\ in 2 $b$-tag $150\,\GeV < \MET < 200\,\GeV$, $200\,\GeV < \MET < 350\,\GeV$, 10~\GeV\ in 2 $b$-tag $350\,\GeV < \MET < 500\,\GeV$, $\geq$3 $b$-tag $150\,\GeV < \MET < 200\,\GeV$, $200\,\GeV < \MET < 350\,\GeV$, 20~\GeV\ in 2 $b$-tag $500\,\GeV < \MET < 750\,\GeV$. In 2 $b$-tag $\MET > 750~\GeV$ and $\geq$3 $b$-tag $\MET > 500~\GeV$, the \mh\ binning is customized as [50, 90, 150, 270) \GeV, while in $\geq$3 $b$-tag $350\,\GeV < \MET < 500\,\GeV$, the \mh\ binning is customized as [50, 110, 150, 280) \GeV.

In the model-independent scenario, each signal region is assigned a separate signal strength $\mu_i$ in order to measure the visible cross-section, degined as:

\begin{equation}
\sigma_{\mathrm{vis},h(b\bar{b})+\mathrm{DM}} \equiv \sigma_{h+\mathrm{DM}} \times \mathscr{B}(h \to b\bar{b}) \times (\mathscr{A} \times \varepsilon)
\end{equation}

where $(\mathscr{A} \times \varepsilon)$ with the acceptance $\mathscr{A}$ and the reconstruction efficiency $\varepsilon$ quantifies the probability for a certain event to be reconstructed within a window around the Higgs boson mass in a given signal region. The visible cross-section is obtained from the number of signal events in each signal region extracted from a fit to the $m(b\bar{b})$ distribution as described below, divided by the integrated luminosity. In contrast to the model-specific scenario, we assume that a signal resonance is produced with a mass close to 125~\GeV\ and decays into a pair of $b$ quarks in association with \MET. For this purpose, the binning in $m_{h}$ in the signal regions is modified from the model-specific scenario such that all bins under the Higgs boson peak in a range from 90~\GeV\ to 150~\GeV\ are merged into one bin (Higgs window). This Higgs window includes the Higgs boson peak, but excludes important parts of the $Z$ boson peak.

\subsection{Signal modeling}

In the model-specific scenario, the signal models considered for the statistical interpretation include \zpthdm\ and \thdma\ as described in section~\ref{chap:higgs:sec:extended}. The MC samples for each of the models are used to model the signal shape in the spectra of the fit discriminants. In the model-independent scenario, the signal is assumed to be present only in the Higgs window.

\subsection{Background modeling}
\label{chap:monoHbb:sec:stat:ssec:signal}

The dominant backgrounds in the signal regions are \ttbar\ and $W/Z$ processes. The $W/Z$ backgrounds are subdivided according to the true flavour of the associated jets that constitute the Higgs boson candidate during reconstruction. Simulated jets are labelled according to which hadrons with $\pT > 5$~\GeV\ are found within a cone of size $\Delta R = 0.3$ around the jet axis. If a $b$-hadron is found the jet is labelled as a $b$-jet. If no $b$-hadron is found, but a $c$-hadron is present, then the jet is labelled as a $c$-jet. Otherwise the jet is labelled as a light jet. The flavour of the two leading $b$-tagged track-jets is used in the merged region. If the flavour of either (or both) those two jets is a $b$-quark, the
event is considered to be $W/Z$+HF background, where HF stands for Heavy Flavour. In the 2 $b$-tag resolved regions the dominant backgrounds are \ttbar\ and $Z$+HF, with the latter becoming more important as the \MET\ increases. The 2 $b$-tag merged regions are dominated by $Z$+HF.
Both the resolved and merged $\geq$3 $b$-tag regions are dominated by \ttbar, where the extra $b$-jet typically is a mis-tagged jet originating from a hadronic $W$ boson decay.
At higher \MET\ values the $Z$+HF background becomes important again.

The kinematic shapes of all background processes are modeled using simulation samples. We use free parameters (i.e. without prior constraint in the likelihood model) to model the normalizations of the dominant backgrounds. The normalizations of \ttbar, $W$+HF are determined by a single free parameter each, while the normalization of $Z$+HF is determined by two free parameters where one is assigned to the events in 2 $b$-tag regions and the other one is assigned to the events in $\geq$3 b-tag regions.

The normalizations of smaller backgrounds are taken directly from simulation.
These include the production of a $W$ or $Z$ boson in association with light jets or at most one jet containing a $c$-hadron, single top-quark production (dominated by production in association with a $W$ boson) and diboson processes.
Small contributions also arise from \ttbar\ processes in association with vector bosons or a Higgs boson.
Another background contribution stems from vector-boson production in association with a Higgs boson ($Vh$), which mimics the signal due to the presence of a Higgs boson peak in association with jets.
In the case where the vector boson is a $Z$ boson decaying to neutrinos this is an irreducible background.
Similarly, the diboson decay $ZZ \rightarrow b\overline{b} \nu \nu$ is nearly irreducible due to small difference in $Z$ and $h$ mass peaks (compared to the Higgs candidate mass resolution).
The multijet background is negligible in all regions after the requirements on the object-based \MET\ significance $S$ and on $\Delta \phi(\text{jet},\MET)$, and thus not further considered.

\subsection{Systematic uncertainties}

Signal and background expectations are subject to statistical, detector-related and theoretical uncertainties, which are all included in the likelihood as nuisance parameters. 
Detector-related and theoretical uncertainties may affect the overall normalisation and/or shape of  the simulated background and signal event distributions.

Detector-related uncertainties are dominated by contributions from the jet reconstruction. 
Uncertainties in the jet energy scale (JES) for small-$R$ jets~\cite{ATLAS:2020cli} arise from the calibration of the scale of the jet and are derived as function of the jet \pt, and also $\eta$. Further contributions emerge from the jet flavour composition and the pile-up conditions. The `category reduction' scheme as described in Ref.~\cite{ATLAS:2020cli} with 29 nuisance parameters is used. Uncertainties in the jet energy resolution (JER) depend on the jet \pt\ and $\eta$ and arise both from the method used to derive the jet resolution and from the difference between simulation and data~\cite{ATLAS:2020cli}, and they are included with eight nuisance parameters. Similarly, uncertainties in the jet energy resolution for large-$R$ jets arise from the calibration, the flavour composition and the topology dependence~\cite{ATLAS:2018bip}. Further uncertainties are considered for the large-$R$ jet mass scale~\cite{ATLAS:2018bip} and resolution~\cite{ATLAS-CONF-2017-063}.

Uncertainties due to the $b$-tagging efficiency for heavy-flavour jets, including $c$-flavour jets, are derived from \ttbar\ data~\cite{ATLAS:2019bwq,ATLAS-CONF-2018-001} and are represented by four nuisance parameters. Uncertainties are also considered for mistakenly $b$-tagging a light-flavour jet, with nine nuisance parameters. These are estimated using a method similar to that in Ref.~\cite{ATLAS-CONF-2018-006}.

Uncertainties in the modeling of \MET\ are evaluated by considering the uncertainties affecting the jets included in the calculation and the uncertainties in soft term's scale and resolution~\cite{ATLAS:2018txj}. The pile-up in simulation is matched to the conditions in data by a reweighting factor. An uncertainty of 4\% is assigned to this reweighting factor. 
The uncertainty in the combined 2015--2018 integrated luminosity is 1.7\% \cite{ATLAS-CONF-2019-021}, obtained using the LUCID-2 detector \cite{Avoni_2018} for the primary luminosity measurements.

Scale factors, including their uncertainties, are calculated specifically for this analysis to correct the efficiency of \MET\ triggers in simulation to that in data. The uncertainties in the scale factors are at most 1\%--2\% for low \MET\ values. 

In the regions requiring the presence of leptons, uncertainties in the lepton identification and lepton energy/momentum scale and resolution are included. These are derived using simulated and measured events with $Z \rightarrow \ell^{+}\ell^{-}$, $J/\psi \rightarrow \ell^{+}\ell^{-}$ and $W \rightarrow \ell\nu$ decays~\cite{ATLAS:2019qmc,ATLAS:2020auj}. 

Modelling uncertainties impact the shape of the $m_h$ distribution, the relative acceptance between different \MET\ and $b$-tag multiplicity bins and between signal and control regions as well as the overall normalisation of the samples that are not freely floating in the fit.

The theoretical uncertainties are dominated by modelling uncertainties in the \ttbar\ and $Z$+HF backgrounds. 
For the \ttbar\ and $Wt$ processes, the impact of the choice of parton shower and hadronisation model is evaluated by comparing the sample from the nominal generator set-up 
with a sample interfaced to \textsc{Herwig}7.04~\cite{Bahr:2008pv,Bellm:2015jjp}. 
To assess the uncertainty in the matching of NLO matrix elements to the parton shower, the \textsc{Powheg}-\textsc{Box} sample is compared with a sample of events generated with \textsc{Mad}\textsc{Graph}5\_aMC@NLO~v2.6.2. 
For the $Wt$ process, the nominal sample is compared with an alternative sample generated using the diagram subtraction scheme~\cite{Frixione:2008yi,ATL-PHYS-PUB-2016-020} instead of the diagram removal scheme to estimate the uncertainty arising from the interference with \ttbar\ production. 

For the $V$+jet processes, uncertainties arising from the modelling of the parton shower and the matching scheme are evaluated by comparing the nominal samples with samples generated with \textsc{Mad}\textsc{Graph}5\_aMC@NLO~v2.2.2. 
For the diboson processes, the uncertainties associated with the modelling of the parton shower, the hadronisation and the underlying event are derived using alternative samples generated with \textsc{Powheg}-\textsc{Box}~\cite{Nason:2004rx,Frixione:2007vw,Alioli:2010xd} and interfaced to \textsc{Pythia}8.186~\cite{Sjostrand:2007gs} or \textsc{Herwig++}. 

For all MC samples, the uncertainties due to missing higher orders are estimated by a variation of the renormalisation and factorisation scales by a factor of two, while the PDF and $\alpha_s$ uncertainties are calculated using the PDF4LHC prescription~\cite{Butterworth:2015oua}.

Tab.~\ref{syst:table:impact} gives the impact of the different sources of systematic uncertainties for selected signal models as evaluated in different model-dependent fits. The signal models with lower masses illustrate the impact of the systematic uncertainties in the resolved regions, while the models with larger mediator masses are more impacted by the merged regions. The theoretical uncertainties in the modelling of the \ttbar\ background, the experimental uncertainties in the calibration of jets and the limited MC sample size show the largest impact. 

\begin{table}
 \caption{\label{syst:table:impact} Relative importance of the different sources of uncertainty for different $Z'$-2HDMs, with the masses of the $Z'$ boson and the $A$ boson given in the second row, expressed as fractional impact on the signal strength parameter. The fractional impact is calculated by considering the square of the uncertainty in the signal strength parameter arising from a given group of uncertainties (as listed in the left column of the table), divided by the square of the total uncertainty in the signal strength parameter. Due to correlations, the sum of the different impacts of systematic uncertainties might not add up to the total impact of all systematic uncertainties.}
 \vspace{0.2cm}
 \begin{center}
 \begin{tabular}{l|ccc}
 \toprule
\multirow{3}{*}{Source of uncertainty} & \multicolumn{3}{c}{Fractional squared uncertainty in $\mu$} \\\cline{2-4}
 & \multicolumn{3}{c}{$Z'$-2HDM signals, $(m_Z',m_A)$ [\GeV]} \\
 & (800, 500) & (1400, 1000) & (2800, 300) \\
\midrule
$Z$+HF normalisation & 0.11 & 0.03 & $<0.01$~~~\\
$W$+HF normalisation & 0.02 & 0.01 & $<0.01$~~~\\
\ttbar\ normalisation & 0.16 & 0.04 & $<0.01$~~~\\
$Z$ modelling uncertainties & 0.02 & 0.07 & $<0.01$~~~\\
$W$ modelling uncertainties & $<0.01$~~~& 0.01 & $<0.01$~~~\\
\ttbar\ modelling uncertainties & 0.13 & 0.05 & $<0.01$~~~\\
Single-$t$ modelling uncertainties& 0.18 & 0.02 & $<0.01$~~~\\
Other modelling uncertainties & 0.05 & 0.01 & $<0.01$~~~\\
Jets & 0.20 & 0.06 & 0.01 \\
$b$-tagging & 0.01 & 0.01 & 0.04 \\
\MET\ soft term and pile-up & $<0.01$~~~& $<0.01$~~~& $<0.01$~~~\\
Other experimental systematic uncertainties & 0.01 & $<0.01$~~~& $<0.01$~~~\\
Signal systematic uncertainties & $<0.01$~~~& $<0.01$~~~& $<0.01$~~~\\
MC sample size & 0.08 & 0.07 & 0.11 \\
\midrule
Statistical uncertainty & 0.27 & 0.61 & 0.79 \\
Total systematic uncertainties & 0.73 & 0.39 & 0.21 \\\bottomrule
 \end{tabular}
 \end{center}
\end{table}

\subsection{Results}

\begin{figure}[h!]
  \centering
  \subfigure[]{\includegraphics[width=0.7\textwidth]{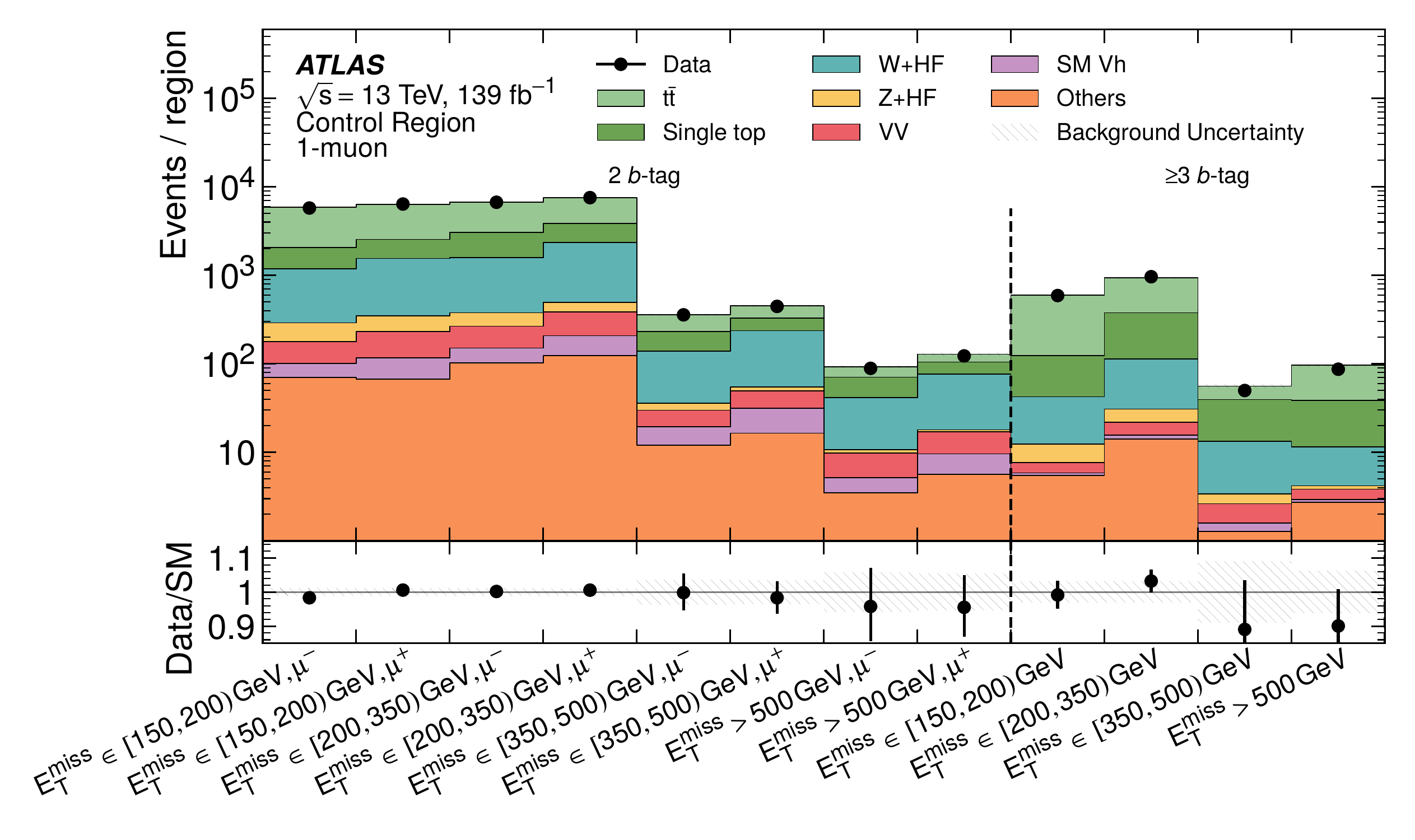}}
  \subfigure[]{\includegraphics[width=0.7\textwidth]{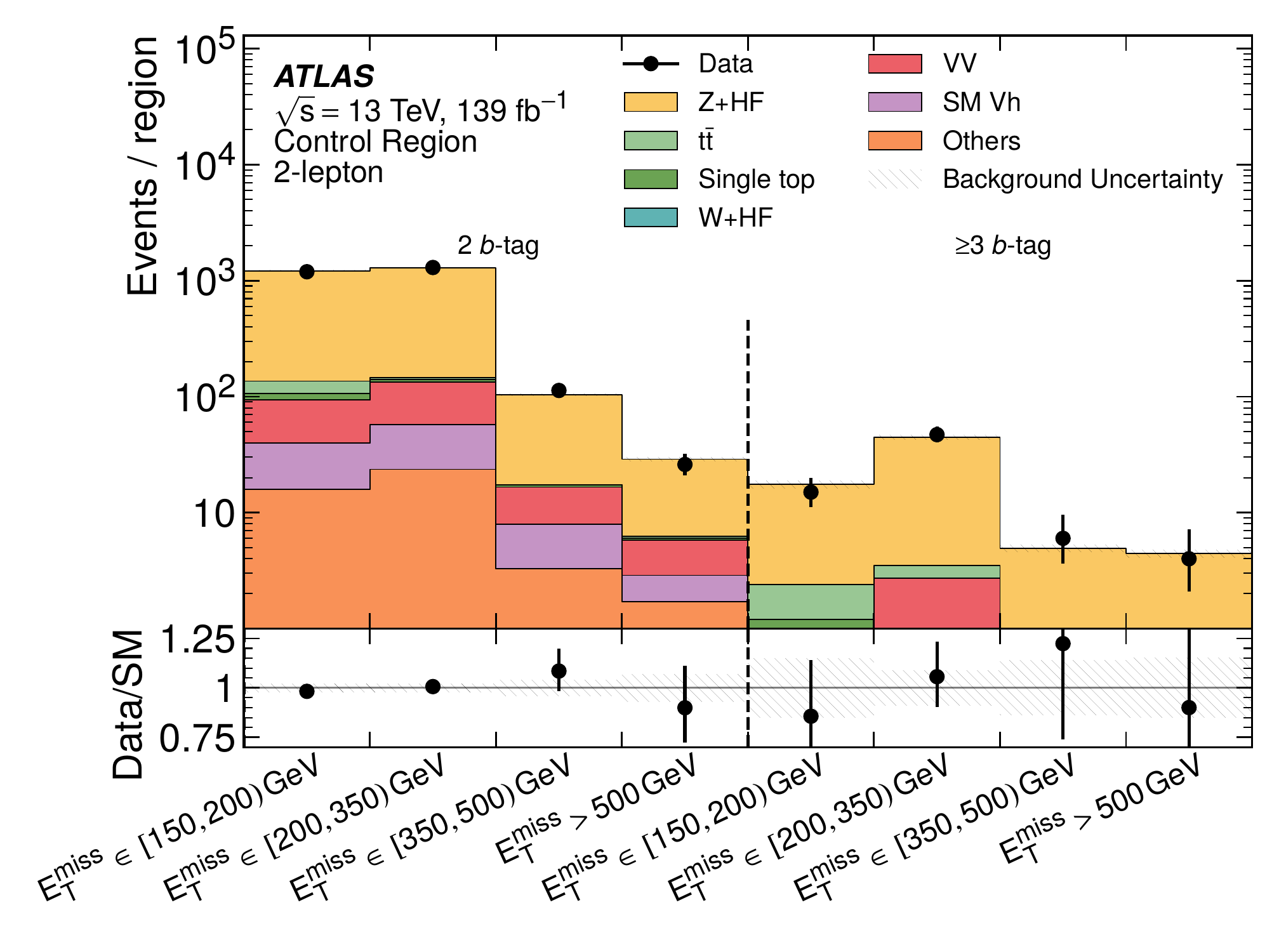}}
  \caption{Yields in the resolved and merged (a)~1-muon control regions and (b)~2-lepton control regions. The top panel compares the fitted background yields with data, while the bottom panel indicates the ratio of the observed data to the predicted Standard Model backgrounds. The different control region bins included in the fit are indicated on the $x$-axis by first giving the range in \MET\ and then the sign of the muon charge (where applicable). }
  \label{fit:fig:CR}
\end{figure}

The post-fit background yields are determined in a background-only profile likelihood fit to data in all regions.
Fig.~\ref{fit:fig:CR} shows the yields in the 1-muon and 2-lepton control regions.
The post-fit normalisation factors for \ttbar\ and for $W$+HF are found to be $0.93 \pm 0.08$ and $0.95 \pm 0.14$, respectively. For $Z$ boson production in association with two (at least three) heavy-flavour jets the normalisation factors are determined to $1.41 \pm 0.09$ ($1.85 \pm 0.24$). An upward scaling of the $Z$+HF background relative to the simulation was also observed in other studies~\cite{ATLAS:2020juj}, and was attributed to an underestimation of the $g \rightarrow b\bar{b}$ rate in \textsc{Sherpa}. A larger scaling is observed in the region with $\geq$3 $b$-tagged jets, dominated by processes with more $g \rightarrow b\bar{b}$ splittings. The uncertainty on the $Z$+HF normalisation factor increases in the $\geq$3 $b$-tag region due to lower statistical precision and the smaller contribution of the $Z$+HF background.

\begin{figure}[h!]
  \centering
  \subfigure[]{\includegraphics[width=0.4\textwidth]{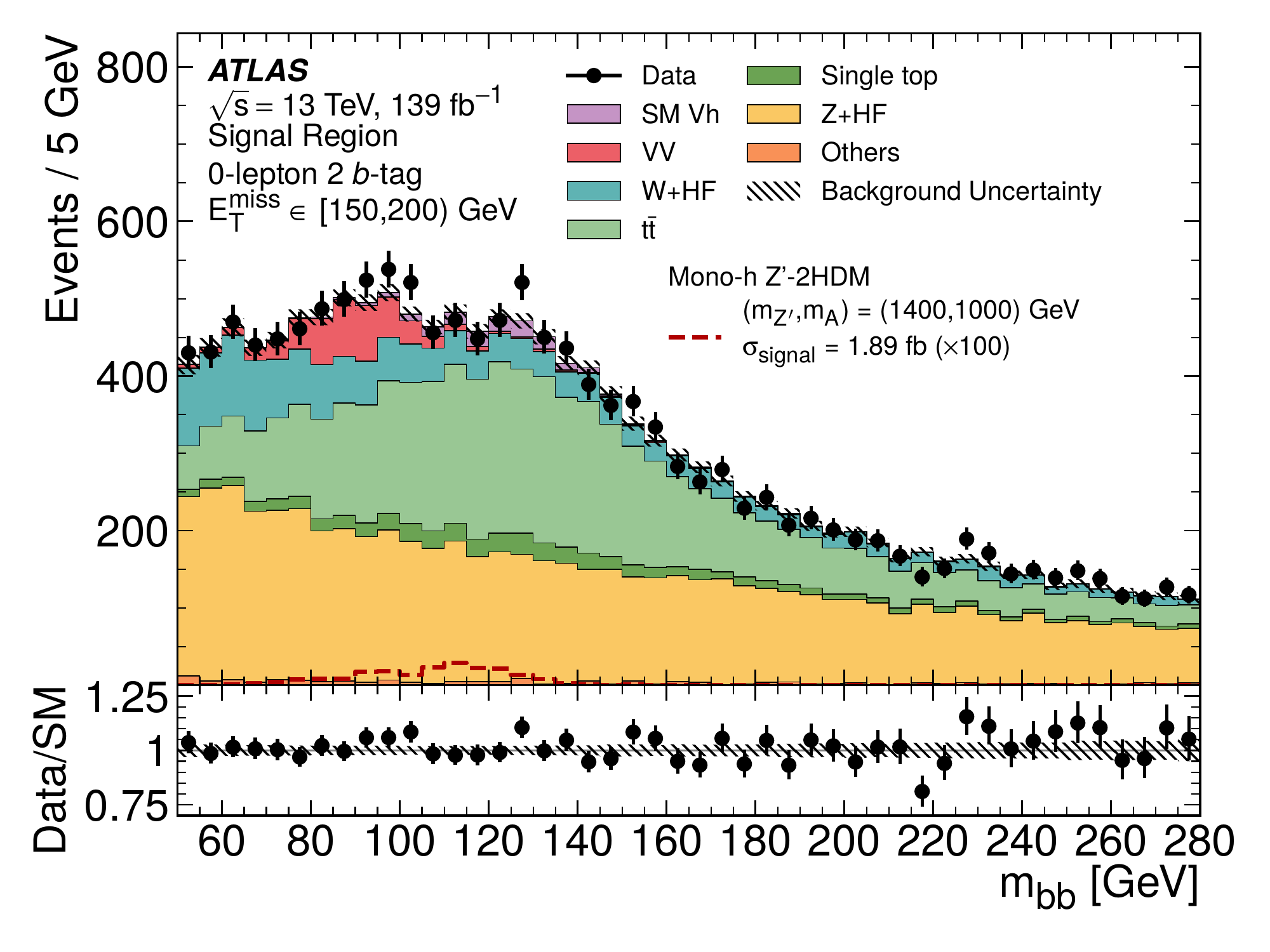}}
  \subfigure[]{\includegraphics[width=0.4\textwidth]{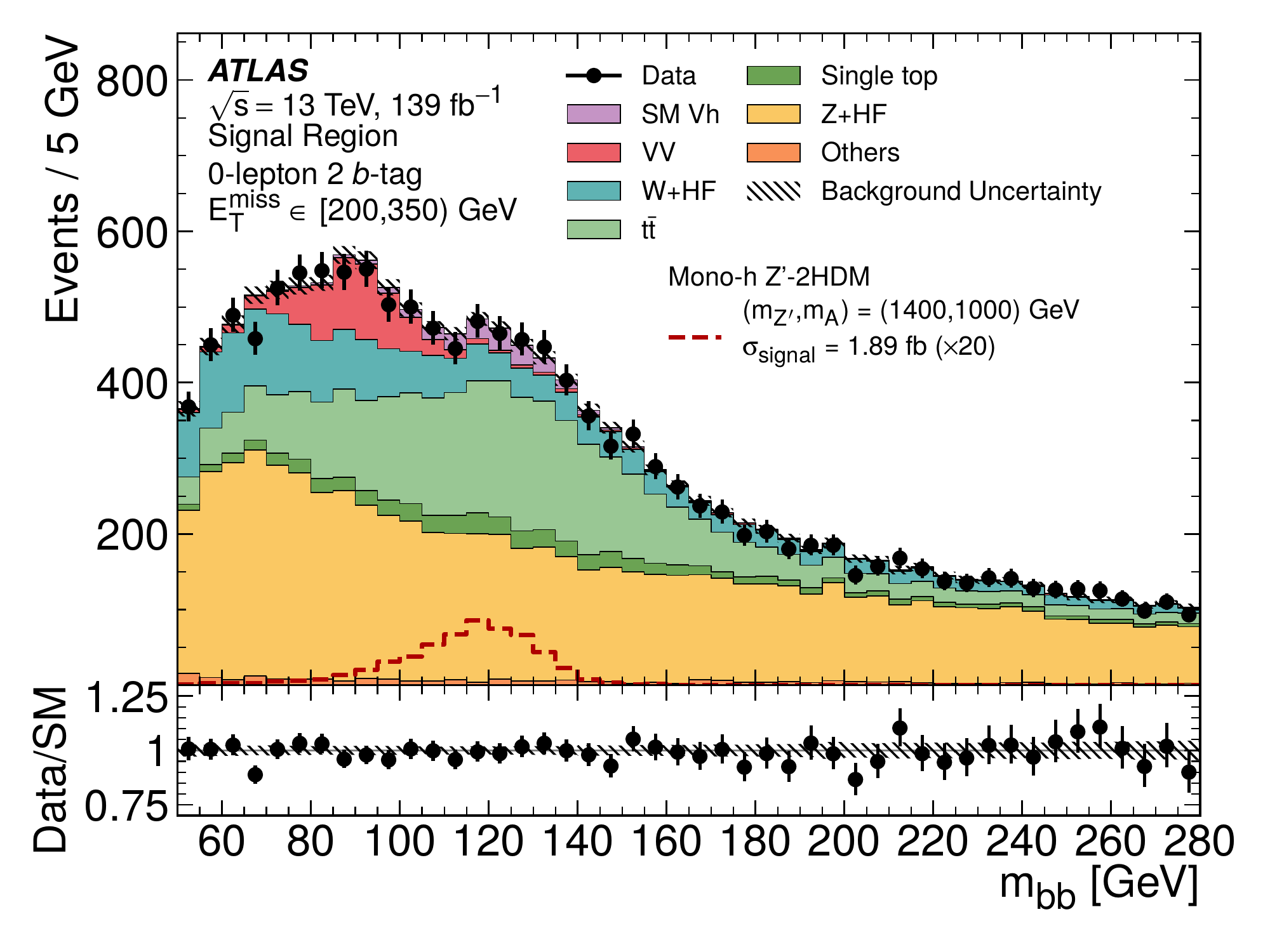}}
  \subfigure[]{\includegraphics[width=0.4\textwidth]{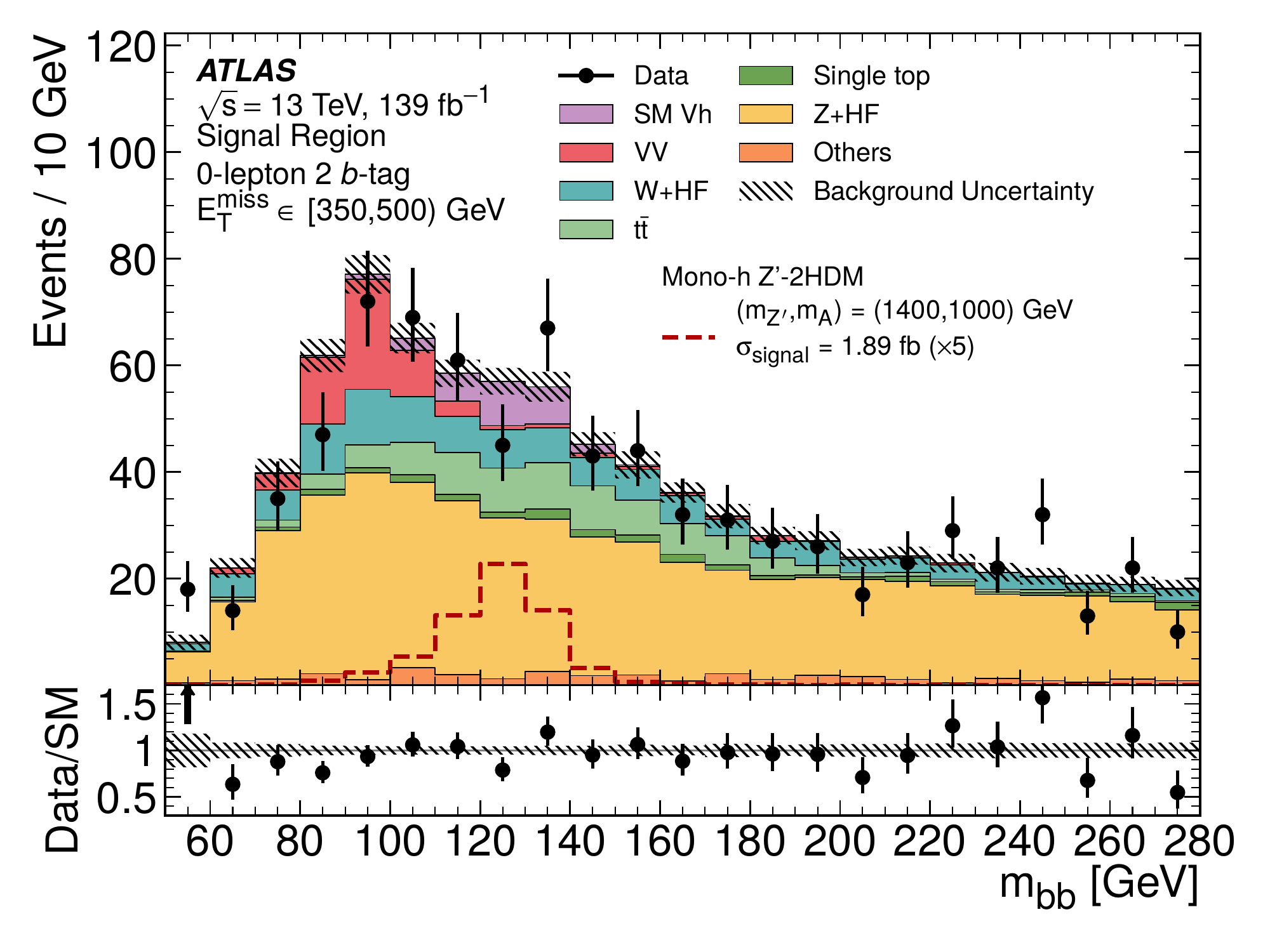}}
  \subfigure[]{\includegraphics[width=0.4\textwidth]{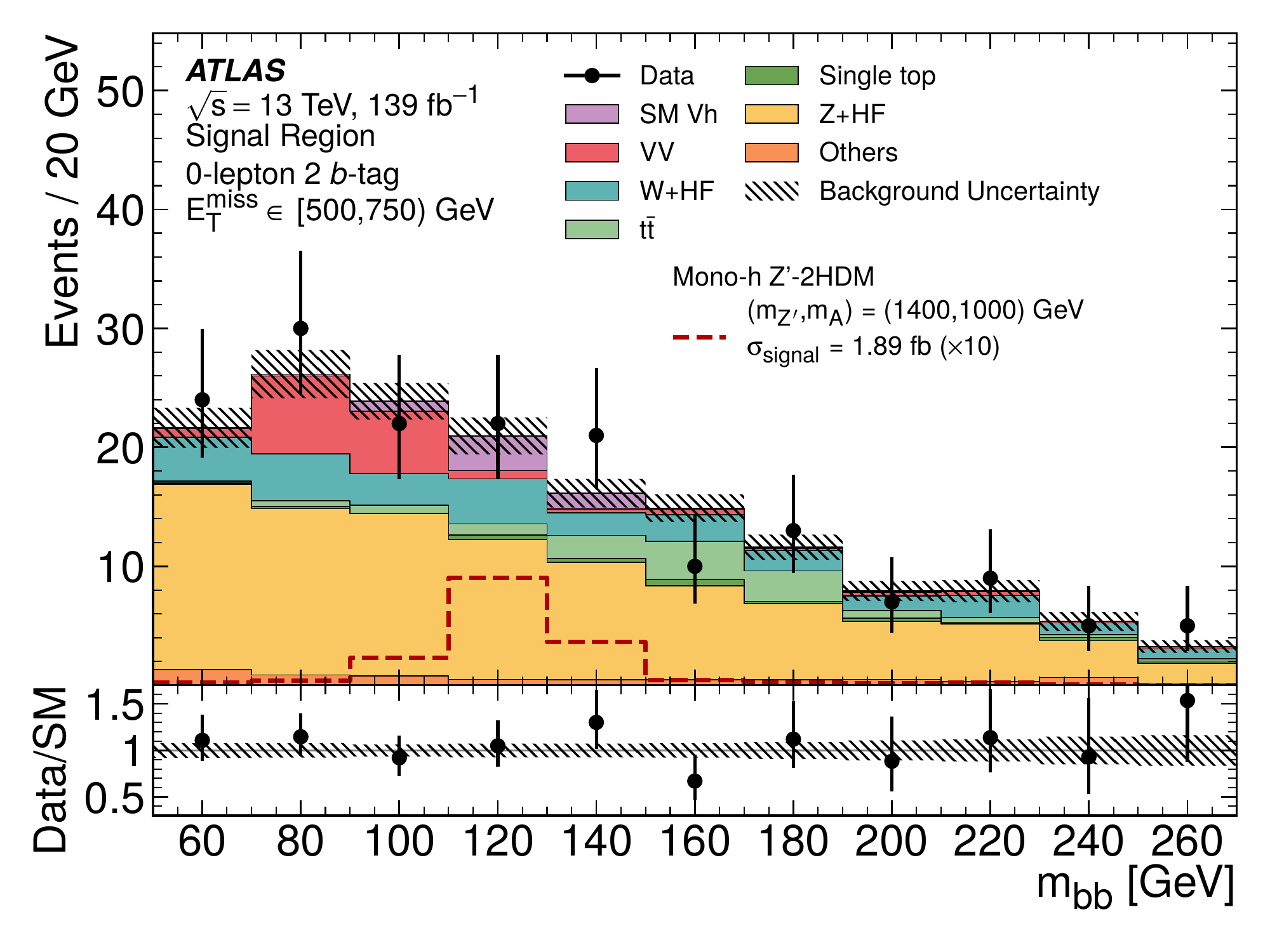}}
  \subfigure[]{\includegraphics[width=0.4\textwidth]{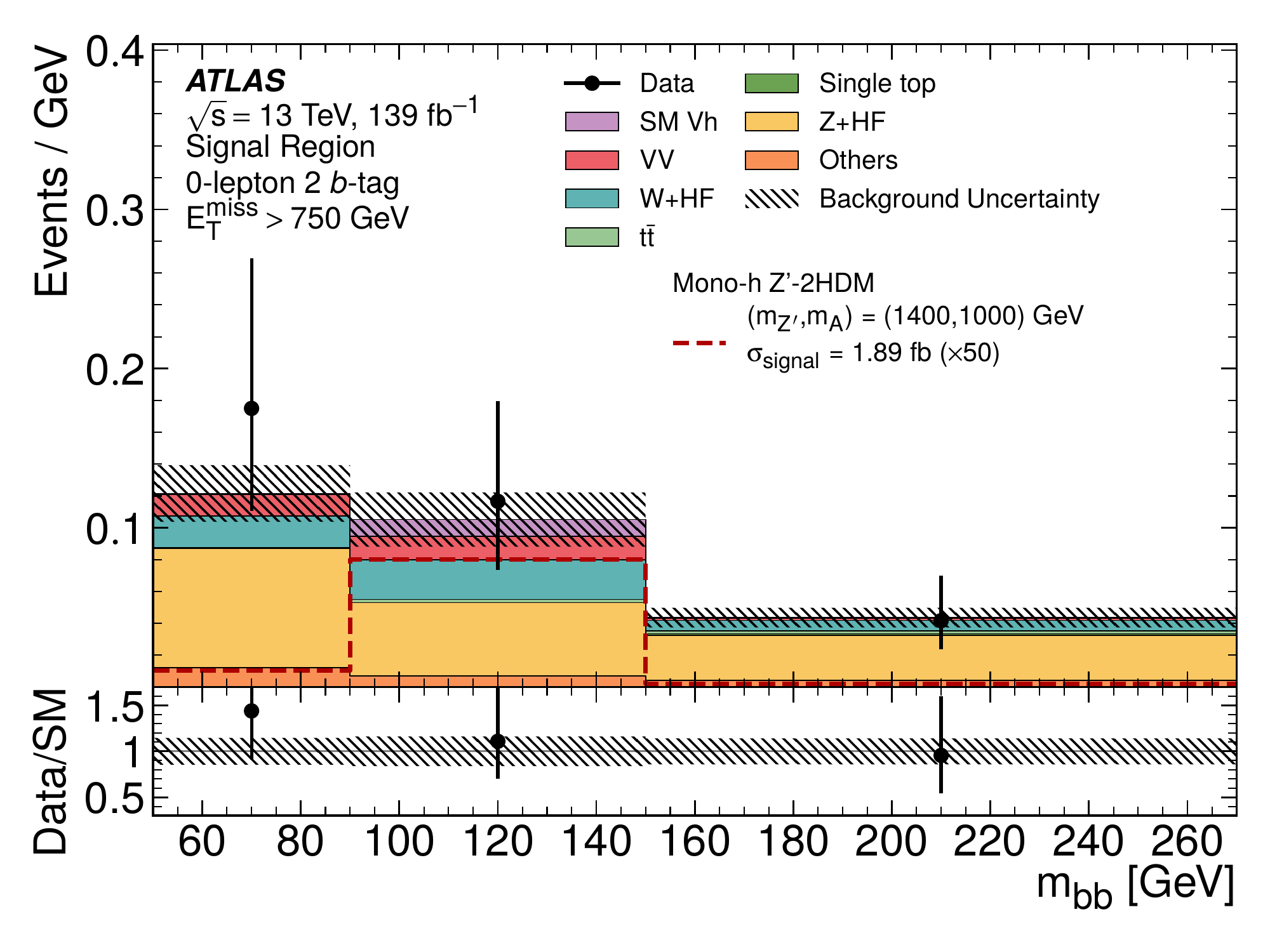}}
  \caption{Distributions of the Higgs boson candidate mass in the 2 $b$-tag signal regions for different \MET\ ranges. The top panel comparies the fitted background yields with data, while the bottom panel indicates the ratio of the observed data to the predicted Standard Model backgrounds. }
  \label{results:fig:SRplots1}
\end{figure}

\begin{figure}[h!]
  \centering
  \subfigure[]{\includegraphics[width=0.45\textwidth]{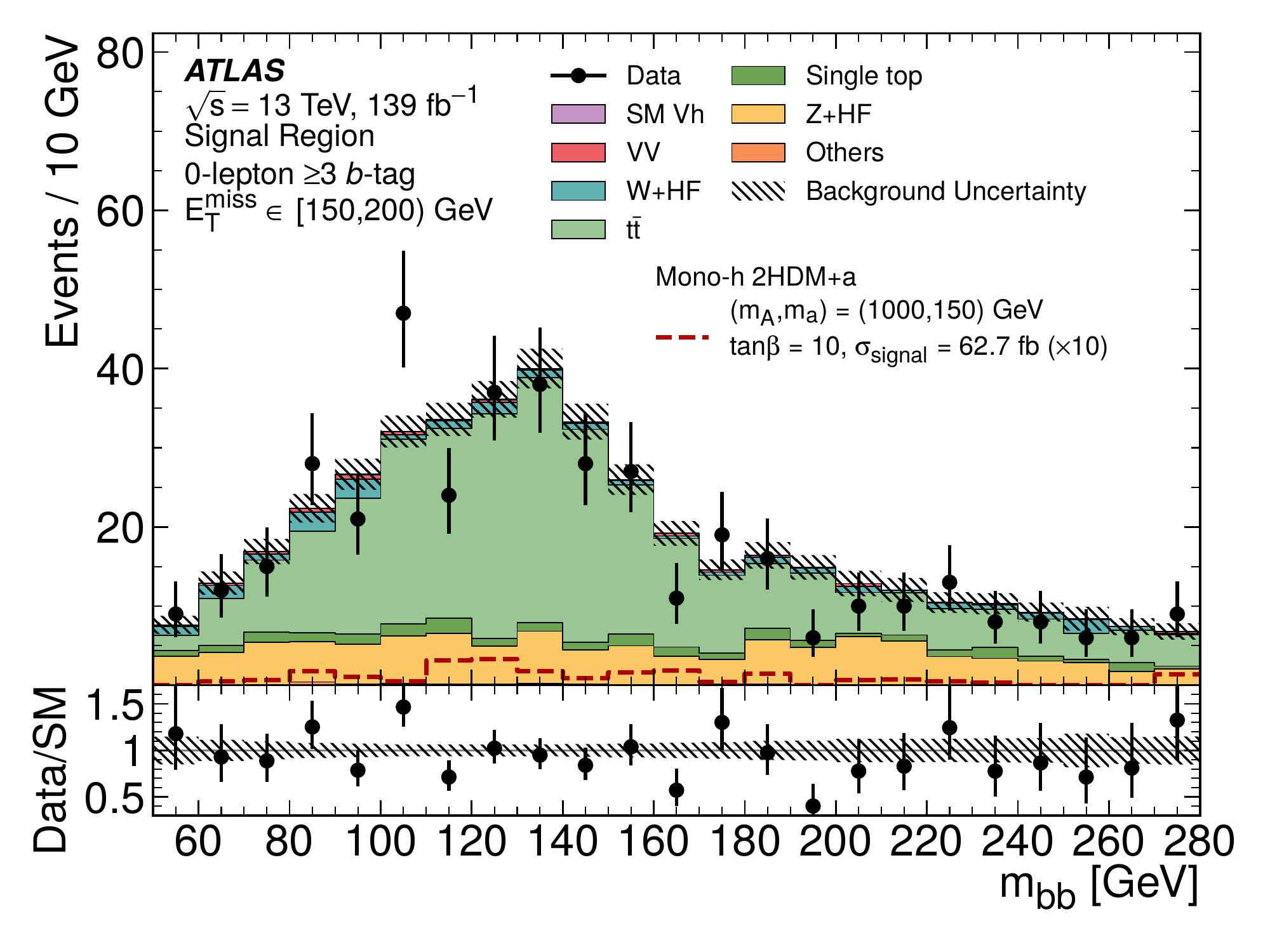}}
  \subfigure[]{\includegraphics[width=0.45\textwidth]{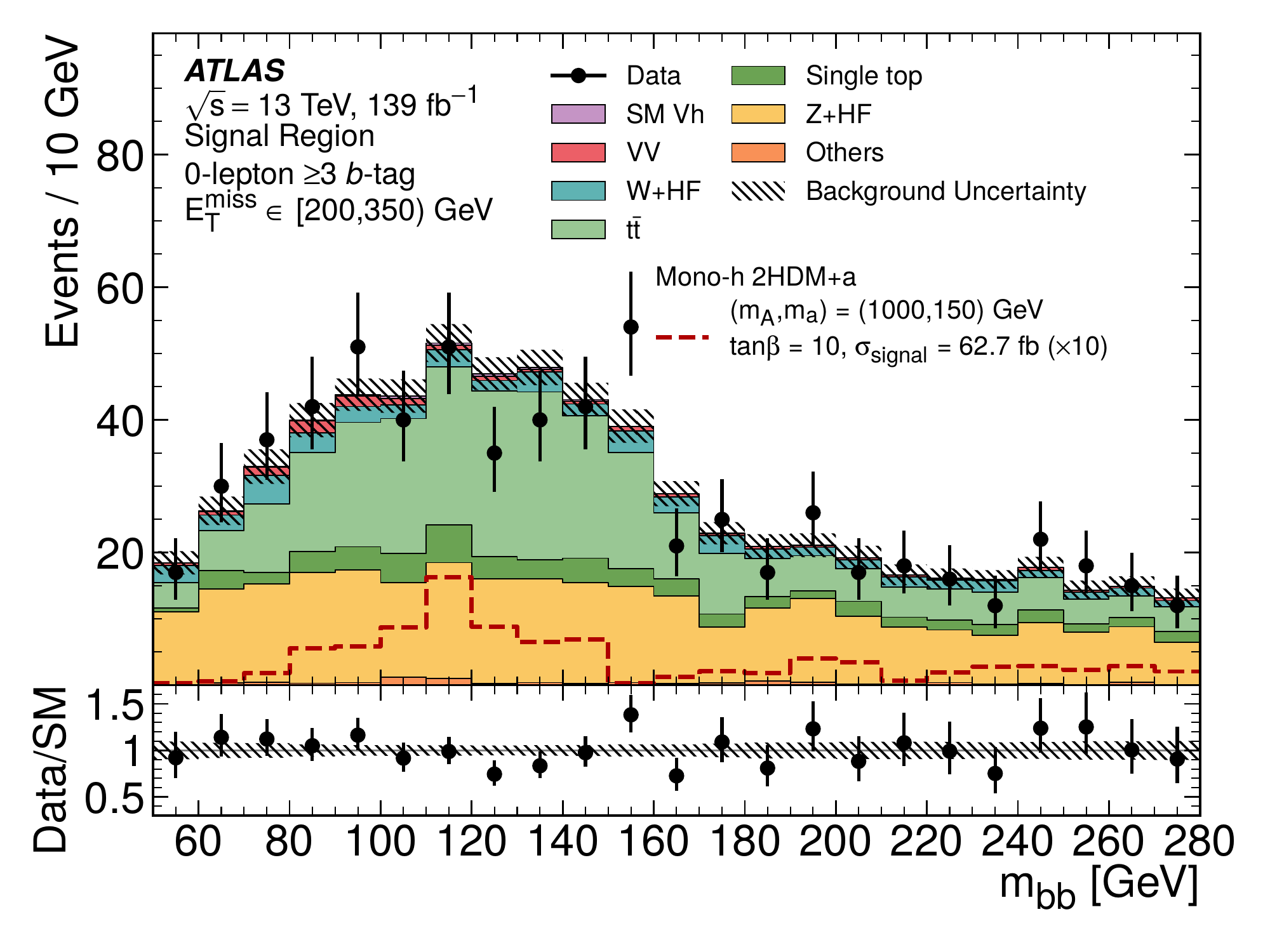}}
  \subfigure[]{\includegraphics[width=0.45\textwidth]{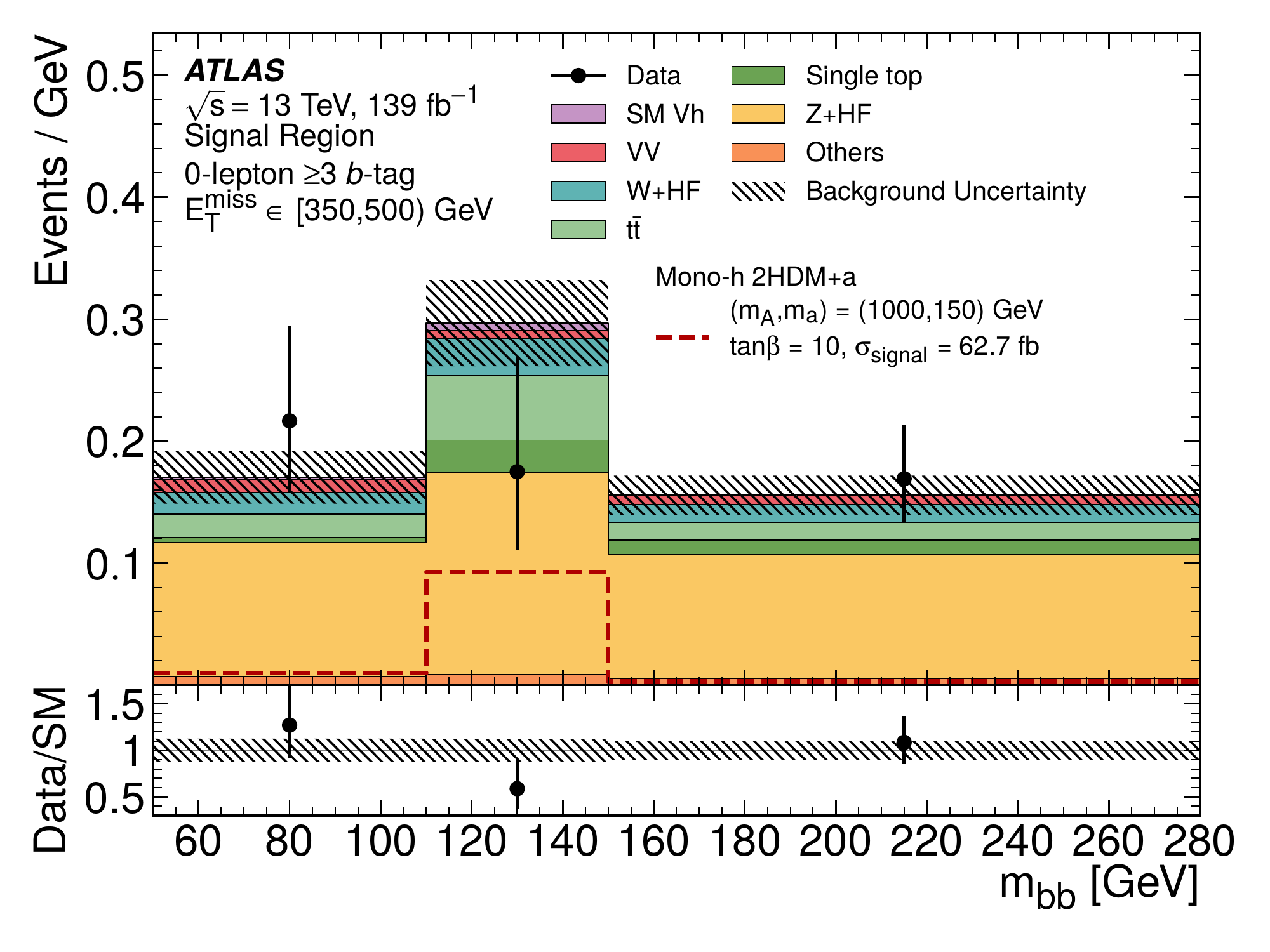}}
  \subfigure[]{\includegraphics[width=0.45\textwidth]{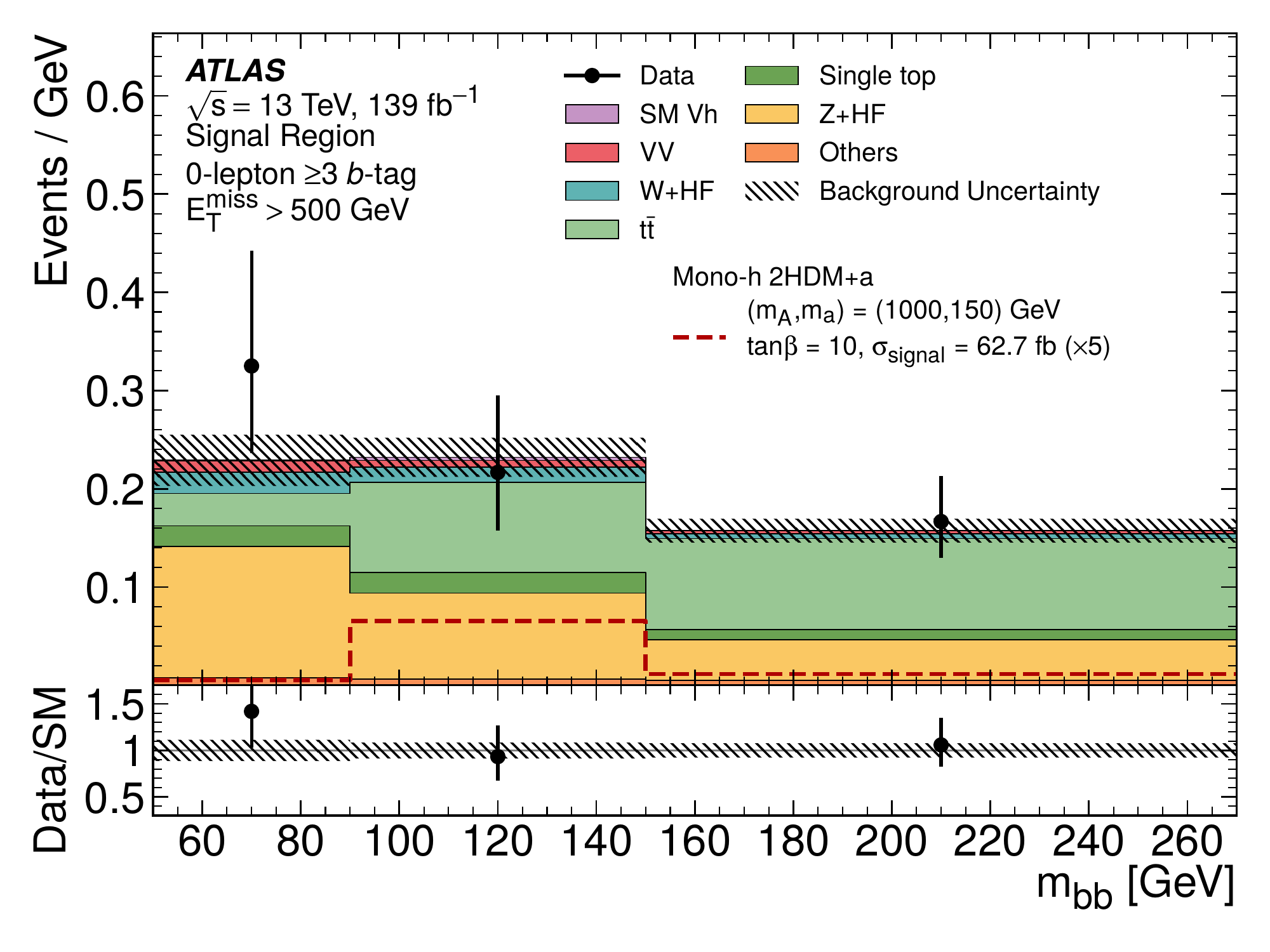}}
  \caption{Distributions of the Higgs boson candidate mass in the $\geq$3 $b$-tag signal regions for different \MET\ ranges. The top panel comparies the fitted background yields with data, while the bottom panel indicates the ratio of the observed data to the predicted Standard Model backgrounds. }
  \label{results:fig:SRplots2}
\end{figure}

\begin{figure}[h!]
  \centering
  \subfigure[]{\includegraphics[width=0.45\textwidth]{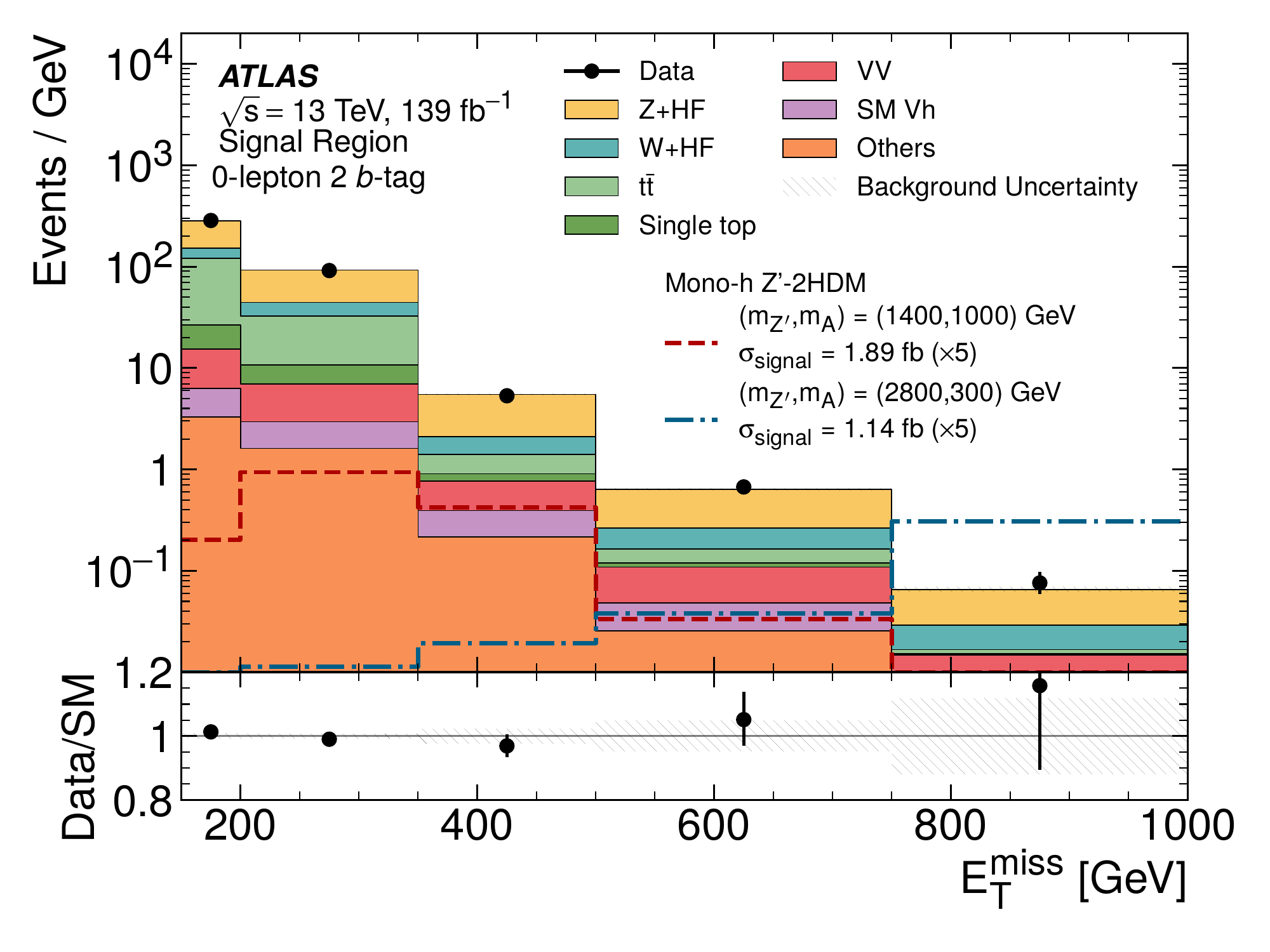}}
  \subfigure[]{\includegraphics[width=0.45\textwidth]{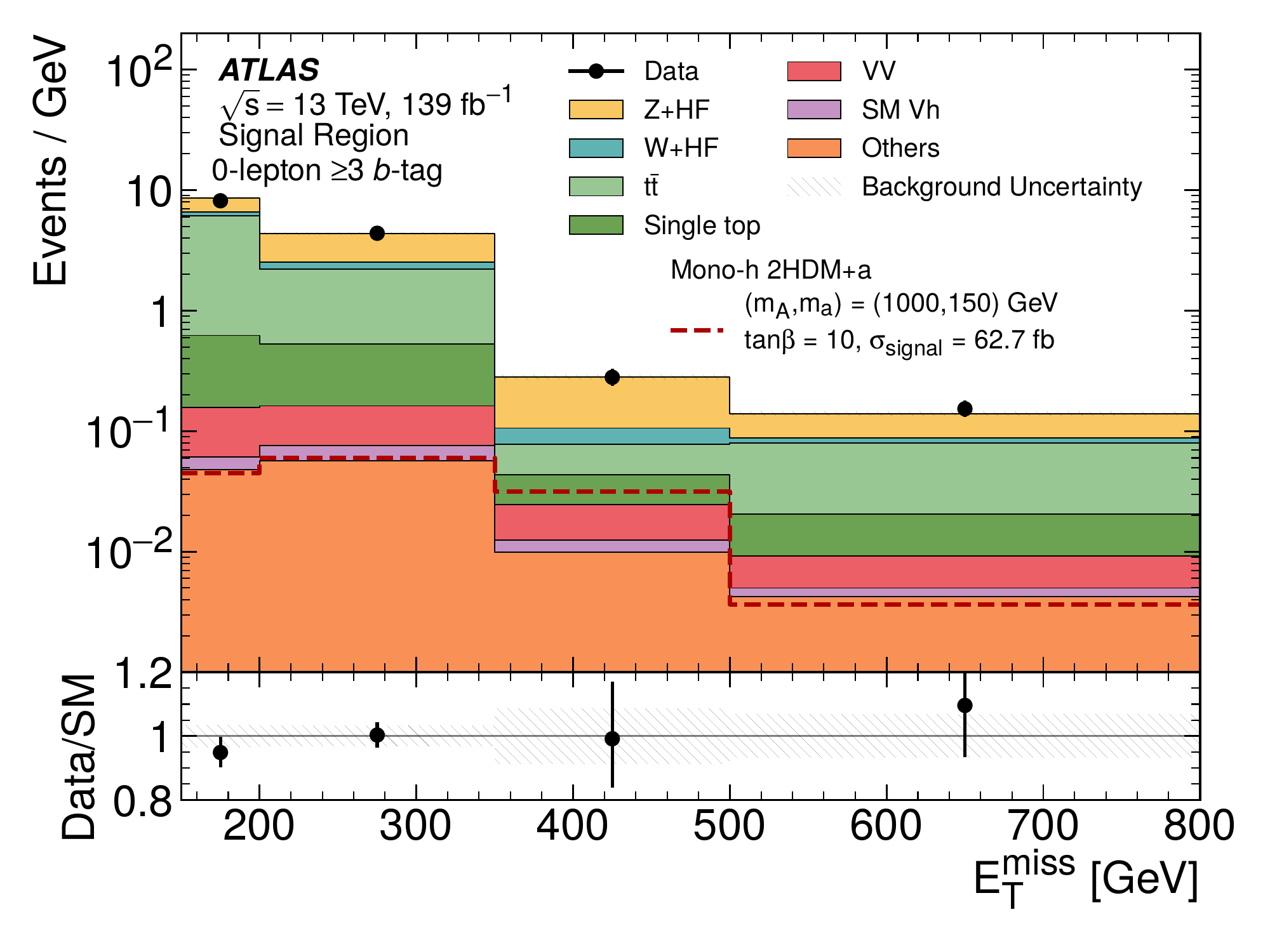}}
  \caption{\MET\ distributions in the (a) 2 $b$-tag and (b) $\geq$3 $b$-tag signal regions for different \MET\ ranges. The top panel comparies the fitted background yields with data, while the bottom panel indicates the ratio of the observed data to the predicted Standard Model backgrounds. }
  \label{fig:results:metdist}
\end{figure}

The distributions of the Higgs boson candidate mass $m_{h}$ after the background-only fit are shown in Fig.~\ref{results:fig:SRplots1} and \ref{results:fig:SRplots2}. The signal to background ratio is higher in the $\geq$3 $b$-tag signal regions, because the 2HDM+$a$ signal model is shown with $\tan\beta=10$, where $b$-associated
production dominates. Tab.~\ref{results:tables:SRyields1} and~\ref{results:tables:SRyields2} present the background estimates in comparison with the observed data. No significant deviation from SM expectations is observed, with the largest deficit corresponding to a local significance of $2.3 \sigma$, and the largest excess amounting to $1.6 \sigma$. Fig.~\ref{fig:results:metdist} summarises the total yields in the signal regions as a function of \MET. The background prediction from simulation is scaled upwards in the fit for lower \MET\ values, while the simulation agrees better with the data for large \MET\ values.

\begin{table}
\centering
\caption{\label{results:tables:SRyields1}Background yields in comparison with data in the 2 $b$-tag signal regions for different \MET\ ranges after a background-only fit to data. Statistical and systematic uncertainties are reported together.}
\vspace{0.2cm}
\footnotesize
\begin{tabular}{l|c|c|c|c|c}
\toprule
2 $b$-tag signal regions & \multirow{2}{*}{$[150,200)~\GeV$} & \multirow{2}{*}{$[200,350)~\GeV$} & \multirow{2}{*}{$[350,500)~\GeV$} & \multirow{2}{*}{$ [500,750)~\GeV$} & \multirow{2}{*}{$ > 750~\GeV$}\\
\MET\ range &  &   &  &   &  \\
\midrule
$Z$+HF & 6470 $\pm$ 310 & 7200 $\pm$ 310 & 507 $\pm$ 26 & 94 $\pm$ 7 & 9.2 $\pm$ 1.8\\
$Z$+light jets & ~~72 $\pm$ 15 & 137 $\pm$ 29 & 18 $\pm$ 4 & ~~4.5 $\pm$ 1.0 & 1.17 $\pm$ 0.30\\ 
$W$+HF & 1590 $\pm$ 210 & 1760 $\pm$ 230 & 106 $\pm$ 14 & 25 $\pm$ 4 & 3.1 $\pm$ 0.6\\
$W$+light jets & ~~86 $\pm$ 35 & ~~92 $\pm$ 35 & 14 $\pm$ 5 & ~~1.6 $\pm$ 0.6 & 0.21 $\pm$ 0.09\\
Single top-quark & ~~570 $\pm$ 260 & ~~570 $\pm$ 260 & ~~21 $\pm$ 10 & ~~2.6 $\pm$ 1.9 & 0.10 $\pm$ 0.16\\
\ttbar\ & 4680 $\pm$ 290 & 3280 $\pm$ 240 & 76 $\pm$ 9 & 11.4 $\pm$ 1.6 & 0.38 $\pm$ 0.08\\
Diboson & 450 $\pm$ 50 & 600 $\pm$ 60 & 56 $\pm$ 7 & 15.2 $\pm$ 1.9 & 1.61 $\pm$ 0.29\\
$Vh$ & 151 $\pm$ 10 & 202 $\pm$ 12 & 26.6 $\pm$ 1.8 & ~~5.6 $\pm$ 0.5 & 0.68 $\pm$ 0.12\\
\ttbar+$V/h$ & ~~7.6 $\pm$ 0.4 & 11.8 $\pm$ 0.5 & ~~0.45 $\pm$ 0.06 & ~~0.286 $\pm$ 0.029 & 0.035 $\pm$ 0.006\\
\midrule
Total background & 14070 $\pm$ 110~~ & 13860 $\pm$ 100~~ & 825 $\pm$ 19 & 160 $\pm$ 8~~ & 16.7 $\pm$ 1.9~~ \\
\midrule
Data & 14259 & 13724 & 799 & 168 & 19\\ \bottomrule
\end{tabular}
\end{table}

\begin{table}
\centering
\caption{\label{results:tables:SRyields2}Background yields in comparison with data in the $\geq$3 $b$-tag signal regions for different \MET\ ranges after a background-only fit to data. Statistical and systematic uncertainties are reported together.}
\vspace{0.2cm}
\small
\begin{tabular}{l|c|c|c|c}
\toprule
$\geq$3 $b$-tag signal regions & \multirow{2}{*}{$[150,200)~\GeV$} & \multirow{2}{*}{$[200,350)~\GeV$} & \multirow{2}{*}{$[350,500)~\GeV$} & \multirow{2}{*}{$ > 500~\GeV$}\\
 \MET\ range & & & &\\
\midrule
$Z$+HF & 102 $\pm$ 15~~ & 278 $\pm$ 28~~ & 26.4 $\pm$ 3.5~~ & 15.6 $\pm$ 1.9~~ \\
$Z$+light jets & 0.6 $\pm$ 0.4 & 2.9 $\pm$ 0.8 & 0.34 $\pm$ 0.12 & 0.46 $\pm$ 0.12\\
$W$+HF & 21 $\pm$ 4~~ & 47 $\pm$ 9~~ & 4.2 $\pm$ 0.9 & 2.4 $\pm$ 0.4\\
$W$+light jets & 0.01 $\pm$ 0.04 & 1.7 $\pm$ 0.9 & 0.8 $\pm$ 0.4 & 0.031 $\pm$ 0.026\\
\ttbar\ & 276 $\pm$ 19~~ & 252 $\pm$ 22~~ & 5.1 $\pm$ 0.7 & 17.9 $\pm$ 1.8~~ \\
Single top-quark & 23 $\pm$ 11 & 55 $\pm$ 25 & 2.9 $\pm$ 1.4 & 3.4 $\pm$ 1.7\\
Diboson & 4.8 $\pm$ 1.4 & 12.9 $\pm$ 2.2~~ & 1.8 $\pm$ 0.4 & 1.26 $\pm$ 0.31\\
$Vh$ & 0.65 $\pm$ 0.28 & 2.9 $\pm$ 0.5 & 0.40 $\pm$ 0.08 & 0.230 $\pm$ 0.025\\
\ttbar+$V/h$ & 1.78 $\pm$ 0.17 & 3.89 $\pm$ 0.26 & 0.371 $\pm$ 0.035 & 0.78 $\pm$ 0.08\\
\midrule
Total background & 430 $\pm$ 15~~ & 656 $\pm$ 21~~ & 42 $\pm$ 4~~ & 42.0 $\pm$ 2.8~~ \\
\midrule
Data & 408 & 658 & 42 & 46\\ \bottomrule
\end{tabular}
\end{table}

The results are interpreted as exclusion limits at 95\% CL in the $Z'$-2HDM and the 2HDM+$a$ scenarios
in Fig.~\ref{results:fig:contourplots2} and \ref{results:fig:contourplots}. Considering the $Z'$-2HDM case, $Z'$ masses up to 3 \TeV\ are excluded for $A$ masses of 300~\GeV\ at 95\% CL. The exclusion boundaries for the 2HDM+$a$ scenario extend up to $m_a = 520~\GeV$ for $m_A = 1.25~\TeV$ for \ggF\ production and $\tan \beta = 1$. This is an improvement of about 200~\GeV\ in $m_a$ on previous results~\cite{ATLAS:2019wdu}, which reinterpreted the earlier $h(\rightarrow b\bar{b})$ + \MET\ analysis using 36.1~\ifb~\cite{ATLAS:2017uis}. 

The higher exclusion limit at high $m_{A}$, low $m_{a}$, is due to an increase of the cross-section of the $a \rightarrow ah$ process, without resonant $A$ production. It should be noted that with the exact parameter choices adopted in this analysis, the $aah$ coupling becomes larger than $4\pi$ for $m_A \gtrsim 1750\,\GeV$. Moreover, as discussed in Refs.~\cite{LHCDarkMatterWorkingGroup:2018ufk,2hdma} values of $m_A \gtrsim 1250\,\GeV$ (for $\tan \beta=1$) or $m_A \gtrsim 2150\,\GeV$ (for $\tan \beta=10$) would not be consistent with the requirement of having a bounded-from-below scalar potential, given the parameter choices discussed in this paper. These constraints can be relaxed substantially if the quartic couplings assume a value closer to the perturbativity limit and also in more general 2HDMs containing additional couplings as discussed in Refs.~\cite{Bauer:2017fsw,Haisch:2018djm}. Therefore, the above should not be considered as a strong requirement for the validity of the model predictions. At high $m_{A}$ the width of the additional Higgs bosons grows substantially and the theoretical predictions are subject to additional theoretical uncertainties associated with the treatment of the width. Exclusion limits are therefore not shown in the region of very large widths ($m_A>2200$ GeV). 

In the case of $bbA$ production and $\tan \beta = 10$, the exclusion limits extend up to $m_a = 240~\GeV$ for $m_A = 900~\GeV$. The inclusion of the $\geq$3 $b$-tag regions helps to increase the sensitivity relative to the 2 $b$-tag region by about 30--70\%. The difference between observed and expected limits arises from data deficits in the $\geq$3 $b$-tag region, especially the deficit around the Higgs boson peak in the $\MET \in [350,500)~\GeV$ region, as shown in Fig.~\ref{results:fig:SRplots2}. The 2HDM+$a$ scenario with $bbA$ production and $\tan \beta = 10$ is considered for the signatures discussed in this paper for the first time.

\begin{figure}[h!]
  \centering
  \includegraphics[width=0.7\textwidth]{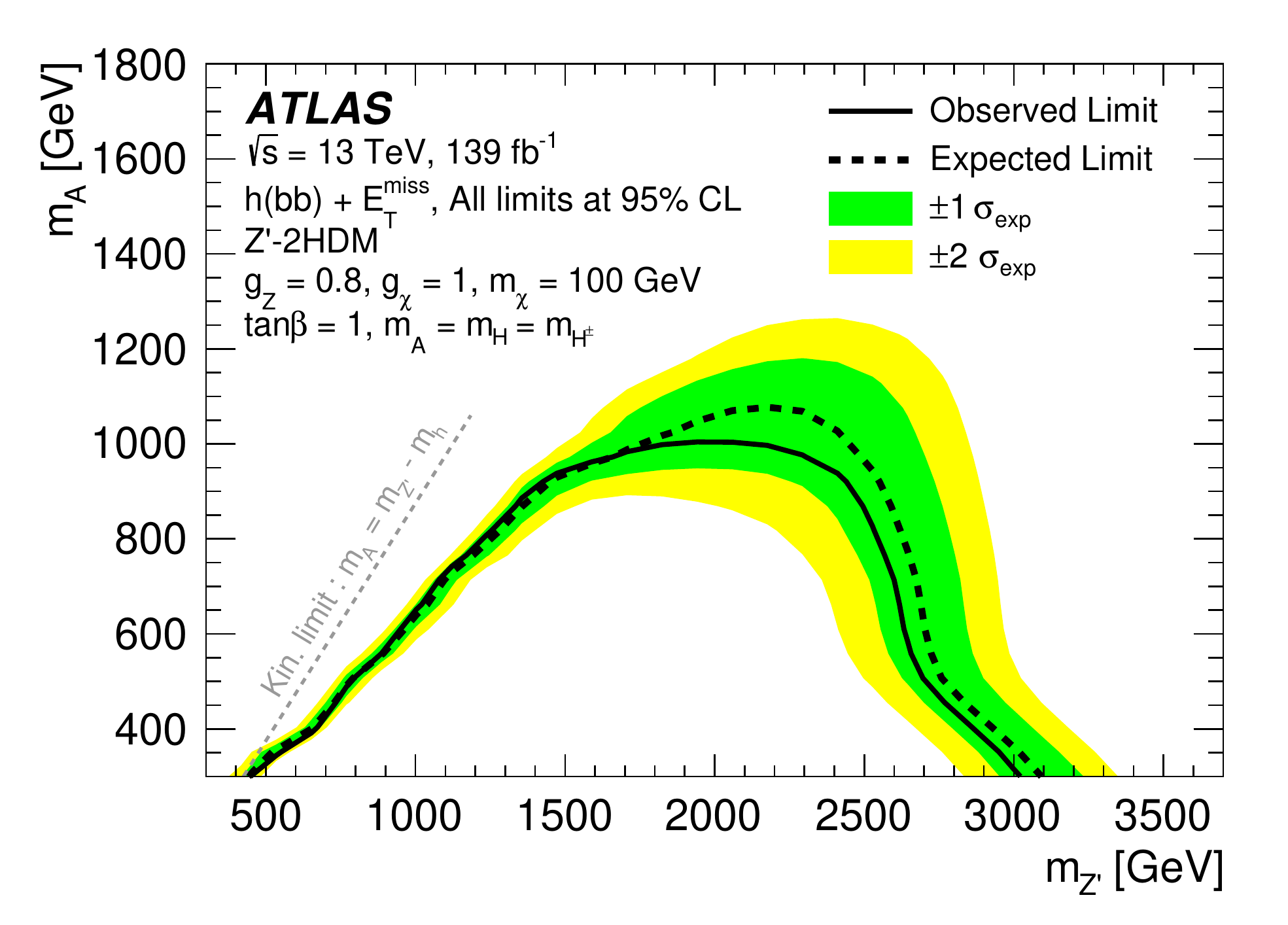}
  \caption{Exclusion limits in the $Z^{'}$-2HDM model. The solid black line shows the observed limit at 95\% CL, the dashed black line the expected limit. The green band gives the $\pm 1 \sigma$ uncertainties of the expected limit, the yellow band the $\pm 2 \sigma$ uncertainties. }
  \label{results:fig:contourplots2}
\end{figure}

\begin{figure}[h!]
  \centering
  \subfigure[]{\includegraphics[width=0.45\textwidth]{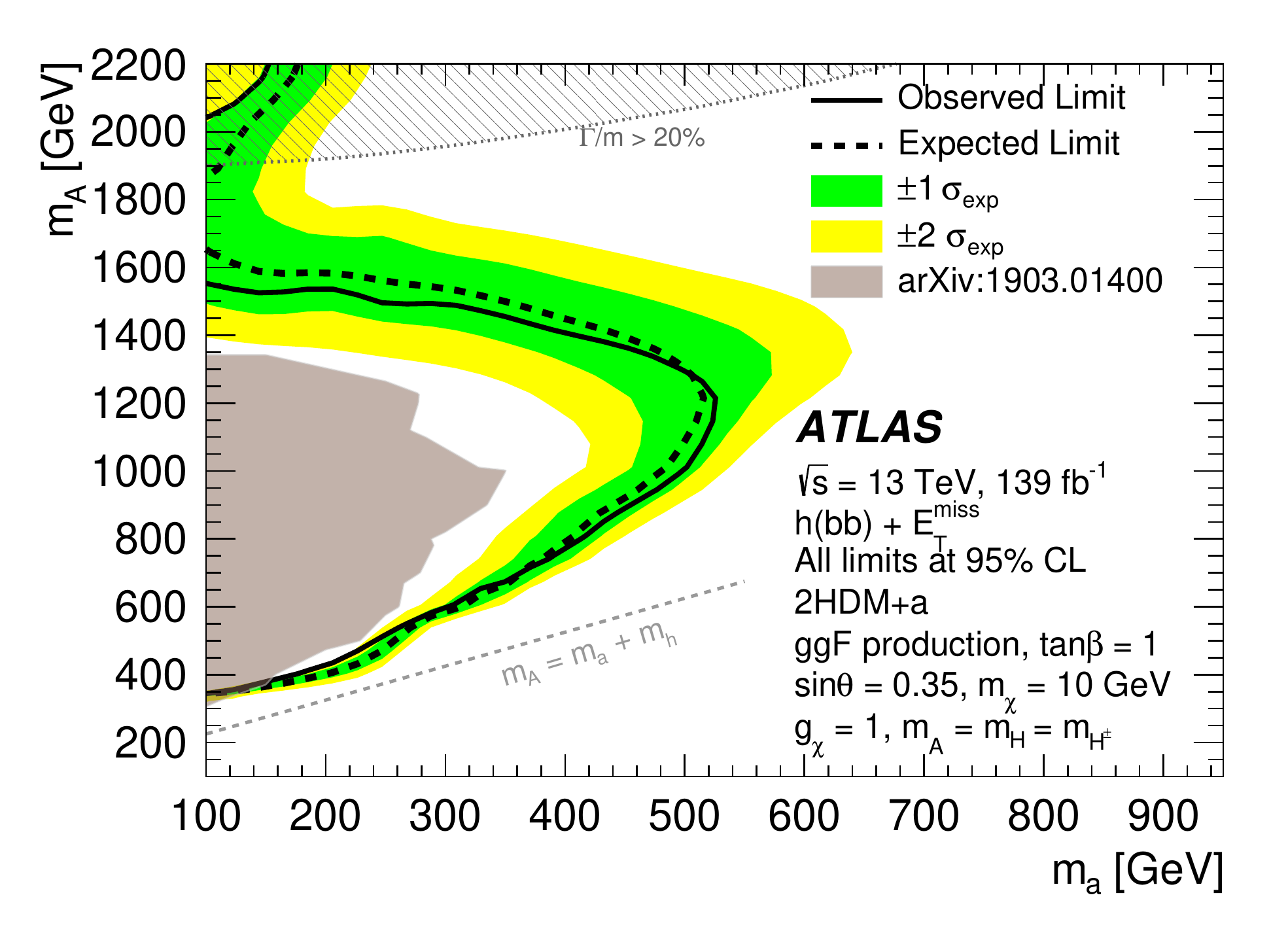}}
  \subfigure[]{\includegraphics[width=0.45\textwidth]{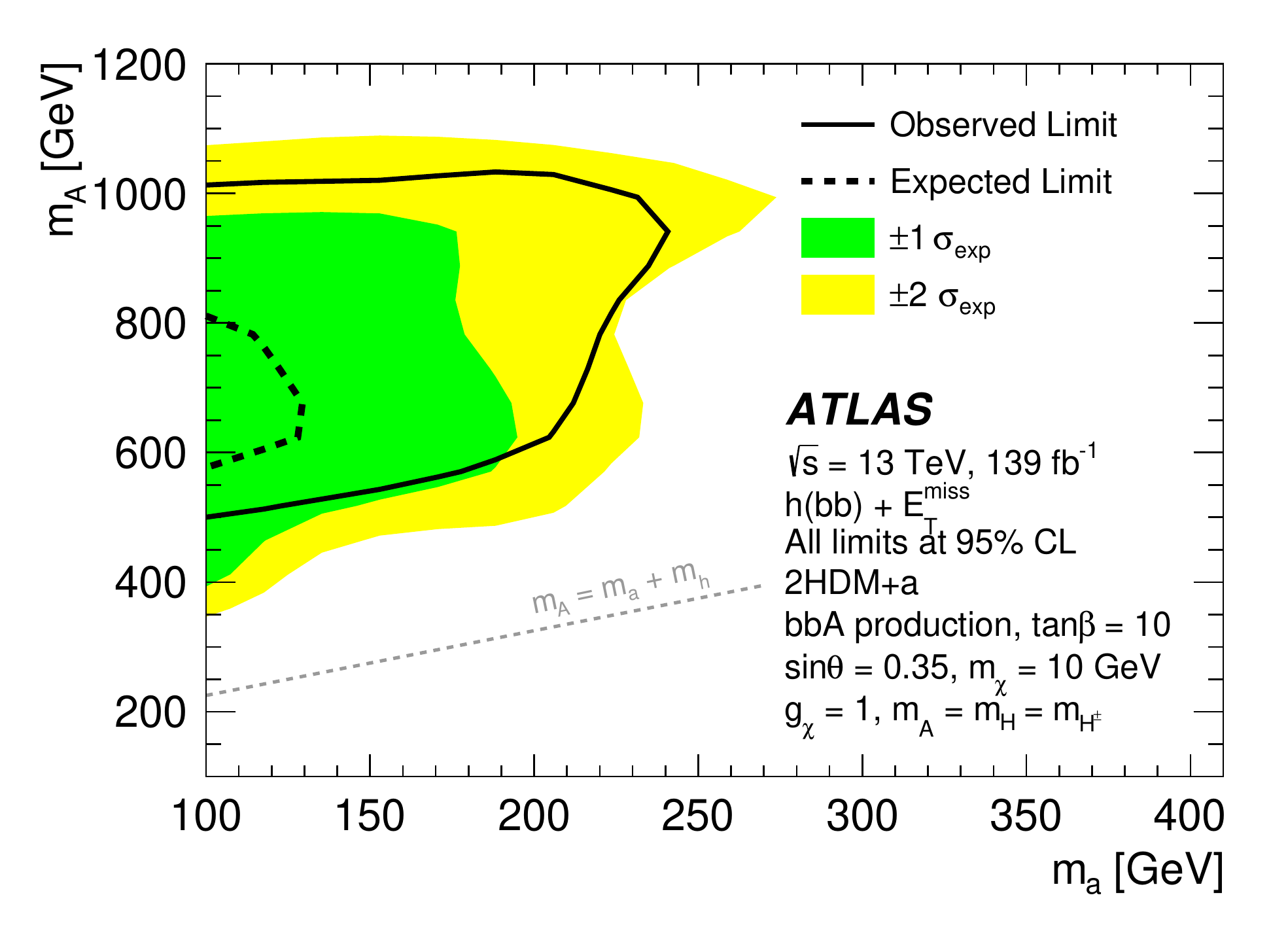}}
  \caption{Exclusion limits for the 2HDM+$a$ signal with (a)~$\tan \beta = 1$ and $gg$F production and with (b)~$\tan \beta = 10$ and $bbA$ production. The solid black line shows the observed limit at 95\% CL, the dashed black line the expected limit. The green band gives the $\pm 1 \sigma$ uncertainties of the expected limit, the yellow band the $\pm 2 \sigma$ uncertainties.
 }
  \label{results:fig:contourplots}
\end{figure}

Fig.~\ref{results:fig:upperlimits} displays the upper limits on the visible cross-section, which are derived from the fit with the the model-independent setup. These model-independent limits are particularly useful for the reinterpretation with any additional interesting signal models.

\begin{figure}[h!]
  \centering
  \includegraphics[width=0.7\textwidth]{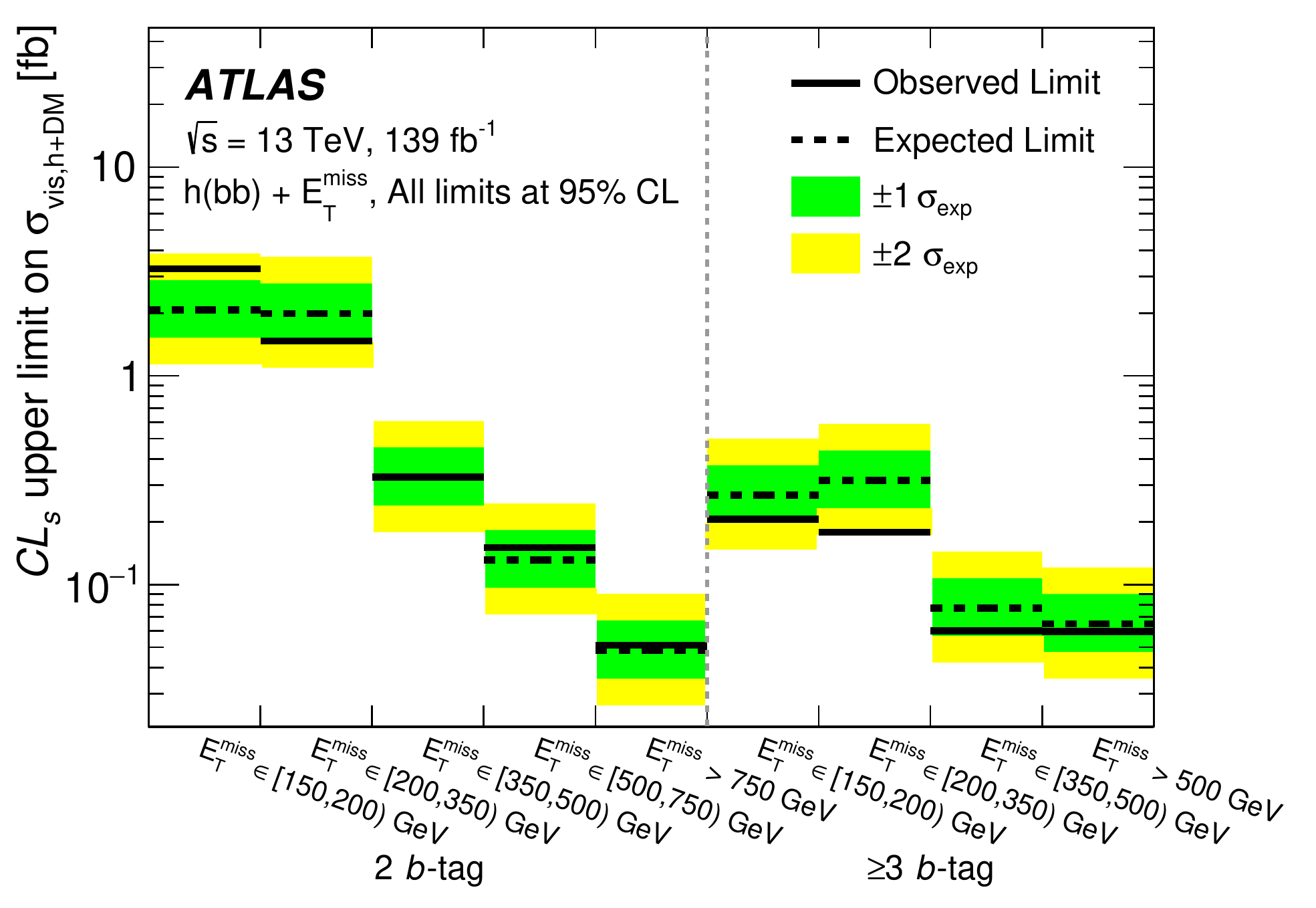}
  \caption{Model-independent upper limits on the visible cross-section $\sigma_{\mathrm{vis},h(b\bar{b})+\mathrm{DM}} \equiv \sigma_{h+\mathrm{DM}} \times \mathscr{B}(h \to b\bar{b}) \times \mathscr{A} \times \varepsilon$ in the different signal regions. }
  \label{results:fig:upperlimits}
\end{figure}

\chapter{Unbinned profiled unfolding}
\label{chap:upu}

\section{Motivation}

One of the most common analysis goals in particle and nuclear physics is the measurement of differential cross sections.  These quantities encode the rate at which a particular process occurs as a function of certain observables of interest.  From measured cross sections, a number of downstream inference tasks can be performed, including the estimation of fundamental parameters, tuning simulations, and searching for physics beyond the Standard Model.  The key challenge of cross section measurements is correcting the data for detector distortions, a process called deconvolution or \textit{unfolding}.  See Refs.~\cite{Cowan:2002in,Blobel:2203257,doi:10.1002/9783527653416.ch6,Balasubramanian:2019itp} for recent reviews on unfolding and Refs.~\cite{DAgostini:1994fjx,Hocker:1995kb,Schmitt:2012kp} for the most widely-used unfolding algorithms.

Until recently, all cross section measurements were performed with histograms.  In particular, the target spectra and experimental observations were binned and the unfolding problem is recast in the language of linear algebra. That is, one would like to determine the signal strength, defined as the ratio of the observed signal yield to the theoretical prediction, for each bin based on the measurements from experimental observations. This approach comes with the limitation that the binning must be determined beforehand. This makes it difficult to compare measurements with different binning. Furthermore, the optimal binning depends on the downstream inference task.

Modern machine learning (ML) has enabled the creation of unfolding methods that can process unbinned data~\cite{Arratia:2021otl}.  Deep generative models such as Generative Adversarial Networks (GAN)~\cite{Goodfellow:2014:GAN:2969033.2969125,Datta:2018mwd,Bellagente:2019uyp} and Variational Autoencoders (VAE)~\cite{kingma2014autoencoding,Howard:2021pos} produce implicit models that represents the probability density of the unfolded result and allow to sample from the probability density. Methods based on Normalizing Flows (NF)~\cite{10.5555/3045118.3045281,Bellagente:2020piv,Vandegar:2020yvw,Backes:2022vmn} allow for both sampling and density estimation.  In contrast, the classifier-based method OmniFold Refs.~\cite{Andreassen:2019cjw,Andreassen:2021zzk} iteratively reweights a simulated dataset.   A summary of machine learning-based unfolding methods can be found in Ref.~\cite{Arratia:2021otl} and recent applications of these techniques (in particular, of OmniFold) to experimental data are presented in Refs.~\cite{H1:2021wkz,H1prelim-22-031,H1prelim-22-034,LHCb:2022rky}.  While powerful, none of these approaches can simultaneously estimate cross sections and fit (nuisance) parameters.  This can be a significant shortcoming when the phase space region being probed has non-trivial constraining power for systematic uncertainties.

Unfolding methods that can also profile have been proposed.  One possibility is to treat the cross section in each region of particle-level phase space (i.e. in a histogram bin) as a free parameter and then perform a likelihood fit as for any set of parameters of interest and nuisance parameters.  For example, this is the setup of the the Simplified Template Cross Section (STXS) (e.g. Refs.~\cite{LHCHiggsCrossSectionWorkingGroup:2016ypw, Andersen:2016qtm, Berger:2019wnu, Amoroso:2020lgh}) measurements for Higgs boson kinematic properties. Another possibility is Fully Bayesian Unfolding (FBU) \cite{https://doi.org/10.48550/arxiv.1201.4612}, which samples from the posterior probability over the cross section in each bin of the particle-level phase space and over the nuisance parameters.  All of these methods require binning.

We are thus motivated to propose a new machine learning-based unfolding method that is both unbinned at particle level and can profile, referred to as Unbinned Profiled Unfolding (UPU). UPU reuses all the standard techniques used in binned maximum likelihood unfolding and combines them with ML methods that allow for unbinned unfolding. Specifically, we use the binned maximum likelihood at detector level as the metric to optimize the unfolding, while the unfolding takes unbinned particle-level simulations as inputs.

The rest of this chapter is organized as follows. In Sec.~\ref{sec:UPU}, we describe the procedure and implementation details of UPU. We then present simple Gaussian examples to demonstrate the usage of UPU in Sec.~\ref{sec:gaussian}. In Sec.~\ref{sec:higgs}, we apply UPU to a simulated Higgs boson cross section measurement at the Large Hadron Collider (LHC). The conclusions and outlook are then given in Sec.~\ref{sec:conclusion}.

\section{Unbinned Profiled Unfolding}
\label{sec:UPU}

\subsection{Statistical Setup}

UPU generalizes binned maximum likelihood unfolding to the unbinned case.  Binned maximum likelihood unfolding can be described by the following optimization setup:

\begin{align}
\label{eq:binned}
    (\hat{k},\hat{\theta})=\text{argmax}_{(k,\theta)}\Pr(m|k,\theta)\,p_0(\theta)\,,
\end{align}
where $m\in\mathbb{R}^{N_m}$ is a vector representing the counts in each of the $N_m$ bins at detector level, $k\in\mathbb{R}^{N_k}$ is a vector representing the counts in each of the $N_k$ bins at particle level (usually $N_m\geq N_k$), $\theta$ are the nuisance parameters, and $p_0$ is the prior on $\theta$.  Our idea is to keep the structure of Eq.~\ref{eq:binned}, but replace $k$ with an unbinned estimator of the particle-level spectrum.  Suppose that the particle-level phase space is\footnote{Assuming the space is suitably standardized to remove units.} $\mathbb{R}^N$ and let\footnote{We will use $[\cdot]$ to denote the parameters of the function and $(\cdot)$ to denote the inputs of the function, e.g. $f[\theta](x)$ is a functional in $\theta$ and a function in $x$.} $\tau[\omega]\in \mathbb{R}^{\mathbb{R}^N}$ parameterize the probability density over this space for parameters $\omega$. The goal of UPU is then to optimize

\begin{align}
\label{eq:unbinned1}
    (\hat{\omega},\hat{\theta})= \text{argmax}_{(\omega,\theta)}\Pr(m|\tau[\omega],\theta)\,p_0(\theta)\,,
\end{align}
where the final result would be given by $\tau[\hat{\omega}]$.  The challenge with the construction in Eq.~\ref{eq:unbinned1} is that for a given truth spectrum $\tau[\omega]$, we need to know the predicted detector-level distribution.  In the binned case, this is readily computed by multiplying $k$ by the response matrix $R_{ij}=\Pr(\text{measure in bin $i$}|\text{truth is bin $j$})$.  When the truth are unbinned, we need the full detector response. This is never known analytically and would be challenging to estimate numerically with a surrogate density estimator\footnote{Note that Eq.~\ref{eq:unbinned1} is a probability distribution over probability distributions so building it from the per-event detector response is non-trivial.}.  To address this challenge, we make use of the fact that simulated events come in pairs, with a matching between particle-level and detector-level events.  Instead of estimating $\tau$ directly, we use a fixed simulation (with particle-level spectrum $\tau[\omega_0]$) and then learn a reweighting function $w_0[\lambda]$ to estimate the likelihood ratio between the unfolded result and the fixed simulation at particle level.  Schematically:
\begin{align}
\label{eq:unbinned}
    (\hat{\lambda},\hat{\theta})= \text{argmax}_{(\lambda,\theta)}\Pr(m|\tau [\omega_0]w_0[\lambda]),\theta)\,p_0(\theta)\,,
\end{align}
where in practice, we only have samples from $\tau[\omega_0]$ and $w_0$ is a surrogate model.  The number of predicted events in a given bin $i$ is then a sum over weights $w_0[\hat{\lambda}]$ (evaluated at particle-level) for simulated events with a detector-level value in bin $i$.  The probability over values $m$ is then a product over Poisson probability mass functions, since the bins are statistically independent.  The fact that the probability mass is known is crucial and means that UPU does not readily generalize the case where the detector-level phase space is also unbinned.

\subsection{Machine Learning Approach}

For particle-level features $T$ and detector-level features $R$, the main goal is to train the likelihood ratio estimator $w_0\left(T\right)$, which reweights the simulated particle-level spectrum.  In the absence of profiling, this corresponds to the following loss function:

\begin{align}
\label{eq:lossnoprofile}
    L &= \prod_{i=1}^{n_\text{bins}}\Pr(n_i\Bigg|\sum_{j=1}^{n_\text{MC}}w_0(T_j)\mathbb{I}_i(R_j))\,,
\end{align}
where $n_i$ is the number of observed events in bin $i$, $n_\text{MC}$ is the number of simulated events, and $\mathbb{I}_i(\cdot)$ is the indicator function that is one when $\cdot$ is in bin $i$ and zero otherwise.  When $w_0$ is parameterized as a neural network (see Sec.~\ref{sec:implement}), then the logarithm of Eq.~\ref{eq:lossnoprofile} is used for training:

\begin{align}
\label{eq:lossnoprofile2}
    &\log L = \\\nonumber
    &\hspace{2mm}\sum_{i=1}^{n_\text{bins}}\left[n_i\log\left(\sum_{j=1}^{n_\text{MC}}w_0(T_j)\mathbb{I}_i(R_j)\right)-\sum_{i=1}^{n_\text{MC}}w_0(T_j)\mathbb{I}_i(R_j)\right]\,,
\end{align}
where we have dropped constants that do not affect the optimization.  Experimental nuisance parameters modify the predicted counts in a particular bin given the particle-level counts.  We account for these effects with a second reweighting function:

\begin{align}
    w_1(R|T,\theta)=\frac{p_{\theta}(R|T)}{p_{\theta_0}(R|T)}\,,
\end{align}
where $p_\theta(R|T)$ is the conditional probability density of $R$ given $T$ with nuisance parameters $\theta$.  Importantly, $w_1$ does not modify the target particle level distribution.  Incorporating $w_1$ into the log likelihood results in the full loss function:


\begin{equation}
\begin{split}
    \log L&=\sum_{i=1}^{n_\text{bins}}\Biggl[n_i\log\left(\sum_{j=1}^{n_\text{MC}}w_0(T_j)w_1(R_j|T_j,\theta)\mathbb{I}_i(R_j)\right) \\
     &-\sum_{j=1}^{n_\text{MC}}w_0(T_j)w_1(R_j|T_j,\theta)\mathbb{I}_i(R_j)\Biggr]+\log p_0(\theta)\,.
\end{split}
\label{eq:lllall}
\end{equation}

Since $w_1$ does not depend on the particle-level spectrum, it can be estimated prior to the final fit and only the parameters of $w_0$ and the value(s) of $\theta$ are allowed to float when optimizing Eq.~\ref{eq:lllall}.

\subsection{Machine Learning Implementation}
\label{sec:implement}

In our subsequent case studies, the reweighting functions $w_0$ and $w_1$ are parametrized with neural networks.  The $w_0$ function is only constrained to be non-negative and so we choose it to be the exponential of a neural network.

The pre-training of $w_1$ requires neural conditional reweighting~\cite{2107.08979}, as a likelihood ratio in $R$ conditioned on $T$ and parameterized in $\theta$.  While there are multiple ways of approximating conditional likelihood ratios, the one we found to be the most stable for the examples we have studied for UPU is the product approach:

\begin{align}
    w_1(R|T,\theta)=\left(\frac{p_\theta(R,T)}{p_{\theta_0}(R,T)}\right)\left(\frac{p_{\theta_0}(T)}{p_{\theta}(T)}\right)\,,
\end{align}
where the two terms on the righthand side are separately estimated and then their product is $w_1$.  For a single feature $T$, a likelihood ratio between samples drawn from a probability density $p$ and samples drawn from a probability density $q$ is estimated using the fact that machine learning-classifiers approximate monotonic transformations of likelihood ratios (see e.g. Ref.~\cite{hastie01statisticallearning,sugiyama_suzuki_kanamori_2012}).  In particular, we use the standard binary cross entropy loss function 

\begin{align}
L_\text{BCE}[f]=-\sum_{Y\sim p}\log(f(Y))-\sum_{Y\sim q}\log(1-f(Y))\,,
\end{align}
and then the likelihood ratio is estimated as $f/(1-f)$.  The last layer of the $f$ networks are sigmoids in order to constrain their range to be between 0 and 1.  The function $f$ is additionally trained to be parameterized in $\theta$ by training with pairs $(Y,\Theta)$ instead of just $Y$, where $\Theta$ is a random variable corresponding to values $\theta$ sampled from a prior.  We will use a uniform prior when training the parameterized classifiers.  

All neural networks are implemented using PyTorch \cite{NEURIPS2019_9015} and optimized with Adam \cite{https://doi.org/10.48550/arxiv.1412.6980} with a learning rate of 0.001 and consist of three hidden layers with 50 nodes per layer.  All intermediate layers use ReLU activation functions.  Each network is trained for 10000 epochs with early stopping using a patience of 10.  The $w_1$ training uses a batch size of 100,000.  The $w_0$ network is simulataneously optimized with $\theta$ and uses a batch size that is the full dataset, which corresponds to performing the fit in Eq.~\ref{eq:lllall} over all the data.

\section{Gaussian Examples}
\label{sec:gaussian}

\subsection{One-dimension in both particle and detector level}
\label{sec:ssec:gaussian1D}

We now demonstrate the proposed method with simple numerical examples. Here, each data set represents one-dimensional Gaussian random variables in both the particle and detector level. The particle-level random variable $T$ is described by mean $\mu$ and standard deviation $\sigma$, while the detector-level variable is given by
\begin{equation}
    R = T + Z,
\end{equation}
where $Z$ is a Gaussian random variable with mean $\beta$ and standard deviation $\epsilon$.

In a first example, $\epsilon$ is considered to be the only nuisance parameter, and $\beta$ is fixed to 0. Three data sets are prepared for the full training procedure. The first data set $D_\mathrm{sim}^{1.0}$ is used as the nominal simulation sample, which contains 200,000 events with $\mu=0$, $\sigma=1$ and $\epsilon=1$. The second data set $D_\mathrm{obs}$ is used as the observed data, which contains 100,000 events with $\mu=0.2$, $\sigma=1$ and $\epsilon=1.2$. To train the $w_1$ reweighter, the third data set $D_\mathrm{sim}^*$, which contains 200,000 events with $\mu=0$, $\sigma=1$ and $\epsilon$ uniformly distributed from 0.2 to 1.8, is prepared. In addition, another data set $D_\mathrm{sim}^{1.2}$ of 100,000 events with $\mu=0$, $\sigma=1$ and $\epsilon=1.2$ is produced for validating the $w_1$ reweighter. All data sets used in the training procedure are split to 50\% for training and 50\% for validating.

\begin{figure}[htbp]
\begin{center}
\includegraphics[width=0.43\textwidth]{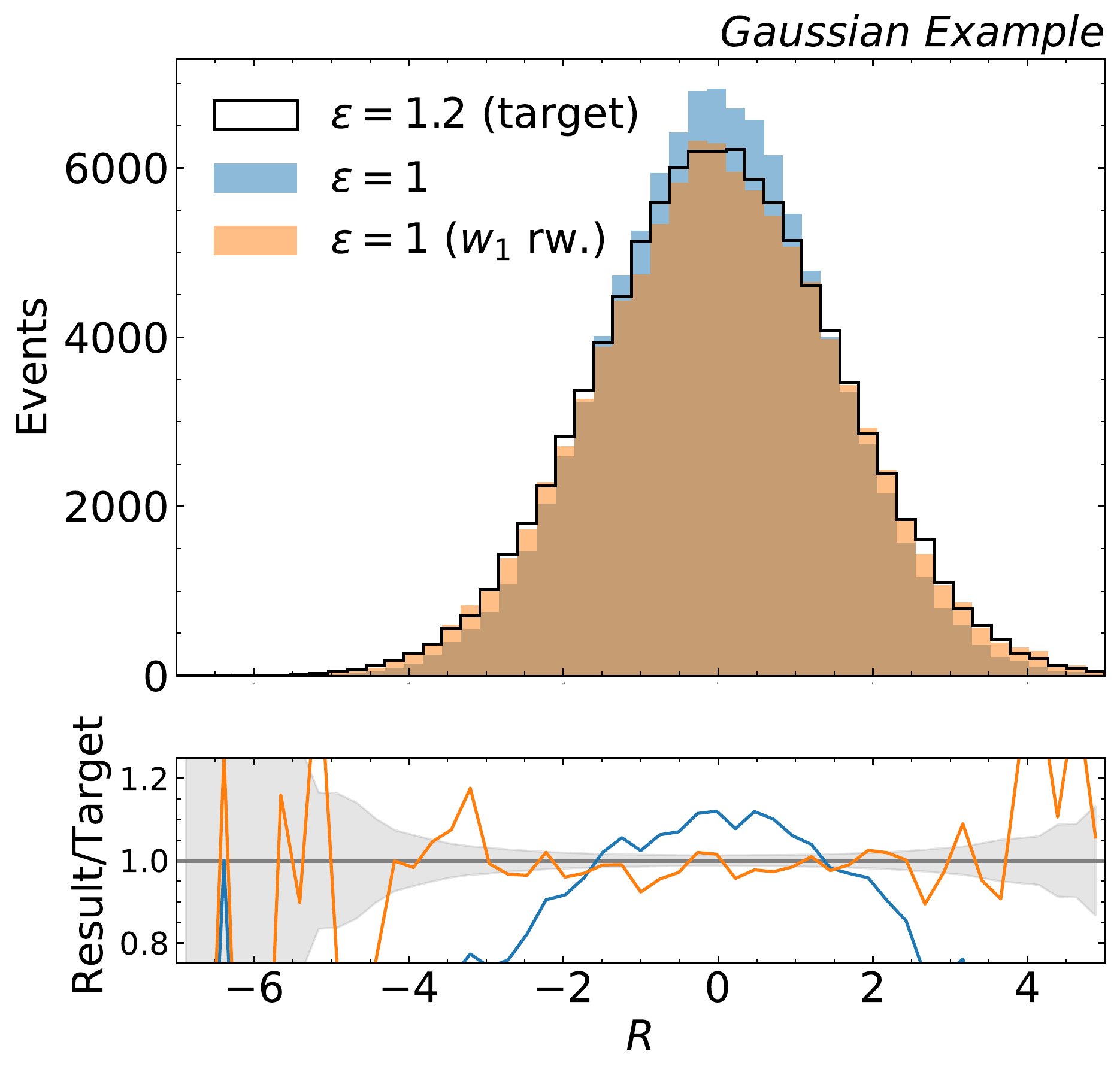}
\caption{Gaussian 1D example: the nominal $R$ distribution ($\epsilon = 1$) in the reweighted by the trained $w_1$ conditioned at $\epsilon = 1.2$ and compared to $R$ distribution with $\epsilon = 1.2$.}
\label{fig:gaussian1D_w1}
\end{center}
\end{figure}

A $w_1$ reweighter is trained to reweight $D_\mathrm{sim}^{1.0}$ to $D_\mathrm{sim}^*$. The trained $w_1$ is then tested with the nominal $R$ distribution ($D_\mathrm{sim}^{1.0}$) reweighted to $\epsilon = 1.2$ ($w_1\left(R|T,\epsilon=1.2\right)$) and compared to the $R$ spectrum with $\epsilon = 1.2$ ($D_\mathrm{sim}^{1.2}$). As shown in Fig. \ref{fig:gaussian1D_w1}, the trained $w_1$ reweighter has learned to reweight the nominal $R$ spectrum to match the $R$ spectrum with $\epsilon$ at 1.2.

With this trained $w_1$ reweighter, a $w_0$ reweighter is trained using $D_\mathrm{sim}^{1.0}$ as the simulation template with $D_\mathrm{obs}$ as the observed data used in Eq.~\ref{eq:lllall}. In the first scenario, the nuisance parameter $\epsilon$ for the $w_1$ reweighter is fixed to 1.2, and the penalty term in Eq.~\ref{eq:lllall} $\log(\theta)$ is set to 0 (no constraint). As shown in Fig.~\ref{fig:gaussian1D_w0_noepsilon}, the $w_0$ reweighter is able to learn to reweight the particle-level spectrum $T$ by matching the detector-level spectrum $R$ to the observed spectrum. In the second scenario, the nuisance parameter $\epsilon$ is trained together with the $w_0$ reweighter. The prior in the penalty term in Eq.~\ref{eq:lllall} is set to be a Gaussian probability density with a 80\% uncertainty. As shown in Fig. \ref{fig:gaussian1D_w0_epsilon}, the trained $w_0$ and optimized $\epsilon$ are tested. The fitted $\epsilon$ is $1.03 \pm 0.016$ \footnote{The fitted value is averaged over five different $w_0$ reweighters which are trained in the same way, but with different random initializations. The standard deviation of the fitted values is taken as the error.} (true value is 1.2). The reweighted distribution matches well with observed data in the detector-level spectrum but the particle-level spectrum has a large non-closure. This is because of the degeneracy between the $w_0$ and $w_1$ reweighters in the effect on the detector-level spectrum.  In other words, detector effects can mimic changes in the particle-level cross section, so the data cannot distinguish between these two scenarios. This is a common issue which also exists in the standard binned maximum likelihood unfolding. For comparison, we also perform the standard binned maximum likelihood unfolding. As shown in App.~\ref{chap:upu_app}, the unfolded $T$ spectrum in this case also fails to represent the true $T$ spectrum. An 80\% uncertainty is highly exaggerated from typical scenarios, but it clearly illustrates the challenge of profiling and unfolding at the same time (see Sec.~\ref{sec:conclusion} for a discussion about regularization).

\begin{figure}[htbp]
\begin{center}
\includegraphics[width=0.43\textwidth]{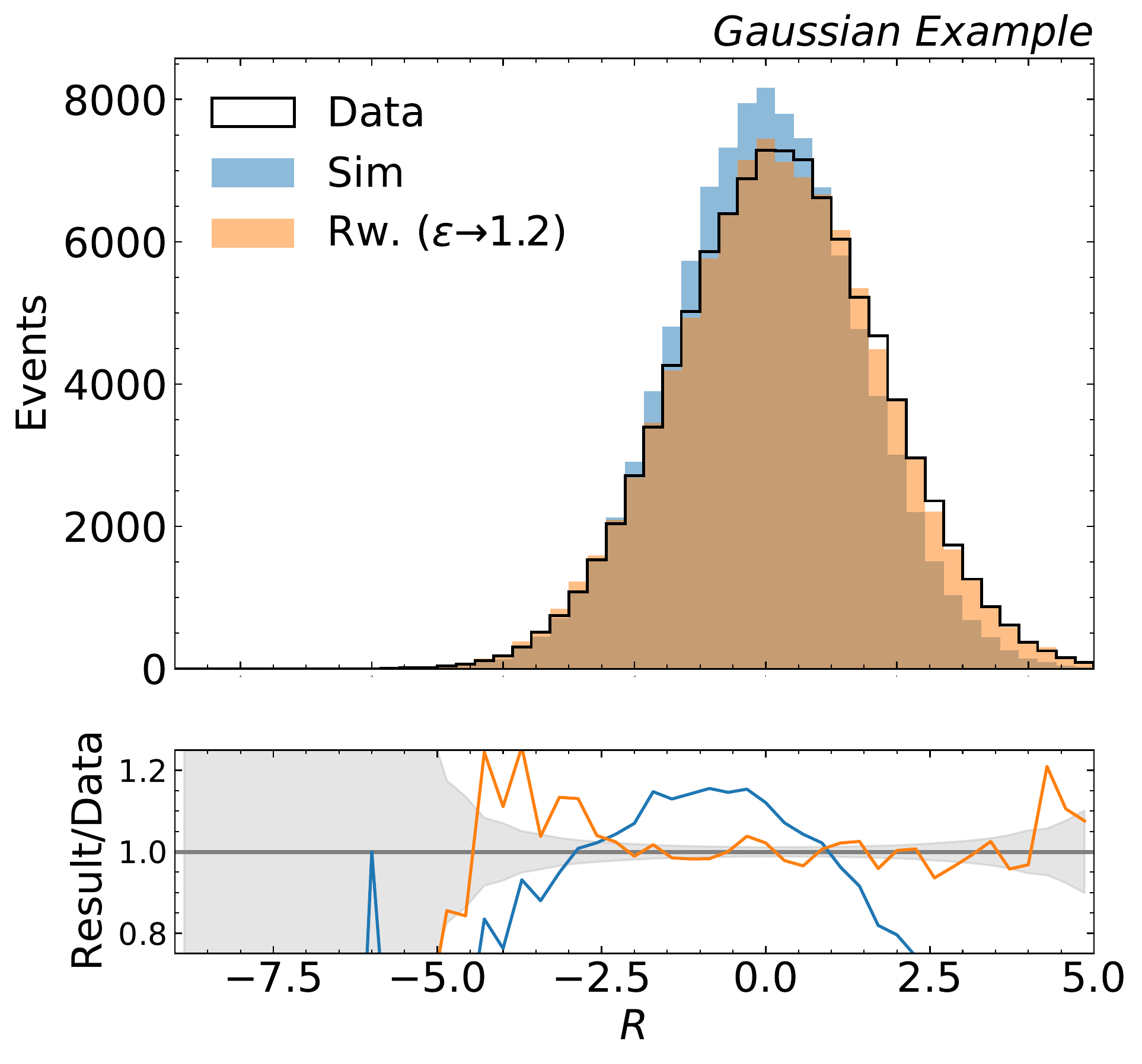}
\includegraphics[width=0.43\textwidth]{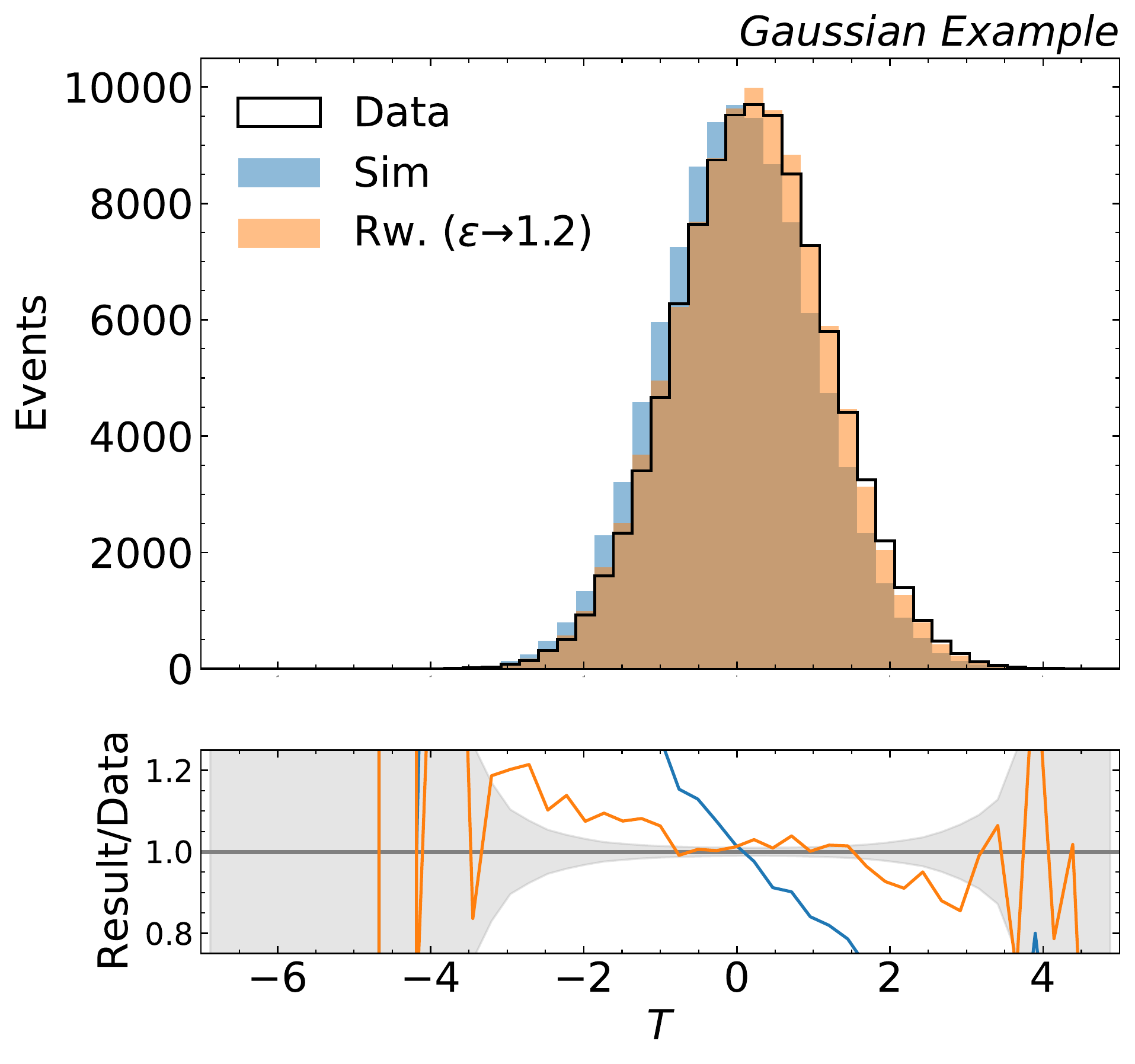}
\caption{Gaussian 1D example: results of the $w_0$ optimization. The nuisance parameter $\epsilon$ is fixed to 1.2, and the the penalty term in Eq.~\ref{eq:lllall} is set to 0. (Top) The detector-level spectrum $R$ of the simulation template $D_\mathrm{sim}$ reweighted by the trained $w_0 \times w_1$, compared to the $R$ spectrum of the observed data $D_\mathrm{obs}$. (Bottom) The particle-level spectrum $T$ of the simulation template $D_\mathrm{sim}$ reweighted by the trained $w_0$, compared to the $T$ spectrum of the observed data $D_\mathrm{obs}$.}
\label{fig:gaussian1D_w0_noepsilon}
\end{center}
\end{figure}

\begin{figure*}[htbp]
\begin{center}
\includegraphics[width=0.43\textwidth]{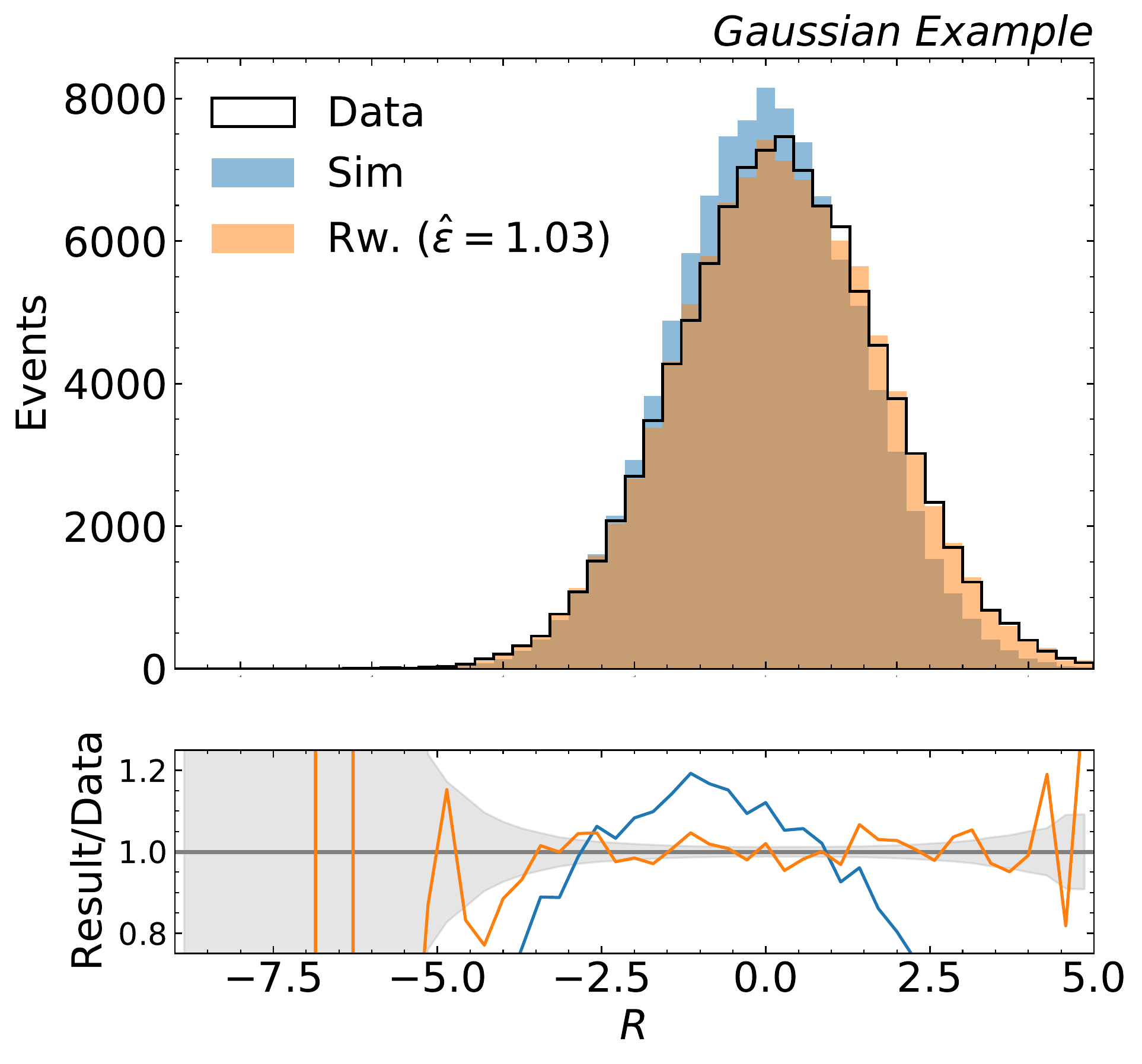}
\includegraphics[width=0.43\textwidth]{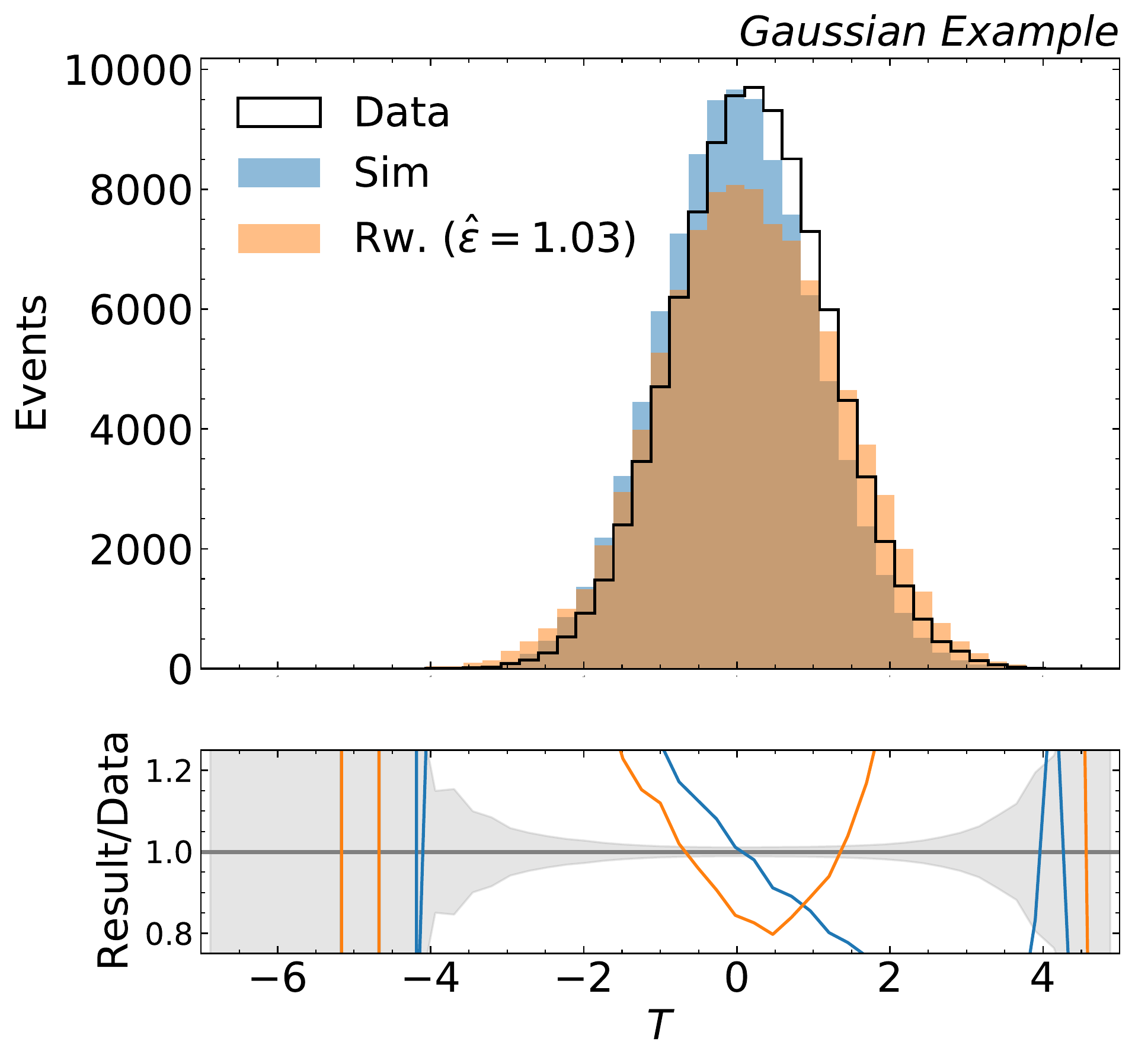}
\caption{Gaussian 1D example: results of the $w_0$ optimization. The nuisance parameter $\epsilon$ is optimized simultaneously with $w_0$ and the best-fit value is $\hat{\epsilon} = 1.03 \pm 0.016$. (Left) The detector-level spectrum $R$ of the simulation template $D_\mathrm{sim}$ reweighted by the trained $w_0 \times w_1$, compared to the $R$ spectrum of the observed data $D_\mathrm{obs}$. (Right) The particle-level spectrum $T$ of the simulation template $D_\mathrm{sim}$ reweighted by the trained $w_0$, compared to the $T$ spectrum of the observed data $D_\mathrm{obs}$.}
\label{fig:gaussian1D_w0_epsilon}
\end{center}
\end{figure*}

\begin{figure*}[htbp]
\begin{center}
\includegraphics[width=0.43\textwidth]{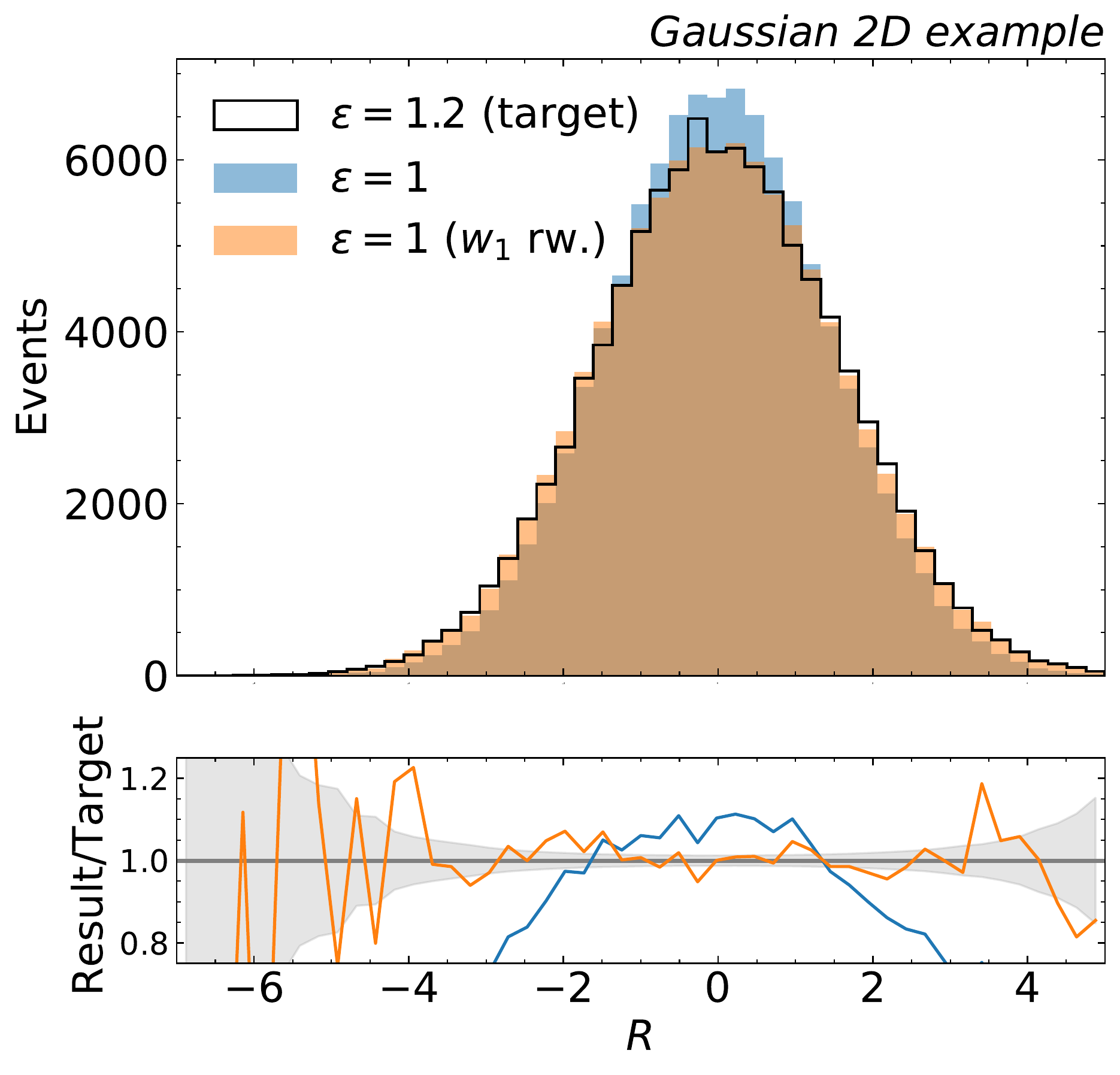}
\includegraphics[width=0.43\textwidth]{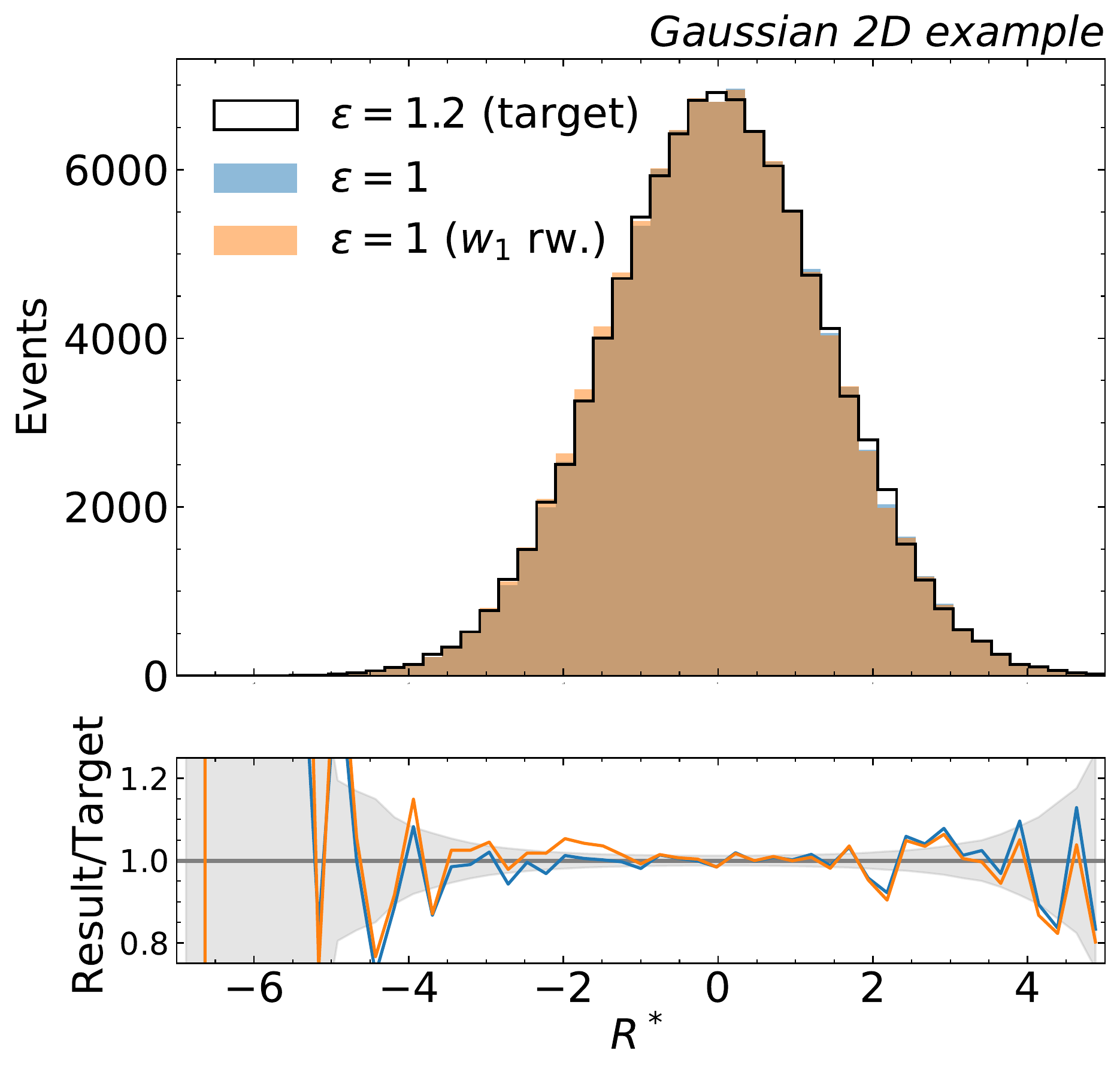}
\caption{Gaussian 2D example: the nominal detector-level spectra $R$ (left) and $R^*$ (right) with $\epsilon = 1$ reweighted by the trained $w_1$ conditioned at $\epsilon = 1.2$ and compared to the spectra with $\epsilon = 1.2$.}
\label{fig:gaussian_2D_w1}
\end{center}
\end{figure*}

\subsection{One-dimension in particle level and two-dimension in detector level}
\label{sec:ssec:gaussian2D}

To break the degeneracy between the $w_0$ and $w_1$ reweighters, we now consider a two-dimension distribution in the detector level, which is given by
\begin{align}
    R & = T + Z,\\
    R^* & = T + Z^*,
\end{align}
where $Z$ ($Z^*$) is a Gaussian random variable with mean $\beta$ ($\beta^*$) and standard deviation $\epsilon$ ($\epsilon^*$). $\epsilon$ is considered to be the only nuisance parameter, and $\beta$, $\beta^*$ are fixed to 0, and $\epsilon^*$ is fixed to 1. In this case, the nuisance parameter $\epsilon$ only has effect on the $R$ spectrum and the $R^*$ spectrum depends purely on the particle-level spectrum $T$.

Similar to the previous example, $D_\mathrm{sim}^{1.0}$ is used as the nominal simulation sample, which contains 200,000 events with $\mu=0$, $\sigma=1$ and $\epsilon=1$. $D_\mathrm{obs}$ is used as the observed data, which contains 100,000 events with $\mu=0.8$, $\sigma=1$ and $\epsilon=1.2$. To train the $w_1$ reweighter, $D_\mathrm{sim}^*$, which contains 200,000 events with $\mu=0$, $\sigma=1$ and $\epsilon$ uniformly distributed from 0.2 to 1.8, is prepared. In addition, another data set $D_\mathrm{sim}^{1.2}$ of 100,000 events with $\mu=0$, $\sigma=1$ and $\epsilon=1.2$ is produced for validating the $w_1$ reweighter. All data sets used in the training procedure are split to 50\% for training and 50\% for validating.

A $w_1$ reweighter is trained to reweight $D_\mathrm{sim}^{1.0}$ to $D_\mathrm{sim}^*$. The trained $w_1$ is tested with the nominal $R$ and $R^*$ spectra ($D_\mathrm{sim}^{1.0}$) reweighted to $\epsilon = 1.2$ and compared to the $R$ and $R^*$ spectra with $\epsilon = 1.2$. As shown in Fig. \ref{fig:gaussian_2D_w1}, the trained $w_1$ reweighter has learned to reweight the nominal $R$ spectrum to match the $R$ spectrum with $\epsilon$ at 1.2, and $R^*$ is independent of the $w_1$ reweighter.

Based on the trained $w_1$ reweighter, a $w_0$ reweighter and the nuisance parameter $\epsilon$ are optimized simultaneously using $D_\mathrm{sim}$ as the simulation template with $D_\mathrm{obs}$ as the observed data used in Eq.~\ref{eq:lllall}. As before, the prior in the penalty term in Eq.~\ref{eq:lllall} is configured with an uncertainty of 80\%. The fitted $\epsilon$ is $1.20 \pm 0.004$ (correct value is 1.2). As shown in Fig. \ref{fig:gaussian2D_w0_epsilon_0.8}, the reweighted spectra match well with observed data in both detector and particle level. For more realistic uncertainties (so long as the simulation is close to the right answer), the fidelity is even better.


\begin{figure*}[htbp]
\begin{center}
\includegraphics[width=0.43\textwidth]{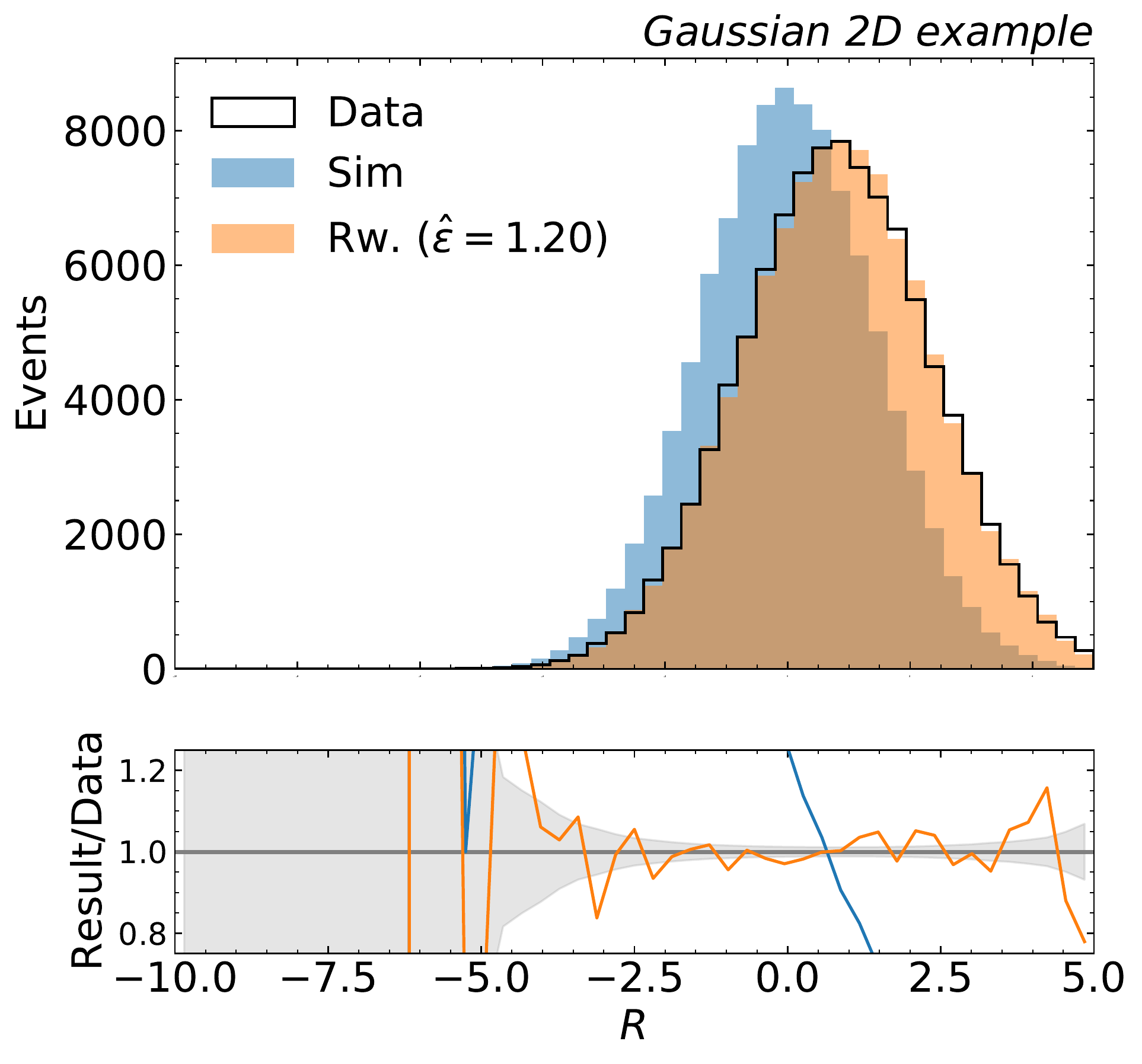}
\includegraphics[width=0.43\textwidth]{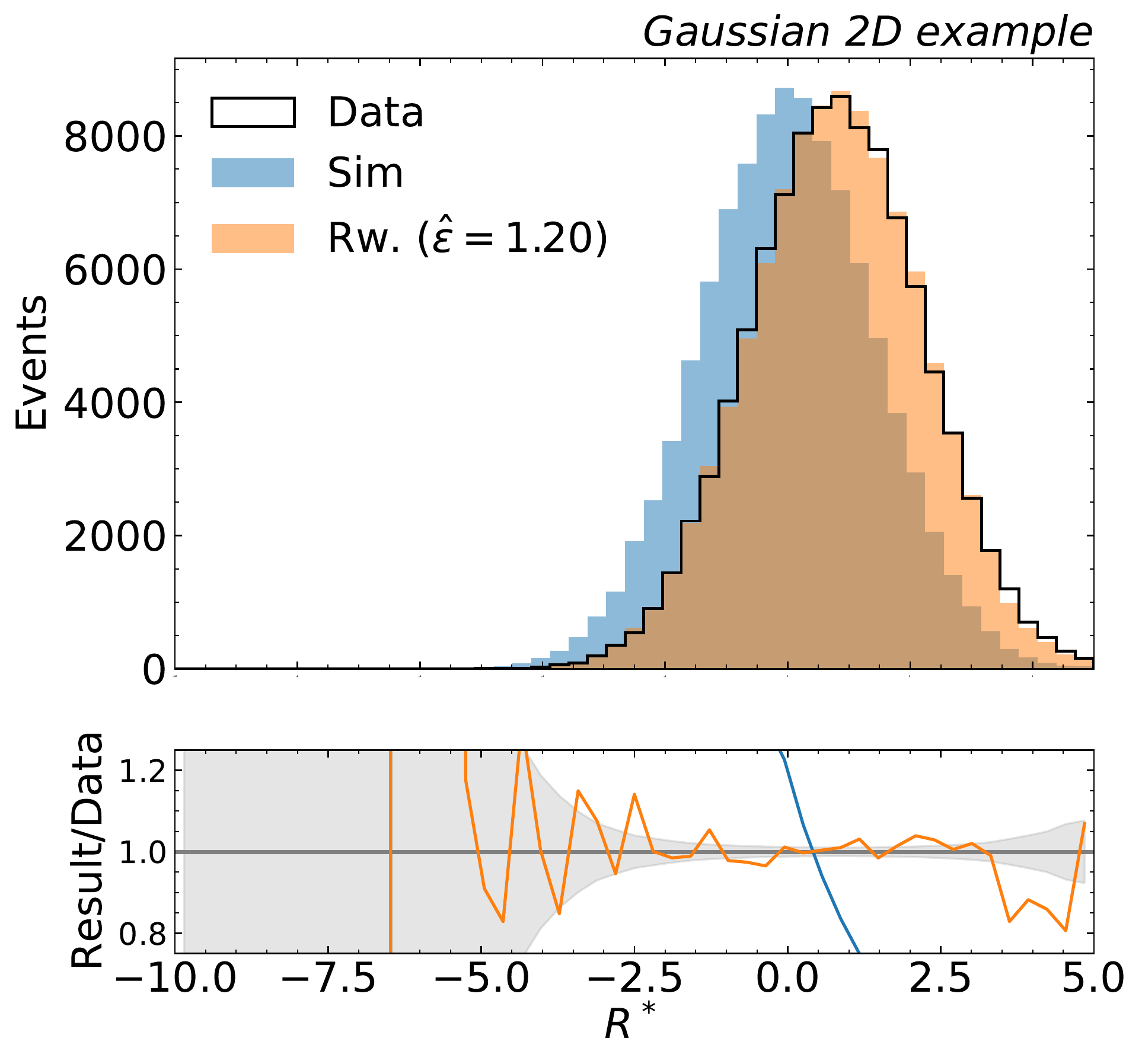}\\
\includegraphics[width=0.43\textwidth]{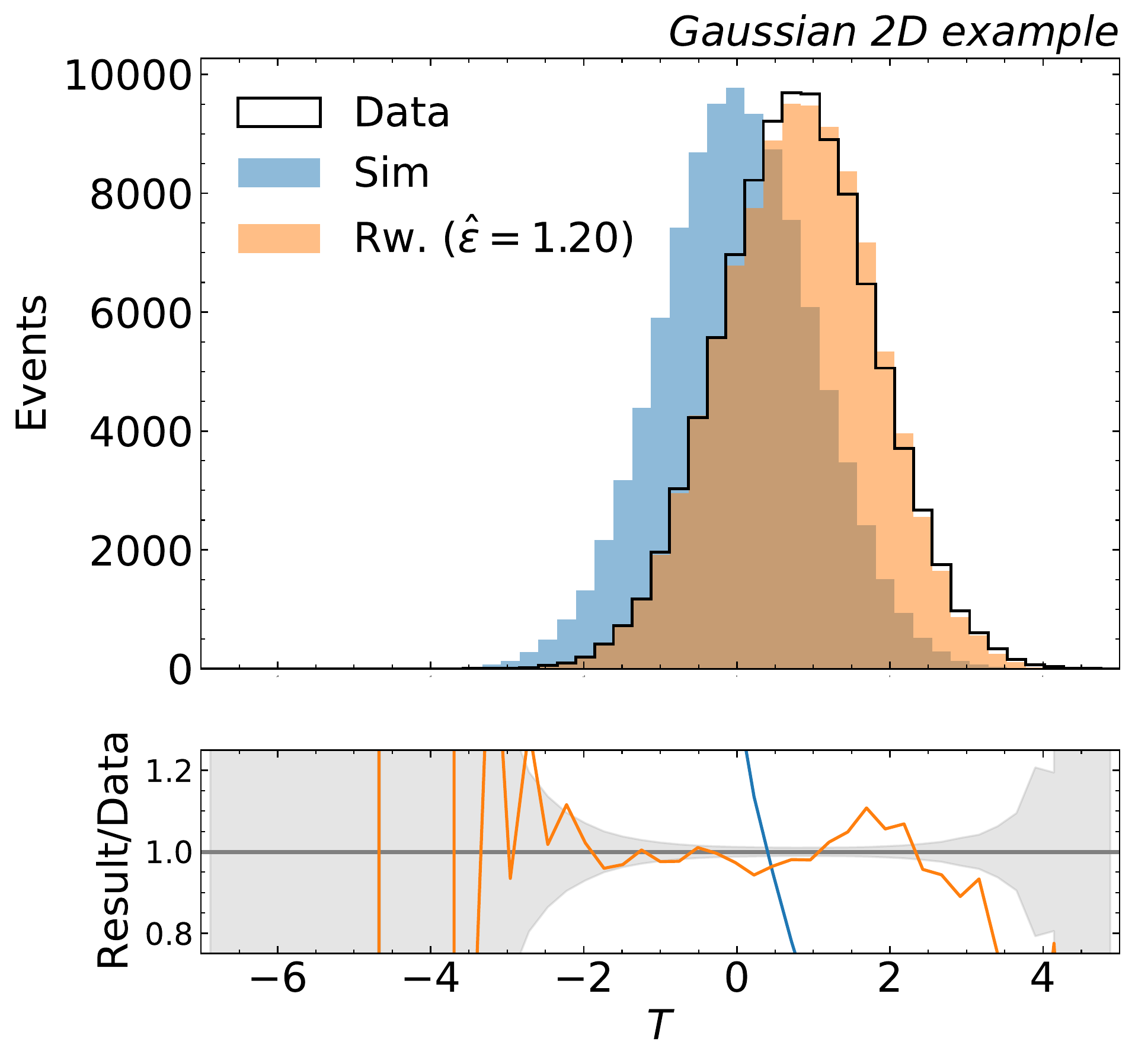}
\caption{Gaussian 2D example: results of the $w_0$ optimization. The nuisance parameter $\epsilon$ is optimized simultaneously with $w_0$ with the prior constraint set to 80\%. The fitted $\epsilon$ is $1.20 \pm 0.004$. (Top-left) The detector-level spectrum $R$ of the simulation template $D_\mathrm{sim}$ reweighted by the trained $w_0 \times w_1$, compared to the $R$ spectrum of the observed data $D_\mathrm{obs}$. (Top-right) The detector-level spectrum $R^\prime$ of the simulation template $D_\mathrm{sim}$ reweighted by the trained $w_0 \times w_1$, compared to the $R^*$ spectrum of the observed data $D_\mathrm{obs}$. (Bottom) The particle-level spectrum $T$ of the simulation template $D_\mathrm{sim}$ reweighted by the trained $w_0$, compared to the $T$ spectrum of the observed data $D_\mathrm{obs}$.}
\label{fig:gaussian2D_w0_epsilon_0.8}
\end{center}
\end{figure*}

\section{Higgs Boson Cross Section}
\label{sec:higgs}

We now demonstrate the unfolding method in a physics case --- a Higgs boson cross section measurement. Here, we focus on the di-photon decay channel of the Higgs boson. The goal is then to measure the transverse momentum spectrum of the Higgs boson $p^\mathrm{T}_H$ using the transverse momentum of the di-photon system $p^\mathrm{T}_{\gamma\gamma}$ at detector level. The photon resolution $\epsilon_{\gamma}$ is considered as a nuisance parameter. In this case, the $p^\mathrm{T}_{\gamma\gamma}$ spectrum is minimally affected by $\epsilon_{\gamma}$. Therefore, we also consider the invariant mass spectrum of the di-photon system $m_{\gamma\gamma}$ at detector level, which is highly sensitive to $\epsilon_{\gamma}$. In addition, In order to have a large spectrum difference between different data sets for demonstration purpose, we consider only events that contain at least two reconstructed jets, where the leading-order (LO) calculation would significantly differ from next-to-leading-order calculation (NLO)

Similar to the Gaussian examples, we prepare the following data sets:

\begin{itemize}
    \item $D_\mathrm{obs}$: used as the observed data.
    \item $D_\mathrm{sim}^{1.0}$: used as the nominal simulation sample.
    \item $D_\mathrm{sim}^{1.2}$: used as the simulation sample with a systematic variation.
    \item $D_\mathrm{sim}^{*}$: simulation sample with various $\epsilon_\gamma$ values for training the $w_1$ reweighter.
\end{itemize}

$D_\mathrm{obs}$ is generated at NLO using the \textsc{Powheg}\textsc{Box} program~\cite{Oleari:2010nx, Alioli:2008tz}, while the rest are generated at LO using \textsc{Mad}\textsc{Graph}5\_aMC@LO v2.6.5~\cite{Alwall:2014hca}. For all samples, the parton-level events are processed by \textsc{Pythia} 8.235~\cite{Sjostrand:2006za,Sjostrand:2014zea} for the Higgs decay, the parton shower, hadronization, and the underlying event. The detector simulation is based on \textsc{Delphes} 3.5.0~\cite{deFavereau:2013fsa} with detector response modified from the default ATLAS detector card. For both $D_\mathrm{obs}$ and $D_\mathrm{sim}^{1.2}$, the photon resolution $\epsilon$ is multiplied by a factor of 1.2. For $D_\mathrm{sim}^{*}$, the multiplier of $\epsilon$ is uniformly scanned between 0.5 and 1.5 with a step size of 0.01. $D_\mathrm{sim}^{1.0}$ uses the default ATLAS detector card. 

Each of the spectra of particle-level $p^\mathrm{T}_{\gamma\gamma}$, detector-level $p^\mathrm{T}_{\gamma\gamma}$ and detector-level $m_{\gamma\gamma}$ is standardized to the spectrum with a mean of 0 and a standard deviation of 1 before being passed to the neural networks. A $w_1$ reweighter is then trained to reweight $D_\mathrm{sim}^{1.0}$ to $D_\mathrm{sim}^*$. The trained $w_1$ is tested with the nominal detector level $p^\mathrm{T}_{\gamma\gamma}$ and $m_{\gamma\gamma}$ spectra ($D_\mathrm{sim}^{1.0}$) reweighted to $\epsilon_{\gamma}=1.2$ and compared to the detector level $p^\mathrm{T}_{\gamma\gamma}$ and $m_{\gamma\gamma}$ spectra with $\epsilon_{\gamma} = 1.2$. As shown in Fig. \ref{fig:higgs_w1}, the trained $w_1$ reweighter has learned to reweight the nominal detector level $m_{\gamma\gamma}$ spectrum to match the detector level $m_{\gamma\gamma}$ spectrum with $\epsilon_{\gamma}$ at 1.2, and the detector level $p^\mathrm{T}_{\gamma\gamma}$ variable is independent of the $w_1$ reweighter.

\begin{figure*}[htbp]
\begin{center}
\includegraphics[width=0.43\textwidth]{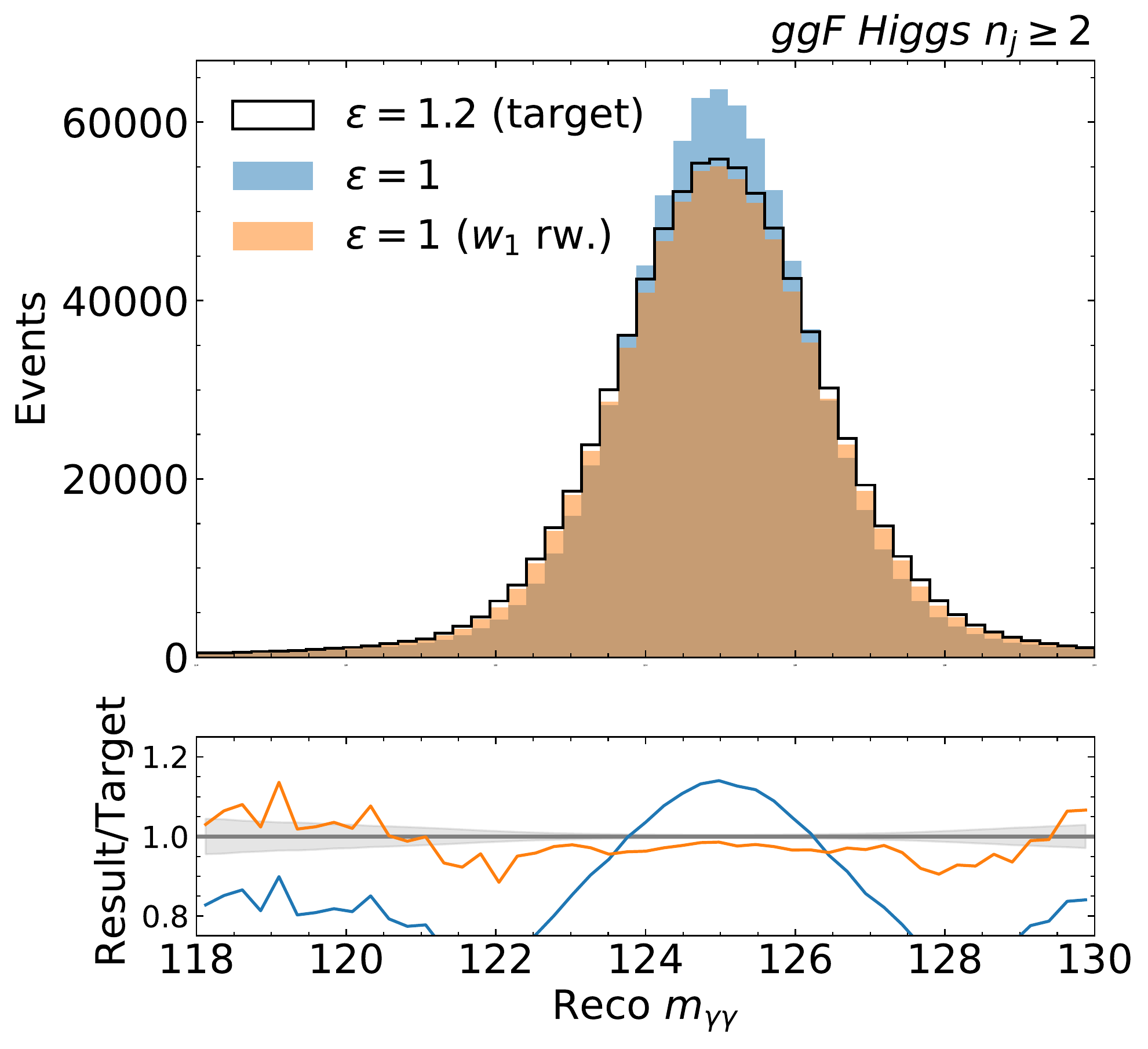}
\includegraphics[width=0.43\textwidth]{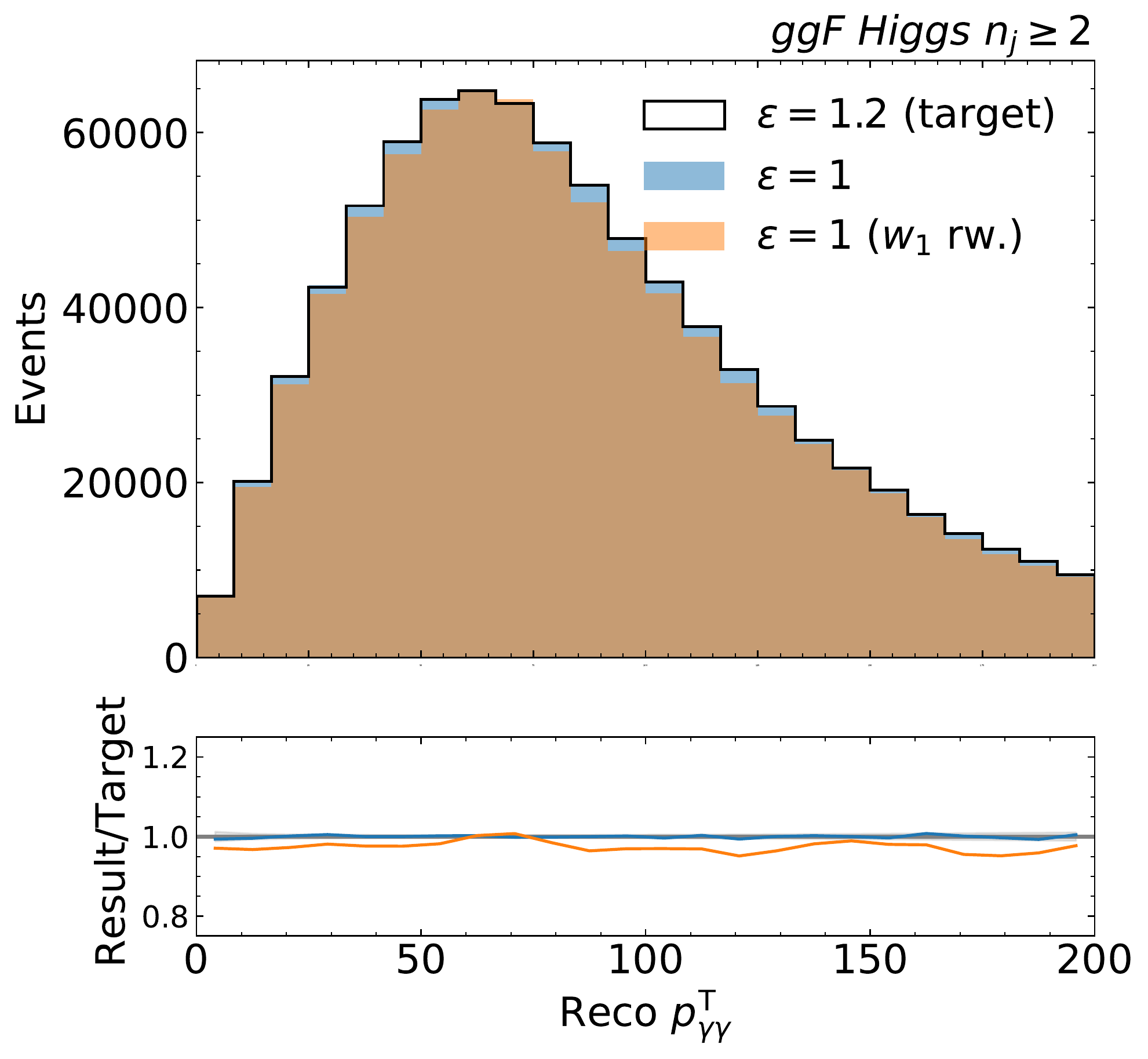}
\caption{Higgs boson cross section: the nominal detector-level spectra $m_{\gamma\gamma}$ (left) and $p^\mathrm{T}_{\gamma\gamma}$ (right) with $\epsilon_\gamma = 1$ reweighted by the trained $w_1$ conditioned at $\epsilon_\gamma = 1.2$ and compared to the spectra with $\epsilon_\gamma = 1.2$.}
\label{fig:higgs_w1}
\end{center}
\end{figure*}

\begin{figure*}[htbp]
\begin{center}
\includegraphics[width=0.43\textwidth]{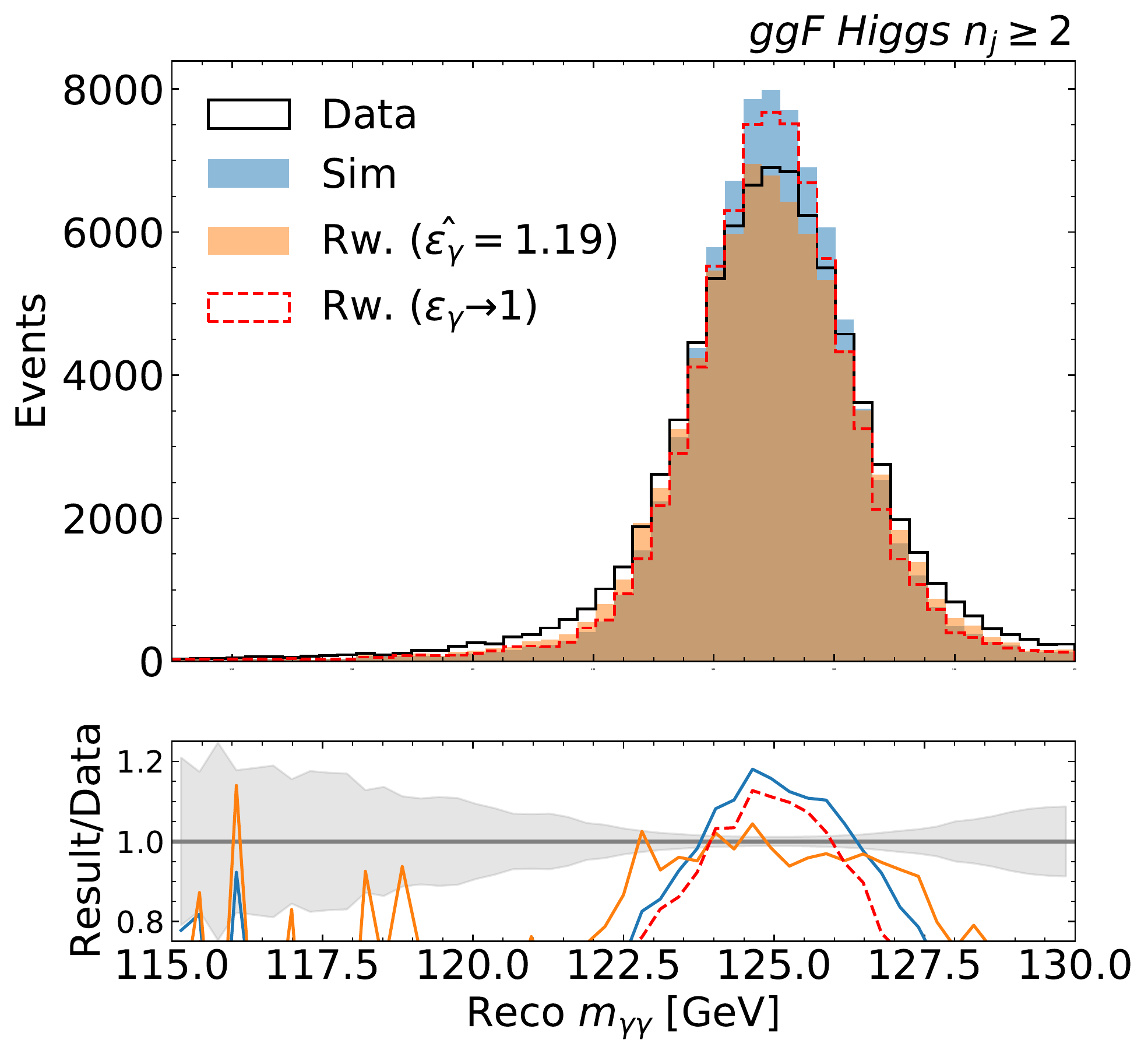}
\includegraphics[width=0.43\textwidth]{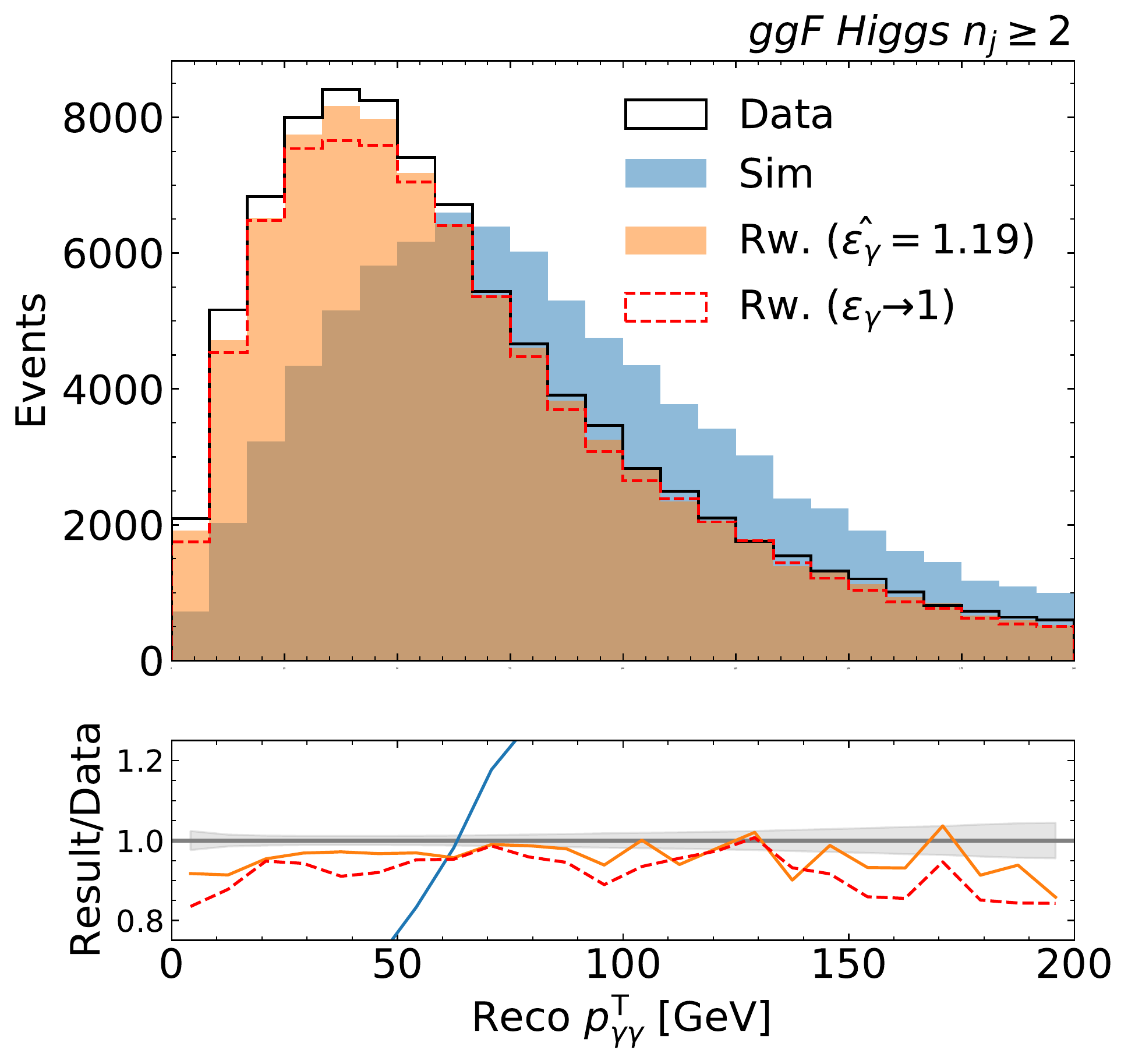}\\
\includegraphics[width=0.43\textwidth]{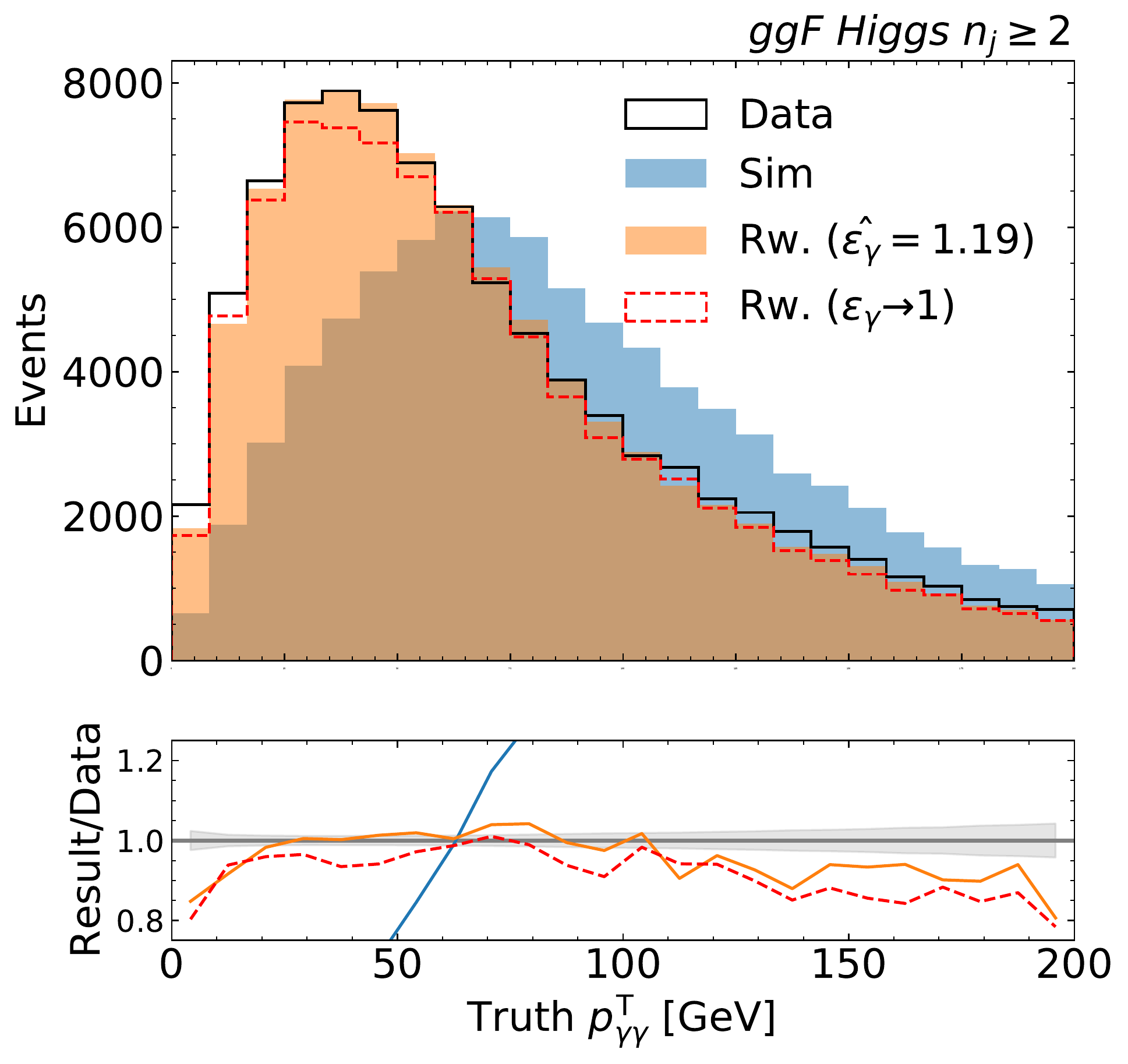}
\caption{Higgs boson cross section: results of the $w_0$ optimization. The nuisance parameter $\epsilon_\gamma$ is optimized simultaneously with $w_0$ with the prior constraint set to 50\% (orange) or fixed to 1 for comparison (red). The fitted $\epsilon_\gamma$ is $1.19 \pm 0.007$. (Top-left) The detector-level spectrum $m_{\gamma\gamma}$ of the simulation template $D_\mathrm{sim}$ reweighted by the trained $w_0 \times w_1$, compared to the $m_{\gamma\gamma}$ spectrum of the observed data $D_\mathrm{obs}$. (Top-right) The detector-level spectrum $p^\mathrm{T}_{\gamma\gamma}$ of the simulation template $D_\mathrm{sim}$ reweighted by the trained $w_0 \times w_1$, compared to the $p^\mathrm{T}_{\gamma\gamma}$ spectrum of the observed data $D_\mathrm{obs}$. (Bottom) The particle-level spectrum $p^\mathrm{T}_{\gamma\gamma}$ of the simulation template $D_\mathrm{sim}$ reweighted by the trained $w_0$, compared to the $p^\mathrm{T}_{\gamma\gamma}$ spectrum of the observed data $D_\mathrm{obs}$.}
\label{fig:higgs_w0_theta}
\end{center}
\end{figure*}

The $w_0$ reweighter and $\epsilon$ are optimized simultaneously based on the pre-trained $w_1$ reweighter. The prior of $\epsilon_\gamma$ is 50\%. The fitted $\epsilon_\gamma$ is $1.19 \pm 0.007$. As shown in Fig.~\ref{fig:higgs_w0_theta}, the reweighted spectra match well with observed data in both detector and particle level. This means that the observed data $p^\mathrm{T}_{H}$ spectrum is successfully unfolded with nuisance parameter $\epsilon_{\gamma}$ properly profiled. For comparison, we also perform UPU with $\epsilon_\gamma$ fixed at 1. As shown in Fig.~\ref{fig:higgs_w0_theta}, the unfolded $p^\mathrm{T}_{H}$ spectrum in this case has a larger non-closure with the observed data due to the lack of profiling.


\section{Summary}
\label{sec:conclusion}

UPU uses the binned maximum likelihood as the figure of merit to optimize the unfolding reweighting function $w_0\left(t\right)$, which takes unbinned particle-level spectra as inputs. $w_0\left(t\right)$ and the nuisance parameters $\theta$ are optimized simultaneously, which also requires to learn a conditional likelihood ratio $w_1(t,r|\theta)$ that reweights the detector-level spectra based on the profiled values of nuisance parameters and is taken as an input for the optimization of $w_0\left(t\right)$ and $\theta$.

In Gaussian examples, we demonstrated the optimization of $w_1$ and the optimization of $w_0$ and $\theta$. A limitation of this method can be seen when the considered detector-level observable, a one-dimension Gaussian distribution, is not able to distinguish between effects from particle level and effects from detector level with $\theta$. This limitation also exists in the standard binned maximum likelihood unfolding, and the problem can be resolved when we consider another observable which does not depend on $\theta$ and thus breaks the degeneracy between particle-level and detector-level effects.  Regularization can also help mitigate these effects.  Our approach has implicit regularization from the smoothness of neural networks.  We leave additional studies of regularization to future work.

We also applied UPU to the Higgs boson cross section measurement. We considered one dimension at particle level and two dimensions at detector level. With one detector-level variable sensitive to the target particle-level observable and one sensitive to the effect of nuisance parameters, the data are successfully unfolded and profiled. The impact of profiling is also demonstrated by comparing with the result of nuisance parameter fixed to the nominal value. This can be readily extended to higher dimensions in either particle level or detector level, provided all particle-level and detector-level effects are distinguishable in the considered detector-level spectra. In the case of more than one nuisance parameters, one can either train multiple $w_1$ for each nuisance parameter separately or train a single $w_1$ which takes all nuisance parameters as inputs.  As the effects of multiple nuisance parameters are usually assumed independent, one could take a product of individually trained reweighters.

As with any measurement, quantifying the uncertainty is critical to interpret UPU results.  Just as in the binned case, one can calculate the uncertainty on the nuisance parameters which can be determined by fixing a given parameter to target values and then simultaneously re-optimizing $w_0$ and the rest of the nuisance parameters.  A new feature of UPU is that the likelihood (ratio) itself is only an approximation, using neural networks as surrogate models.  This is a challenge for all machine learning-based unfolding, and uncertainties can be probed by comparing the results with different simulations.  Future extensions of UPU may be able to also use machine learning to quantify these model uncertainties as well as process unbinned data also at detector level.

\clearpage

\chapter{Conclusion}
\label{chap:conclusion}

With 139 fb$^{-1}$ of $pp$ collision data collected at 13 TeV from 2015 to 2018 by the ATLAS detector at the LHC, two searches involving Higgs boson which use the proton-proton ($p-p$) collision data collected by the ATLAS detector are conducted. The first search is a search for dimuon decay of the SM Higgs boson ($H\to\mu\mu$), which is the most promising channel for probing the couplings between SM Higgs Boson and the second generation fermions at the LHC. Boosted decision trees are trained to separate signal and background. The observed (expected) significance over the background-only hypothesis for a Higgs boson with a mass of 125.09 GeV is 2.0$\sigma$ (1.7$\sigma$), and the observed upper limit on the cross section times branching ratio for $pp\to H\to\mu\mu$ is 2.2 times the SM prediction at 95\% confidence level, while the expected limit on a $H\to\mu\mu$ signal assuming the absence (presence) of a SM signal is 1.1 (2.0). The best-fit value of the signal strength parameter, defined as the ratio of the observed signal yield to the one expected in the SM, is $\mu = 1.2 \pm 0.6$. This result represents an improvement of about a factor of 2.5 in expected sensitivity compared with the previous ATLAS publication. A factor of two from the improvement is due to the larger analysed dataset and the additional 25\% improvement can be attributed to more advanced analysis techniques.

The second search targets events that contain large missing transverse momentum and either two $b$-tagged small-radius jets or a single large-radius jet associated with two $b$-tagged subjects, which is sensitive to Dark Matter models in the extended Higgs sector.
Events are split into multiple categories which target different phase space of the DM signals. No significant excess from the SM prediction is observed.
The results are interpreted with two benchmark models with two Higgs doublets extended by either a heavy vector boson $Z^\prime$ ($Z^{\prime}$-2HDM) or a pseudoscalar singlet $a$ (2HDM+$a$) and which provide a Dark Matter candidate $\chi$.
In the case of $Z^{\prime}$-2HDM, the observed limits extend up to a $Z^{\prime}$ mass of 3.1 TeV at 95 \% confidence level for a mass of 100 GeV for the Dark Matter candidate.
For 2HDM+$a$, masses of $a$ are excluded up to 520 GeV and 240 GeV for $\tan\beta=1$ and $\tan\beta=10$ and a Dark Matter mass of 10 GeV, respectively.
This analysis achieved a large imporvement in the sensitivity for \zpthdm\ as well as \thdma\ with the \ggF\ production mode due to the \MET\ binning in the merged region. This analysis also estasblished the exclusion limits for \bbA\ production of \thdma\ for the first time thanks to the inclusion of the $\geq$3 $b$-tag regions.
In addition, limits on the visible cross sections in the model-independent context are set, which will be useful for the future reinterpretation with additional signal models.

Both searches will benefit from the additional data that will be collected in the next run of LHC, which could increase statistical power and enhance the sensitivity. In addition to the higher statistics, there are many developments and potential tasks which will bring further improvements in the future analyses. For example, there have been a significant improvement in the LAr calorimeter, which increases the granularity of information used by the trigger and will increase the \MET\ trigger efficiency. This will enhance the signal efficiency of the \monoHbb\ search and thus increase the change of finding signals.  \monoHbb\ can also benifit from adopting an ML-based categorization. In particular, it is worth applying a parametrized neural network to accomadate to different signals or signals in different parameter phase space. On the other hand, we expect further improvements in \Hmm\ replacing the BDT with deep neural network for the categorization. In particular, deep sets architecture can preserve jet symmetry during the classifier training, and adverial neural network can allow fully exploiting powerful training variables (e.g. kinematics of individual muons) while maintaining mass independence. All of these items will be crucial for pushing the analyses towards potential discovery.

In addition to the two searches presented, the unbinned profiled unfolding (UPU), a newly proposed unfolding method, is introduced. This method combines the standrad binned maximum likelihood unfolding and the machine learning-based methods, resulting in unbinned differential cross-section and is also able to profile nuisance parameters.
This method singificantly increases the flexibility and reusability of the unfolding result and will be particularly useful for model intepretations with any physics search channel in the future.

\begin{appendix}
  \chapter{Binned maximum likelihood unfolding with Gaussian examples}
\label{chap:upu_app}

  In this appendix, we present results of the standard binned maximum likelihood unfolding (BMLU) with Gaussian examples. The scenarios are:
  
  \begin{itemize}
      \item One-dimension in both particle and detector
level: this is the same example as described in Sec.~\ref{sec:ssec:gaussian1D}. The prior constraint for $\epsilon$ is set to 80\%. The result is shown in Fig.~\ref{fig:gaussian1D_bmlu} with $\epsilon$ fitted to $1.08 \pm 0.02$, which also indicates a degeneracy problem between particle and detector levels.
       \item One-dimension in particle level and
two-dimension in detector level: this is the same example as described in Sec.~\ref{sec:ssec:gaussian2D}. The prior constraint for $\epsilon$ is set to 80\%. The result is shown in Fig.~\ref{fig:gaussian2D_bmlu} with $\epsilon$ fitted to $1.19 \pm 0.003$. The degeneracy problem is resolved after considering an additional spectrum in the detector level.

All the maximum likelihood fittings are performed using pyhf \cite{pyhf, pyhf_joss}.
  \end{itemize}
  
\begin{figure*}[htbp]
\begin{center}
\includegraphics[width=0.43\textwidth]{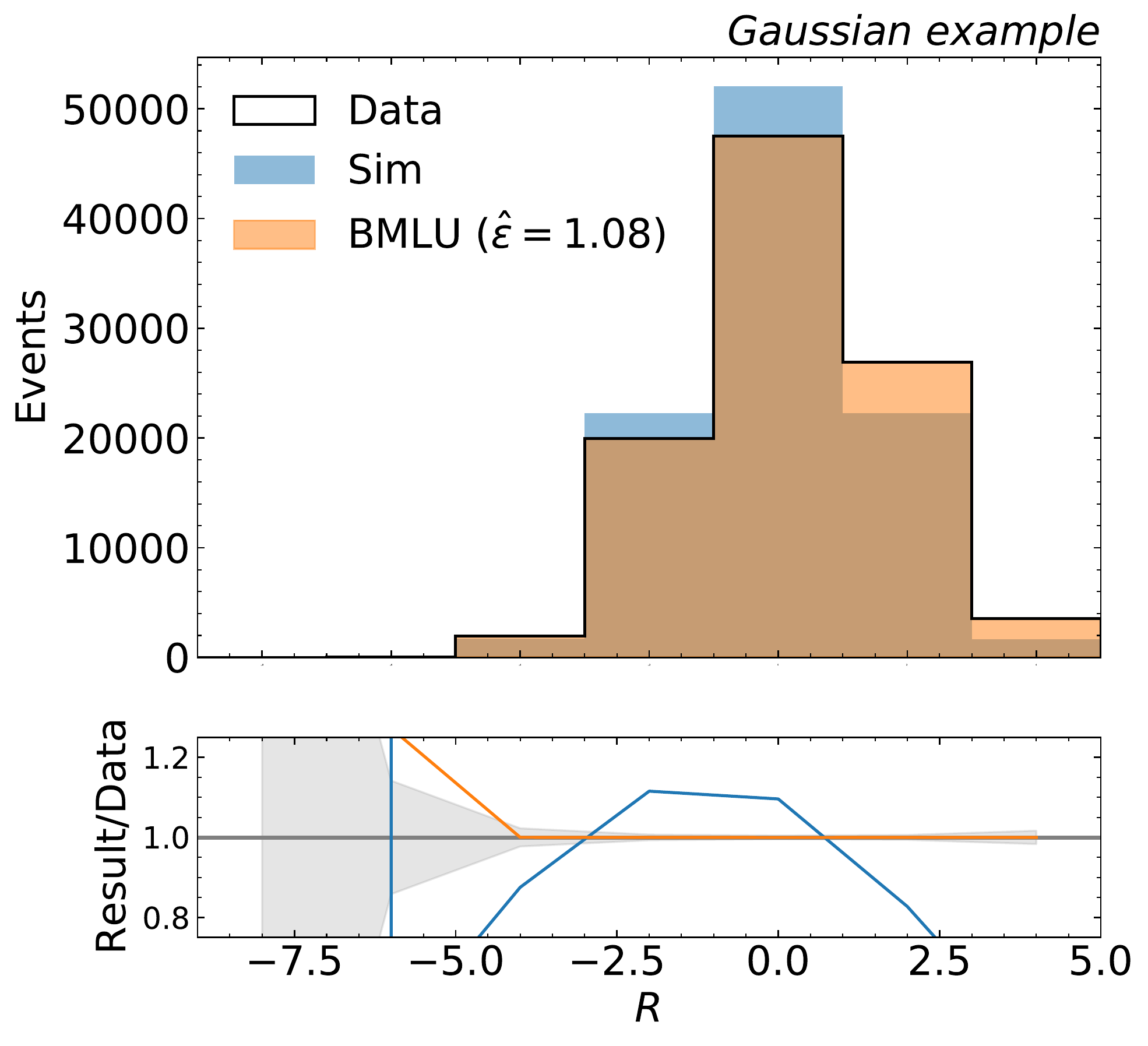}
\includegraphics[width=0.43\textwidth]{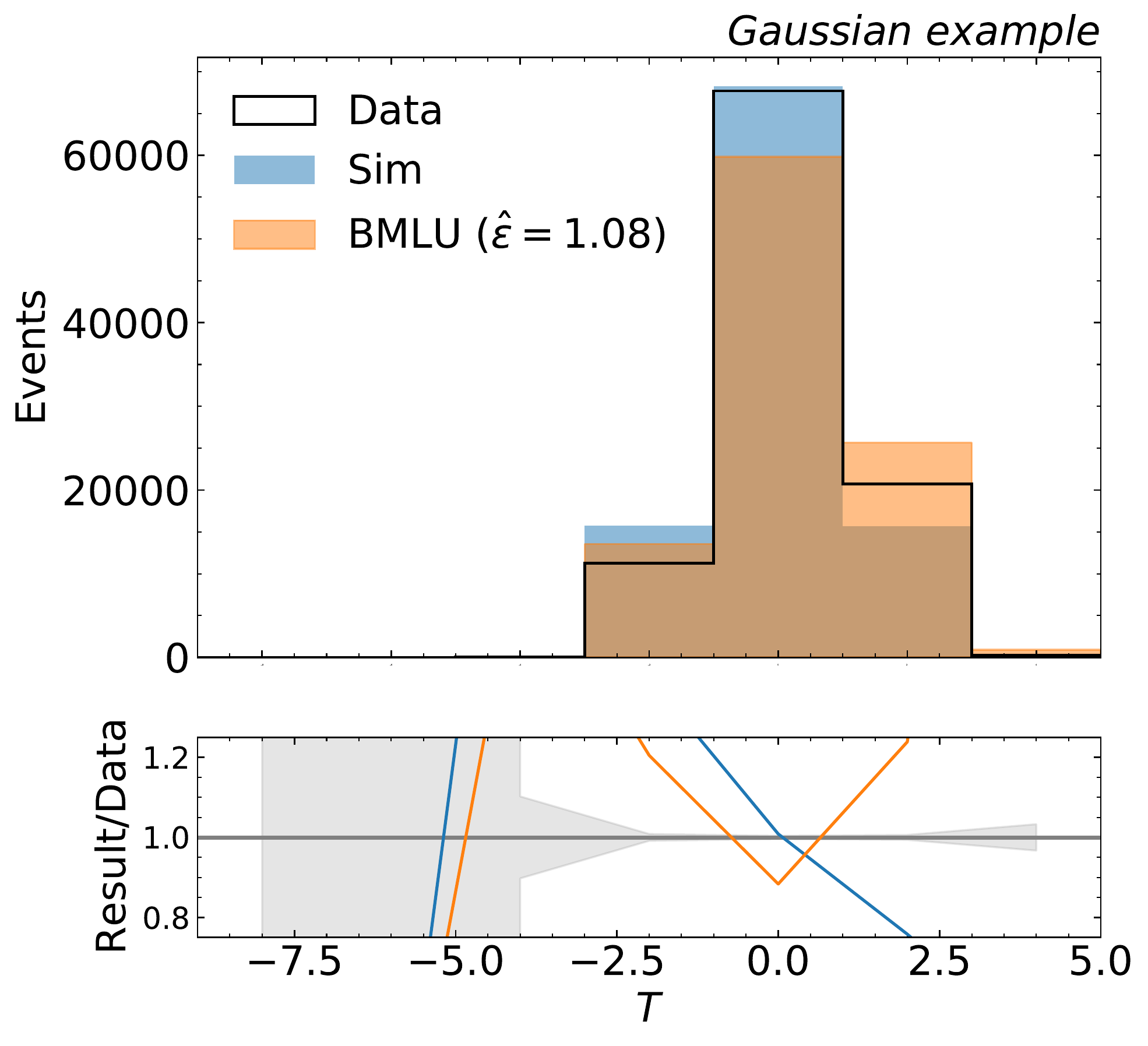}
\caption{Gaussian 1D example: results of the binned maximum likelihood unfolding. The prior constraint for $\epsilon$ is set to 80\% and the fitted $\epsilon$ is $1.08 \pm 0.02$. (Left) The fitted detector-level spectrum $R$ of the simulation template $D_\mathrm{sim}$, compared to the $R$ spectrum of the observed data $D_\mathrm{obs}$. (Right) The unfolded particle-level spectrum $T$ of the simulation template $D_\mathrm{sim}$, compared to the $T$ spectrum of the observed data $D_\mathrm{obs}$.}
\label{fig:gaussian1D_bmlu}
\end{center}
\end{figure*}

\begin{figure*}[htbp]
\begin{center}
\includegraphics[width=0.43\textwidth]{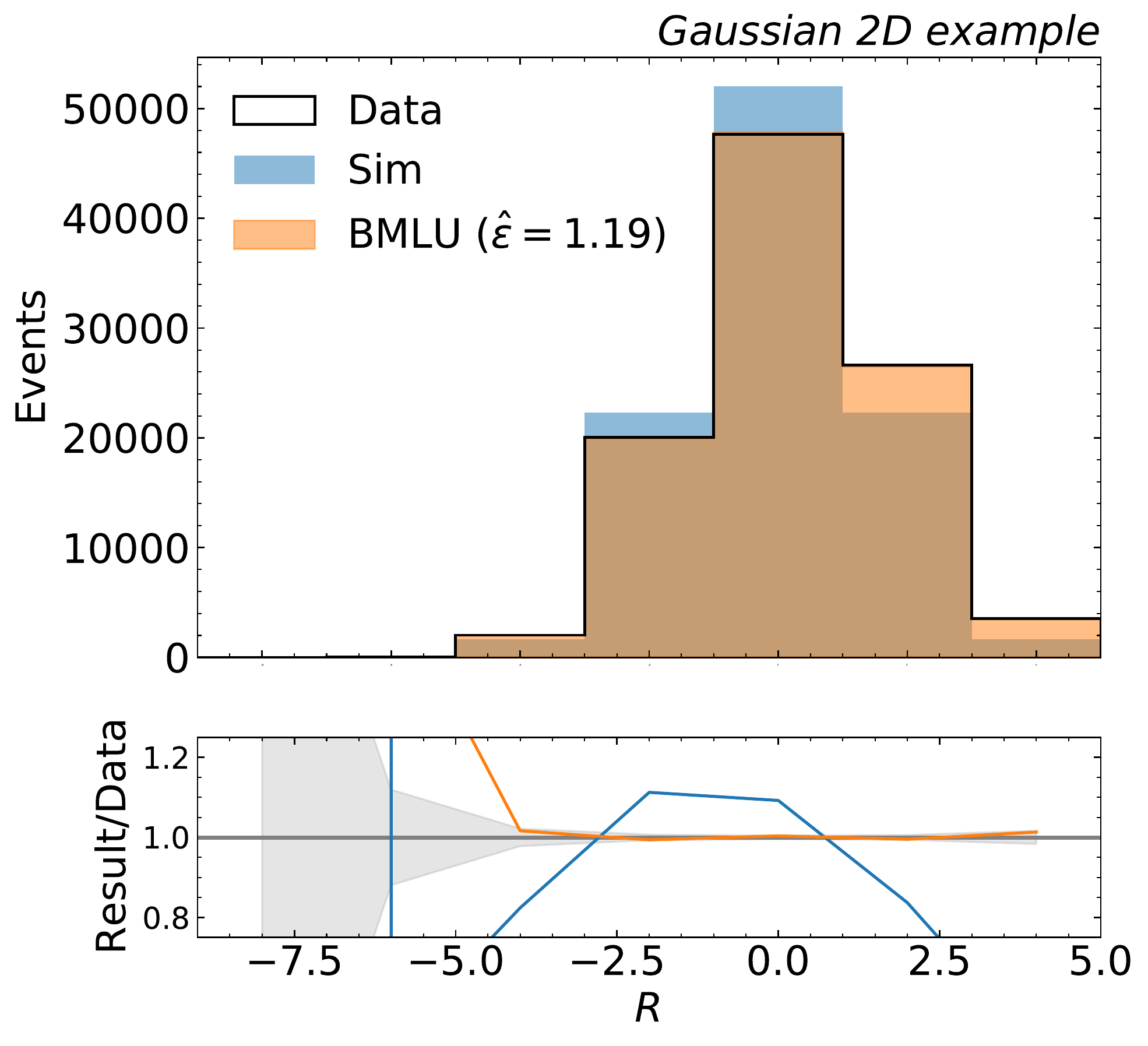}
\includegraphics[width=0.43\textwidth]{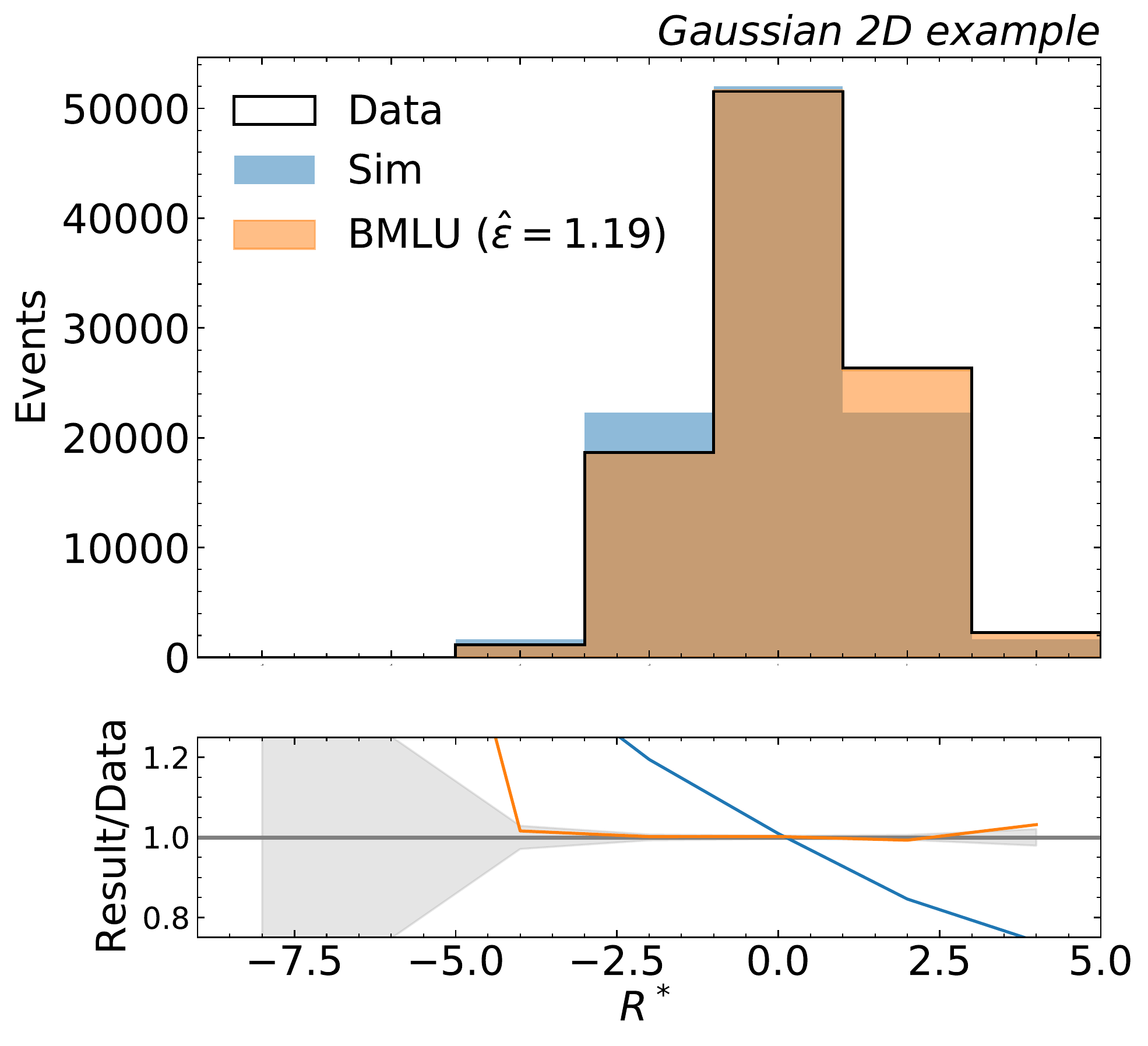}\\
\includegraphics[width=0.43\textwidth]{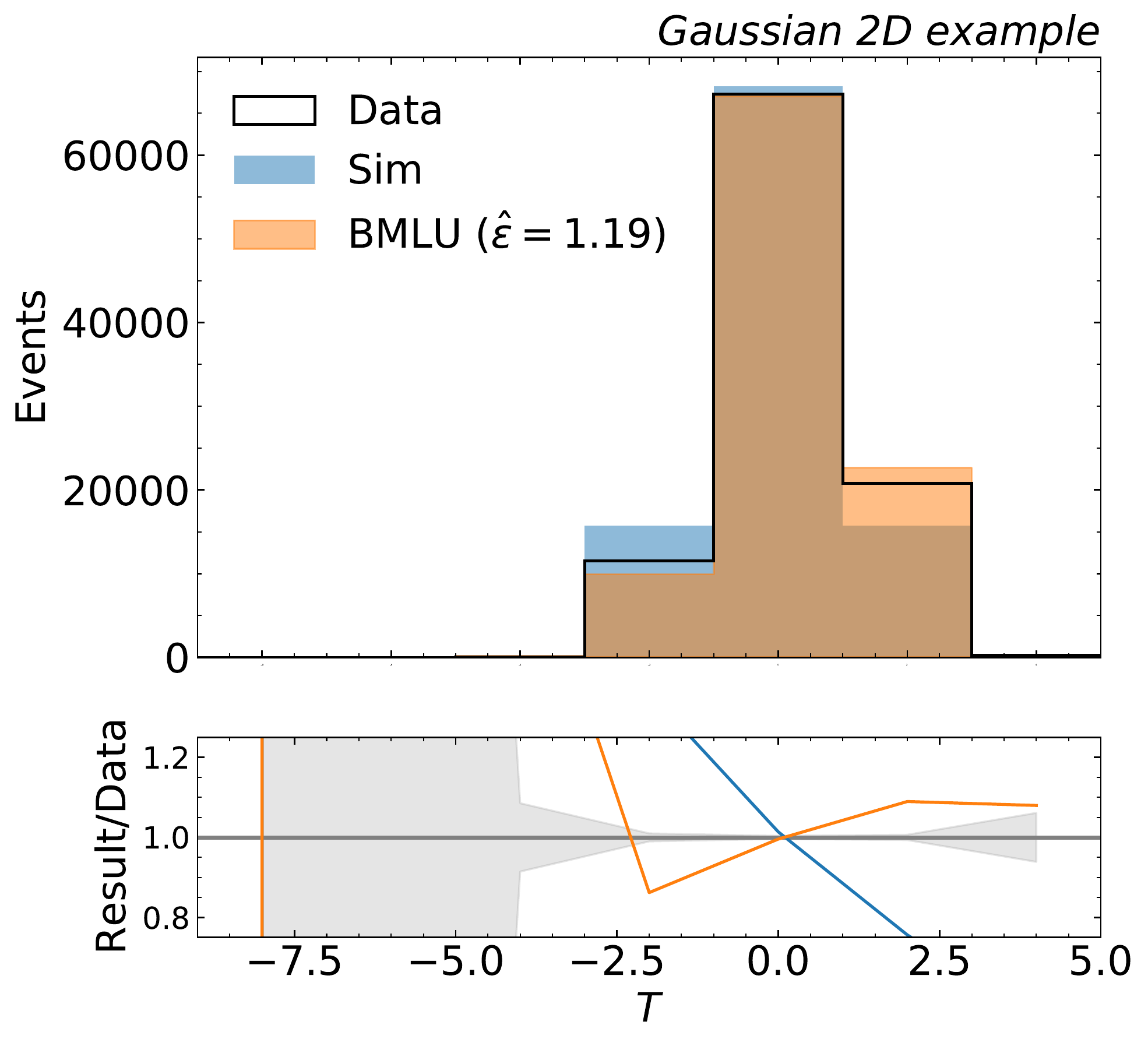}
\caption{Gaussian 2D example: results of the binned maximum likelihood unfolding. The prior constraint for $\epsilon$ is set to 80\% and the fitted $\epsilon$ is $1.19 \pm 0.003$. (Top-left) The fitted detector-level spectrum $R$ of the simulation template $D_\mathrm{sim}$, compared to the $R$ spectrum of the observed data $D_\mathrm{obs}$. (Top-right) The fitted detector-level spectrum $R^*$ of the simulation template $D_\mathrm{sim}$, compared to the $R^\prime$ spectrum of the observed data $D_\mathrm{obs}$. (Bottom) The unfolded particle-level spectrum $T$ of the simulation template $D_\mathrm{sim}$, compared to the $T$ spectrum of the observed data $D_\mathrm{obs}$.}
\label{fig:gaussian2D_bmlu}
\end{center}
\end{figure*}

\clearpage

\end{appendix}

\begin{singlespace}
\bibliographystyle{cms_unsrt}
\bibliography{main}              
\end{singlespace}

\end{document}